\documentclass[twoside]{Style/uwaterloothesis}%
\usepackage[list]{Style/artratex}
\usepackage{Style/artracom}
\usepackage{mathtools}  
\DeclareMathOperator{\Tr}{Tr}
\usepackage{tikz}
\usepackage{caption}
\usepackage{subcaption}
\usepackage{nccmath}
\usepackage{braket}
\usepackage{afterpage}
\usepackage{bm}
\usepackage{mathtools}
\usetikzlibrary{arrows.meta}

\usetikzlibrary{calc,arrows}
\newcommand{\tikzmark}[1]{%
  \tikz[overlay,remember picture] \coordinate (#1) {};}

\usetikzlibrary{arrows,positioning} 
\tikzset{
    >=stealth',
    punkt/.style={
           rectangle,
           rounded corners,
           draw=black, very thick,
           text width=6.5em,
           minimum height=1em
           ,text centered
           },
    pil/.style={
           ->,
           thick,
           shorten <=2pt,
           shorten >=2pt,}
}

\title[]{Correlations and Work Statistics in Critical Quantum Systems}
\author{Zakaria Mzaouali}
\degree{Doctor of Philosophy}
\discipline{Theoretical Physics}
\begin{document}
\frontmatter
\maketitle
\intotoc*{\cleardoublepage}{Examining Committee Membership}
\begin{committee}
    \noindent
    The following served on the Examining Committee for this thesis. The decision of the Examining Committee is by majority vote.

    \begin{center}
        \setlength{\tabcolsep}{10pt}
        \renewcommand{\arraystretch}{3}
        \begin{tabular}{lc}
            \textbf{Committee President:} & \parbox[t]{10cm}{Prof. Yassine HASSOUNI\\\it Mohammed V University in Rabat}\\
            
            \textbf{Supervisors:} & \parbox[t]{10cm}{Prof. Morad EL BAZ\\\it Mohammed V University in Rabat}\\
             & \parbox[t]{10cm}{Prof. Steve CAMPBELL\\\it University College Dublin in Ireland}\\
        
            \textbf{Reporters:} & \parbox[t]{10cm}{Prof. Adil BELHAJ\\ \it Mohammed V University in Rabat}\\
             & \parbox[t]{10cm}{Prof. Hamid EZ-ZAHRAOUY\\\it Mohammed V University in Rabat} \\
             & \parbox[t]{10cm}{Prof. Mohamed AZZOUZ\\ \it Al Akhawayn University in Ifrane}\\
            \textbf{Examiners:} & \parbox[t]{10cm}{Prof. Bartłomiej GARDAS\\ \it Institute of Theoretical and Applied Informatics,\\ Polish Academy of Sciences, Poland}\\
            & \parbox[t]{10cm}{Prof. Mustapha FAQIR\\\it International University of Rabat}\\
        \end{tabular}
    \end{center}
\end{committee}
\intotoc*{\cleardoublepage}{Author's Declaration}
\begin{declaration}
    \noindent
    I hereby declare that I am the sole author of this thesis. This is a true copy of the thesis, including any required final revisions, as accepted by my examiners.

    \bigskip

    \noindent
    I understand that my thesis may be made electronically available to the public.
\end{declaration}
\intotoc*{\cleardoublepage}{Abstract}
\begin{abstract}
    We analyze equilibrium and out-of-equilibrium quantum phase transitions present in quantum spin-$\frac{1}{2}$ models, from the quantum information and the quantum thermodynamics perspective. We provide an overview on the literature of classical and quantum phase transitions. Additionally, we span over classical and quantum information theory, in order to build the necessary background for studying the equilibrium properties of critical quantum spin systems, using quantum-information theoretic measures. We apply this formalism to the critical Heisenberg $XX$ model, where we study the long range behavior of the entanglement, quantum discord, and quantum coherence, with respect to magnetic and thermal perturbations. We show the weakness of the entanglement, the robustness of general forms of correlations to thermal fluctuations, and the versatility of these measures in revealing the second-order quantum phase transition. Furthermore, we extend the study toward detecting equilibrium quantum phase transitions in phase space, through the Wigner function. First we discuss and review the Wigner function for infinite-dimensional systems, with special focus on the finite-dimensional case, where we introduce the Wootters and Stratonovich constructions of the Wigner function. Then, we apply this formalism to detect the critical features of the $XY$ and $XXZ$ model, where we show a general formula relating Wootters', Stratonovich's Wigner function and the thermodynamical quantities of spin models, which can be applied to single, bipartite and multipartite systems. This approach allows us to introduce a novel way to represent, detect, and distinguish first-, second-, and infinite-order quantum phase transitions. Furthermore, we show that the factorization phenomenon in the $XY$ model is only directly detectable by quantities based on the square root of the bipartite reduced density matrix. We establish that phase space techniques provide a simple, experimentally promising tool in the study of many-body systems and we discuss their relation with measures of quantum correlations and quantum coherence. Finally, we focus on the quantum thermodynamics of out-of-equilibrium quantum phase transitions. After presenting the classical theory of thermodynamics, we discuss its extension to non-equilibrium and quantum settings; as well as the resource theory of thermodynamics and its impact on the theory of computation at the classical and quantum level. Then, we focus on the work statistic across an excited state quantum phase transition manifesting in the dynamics of the Lipkin-Meshkov-Glick model, initialized in the ground state of the ferromagnetic phase, and subject to a sudden quench. We demonstrate that the work probability distribution displays non-Gaussian behavior for quenches in the vicinity of the excited state critical point. Furthermore, we show that the entropy of the diagonal ensemble is highly susceptible to critical regions, making it a robust and practical indicator of the associated spectral characteristics. We assess the role that symmetry breaking has on the ensuing dynamics, highlighting that its effect is only present for quenches beyond the critical point. Finally, we show that similar features persist when the system is initialized in an excited state and briefly explore the behavior for initial states in the paramagnetic phase.
 
\end{abstract}
\intotoc*{\cleardoublepage}{Acknowledgements}
\begin{acknowledgements}
    \noindent 
    This dissertation could not have been completed without the support of a number of people. I would like to express my sincere gratitude for my family for believing in me, and for providing me with all kinds of support during this journey. Your prayer for me was what sustained me so far.\\
    \noindent
    I am highly indebted to my main supervisor, Prof. Morad El Baz, for accepting me in his Master's program, and for integrating me in his research team, at Mohammed V University in Rabat, as a PhD student. His kindness, support, and guidance through the last six and half years impacted me greatly on a personal and scientific level.  I would like also to thank my second supervisor, Prof. Steve Campbell, for the strong collaboration we built in the last three and half years, which opened many research horizons and opportunities for me, especially in the field of quantum thermodynamics.\\
    \noindent
    Along this journey, I have been very fortunate to participate in several international scientific events which was possible thanks to the travel grants offered by the organizers of the events. I would like to thank the International Centre for Theoretical Physics in Trieste-Italy, The Okinawa Institute of Science and Technology in Japan, The University of Evora in Portugal, and The University of Vienna in Austria, for the financial assistance during my research visits. These visits allowed me to present my research and interact with many internationally leading experts in the field of quantum information science and condensed matter physics, which impacted the quality of my published works. \\
    \noindent
    I have benefited from discussions and exchanges with many senior researchers, which allowed me to grow personally and scientifically. In particular, I would like to thank Abderrahman Maaouni, Fabio Benatti, Rosario Fazio, Antonello Scardicchio, Maurizio Fagotti, Pei Wang, Markus Heyl, Thomas Busch, Ugo Marzolino, Marcello Dalmonte, Mauro Paternostro, Sebastian Deffner, John Goold, Ricardo Puebla, Nicolas Gigena, and Tiago Mendes-Santos. \\
    \noindent 
    Finally, I would like to thank the jury members for accepting to judge my work, and for providing me with remarks and propositions to improve the quality of the dissertation.

\end{acknowledgements}
\intotoc*{\cleardoublepage}{Dedication}
\begin{dedication}
    \it To my ancestors, and successors.
\end{dedication}
\intotoc*{\cleardoublepage}{List of Publications}
Parts of the dissertation are based on the following publications:
\begin{enumerate}
    \item \textbf{Z. Mzaouali}, R. Puebla, J. Goold, M. El Baz, S. Campbell. “Work statistics and symmetry breaking in an excited-state quantum phase transition”. \textbf{Physical Review E 103 (3), 032145.}
    
    \item \textbf{Z. Mzaouali}, S. Campbell, M. El Baz. “Discrete and generalized phase space techniques in critical quantum spin chains”. \textbf{Physics Letters A 383 (30), 125932.}
    
    \item \textbf{Z. Mzaouali}, M. El Baz. “Long range quantum coherence, quantum \& classical correlations in Heisenberg XX chain”. \textbf{Physica A: Statistical Mechanics and its Applications 518, 119-130.}
\end{enumerate}
{
\linespread{1.1}
\intotoc*{\cleardoublepage}{\contentsname}
\tableofcontents
\intotoc*{\cleardoublepage}{\listfigurename}
\listoffigures
\intotoc*{\cleardoublepage}{\listtablename}
\listoftables
}
\mainmatter

\intotoc*{\cleardoublepage}{Prologue}
\chapter*{Prologue}
\noindent The history of life, humankind, and society is marked with prominent transitions in lifeforms, ecosystems, and civilization~\cite{harari2014sapiens}. The evolution of the universe from creation to its today's shape, is argued to have gone through several phase transitions, from the high-temperature plasma, due to the Big Bang, to the cooled form we know nowadays. Therefore, phase transitions are a cornerstone in nature and are of great importance regardless of the field one might be interested in~\cite{chappin2012review, mathis2017emergence}. In physics, complex systems display macroscopic phase transitions as a consequence of the interactions between numerous microscopic components. This makes phase transition an emergent phenomenon~\cite{goldenfeld2018lectures, sachdev2011quantum, vojta2003quantum}, and throughout history many advances have been made in order to capture the essence of how the emergence of states of matter takes place. In the late $19^{\text{th}}$ century, Pierre Curie discovered the mechanism by which metals transform between ferromagnetism and paramagnetism~\cite{curie1895proprietes, kittel1949physical}. The phenomenon of superconductivity was discovered by Onnes in 1911, which describes how the ability of the material to conduct electricity changes with the temperature~\cite{onnes1913investigations}. In order to approximate the behavior of real-world materials, physicists work with mathematical models. A prominent model for complex systems is the Ising model, which was proposed by Lenz and solved by Ising in 1925, in order to describe magnetic materials~\cite{Ising:1925em, magnetismhandbook, physicsmagnetism}. The advent of the theory of quantum mechanics revolutionized the way physicists think about and approach the atomic realm~\cite{cohen1986quantum, cohen1986quantum2, cohen2019quantum3}. As a consequence, a series of discoveries have been achieved, i.e. the quantum theory of solids by Felix Bloch in 1928~\cite{bloch1928quantum}, the quantum theory of ferromagnetism by Paul Dirac and Werner Heisenberg in 1929~\cite{dirac1928quantum}, as well as the discovery of antiferromagnetism and diamagnetism by N\'eel~\cite{neel} and Meissner~\cite{diamagnetism}, respectively, in the early $30$'s. Following these developments, Lev Landau worked on a universal theory for criticality in physical systems between 1933-1937, after which he presented his phenomenological theory of phase transitions, where he provided a classification and description of first-, and second-order phase transitions~\cite{landau1936theory}. Shortly after, Landau explained the state of superconductivity~\cite{landausuperc} and superfluidity, where a fluid has zero viscosity~\cite{landausuperf}. Building on the discovery of Onnes and using quantum mechanics, the microscopic theory of superconductivity was developed by Bardeen, Cooper, and Schrieffer (BCS theory) in 1957~\cite{bcs1, bcs2, bcs3}. Then, in the early 60's the concept of localization was predicted by Anderson in disordered systems~\cite{andresonlocal}. Numerical methods were also subject to important developments in order to simulate, efficiently, the salient properties of physical systems and in particular, complex critical systems. In this regard, Wilson introduced in 1974 the renormalization group technique for treating phase transitions~\cite{wilson1974renormalization, wilson1983}.\\
\noindent By the 80's a new field started emerging, that is quantum information theory, which characterizes information and how it can be manipulated at the quantum level~\cite{bennett1998quantum, wilde2013quantum}. Two decades later, the connection between the field of condensed matter theory, statistical mechanics, and quantum information has become more evident, and physicists started to study the features of condensed matter architectures using quantum information tools~\cite{preskill2000quantum, nielsen2002quantum, petz2007quantum}. In fact, at zero temperature the features of many-body systems are dictated by a complex ground state wave function, which contains all the necessary information on the correlations that give rise to the various phases of matter, such as: superfluidity, ferromagnetism, and superconductivity~\cite{fazioreview}. In contrast, these correlations can be captured by quantum-information theoretic measures~\cite{dutta2015quantum}. As a consequence, a fertile line of research emerged where figures of merit in quantum information, such as: the entanglement, quantum discord, quantum coherence, and Wigner functions, are used to spotlight quantum phase transitions and extract their critical exponents, which cemented the important role that such figures of merit play in unraveling the curious properties of many-body systems~\cite{niel_osb, rozario, QPT2004, qptdiscord, CampbellPRA2013, AmicoPRB, SarandyPRA2009, Werlang2010, CakmakPRB2014, qptcoherence, TonySciRep, CampbellPRB2015, CakmakPRB2016, RogersPRA2014, HofmannPRB, GiampaoloPRA, NJPCampbell, BayatPRL2017, BellIneqPRA2012, JafariPRA2017, JafariPRA2008, RulliPRA2010, Zakaria2019, mzaouali2019, ahami2021thermal, abaach2021pairwise, mansour2020quantum, el2020quantum}. Indeed, while quantum phase transitions only strictly occur at zero-temperature, approaches based on these quantum information theoretic tools have revealed that signatures of these phenomena persist even at finite temperatures and can be rigorously studied~\cite{DeChiaraReview}. The cross-fertilization of the experience built up over the last decades in condensed matter physics, with the new tools offered by quantum information theory, has paved the way to many discoveries of new properties of critical quantum systems, and the development of new protocols for quantum computation and communications~\cite{nishimori, heim2015annealer, yan2021annealer, bose2007quantumcommu}. A prominent example is the quantum computer, which can be represented as a many-body system and can be controlled and manipulated through protocols. Numerical methods were also re-visited under the tools of quantum information science in order to design new efficient simulation methods for quantum many-body systems, e.g. the density matrix renormalization group, and tensor networks~\cite{dmrgtensor}. \\
\noindent Information is physical, as a consequence of the intimate relationship between information and thermodynamics, through the entropy. Einstein describes thermodynamics as: \textit{``the only physical theory of universal content, which I am convinced, that within the framework of applicability of its basic concepts will never be overthrown''}. Indeed, thermodynamics is the most basic building block of modern science, and provides a framework to understand various range of natural phenomenons in physics, chemistry, and biology~\cite{perrot1998z}. Historically, the motivation behind the invention of thermodynamics is to enhance the efficiency of heat engines, through the study of the relationship between work and heat~\cite{carnot1978reflexions}. However, with the advances of statistical mechanics~\cite{reif}, the theory of thermodynamics attained its mathematically rigorous form, which paved the way to apply it to quantum settings~\cite{thermo_quantum_regime, booksteve}. Thermodynamics benefited heavily from the developments of the field of classical and quantum information theory, which allowed extending the theory to atomic systems and solving paradoxes, such as: Maxwell's demon~\cite{bennett2003notes}. Additionally, the relationship between information and thermodynamics was also inspected in the context of the Szillard engine~\cite{szilard, szilardexp}, and for finding the thermodynamic cost of classical (quantum) information processing, via the Landauer's principle~\cite{Bennett1982TheTO}. The resource theory for thermodynamics was developed, by inspiration of its analogue in quantum information theory, which sets the boundaries on what can and cannot be achieved while performing thermodynamic state transformations~\cite{gooldreview}. Quantum thermodynamics has been very successful in capturing and exploiting the features of quantum systems in order to design efficient quantum devices~\cite{CampbellPRB2016, mzaouali2021}. In particular, entanglement and coherence have been used as resources in order to design quantum-analogues of thermal machines and batteries~\cite{Hammam_2021, Giorgi_2015iop, batt2020, batt2021}. Quantum criticality has been harnessed in order to increase the efficiency of quantum heat engines, where the working medium is a quantum spin chain driven across its equilibrium (non-equilibrium) quantum phase transition~\cite{topo_qhe, Fogarty_2020qhe, qhe_campsi2020, myers_qhe, qpt_qhe2017, lmg_qhe2016}. 

\noindent The scope of the present thesis is to provide a theoretical study on the equilibrium and out-of-equilibrium critical features of quantum systems. Therefore, to build the necessary background, we start in \autoref{chap1} with an overview on the critical phenomenons at the classical and quantum levels. In particular, we discuss how a classical phase transition emerges, their classifications, and the mathematical theories that describes them, i.e. Weiss' mean field and Landau's theory. Furthermore, we extend the overview toward the quantum level in order to discuss quantum phase transitions, and the difference with their classical counterparts. \\
\noindent The first goal in this thesis is to analyze critical quantum systems, using the quantum information formalism. Hence, in \autoref{chap2} we review the topic of classical and quantum information theory, where we discuss the notion of information, how it can be measured, and how it can be manipulated in a classical and quantum manner. Additionally, we present in \autoref{chap3} an application of the formalism of quantum correlations from quantum information theory, to the study of the second-order quantum phase transition in the critical Heisenberg $XX$ model.\\
\noindent The second goal is to study quantum phase transition in phase space. For this reason,  we discuss in \autoref{chap4} the phase space formulation of quantum mechanics, where we present the traditional Wigner function for infinite-dimensional systems, its properties, and some experimental procedures to measure phase space quantities. Through the rest of the chapter, special focus will be on the case of finite-dimensional systems, where we introduce two formulations of the Wigner function for discrete systems, i.e. Wootters' and Stratonovich Wigner function; and we discuss their features, as well as some experimental protocols to measure them. In the end of this part, \autoref{chap5} is devoted to applying the Wootters and Stratonovich Wigner function in order to detect first-, second-, and infinite-order quantum phase transitions in the $XY$ and $XXZ$ model. \\
\noindent The final goal in this thesis is to inspect the non-equilibrium aspect of critical quantum systems, using the quantum thermodynamics approach. As such, we explore in \autoref{chap6} the phenomenological theory of thermodynamics, its extension to out-of-equilibrium and quantum settings. We discuss in parallel the resource theory for thermodynamics, and the interplay between classical and quantum computing with thermodynamics. We provide in \autoref{chap7} an analysis on the dynamics and thermodynamics of the Lipkin-Meshkov-Glick model. In particular, we present the effect of the presence of the excited state quantum phase transition on the statistics of the work, when the system is initialized in the ferromagnetic phase, and subject to a quench. Furthermore,  We assess the role of symmetry breaking has on the ensuing dynamics, highlighting that its effect is only present for quenches beyond the critical point. Moreover, we show that similar features persist when the system is initialized in an excited state and briefly explore the behavior for initial states in the paramagnetic phase. In the end, we conclude the thesis by presenting the short-term and long-term future projects. 
\part{Quantum Information meets Quantum Phase Transitions \label{part1}}
\chapter{Overview on Phase Transitions \label{chap1}}
\section{Classical phase transitions}
\subsection{What is a phase transition?}
A phase transition is a sudden and discontinuous change in the properties of a system, driven by an external parameter. When this change is done by temperature we call the phenomenon classical phase transition~\cite{goldenfeld2018lectures}. Examples of classical phase transitions are ubiquitous in nature, from the melting of ice, evaporation of water to the loss of magnetisation in metals. The study of phase transitions lie at the heart of condensed matter physics and translates into numerous applications: semiconductors, transistors and microchips. The theoretical framework to analyze phase transitions is via statistical mechanics, which is a theory that aims to provide a microscopic description of macroscopic phenomena. Mathematically, the goal of statistical physics is to calculate the partition function $``\mathcal{Z}"$ of the system. It is given by
\begin{equation}
    \mathcal{Z}_{\Omega}=\Tr e^{-\beta \mathcal{H}_{\Omega}},
    \label{partition_function}
\end{equation}
where $\beta=\frac{1}{k_b T}$ is the Boltzmann factor, $\mathcal{H}_{\Omega}$ is the Hamiltonian of the system and $\Omega$ is the microstates sample region. All the thermodynamical quantities follow from the partition function ~\eqref{partition_function}, such as: the mean energy $\langle E\rangle$, the variance in the energy $\langle (\Delta E)^2\rangle$, the heat capacity $C_v$ and the entropy $S$. The most prominent figure of merit for analyzing phase transitions is the free energy ``$\mathcal{F}"$, given by
\begin{equation}
    \mathcal{F}_{\Omega}[K]=-k_b T \log(\mathcal{Z}_{\Omega}[K]).
    \label{classical_free_energy}
\end{equation}
It provides access to the information on the thermodynamics of the system $\Omega$, through the derivatives $\frac{\partial F_{\Omega}}{\partial K_n}$ and $\frac{\partial^2 F_{\Omega}}{\partial K_n \partial K_m}$, where the $K_n$ are external parameters, such as: the magnetic field, the temperature or the exchange interactions. The nature of the phase transition present in the system can be identified through the discontinuity present in the derivatives of $\mathcal{F}_{\Omega}$~\eqref{classical_free_energy}. However, when $\Omega$ is finite, the partition function is a sum of terms each of which is the exponential of an analytic function of the parameter of the Hamiltonian, which cannot give rise to non-analytic behaviour. Only at the limit $\Omega \to \infty$, where non-analytical behaviour arise and phase transitions can effectively be studied. For a very large system of $N$ sites, which is the case studied in this thesis, the bulk free energy can be defined as:

\begin{equation}
    f_b[K]=\lim_{N \to \infty} \frac{\mathcal{F}_{\Omega}[K]}{N}.
    \label{thermo_limit}
\end{equation}

\noindent The limit in~\eqref{thermo_limit}, is known as ``the thermodynamical limit''. Its existence allows to give, through $f_b[K]$, a precise definition of a phase boundary and the nature of the phase transition. However, the existence of the thermodynamical limit is not trivial, and some systems may fail to show bulk behaviour because of the nature of the interactions governing the system. For example, a charged system following inverse square law forces fail to show bulk behaviour, which is due to the fact that inverse square law forces are very long ranged to permit bulk behaviour in the thermodynamic limit~\cite{stab_matter}. 
\subsection{Classification of phase transitions \label{1.1.2}}
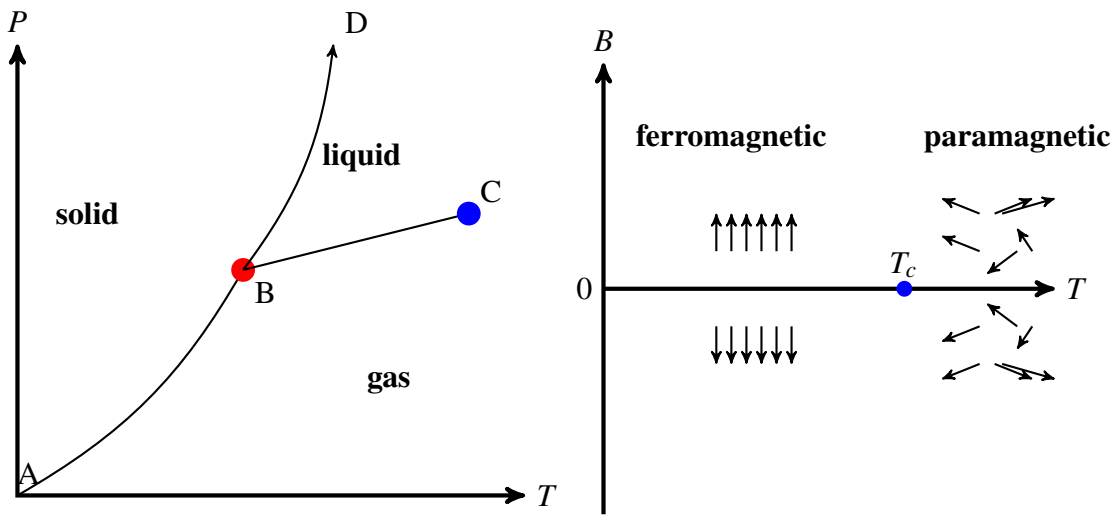
\begin{figure}[t!]
    \centering
    \begin{subfigure}[t]{0.49\textwidth}
    	\begin{tikzpicture}[scale=1.5]
    	\draw [<->,ultra thick] (0,4) node (yaxis) [above] {$P$}
    	|- (4.5,0) node (xaxis) [right] {$T$};
    	\node[left] (A) at (1,2.5) {\textbf{solid}};
    	\node[left] (A) at (3.5,3) {\textbf{liquid}};
    	\node[right] (A) at (3,1) {\textbf{gas}};
    	\coordinate (O) at (0,0);
    	\coordinate (A) at (2,2);
    	\coordinate (B) at (2.8,4);
    	\coordinate (C) at (4,2.5);
    	\draw (0.3,0) node[anchor=south east]{A};
    	\draw[thick,color=black] (O) to [bend right=15] (A) node[anchor=north west]{B};
    	\fill[red] (2,2) circle (3pt);
    	\draw[->,thick,color=black] (A) to [bend right=15] (B) node[anchor=south west]{D};
    	\draw[thick,color=black] (A) to [bend right=0] (C) node[anchor=south west]{C};
    	\fill[blue] (4,2.5) circle (3pt);
    	\end{tikzpicture}
    \caption{Phase diagram of water, with respect to the temperature $T$ and pressure $P$.}
    \label{water}
    \end{subfigure}
        \begin{subfigure}[t]{0.49\textwidth}
        	\begin{tikzpicture}
        	\node[] (A) at (1.7,2) {\textbf{ferromagnetic}};
        	\node[] (A) at (5.5,2) {\textbf{paramagnetic}};
        	\draw[->,ultra thick] (0,0)--(6,0) node[right]{$T$};
        	\draw[->,ultra thick] (0,-3)--(0,3) node[above]{$B$};
        	\node[left] (A) at (0,0) {$0$};
        	\draw[->,thick] (1.5,0.5)--(1.5,1.);
        	\draw[->,thick] (1.7,0.5)--(1.7,1.);
        	\draw[->,thick] (1.9,0.5)--(1.9,1.);
        	\draw[->,thick] (2.1,0.5)--(2.1,1.);
        	\draw[->,thick] (2.3,0.5)--(2.3,1.);
        	\draw[->,thick] (2.5,0.5)--(2.5,1.);
        	\draw[->, thick] (1.5,-0.5)--(1.5,-1.0);
        	\draw[->,thick] (1.7,-0.5)--(1.7,-1.);
        	\draw[->,thick] (1.9,-0.5)--(1.9,-1.);
        	\draw[->,thick] (2.1,-0.5)--(2.1,-1.);
        	\draw[->,thick] (2.3,-0.5)--(2.3,-1.);
        	\draw[->,thick] (2.5,-0.5)--(2.5,-1.);
        	\node[above] (A) at (4,0) {$T_c$};
        	\draw[->,thick] (5,0.5)--(4.5,0.7);
        	\draw[->,thick] (5.5,0.5)--(5.1,0.2);
        	\draw[->,thick] (5.7,0.5)--(5.5,0.8);
        	\draw[->,thick] (5,1)--(4.5,1.2);
        	\draw[->,thick] (5.2,1)--(5.7,1.2);
        	\draw[->,thick] (5.3,1)--(6,1.2);
        	\fill[blue] (4,0) circle (3pt);
        	\draw[->,thick] (5,-0.5)--(4.5,-0.7);
        	\draw[->,thick] (5.5,-0.5)--(5.1,-0.2);
        	\draw[->,thick] (5.7,-0.5)--(5.5,-0.8);
        	\draw[->,thick] (5,-1)--(4.5,-1.2);
        	\draw[->,thick] (5.2,-1)--(5.7,-1.2);
        	\draw[->,thick] (5.3,-1)--(6,-1.2);
        	\end{tikzpicture}
            \caption{Phase diagram of a ferromagnetic material, with respect to temperature $T$ and the magnetic field $B$.}
            \label{ph_ferr}
        \end{subfigure}
    \caption{Examples of (a) first order phase transition and (b) second order phase transition}
    \label{fig}
\end{figure}

A phase is a region where $f_b[K]$~\eqref{thermo_limit}, is analytic. Phase boundaries are regions of discontinuity of  $f_b[K]$ ~\eqref{thermo_limit} and can be classified into two types
\begin{enumerate}
    \item \textbf{First order phase transition}:\\
    $\partial{f_b}/\partial{K}$ is discontinuous in the boundary between the phases. First order phase transitions are characterized by the release or absorption of energy in the form of latent heat. Figure~\eqref{water} represents the phase diagram of water, which undergoes this kind of phase transition. The diagram represents in the $P$-$T$ plane the different phases of water: solid, liquid and gas, while the solid lines separate the phases and the coexistence zones. The AB curve is the \textbf{``sublimation curve''}, which separates the solid and the vapor phases, while they coexist along the curve. The BD line is called the \textbf{``fusion curve''}, and it represents the coexistence zone of the solid and the liquid phases. The BC branch, is the \textbf{``vaporization curve''}, where the liquid phase coexists with the vapor phase. The point B is called \textit{``the triple point''}, where all the phases coexist at $T=273.16 K$. The point C is a \textit{critical point} where the density of the liquid and the gas are equal, and are not distinguishable. After crossing C, water transits continuously from liquid to gas.
    \item \textbf{Second order phase transition}: \\ $\partial{f_b}/\partial{K}$ is smooth, while $\partial^2{f_b}/\partial{K}^2$ is divergent. Second order phase transitions are also referred to as continuous phase transition. A prominent example of a second order phase transition is the ferromagnetic-paramagnetic transition in metals. The phase diagram of a ferromagnetic metal is represented in Figure~\eqref{ph_ferr}. The ferromagnetic-paramagnetic transition emerges, experimentally, when we heat up a ferromagnetic material above a critical temperature $T_c$ (which depends on the nature of the metal), where order (ferromagnetism) is destroyed by thermal fluctuations and disorder (paramgnetism) emerges. 
\end{enumerate}

\subsection{The Ising model \label{1.1.3}}
Nature is abundant with phenomena where complex and large number of microscopic constituents interact to give rise to a macroscopic behaviour. For example, magnets and liquids are composed, respectively, of many interacting magnetic dipoles and molecules. In order to investigate the features of real world materials, physicists make use of mathematical models from statistical mechanics in order to simulate systems exhibiting global behaviour due to local interactions between large number of entities. The success of this approach is tightly connected to the theoretical assumption on how the particles of the system interact among themselves. In general, it is very difficult to design a realistic model that is mathematically tractable and provides comprehensive description of all macroscopic features. Among the class of exactly solvable models~\cite{baxterbook, franchini2017}, the Ising model is one of the touchstones of modern physics as it is the most simple and versatile model for magnetic materials.

\noindent The Ising model was introduced by Lenz and Ising in 1925 to describe how temperature can transform a metal from a paramagnet to a permanent magnet (ferromagnet)~\cite{Ising:1925em}. It is defined on a regular lattice, where each site contains a magnetic dipole, i.e. spin, which can take $\pm 1$ values. Table~\eqref{isingtab} represents a sketch of the model in a two dimensional square lattice. 
    \begin{table}
        \centering
        \begin{tabular}{| c c c c|}
        \hline
         + & + & + & + \\ 
         - & - & - & - \\
         - & + & + & - \\  
         - & + & + & + \\
         \hline
        \end{tabular}
        \caption{Possible configuration of an $N=4^4=16$ Ising lattice~\eqref{ising1}.}
        \label{isingtab}
    \end{table}
The Hamiltonian of the Ising model can be written as follows
    \begin{ceqn}
    	\begin{equation}
    		H_{\Omega} = \sum_{ij} J_{ij} S_iS_j+\sum_{i\in \Omega} h_iS_i,
    	\label{ising1}
    	\end{equation}
    \end{ceqn}
where the $S_i$'s are classical spin operators and $J_{ij}$ is the exchange interaction between two classical spins. The spins also interact with an external magnetic field $h_i$ which, in general, varies from site to site. The Ising model~\eqref{ising1}, is exactly solvable in one-dimension for arbitrary values of the magnetic field, while in two-dimensions it is only solvable at zero magnetic field~\cite{onsager1, onsager2, onsager3}. The three dimensional Ising model has not yet been solved. The simplicity of the Ising model allows for the study of phase transitions. For instance, it was exploited in order to describe the critical behaviour in ferromagnetic films, binary alloys and liquid-gas phase transitions. Critical phenomena arise only in the thermodynamic limit, for this limit to exist in the Ising model~\eqref{ising1}, the exchange interaction $J_{ij}$ need to satisfy the following property~\cite{griffith} 
\begin{equation}
    \sum_{j\neq i} |J_{ij}| < \infty.
\end{equation}
The phase diagram of physical models, and in particular the Ising model, can be constructed by following the energy-entropy approach which consists of analysing and comparing the behaviour of the free energy $\mathcal{F}$~\eqref{classical_free_energy}, at high and low temperature. For a given system, the change of the free energy is given by
\begin{equation}
    F= U - TS,
    \label{fe}
\end{equation}
where $U$ is the internal energy, $T$ is the temperature and $S$ is the entropy of the system. At high temperature, the free energy is dominated by the entropy $S$ and the free energy ~\eqref{fe}, is minimized by maximizing the entropy $S$. At low temperature, the internal energy $U$ has a significant chance of dominating the free energy $F$~\eqref{fe}. Therefore, the free energy is minimized by minimizing $U$. If the resulting macroscopic states at high and low temperatures are different, we conclude that at least one phase transition occurred at some critical temperature $T_c$.
\subsection{One-dimensional phase diagram}
We consider a one-dimensional chain of $N$ spins, described by the Ising model~\eqref{ising1}. At $T=0$ and zero magnetic field, the spins can point either all up or all down. Consider the spin up configuration, the free energy of the system is $F=-NJ$. Switching on the temperature has the effect of creating a domain of flipped spins as sketched below
$$
\uparrow \uparrow \uparrow \uparrow \dots \uparrow \uparrow \uparrow \uparrow \uparrow \uparrow \uparrow | \downarrow \downarrow \downarrow \downarrow \downarrow \downarrow \dots \downarrow \downarrow \downarrow \downarrow \downarrow \downarrow 
$$
The energetic cost of the creation of this domain of flipped spins is: $E=-NJ+2J$. The domain can be placed at any of the $N$ sites, therefore the entropy is: $S=k_b\log N$. Thus, the change in the free energy is
\begin{equation}
    \Delta F =\Delta E-T\Delta S= 2J -k_b T\log N.
    \label{fe131}
\end{equation}
For finite temperature, the free energy~\eqref{fe131}, diverges as $N \to \infty$. To gain stability, the system lowers its free energy by creating domain walls, a process that gets repeated until there are no domains left. Thus, at zero magnetic field, long-range order (the ferromagnetic phase) does not hold against thermal fluctuations. We conclude that for the one-dimensional Ising model, there are no finite temperature phase transitions at zero field.
\subsection{Two-dimensional phase diagram \label{2d_diagram_ising}}
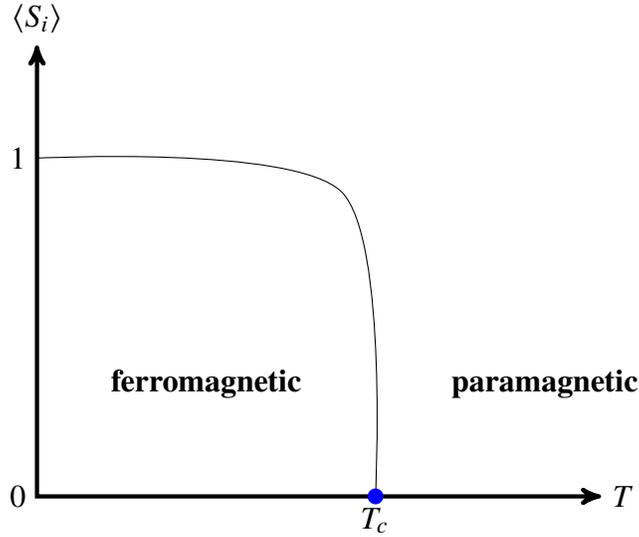
\begin{figure}[t!]
    \centering
    	\begin{tikzpicture}[scale=1.5]
    	\draw [<->,ultra thick] (0,4) node (yaxis) [above] {$\langle S_i \rangle$}
    	|- (5,0) node (xaxis) [right] {$T$};
        \draw plot [smooth] coordinates {(0,3) (2.7,2.7) (3,0)};
        \node[left] (A) at (0,0) {$0$};
        \node[left] (A) at (0,3) {$1$};
        \node[below] (A) at (3,0) {$T_c$};
        \fill[blue] (3,0) circle (2pt);
        \node[] at (1.5,1) {\textbf{ferromagnetic}};
        \node[] at (4.5,1) {\textbf{paramagnetic}};
    	\end{tikzpicture}
    \caption{Net magnetization in the two-dimensional Ising model ~\eqref{ising1}, with respect to the temperature $T$.}
    \label{mag_132}
\end{figure}

In the same spirit, we consider a two-dimensional lattice described by the Ising model~\eqref{ising1}. We switch on the temperature and we analyze the energetic and entropic cost of the domain of flipped spins on the lattice. Since each flipped spin costs an amount of energy of $2J$, then the internal energy change of a domain of length $N$ is $\Delta E = 2JN$. The entropy can be estimated by enumerating the different possibilities of the domain wall, which is proportional to the coordinate number of the lattice $z^N$. The entropy is then $\Delta S = k_b N \log (z-1)$ and the free energy is
\begin{equation}
    \Delta F =(2J-k_b T\log (z-1))N.
    \label{fe132}
\end{equation}
At the thermodynamical limit, i.e. $N \to \infty$, the stability of the free energy~\eqref{fe132}, is dependent on the temperature $T$. If 

\begin{equation}
    T>T_c=\frac{2J}{k_b \log(z-1)},
\end{equation}
the free energy~\eqref{fe132}, diverges and the system is unstable toward the formation of domains. This implies that at $T>T_c$ the system is in a paramagnetic phase where the net magnetization $\langle S_i \rangle=0$. At $T<T_c$, long range order is stable and the net magnetization $\langle S_i \rangle\neq 0$. We conclude that the two-dimensional Ising model, exhibit a finite temperature phase transition at zero magnetic field. Figure~\eqref{mag_132}, represent the net magnetization $\langle S_i \rangle$ with respect to the temperature $T$, in the two-dimensional Ising model. The magnetization is non-zero in the ferromagnetic phase and decreases as we increase the temperature, until it vanishes at the critical point $T_c$, as the system enters the paramagnetic phase.

\subsection{Mean field theory \label{sec1.4}}
Many-body systems are, in general, plagued with exact solvability issues which hinder the computation of their partition function $``\mathcal{Z}$''~\eqref{partition_function}, and therefore affect the access to their critical features, except for few cases like the one-dimensional Ising model. Mean field theory is an approximation method that aims to approximate the local field in the original ``\textit{hard to solve}'' Hamiltonian by an average effective field determined by the entire system. In other words, we look at the chain from the point of view of a given spin and we ignore the fluctuations of the system, while focusing on its mean behaviour~\cite{weiss-curie}.

\noindent According to mean field theory, the Hamiltonian of the Ising model ~\eqref{ising1}, can be written as
\begin{equation}
    H_{\Omega} = -\sum_i S_i H_i,
\end{equation}
where
\begin{equation}
H_i=\underbrace{h}_{\text{magnetic field}}+ \underbrace{\sum_j J_{ij}\langle S_j \rangle}_{\text{the mean field}} + \underbrace{\sum_j J_{ij} \left(S_j-\langle S_j. \rangle\right)}_{\text{the fluctuations}}.
\end{equation}
We ignore the fluctuation term, and we write $H_i$ for a $d$-dimensional hypercubic lattice as
\begin{equation}
    H_i = h + 2dJm,
\end{equation}
where $2d$ is the coordination number of each site of the lattice and $m$ is the mean field sensed by the spin. Then, the magnetization is
\begin{equation}
    M= \tanh\left(\frac{h+2dJm}{k_B T}\right).
    \label{sol_mft}
\end{equation}
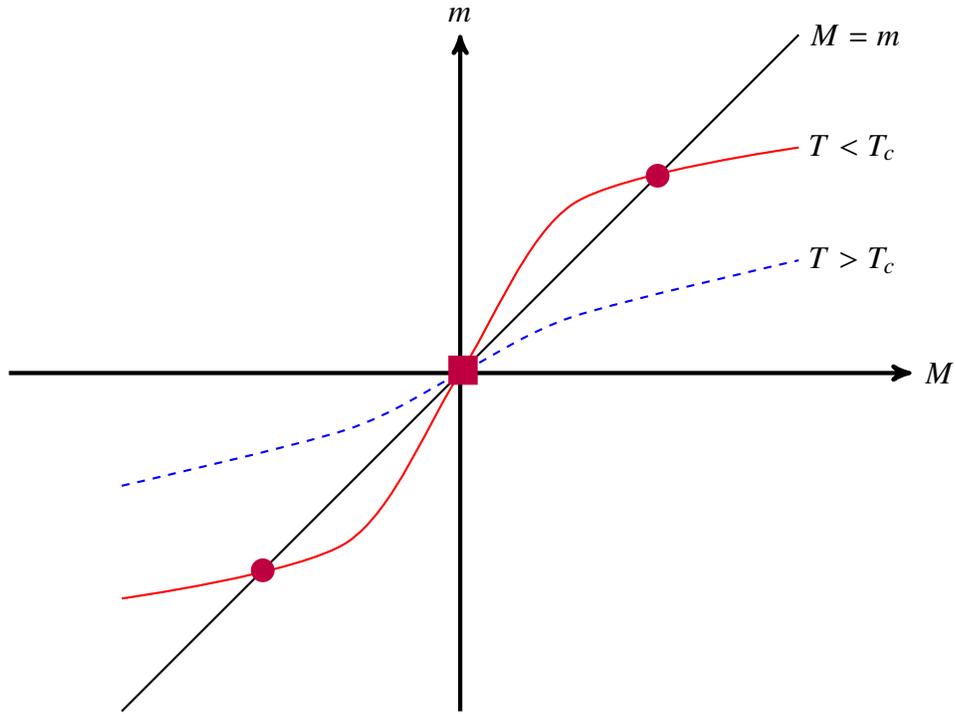
\begin{figure}[t!]
    \centering
    	\begin{tikzpicture}[scale=1.5]
        \draw[->,ultra thick] (-4,0)--(4,0) node[right]{$M$};
        \draw[->,ultra thick] (0,-3)--(0,3) node[above]{$m$};
        \draw [thick] plot [smooth] coordinates {(-3,-3) (0,0) (3,3)};
        \draw [red,thick] plot [smooth] coordinates {(-3,-2) (-1,-1.5) (0,0) (1,1.5) (3,2)};
        \draw [blue,dashed,thick] plot [smooth] coordinates {(-3,-1) (-1,-0.5) (0,0) (1,0.5) (3,1)};
        \node[right] (A) at (3,2) {$T<T_c$};
        \node[right] (A) at (3,3) {$M=m$};
        \node[right] (A) at (3,1) {$T>T_c$};
        \filldraw[purple] (-0.1,-0.1) rectangle (0.15,0.15);
        \fill[purple] (1.75,1.75) circle (3pt);
        \fill[purple] (-1.75,-1.75) circle (3pt);
    	\end{tikzpicture}
    \caption{Graphical solution to mean field approximation~\eqref{sol_mft}.}
    \label{mean_field}
\end{figure}
\noindent At zero magnetic field, i.e. $h=0$, the critical temperature is 
\begin{equation}
    T_c =2dJ/k_B.
\end{equation}
Non-analytical behaviour arise due to the behaviour of the solutions of the magnetization~\eqref{sol_mft}, with respect to the temperature $T$. This is visualised graphically in Figure~\eqref{mean_field}, where we see at $T>T_c$, the $\tanh$ curve lies below $M=m$ and intersect only at $M=0$. Hence, at high temperatures, there is no net magnetization and that the ordered state is destroyed by thermal fluctuations. However, at $T<T_c$ the $\tanh$ intersects at $\pm M_0$ and $M=0$ (which is unstable). Thus, the spins are pointing in the same directions, either up or down.
\subsection{Critical exponents \label{1.1.4.1}}
We can study the behaviour of physical quantities around the critical point $T_c$, by expanding the equation of state~\eqref{sol_mft}, in the vicinity of $T_c$ as
\begin{equation}
    M= \tanh(\frac{h}{k_B T} +m \tau) =
\frac{\tanh(\frac{h}{k_B T})+\tanh m\tau }{1+\tanh(\frac{h}{k_B T}) \tanh m\tau},
\end{equation}
where $\tau=T/T_c$. Then
\begin{ceqn}
	\begin{equation}
	\tanh(\frac{h}{k_B T}) = \frac{M-\tanh m \tau}{1-M \tanh m \tau},
	\label{criexpomagn}
	\end{equation}
\end{ceqn}
for weak $h$ and small $m$, we can write~\eqref{criexpomagn} as
\begin{ceqn}
	\begin{equation}
	\frac{h}{k_B T} \approx M(1-\tau) + M^3\left(\tau -\tau^2 +\frac{\tau^3}{3} +\dots\right) + \dots
	\label{criexpomagn2}
	\end{equation}
\end{ceqn}
For zero magnetic field and when $T \to T_c^-$, we have
\begin{equation}
    M^2 \approx 3 \frac{T_c - T}{T} + \dots.
\end{equation}
As $M \propto \left(\frac{T- T_c}{T}\right)^{\beta}$ we can extract immediately the critical exponent of the ferromagnetic transition: $\beta=1/2$. The critical isotherm is the curve in the $h$-$M$ plan corresponding to $T=T_c$. Near the critical point, it is described by a critical exponent $\delta$
\begin{equation}
    h \sim M^{\delta}.
\end{equation}
Setting $\tau=1$ in~\eqref{criexpomagn2}, we find $\delta=3$. That is
\begin{ceqn}
	\begin{align}
		\frac{h}{k_B T} \sim M^3.
	\end{align}
\end{ceqn}
The isothermal magnetic susceptibility also diverges near $T_c$:
\begin{equation}
    \chi_T =\frac{\partial M}{\partial h},
\end{equation}
from~\eqref{criexpomagn2}, we get
\begin{equation}
    \frac{1}{k_B T} = \chi_T(1-\tau)+3M^2 \chi_T (\tau -\tau^2+\frac{1}{3} \tau^3).
\end{equation}
At high temperatures, i.e. $T>T_c$, $M=0$. Then
	\begin{ceqn}
		\begin{align}
			\chi_T &=\frac{1}{k_B} \frac{1}{T-T_c} + \dots\\& \sim |T-T_c|^{-\gamma}.
			\label{eq113}
		\end{align}
	\end{ceqn}
The critical exponent that characterizes the divergence in the isothermal susceptibility is $\gamma =1$. At low temperatures, i.e. $T<T_c$, we have
\begin{equation}
    M=\sqrt{3} \left( \frac{T_c - T}{T} \right)^{1/2} + \dots
\end{equation}
	which yields to
\begin{equation}
    	\chi_T =\frac{3}{2k_B} \frac{1}{T-T_c} + \dots
\end{equation}
Below the transition temperature, the isothermal susceptibility diverges with $\gamma=1$.
The critical exponent $\alpha$ of the specific heat can also be calculated, for that we write the free energy in the mean field approximation as
\begin{equation}
    F_m= -k_B T \ln\left[ 2\cosh\left( \beta J2dm \right) \right],
\end{equation}
knowing that $\cosh(x) = 1+\frac{x^2}{2!} + \frac{x^4}{4!} + \dots$ and that $M=0$ for $T>T_c$ while $M=\left(3 \frac{T_c - T}{T} \right)^{1/2}$ for $T<T_c$. Taking the second derivative of the free energy with respect to the temperature leads to
\begin{equation}
    C=\begin{cases}
\displaystyle{\frac{3}{2}} k_B N &T<	T_c,\\
0&T>T_c.
\end{cases}
\end{equation}
Since $C\sim |\frac{T_c - T}{T}|^{-\alpha}$, the critical exponent $\alpha$ must be zero.
An important relationship can be derived between the isothermal magnetic susceptibility and the correlations functions. For that, we start from the partition function of the Ising model~\eqref{ising1}, given by
\begin{equation}
    Z= \Tr \exp \left[ -h \beta \sum_i S_i - \beta J \sum_{\langle ij \rangle} S_iS_{j} \right].
\end{equation}
Then, we have
\begin{ceqn}
	\begin{align}
		\sum_i \langle S_i \rangle &= \frac{1}{\beta Z} \frac{\partial Z}{\partial h}, \\
		\sum_i \langle S_iS_j\rangle &= \frac{1}{\beta^2 Z} \frac{\partial^2 Z}{\partial h^2}.
	\end{align}
\end{ceqn}
On the other side

\begin{ceqn}
	\begin{align}
	\chi_T &=\frac{\partial M}{\partial h}=\frac{1}{\beta N} \frac{\partial^2 \log Z}{\partial h^2},\nonumber\\
	&=\frac{1}{N} k_B T \left[ \frac{1}{Z} \frac{\partial^2  Z}{\partial h^2} - \frac{1}{Z^2} \left(  \frac{\partial  Z}{\partial h}\right)  \right], \nonumber\\
	&=\frac{1}{N} (k_B T)^{-1} \left[ \sum_{ij} \langle S_iS_j\rangle - \left(\sum_i \langle S_i \rangle \right)^2 \right],\nonumber\\
	&=\frac{1}{N} (k_B T)^{-1}  \sum_{ij} G(r_i-r_j),\nonumber\\
	&=(k_B T)^{-1} \sum_i G(x_i),\nonumber\\
	&=(a^d k_B T)^{-1} \int_{\Omega} d^d rG(r).
	\end{align}
\end{ceqn}
Here, $G(r_i-r_j)=\langle S_iS_j \rangle - \langle S_i \rangle \langle S_j \rangle$ is the two point correlation function, where $r_i(r_j)$ is the spatial position of the spin $S_i(S_j)$.
This result relates the divergence in $\chi_T$ with the two point correlation function $G$. Then, $G$ has to reflect the divergence in $\chi_T$, in general we have
\begin{ceqn}
	\begin{align}
		G(r) \sim \frac{e^{-|r|/\xi}}{|r|^{(d-1)/2}\xi^{(d-3)/2}}, && \text{for } |r|>>\xi.
		\label{g(r)}
	\end{align}
\end{ceqn}
Combining this result with~\eqref{eq113} yields to
\begin{ceqn}
	\begin{align}
		\left(\frac{T_c -T}{T}\right)^{-1} &\sim \int \frac{r^{d-1}e^{-r/\xi}}{r^{(d-1)/2}\xi^{(d-3)/2}} dr \\
		&\sim \int \left(z^{(d-1)/2}  e^{-z} dz\right) \xi^2,
	\end{align}
\end{ceqn}
with $z=r/\xi$. Thus
\begin{ceqn}
	\begin{equation}
		\xi \sim \left(\frac{T_c -T}{T}\right)^{-\nu},
	\end{equation}
\end{ceqn}
with $\nu=1/2$. $\xi$ is called the correlation length which describes the extent of spatial correlations between two spins $S_i$ and $S_j$. In the thermodynamical limit, $\xi$ diverges at the critical point. The last critical exponent we mention is $\eta$, which describes how the two points correlation function $G(r)$~\eqref{g(r)}, behave at long distances at the critical point. $G(r)$ for long distances near the critical point is given by: $G(r) \sim r^{-(d-2+\eta)}$, with $\eta=0$. In principle, $\eta$ can be non zero.
\begin{table}[t!]
	\centering
	\begin{tabular}{ccccc }
		\hline
		Exponent  & Mean Field  & Experiment & 2D Ising & 3D Ising  \\ \hline
		$\alpha$ & 0 &0.110-0.116 & 0& 0.110\\
		$\beta$ & 1/2 & 0.316-0.327 & 1/8 & 0.325 $\pm$ 0.0015\\
		$\gamma$ & 1 & 1.23-1.25 & 7/4 & 1.2405 $\pm$ 0.0015\\
		$\delta$ & 3 & 4.6-4.9 & 15 & 4.82\\
		$\nu$ & 1/2 & 0.625 $\pm$ 0.010 & 1 & 0.630 \\
		$\eta$ & 0 & 0.016 - 0.06 & 1/4 & 0.032 $\pm$0.003\\
		\hline
	\end{tabular}
\caption{Critical exponents for the Ising universality class}
\label{cri_exp_tab}
\end{table}

\noindent How accurate is the mean field theory? In Table~\eqref{cri_exp_tab}, we see the discrepancy between the critical exponents obtained by the mean field approximation and the experimental results while the exponents of the three-dimensional Ising model are in accordance with experience. This is due to the mean field approximation, where the exponents in this approach do not depend on the dimension, while from Table~\eqref{cri_exp_tab}, it is clear that the critical exponents depend on the dimensionality of the system. On the other hand, the approximation supposes that the spins do not interact and each spin feels the same field due to all the other spins, and this contradicts the essence of magnetism, which is based on cooperative long range cooperative behavior of spins. The mean field approximation is clearly not a good choice for magnetic systems.\\  
The critical exponents satisfy scaling relations, which can be obtained by thermodynamic considerations, they are given by
\begin{ceqn}
	\begin{align}
		\alpha +2\beta+\gamma=2&,\nonumber\\
		\gamma=\beta(\delta-1)&,\\
		\gamma=\nu(2-\eta).& \nonumber
	\end{align}
\end{ceqn}
The precision of the mean field approximation increases as we increase the dimension of the system. In fact, from the scaling relation $2-\alpha=d\nu$, where $d$ is the dimension of the system, we can deduce the critical dimension $d_c$ at which we get precise results from the mean field approximation. Since $\alpha=0$ and $\nu=1/2$, $d_c$ must be 4.

\subsection{Landau theory of phase transitions}
One of the main challenges in physics is to develop a universal framework, a theory of everything, for all physical aspects of our universe. In condensed matter physics and statistical mechanics, Landau proposed a theory for all phase transitions based on simple hypotheses~\cite{landau1936theory}. Landau theory of phase transitions consists of \textit{guessing} a quantity called the \textit{Landau potential}, where its minima with respect to the order parameter should describe the thermodynamic properties of the system at the critical point.
\subsubsection{The order parameter}
The order parameter $\Psi$ is a quantity used to describe phase transitions, it is zero (non-zero) in the disordered (ordered) phase. The definition of the order parameter is not always trivial as there are systems for which this quantity can not be defined. Depending on the nature of the system, the order parameter can be a scalar, vector or a tensor.\\
The order parameter of the ferromagnetic-paramagnetic transition is, in general, a vector: the net magnetization $M=\sum_i \langle S_i \rangle$. It is zero (non-zero) above (below) the critical temperature. For the liquid-gas transition, the order parameter is a scalar which is computed via the difference between the density of liquid and that of the gas, that is $\rho_l-\rho_g$.
\subsubsection{Landau Theory}
Landau theory of phase transitions consists of writing a function $L$ called \textit{Landau free energy} or \textit{Landau functional} in terms of  the order parameter $\Psi$ and the coupling constants $\left[ K_i\right]$. In this decomposition, we keep only the terms that are compatible with the symmetry of the system and we assume that thermodynamic functions of the system can be calculated by differentiation of $L$. Landau free energy has the following constraints
\begin{enumerate}
	\item $L$ has to follow the symmetries of the system.
	\item Near $T_c$, $L$ is an analytic function of $\Psi$ and $[K]$. We can write
        \begin{equation}
            	L = \sum_{n=0}^{\infty} a_n([K])\Psi^n.
        \end{equation}
	\item $\Psi$=0 in the disordered phase, while it is small and non zero in the ordered phase near $T_c$. Thus, for $T>T_c$ we solve the minimum equation for $L$ by $\Psi=0$ and for $T<T_c$ $\Psi \neq 0$ solves the minimum equation. For a homogeneous system we can write
	\begin{ceqn}
		\begin{equation}
			L = \sum_{n=0}^{4} a_n([K])\Psi^n.
			\label{landau1}
		\end{equation}
	\end{ceqn}
	Where we suppose that all the necessary information about the physics of phase transitions appear at this order as it did for mean field theory in Section~\eqref{sec1.4}.
\end{enumerate}
Using~\eqref{landau1}, we have at equilibrium 
\begin{ceqn}
	\begin{equation}
		\frac{\partial L}{\partial \Psi} = a_1 +2a_2 \Psi+3a_3\Psi^2+4a_4\Psi^3=0.
		\label{dlandau1}
	\end{equation}
\end{ceqn}
For $T>T_c$, the order parameter $\Psi$ is zero. Then, $a_1=0$. In fact, the system is invariant under change of $\Psi$ by $-\Psi$, that is $L$ is an even function : $L(\Psi)=L(-\Psi)$. Thus
\begin{ceqn}
	\begin{equation}
		L=a_0([K],T)+a_2([K],T)\Psi^2+a_4([K],T)\Psi^4.
		\label{landau0}
	\end{equation}
\end{ceqn}
Each term in Landau's free energy serve a role. $a_0([K],T)$ represents the value $L$ in the disordered phase ($\Psi=0$ for $T>T_c$), it describes the degrees of freedom of the system that cannot be understood via the order parameter. For $a_2$ and $a_4$ expanding in temperature near $T_c$, we obtain
\begin{ceqn}
	\begin{align}
		a_4&=a^0_4+\left(T-T_c\right) a^1_4+\dots, \\
		a_2&=a^0_2+\frac{\left(T-T_c\right)}{T_c}a^1_2+O\left(\left(T-T_c\right)^2\right).
	\end{align}
\end{ceqn}
Since $\partial^2L/\partial \Psi^2 =1/\chi=0$ as $T \to T_c$, one has $a^0_2=0$ and
\begin{equation}
    a_2=\frac{\left(T-T_c\right)}{T_c}a^1_2+O\left(\left(T-T_c\right)^2\right).
\end{equation}
The extension to the case $h \neq 0$ for the Ising ferromagnet is immediate. The order parameter in the Ising model is the magnetization $M$, that is $\Psi=M$. The additional energy in the Hamiltonian due to the magnetic field is $-h\displaystyle{\sum_i S_i}=-hNM$. Thus, Landau free energy can be written as
\begin{ceqn}
	\begin{equation}
		L = a\left(\frac{T-T_c}{T_c}\right)\Psi^2+\frac{1}{2}b\Psi^4-h\Psi,
		\label{landau2}
	\end{equation}
\end{ceqn} 
where the coefficient $a$ and $b$ can be obtained from a microscopic analysis.
\subsubsection{Continuous Phase Transitions}
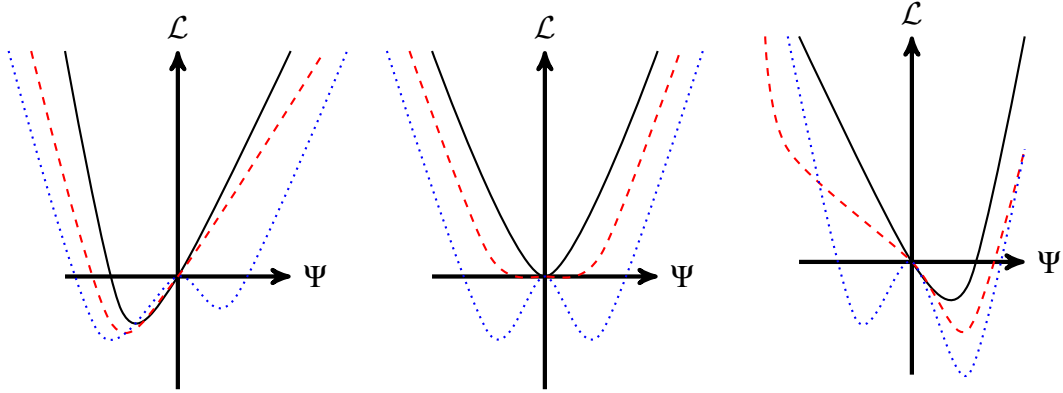
\begin{figure}[t!]
	\centering
	\begin{subfigure}[t]{0.33\textwidth}
    	\begin{tikzpicture}[scale=1.5]
        \draw[->,ultra thick] (-1,0)--(1,0) node[right]{$\Psi$};
        \draw[->,ultra thick] (0,-1)--(0,2) node[above]{$\mathcal{L}$};
        \draw [thick] plot [smooth] coordinates {(-1,2) (-0.5,-0.3) (0,0) (1,2)};
        \draw [dashed,red,thick] plot [smooth] coordinates {(-1.3,2) (-0.6,-0.4) (0,0) (1.3,2)};
        \draw [dotted,blue,thick] plot [smooth] coordinates {(-1.5,2) (-0.7,-0.5) (0,0) (0.5,-0.2) (1.5,2)};
    	\end{tikzpicture}
    \end{subfigure}%
    \begin{subfigure}[t]{0.33\textwidth}
    	\begin{tikzpicture}[scale=1.5]
        \draw[->,ultra thick] (-1,0)--(1,0) node[right]{$\Psi$};
        \draw[->,ultra thick] (0,-1)--(0,2) node[above]{$\mathcal{L}$};
        \draw [thick] plot [smooth] coordinates {(-1,2) (0,0) (1,2)};
        \draw [dashed,red,thick] plot [smooth] coordinates {(-1.2,2) (-0.5,0.2) (0,0) (0.5,0.2) (1.2,2)};
        \draw [dotted,blue,thick] plot [smooth] coordinates {(-1.4,2) (-0.5,-0.5) (0,0) (0.5,-0.5) (1.4,2)};
    	\end{tikzpicture}
    \end{subfigure}%
    \begin{subfigure}[t]{0.33\textwidth}
    	\begin{tikzpicture}[scale=1.5]
        \draw[->,ultra thick] (-1,0)--(1,0) node[right]{$\Psi$};
        \draw[->,ultra thick] (0,-1)--(0,2) node[above]{$\mathcal{L}$};
        \draw [thick] plot [smooth] coordinates {(-1,2) (0,0) (0.5,-0.2) (1,2)};
        \draw [dashed,red,thick] plot [smooth] coordinates {(-1.3,2) (-1.1,1) (0,0) (0.5,-0.6) (1,1)};
        \draw [dotted,blue,thick] plot [smooth] coordinates {(-1.1,2) (-0.5,-0.5) (0.,0.) (0.5,-1) (1,1)};
    	\end{tikzpicture}
    \end{subfigure}
	\caption{Landau free energy~\eqref{landau2}, for different values of $T$ and $H$. From left to right : $H<0$, $H=0$ and $H>0$.}
	\label{plot_l}
\end{figure}

We present in Figure~\eqref{plot_l} the variation of Landau free energy~\eqref{landau2}, with respect to the order parameter $\Psi$. The case $h=0$ is when the continuous (second-order) phase transition occur. For $T>T_c$, $L$ has a minimum at $\Psi=0$. When $T=T_c$ Landau potential has zero curvature at $\Psi=0$ while $\Psi=0$ is still the global minimum. For $T<T_c$, Landau free energy shows two degenerate minima at $\Psi=\pm n_s(T)$.\\
Solving~\eqref{dlandau1} for $\Psi$ we can read off the critical exponent $\beta$. We have
\begin{ceqn}
	\begin{align}
		\Psi=0 && \text{or} && \Psi =\sqrt{-\frac{at}{b}},
	\end{align}
\end{ceqn}
where $t=\frac{T-T_c}{T_c}$. Then, for $T<T_c$  $\beta=1/2$.\\
Landau potential is zero for $t>0$ and for $t<0$ we can write
\begin{equation}
    L=-\frac{1}{2} \frac{a^2t^2}{b}.
\end{equation}
The critical exponent $\alpha$ of the heat capacity can be extracted by writing: $C_V=-T \partial^2 L/\partial T^2$. Hence
\begin{equation}
    C_v = 
	\begin{cases}
	0 & T>T_c,\\
	a^2/bT_c & T<T_c.
	\end{cases}
\end{equation}
which shows that the heat capacity exhibits a discontinuity and that $\alpha=0$.\\
Switching on the magnetic field ``$h$'' allows to calculate the other critical exponents. Taking the derivative with respect to $\Psi$ in~\eqref{landau2} gives:
\begin{ceqn}
	\begin{equation}
	at\Psi +b\Psi^3=\frac{1}{2}h.
	\label{landau3}
	\end{equation}
\end{ceqn}
On the critical isotherm: $t=0$. We have $h\propto\Psi^3$ and we read the critical exponent $\delta=3$. The isothermal susceptibility $\chi_T$ can be computed by taking the derivative of~\eqref{landau3} with respect to $h$. That is
\begin{ceqn}
	\begin{equation}
		\chi_T =\frac{\partial \Psi(h)}{\partial h}
		=\frac{1}{2\left(at+3b\Psi(h)^2\right)},
	\end{equation}
\end{ceqn}
where $\Psi(H)$ is a solution of~\eqref{landau3}. For $t>0$, we have $\Psi=0$, then $\chi_T \propto t^-1$ while for $t<0$, we have $\Psi=\left(-at/b\right)^{1/2}$ and $\chi_T \propto t^{-1}$. Thus, the critical exponent is $\gamma=1$.
\subsubsection{First Order Phase Transitions}
\begin{figure}[t!]
    \centering
    \begin{subfigure}[t]{0.49\textwidth}
    	\begin{tikzpicture}[scale=1.5]
        \draw[->,ultra thick] (-1,0)--(2,0) node[right]{$\Psi$};
        \draw[->,ultra thick] (0,0)--(0,3) node[above]{$\mathcal{L}$};
        \draw [thick] plot [smooth] coordinates {(-1,2) (0,0) (1.5,2)};
        \node[above] (A) at (1.5,2) {$t>t^*$};
    	\end{tikzpicture}
    \end{subfigure}%
    \begin{subfigure}[t]{0.49\textwidth}
        \begin{tikzpicture}[scale=1.5]
            \draw[->,ultra thick] (-1,0)--(2,0) node[right]{$\Psi$};
            \draw[->,ultra thick] (0,0)--(0,3) node[above]{$\mathcal{L}$};
            \draw [thick] plot [smooth] coordinates {(-1.5,2) (0,0) (0.75,1) (1.5,0.5) (2,2)};
            \node[above] (A) at (2,2) {$t<t^*$};
        \end{tikzpicture}
    \end{subfigure}
    \begin{subfigure}[t]{0.49\textwidth}
    	\begin{tikzpicture}[scale=1.5]
        \draw[->,ultra thick] (-1,0)--(2,0) node[right]{$\Psi$};
        \draw[->,ultra thick] (0,0)--(0,3) node[above]{$\mathcal{L}$};
        \draw [thick] plot [smooth] coordinates {(-1,2) (0,0) (0.75,1) (1.5,0.) (2,2)};
        \node[right] (A) at (2,2) {$t=t_1$};
    	\end{tikzpicture}
    \end{subfigure}
    \begin{subfigure}[t]{0.49\textwidth}
        \begin{tikzpicture}[scale=1.5]
            \draw[->,ultra thick] (-1,1)--(2,1) node[right]{$\Psi$};
            \draw[->,ultra thick] (0,0)--(0,3) node[above]{$\mathcal{L}$};
            \draw [thick] plot [smooth] coordinates {(-0.5,2.5) (0,1) (0.5,1.5) (1,0) (2,2)};
            \node[right] (A) at (2,2) {$t<t_1$};
        \end{tikzpicture}
    \end{subfigure}
    \caption{Landau free energy ~\eqref{1PT_landau}, with respect to the order parameter $\Psi$.}
    \label{1PT_landau}
\end{figure}
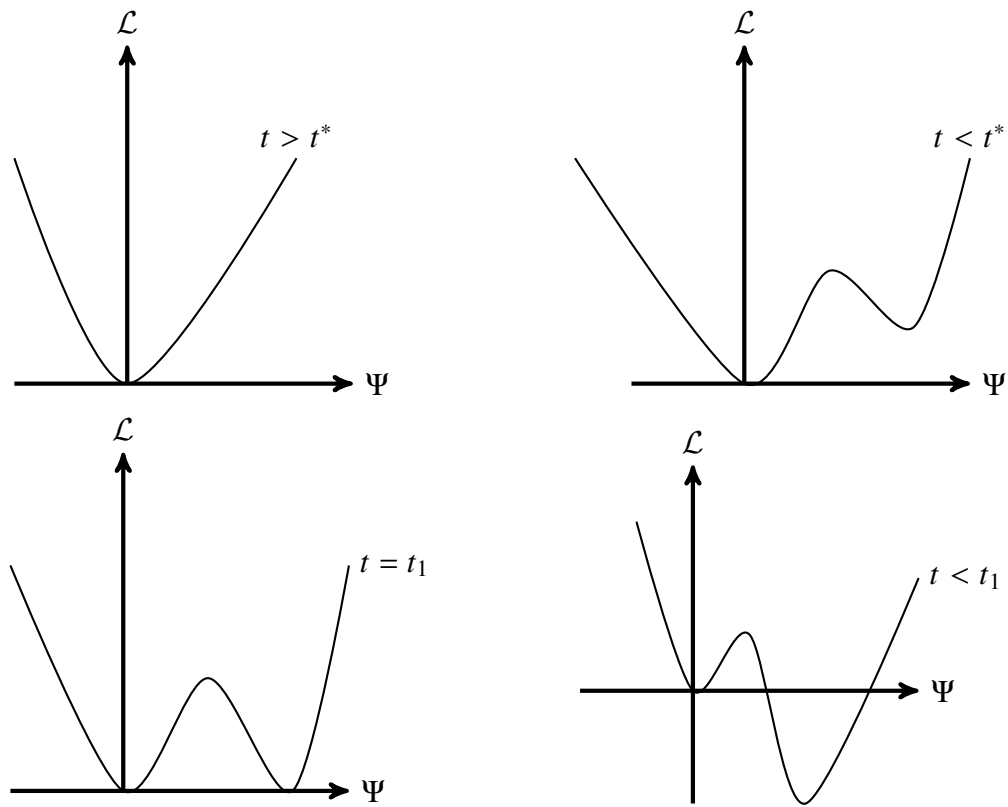

The symmetry requirement prevents the Landau free energy in~\eqref{landau0} of having a cubic term in $\Psi$. In general, what is the effect of this term in Landau potential? We first write $L$ as
\begin{ceqn}
	\begin{equation}
	L=at\Psi^2+\frac{1}{2}b\Psi^4+C\Psi^3-h\Psi.
	\label{1pt}
	\end{equation}
\end{ceqn}
Where $a$ and $ b $ are positive. A derivative of $L$ with respect to $\Psi$ at equilibrium and at zero magnetic field ($ h=0 $) gives:
\begin{ceqn}
	\begin{align}
		\Psi= 
			\begin{cases}
			0,\\
			-c \pm \sqrt{c^2-at/b},&\text{with  } c= 3C/4b.
			\end{cases}
	\end{align}
\end{ceqn}
The solution $\Psi \neq 0$ is real when the argument of the square root is positive, i.e. $ \sqrt{c^2-at/b} > 0$. That is, $ t<t^*=bc^2/a $. When $t<t^*$, in addition to the minimum $\Psi=0$, $L$ develops a secondary minimum and a maximum. Reducing $t$ further, below a value $t_1$, the value of the order parameter which minimizes $L$ \textbf{jumps discontinuously} from $\Psi=0$ to  a non-zero value and the secondary minimum becomes the global minimum of $L$. This is a \textit{first order phase transition}. In Figure~\eqref{1PT_landau} we sketch Landau's free energy~\eqref{1pt}, with respect to the order parameter $\Psi$. A sufficient but not necessary condition of the occurrence of continuous phase transitions is that there are no cubic terms in the potential. In general, the cubic term causes a first order phase transition.

\section{Quantum phase transitions}
\subsection{What is a quantum phase transition?}
The phase transitions studied so far occur at finite temperature, where long-range order is destroyed by thermal fluctuations. Many-body systems can exhibit a different kind of critical phenomena at zero temperature, called quantum phase transitions, where quantum fluctuations are dictating the properties of the system~\cite{sachdev2011quantum, vojta2003quantum}. Quantum phase transitions are driven by non-thermal parameters,  such as pressure, magnetic field or the chemical composition, which creates quantum fluctuations, due to the Heisenberg principle, that destroy the long range order across the quantum critical point. 

\noindent From an experimental point of view, the study of quantum phase transitions appears to be a theoretical marginal problem and reserved only to a class of specialist theorists, due to their occurrence at an impossible to reach temperature in experiments. However, recent theoretical and experimental advances on the field have proved that the presence of quantum critical phenomena at zero temperature, provide a framework to solve problems in condensed matter physics, such as rare-earth magnetic insulators~\cite{bitko1996}, heavy-fermion compounds~\cite{coleman1999, L_hneysen_1996}, high-temperature superconductors~\cite{dagotto1994, orenstein2000} and two-dimensional electron gases~\cite{sachdev2000,krav1995}.

\noindent Consider a quantum system describing a lattice of spins via a Hamiltonian $H(g)$, where $g$ is a dimensionless coupling parameter used to drive the system. The nature of the quantum phase transition can be inspected by tracking the low-lying energies of the system as a function of $g$. A quantum phase transition is essentially a non-analyticity in the ground state, which can manifest into two forms:
\begin{itemize}
    \item \textbf{Level crossing:}\\
    A singularity can manifest in the low-lying energy levels of the quantum system at $g=g_c$, by an excited state becoming the ground state of the system, as depicted in Figure~\eqref{lvl_cross}. Energy level crossing is a feature indicating a ``first-order quantum phase transition", which can occur in both finite and infinite systems
    \item \textbf{Avoided level crossing:}\\
    The non-analyticity can develop by an avoided level crossing of the energy levels between the ground state and the first excited state as shown in Figure~\eqref{avoid_lvl_cross}. Continuous (second-order) quantum phase transitions are characterized by an avoided level crossing in the ground state energy. In the thermodynamical limit, the avoided level crossing tends to be infinitely sharp.
\end{itemize}
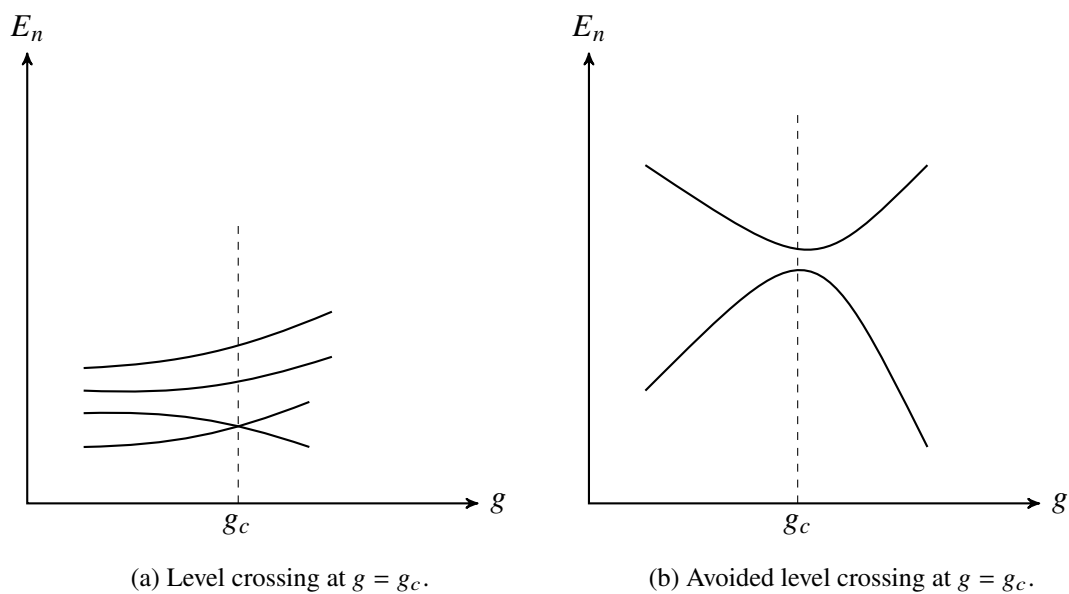
\begin{figure}[t!]
	\centering
	\begin{subfigure}[h]{0.49\textwidth}
         \begin{tikzpicture}[scale=1.5]
	    \draw [<->,thick] (0,4) node (yaxis) [above] {$E_n$}|- (4,0) node (xaxis) [right] {$g$};
	    \coordinate (O) at (0.5,0.5);
	    \coordinate (A) at (2.5,0.9);
	    \coordinate (B) at (0.5,0.8);
	    \coordinate (C) at (2.5,0.5);
	    \coordinate (D) at (0.5,1.0);
    	\coordinate (E) at (2.7,1.3);
    	\coordinate (F) at (0.5,1.2);
    	\coordinate (G) at (2.7,1.7);
    	\draw[thick,color=black] (O) to [bend right=10] (A) node[anchor=south east]{};
    	\draw[thick,color=black] (B) to [bend left=10] (C) node[anchor=south east]{};
    	\draw[thick,color=black] (D) to [bend right=10] (E) node[anchor=south east]{};
    	\draw[thick,color=black] (F) to [bend right=10] (G) node[anchor=south east]{};
    	\draw[dashed] (1.87,0.0)--(1.87,2.5);
    	\node[below] (A) at (1.86,0.0) {$g_c$};
    	\end{tikzpicture}
    	\caption{Level crossing at $g=g_c$.}
    	\label{lvl_cross}
    	 \end{subfigure}%
     \begin{subfigure}[h]{0.49\textwidth}
         \begin{tikzpicture}[scale=1.5]
         \draw [<->,thick] (0,4) node (yaxis) [above] {$E_n$}|- (4,0) node (xaxis) [right] {$g$};
         \draw[dashed] (1.85,0.0)--(1.85,3.5);
        \draw[thick] (0.5,1) .. controls (2,2.5) .. (3,0.5);
        \draw[thick] (0.5,3) .. controls (2.,2.) .. (3,3);
	    \node[below] (A) at (1.85,0.0) {$g_c$};
        \end{tikzpicture}
        \caption{Avoided level crossing at $g=g_c$.}
        \label{avoid_lvl_cross}
     \end{subfigure}
	\caption{Energy level crossing at (a) first-order quantum phase transition, and (b) second-order quantum phase transition, with respect to the dimensionless coupling $g$.}
	\label{energy_crossing}
\end{figure}
The focus of this part will be on second-order quantum phase transitions, which are characterized by a vanishing energy gap $\Delta$, between the ground state and the first excited state as $g$ tends to $g_c$. In general, the energy gap $\Delta$ vanishes as
\begin{equation}
\Delta \sim |g-g_c|^{zv}, 
\label{quantum_delta}
\end{equation}
where $zv$ is a critical exponent, which is usually independent of the microscopic details of the Hamiltonian, in other words, $zv$ is universal.

\noindent Continuous quantum phase transitions are also characterized by a diverging correlation length $\xi$, which describes the length scale of the decay of equal-time correlations in the ground state or the length scale where a crossover between correlations at long distances occur. The equal-time correlation function $G(r)$ of the order parameter $O(r,t)$ between two points distant by $r$, is given by
\begin{equation}
    G(r)=\langle O(0,t)O(r,t) \rangle - \langle O(0,t)\rangle \langle O(r,t) \rangle \propto \frac{e^{-r/\xi}}{r^{d-2+\eta}},
\end{equation}
where $d$ is the dimension of the system and $\eta$ is the Fisher exponent. Approaching the quantum critical point, the correlation length $\xi$ diverges as
\begin{equation}
    \xi^{-1} \sim |g-g_c|^v.
    \label{quantum_xi}
\end{equation}
Unlike classical phase transitions, space and time are connected in a quantum phase transition. This can be seen from~\eqref{quantum_delta} and~\eqref{quantum_xi}, where we can define a time scale $\xi_{\tau}$ for the decay of equal-space correlations, which diverges as
\begin{equation}
    \xi_{\tau} \sim \Delta^{-1} \propto \xi^{-z}.
\end{equation}
At the critical point, fluctuations occur on all length and time scales, due to the divergence of the correlation length and time. In Table~\eqref{quantum_cri_exp_tab}, we list the critical exponents for quantum magnetic systems. Similarly to the classical case, not all the quantum critical exponents are independent. From the seven critical exponents in Table~\eqref{quantum_cri_exp_tab}, we can obtain four scaling relations and only three exponents are independent.
\begin{table}[t!]
	\centering
	\begin{tabular}{ccc }
		\hline
		Quantity  & Exponent & Definition  \\ \hline
		Correlation length $\xi$ & $v$ & $\xi \propto |g-g_c|^{-v}$\\
		Order parameter $O$ &$\beta$ & $O \propto (g_c-h)^{\beta}$\\
		Specific heat $C_v$ &$\alpha$ & $C_v \propto |g-g_c|^{-\alpha}$ \\
		Susceptibility $\chi$ &$\gamma$ & $\chi \propto |g-g_c|^{-\gamma}$ \\
		Critical isotherm &$\delta$ & $g_c \propto |O|^{\delta}$ \\
		Correlation function $G(r)$ &$\eta$ &  $G(r) \propto |r|^{-d+2-\eta}$ \\
		Correlation time $\xi_{\tau}$ &$z$ & $\xi_{\tau}\propto \xi^z$ \\
		\hline
	\end{tabular}
\caption{Critical exponents for the quantum Ising universality class.}
\label{quantum_cri_exp_tab}
\end{table}
\subsection{Quantum-classical crossovers}
The singularities discussed above occur strictly at $T=0$. Therefore, theoretically speaking, quantum phase transitions occur only at zero temperatures. Experiments are always performed at very small, but nonzero temperature. One of the main goals of the theory of quantum phase transitions is to understand the ramifications of the presence of the singularity at $T=0$ on the physical properties of the system at $T>0$.

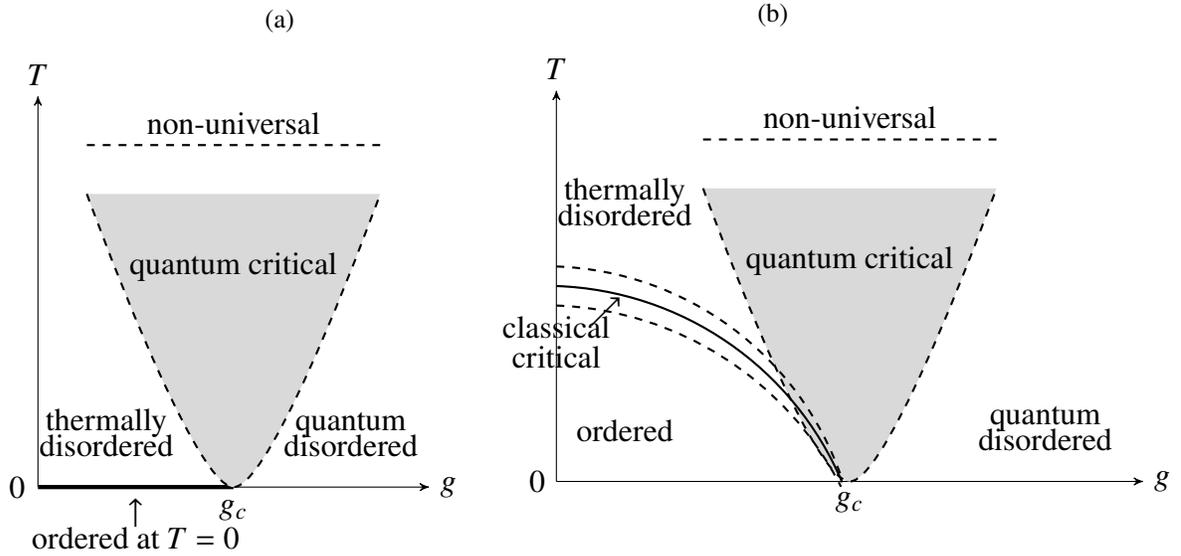
\begin{figure}[t!]
    \begin{subfigure}[t!]{0.49\textwidth}
    \caption{}
    	\begin{tikzpicture}[scale=1.3]
        \draw[-,ultra thick] node[left]{$0$} (0,0)--(2,0) node[below]{$g_c$};
        \draw[->] (2,0)--(4,0) node[right]{$g$};
        \draw[->] (0,0)--(0,4) node[above]{$T$};
        \draw [dashed,thick,fill=gray!30] plot [smooth] coordinates {(0.5,3) (2,0) (3.5,3)};
        \draw[dashed, thick]  (0.5,3.5)--(3.5,3.5);
        \node[above] (A) at (2,3.5) {non-universal};
        \node[above] (A) at (2,2) {quantum critical};
        \node[above] (A) at (3.2,0.4) {quantum};
        \node[above] (A) at (3.2,0.2) { disordered};
        \node[above] (A) at (0.7,0.4) {thermally};
        \node[above] (A) at (0.7,0.2) {disordered};
        \node[below] (A) at (1,-0.3) {ordered at $T=0$};
        \node[below] (A) at (1,0) {\textbf{$\uparrow$}};
    	\end{tikzpicture}
    \label{q_c_mapping_0}
    \end{subfigure}
    \hspace{-2.\baselineskip}
    \begin{subfigure}[t!]{0.49\textwidth}
    \caption{}
        \begin{tikzpicture}[scale=1.3]
        \draw[-] node[left]{$0$} (0,0)--(3,0) node[below]{$g_c$};
        \draw[->] (3,0)--(6,0) node[right]{$g$};
        \draw[->] (0,0)--(0,4) node[above]{$T$};
        \draw [dashed,thick,fill=gray!30] plot [smooth] coordinates {(1.5,3) (3,0) (4.5,3)};
        \draw[thick] (0,2) arc (88:23:3.3);
        \draw[dashed,thick] (0,2.2) arc (88:19:3.2);
        \draw[dashed,thick] (0,1.8) arc (88:27:3.4);
        \draw[dashed, thick]  (1.5,3.5)--(4.5,3.5);
        \node[above] (A) at (3,3.5) {non-universal};
        \node[above] (A) at (3,2) {quantum critical};
        \node[above] (A) at (5,0.4) {quantum};
        \node[above] (A) at (5,0.2) { disordered};
        \node[above] (A) at (0.7,2.7) {thermally};
        \node[above] (A) at (0.7,2.5) {disordered};
        \node[above] (A) at (0.7,0.3) {ordered};
        \node[below] (A) at (0,1.8) {classical};
        \node[below] (A) at (0,1.5) {critical};
        \node[below] (A) at (0.5,2) {\textbf{$\nearrow$}};
    	\end{tikzpicture}
    \label{q_c_mapping_nonzero}
    \end{subfigure}
    \caption{Schematic phase diagram in the vicinity of a quantum critical point $g_c$. In (a) order exists only at $T=0$ and in (b) the order can exist even at finite temperature.}   
    \label{q_c_mapping}
\end{figure}

\noindent At nonzero temperature, we have two energy scales: the thermal energy $k_bT$, and the energy associated to the characteristic frequency belonging to the diverging relaxation time $\hbar\omega_c$, quantum theory will be important as long as $\hbar\omega_c \gg k_bT $. On the other hand, when $\hbar\omega_c \ll k_bT $ the fluctuations follow the laws of classical mechanics. The crossover between quantum and classical theory in the vicinity of the quantum critical point leads to a rich phase diagram as depicted in Figure~\eqref{q_c_mapping}. We distinguish two cases depending on whether long range exists at zero or nonzero temperature.

\noindent The situation where order exists only at $T=0$ is shown in Figure~\eqref{q_c_mapping_0}, a true phase transition cannot be seen experimentally at $T>0$. The nonzero temperature behavior is characterized by three regimes, separated by crossovers. In the thermally disordered regime, the long-range order is destroyed by thermal fluctuations, while in the quantum disordered region the system is dominated by quantum fluctuations. In both regimes, the energy gap $\Delta$ satisfies: $\Delta>k_bT$, which implies that the system reaches local thermal equilibrium after a time $\tau_{\text{eq}}$, satisfying $\tau_{\text{eq}}\gg \frac{\hbar}{k_bT}$. Thus, the dynamics can be effectively described with classical equation of motions. Between the thermally and quantum disordered regimes, is the quantum critical region, the energy gap $\Delta$ satisfies: $\Delta<k_bT$. Implying that both quantum and classical fluctuations are important. In this region, the system looks critical with respect to $g$, but thermal fluctuations drives it away from criticality. Therefore, the physics in the quantum critical region is dictated by thermal excitations of the quantum critical ground state, whose main feature is the absence of quasiparticle excitations.

\noindent The phase diagram can be richer when order can exist at finite temperature, as depicted in Figure~\eqref{q_c_mapping_nonzero}. In this case, a real phase transition is encountered at low temperature while varying $g$. There is an additional line of finite temperature transitions surrounded by a region, referred to as ``classically critical" where classical fluctuations are dominant. The quantum critical point can be seen as the endpoint of a line of finite temperature phase transitions.
\subsection{Quantum-classical mapping}
Deeper insight can be gained on the interplay between the quantum and classical behavior in the vicinity of the quantum critical point, by calling for quantum statistical mechanics. We consider a system described by a Hamiltonian of the form $H=H_{\text{kin}}+H_{\text{pot}}$, where $H_{\text{kin}}$ denotes the kinetic energy term and $H_{\text{pot}}$ represents the potential energy part. In classical mechanics, the kinetic and potential energy commute. Therefore, the partition function factorizes as $Z=e^{-H/{k_bT}}=Z_{\text{kin}}Z_{\text{pot}}$. Thus, in a classical system the statics and dynamics decouple. The kinetic part of the free energy is analytic, since it can be written as a product of simple Gaussian integrals. Therefore, classical critical phenomena can be studied using time-independent theories that live in $d$ dimensions. 

\noindent In contrast, in quantum mechanics the kinetic and potential part of the Hamiltonian $H$, do not commute in general. Therefore, the quantum partition function cannot be written in a simple factorized form. Thus, the statics and dynamics are always coupled. The canonical density operator $e^{-H/{k_bT}}$, can be seen as a time evolution operator with an imaginary time $\tau$, by identifying $1/k_bT=\tau=-i\theta/\hbar$ where $\theta$ is the real time. At zero temperature, the imaginary time acts as an additional space dimension, due to the system being infinite in this direction.  Close to the critical point, the correlation length is the only relevant length scale. Therefore, any re-scale of the system by a factor, does not affect the physical properties while adjusting the external parameters in a way that keeps the correlation length unchanged. This allow us to introduce a scaling law for the singular part in the free energy,
    \begin{align}
    f(t,B)&=b^{-d}f(tb^{1/v},Bb^{yb}), &\text{classical phase transition,}\\
    f(t,B)&=b^{-(d+z)}f(tb^{1/v},Bb^{yb}), &\text{quantum phase transition},
    \label{qc_mapping_}
    \end{align}
where $t=|g-g_c|/g_c$, $b$ is an arbitrary positive number and $(v,y_b)$ are critical exponents. The relation~\eqref{qc_mapping_} implies that a quantum phase transition in $d$ dimension is related to a classical phase transition in $(d+z)$ space dimension. For many phase transitions $z=1$, such as: clean insulators and Ising transition in two-dimensions. The classical phase transition in the two-dimensional model seen in Section~\eqref{2d_diagram_ising} is related to its counterpart in the one-dimensional quantum Ising model.

\noindent At finite temperature, the crossover from quantum to classical corresponds to a dimensional jump. The temperature under the quantum-classical mapping is equivalent to an inverse length of the imaginary time axis. Approaching the transition, c.f. Figure~\eqref{q_c_mapping_nonzero}, the crossover is controlled by $\beta=1/k_bT$ and the correlation time $\tau_c$. When $\tau_c>\beta$, the crossover from quantum to classical occur, as the system is effectively only $d-$dimensional, instead of $(d+z)-$dimensional. The scaling law~\eqref{qc_mapping_}, can be generalized to finite temperatures, and we can write
\begin{equation}
    f(t,B,T)=b^{-(d+z)}f(tb^{1/v},Bb^{yb},Tb^z).
\end{equation}
What is so special about quantum phase transitions, when they can be mapped to their classical counter parts?
Quantum systems have special features and properties that outshine their classical counterparts, such as quantum entanglement, topological Berry phase and long-ranged effective interactions. Calculating the dynamics of quantum systems is never easy and requires sophisticated theories and careful considerations~\cite{sachdev2011quantum}. Therefore the approximation done on the imaginary axis, due to the quantum-classical mapping, is in general not valid for real times. Furthermore, near the critical point the dynamics of quantum systems are characterized by the phase coherence time $\tau_{\phi}$, which does not have a classical counterpart. When $T \to 0$, the phase coherence time diverges for all parameter values. This means a quantum system has perfect phase coherence in both the ordered and disordered phase. This peculiarity is related to the nature of the analytical continuation between imaginary and real times. For many models, it is found that $\tau_{\phi} \propto 1/T$ as $T \to 0$, and diverges faster in stable phases.
\subsection{Quantum Ising model \label{1.2.3}}
The Hamiltonian of the quantum Ising model is given by
\begin{equation}
    H=-J\left(\sum_{i} \sigma_i^z\sigma_{i+1}^z-g\sigma_i^x\right),
    \label{eq_quantum_ising}
\end{equation}
where $J$ is the exchange energy between the spins and $g$ is a dimensionless coupling used to drive the system across the quantum phase transition. At each site $i$, the quantum degrees of freedom are represented by the Pauli matrices $\sigma_i^{\alpha}$, with $\alpha=x,y, z$, and they are given by
\begin{equation}
    \sigma^x=\begin{pmatrix}
     1&0\\
     0&-1
    \end{pmatrix},%
    \sigma^y=\begin{pmatrix}
     0&-i\\
     i&0
    \end{pmatrix},%
    \sigma^z=\begin{pmatrix}
     1&0\\
     0&1
    \end{pmatrix}.
    \label{eq_pauli_matrix}
\end{equation}
The Pauli matrices~\eqref{eq_pauli_matrix}, satisfy the following commutation relations:
\begin{itemize}
    \item At different sites $i$ and $j$: $[\sigma_i^{\alpha},\sigma_j^{\alpha}]=0$, for $i \neq j$.
    \item On the same site $i$: $[\sigma_i^{x},\sigma_i^{y}]=2i\sigma_i^{z}$.
\end{itemize}
The quantum Ising chain~\eqref{eq_quantum_ising}, can take two possible orientations: up $\ket{\uparrow}_i$ and down $\ket{\downarrow}_i$, corresponding to the two eigenvalues of $\sigma^z$: $\pm 1$. Therefore, when $g=0$ the quantum Ising model~\eqref{eq_quantum_ising}, reduces to a diagonal form in the basis of $\sigma^z$ which is equivalent to the classical Ising model~\eqref{ising1}, studied in Section~\eqref{1.1.3}. Switching on the coupling $``g$'' creates quantum fluctuations which are induced by the $\sigma^x$ term being non-diagonal in the $\sigma^z$ basis. These quantum fluctuations create quantum-tunneling in the Hamiltonian~\eqref{eq_quantum_ising}, which is capable of flipping the Ising spin on a site. As a consequence, competition arise between the term proportional to $``J$''  which favors the alignment of spins, either in the up or down orientation, which creates long-range magnetic order, and the term proportional to $``Jg$'' which disrupt and breaks the long-range magnetic order.
The ground state of the quantum Ising model~\eqref{eq_quantum_ising}, is controlled solely by the coupling $``g$''. The phase diagram can be constructed by studying two limiting cases: $g\gg1$ and $g\ll1$. 
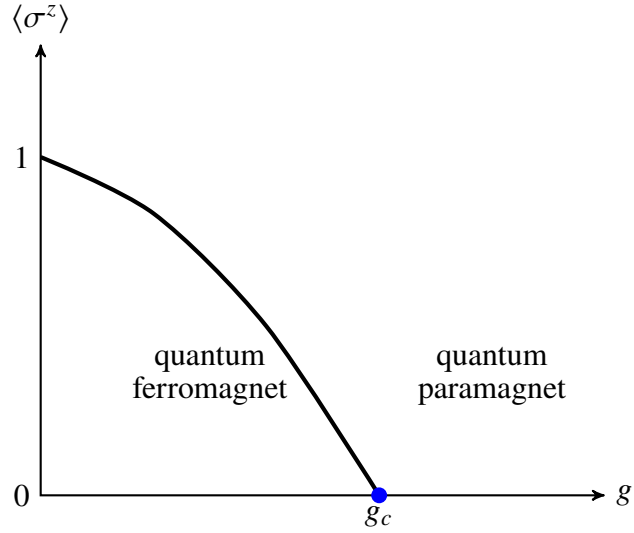
\begin{figure}[t!]
    \centering
    	\begin{tikzpicture}[scale=1.5]
    	\draw [<->, thick] (0,4) node (yaxis) [above] {$\langle \sigma^z \rangle$}
    	|- (5,0) node (xaxis) [right] {$g$};
        \draw[ultra thick] plot [smooth] coordinates {(0,3) (1,2.5) (2,1.5) (3,0)};
        \node[left] (A) at (0,0) {$0$};
        \node[left] (A) at (0,3) {$1$};
        \node[below] (A) at (3,0) {$g_c$};
        \fill[blue] (3,0) circle (2pt);
        \node[above] (A) at (1.5,1) {quantum};
        \node[above] (A) at (1.5,0.7) {ferromagnet};
        \node[above] (A) at (4,1) {quantum};
        \node[above] (A) at (4,0.7) {paramagnet};
    	\end{tikzpicture}
    \caption{Spontaneous magnetization in the ground state~\eqref{gs_quantum_ising}, of the quantum Ising model~\eqref{eq_quantum_ising}, with respect to the coupling $g$.}
    \label{quantum_mag}
\end{figure}

\noindent At $g\gg1$, the $\sigma^x$ term in the Hamiltonian~\eqref{eq_quantum_ising}, dominates and the ground state $\ket{0}$ can be written, at a leading order of $1/g$, as
\begin{equation}
    \ket{0}=\prod_i \ket{\rightarrow}_i,
    \label{gs_quantum_ising}
\end{equation}
where
\begin{align}
    \ket{\rightarrow}_i=(\ket{\uparrow}_i + \ket{\downarrow}_i)/\sqrt{2},\\
    \ket{\leftarrow}_i=(\ket{\uparrow}_i - \ket{\downarrow}_i)/\sqrt{2},
\end{align}
are the eigenstates of $\sigma^x$ with eigenvalues $\pm1$. The spin-spin correlation function between two different sites $i$ and $j$, in the $z$-direction, is totally uncorrelated in the ground state~\eqref{gs_quantum_ising}, i.e. $\bra{0}\sigma_i^z\sigma_j^z \ket{0}=\delta_{ij}$. Using perturbation theory we can add corrections, at the leading order of $1/g$, in the form of correlations in $\sigma^z$. At each order in $1/g$, the build up correlations increase but remain short-ranged, so that for a large distance between two spins $i$ and $j$, the correlations decay as
\begin{equation}
    \bra{0}\sigma_i^z\sigma_j^z \ket{0}\sim e^{-|x_i-x_j|/\xi},
    \label{eq1.2.5}
\end{equation}
where $|x_i-x_j|$ is the distance between the spins $i$ and $j$ and $\xi$ is the correlation length discussed in Section~\eqref{1.1.4.1}.

\noindent In the opposing limit, i.e. $g\ll1$, the first term in the Hamiltonian ~\eqref{eq_quantum_ising}, takes the lead. At $g=0$, the ground state is described via the eigenstates of $\sigma^z$, as it can take two orientations: up $\ket{\uparrow}_i$ or down $\ket{\downarrow}_i$. Consider the up configuration, turning on the coupling $``g"$ flips a fraction of spins. However, in the thermodynamic limit the degeneracy of the ground state survives due to the global spin-flip ($\mathcal{Z}_2$) symmetry that the Hamiltonian~\eqref{eq_quantum_ising}, obeys which maps the two ground states into each other, so that the Hamiltonian~\eqref{eq_quantum_ising}, remain invariant. The system will always chose one of the states as its ground state, this phenomenon is commonly referred to as ``spontaneous symmetry breaking''.

\noindent In the thermodynamic limit and at small-$g$ perturbation theory, the spin spin correlation function between two different sites $i$ and $j$, in the $z$-direction, can be written as
\begin{equation}
    \lim_{|x_i-x_j| \to \infty} \bra{0}\sigma_i^z\sigma_j^z \ket{0} =N_0^2,
    \label{eq1.2.6}
\end{equation}
where $\bra{0}\sigma_i^z\ket{0}=\pm N_0$ is the spontaneous magnetization. At $g=0$, we have $N_0=1$ meaning that all spins are in the up orientations. For small-$g$, quantum fluctuations reduces $N_0$ to a smaller, but non-zero value. 

\noindent The transition of the two-point correlation function between~\eqref{eq1.2.5} and~\eqref{eq1.2.6} does not occur analytically. In the thermodynamic limit and at a critical value of the coupling $g=g_c$, a quantum phase transition occurs between a quantum paramagnet ($g>g_c$) where equation~\eqref{eq1.2.5} holds, and a quantum ferromagnet ($g<g_c$) where equation~\eqref{eq1.2.6} is obeyed. Figure~\eqref{quantum_mag} illustrates the behaviour of the spontaneous magnetization of the ground state, upon crossing the quantum phase transition, with respect to $g$. We see $\langle \sigma^z \rangle$ is non zero in the quantum ferromagnetic phase and it reduces as we approach the quantum critical point at $g=g_c$, where it vanishes in the quantum paramagnet phase.
\begin{table}[t!]
	\centering
	\begin{tabular}{cc }
		\hline
		System  & Examples  \\ \hline
		Order-disorder ferroelectrics & $KH_2PO_4$, $KD_2PO_4$ \\
		Jahn-Teller systems & $DyVO_4,TbVO_4$\\
		proton glasses & $Rd_{1-x}(NH_4)_x H_2PO_4$\\
		quantum spin glasses & $LiHo_x Y_{1-x}F_4$\\
		quasi-one-dimensional Ising systems & $CoNb_2O_6$\\
		\hline
	\end{tabular}
\caption{Systems described by transverse Ising model.}
\label{table_transver_ising}
\end{table}
\subsection{Experimental realizations}
There are a number of experimental verification of quantum phase transitions. In Table~\eqref{table_transver_ising}, we list some materials that can be described by the quantum Ising model~\eqref{eq_quantum_ising}. In the following we discuss some of the experiments of quantum phase transitions~\cite{under_qpt}.
\begin{itemize}
    \item \textbf{Quantum dots:}\\
    Over the last three decades, technological advances in nanofabrication and control of systems at low temperatures, allowed for the design of artificial atomic structures which imitates conventional materials. A prominent example is a quantum dot, which represents an artificial atom and can be used to model a localized magnetic moment~\cite{quantumdot}. A lattice of multiple quantum dots offer rich applications, it can be used to study spin transport~\cite{qdots_spin_trans}, heat transport~\cite{qdot_heat_transp} in low dimensional architectures and the interplay between single spins and conduction electrons~\cite{spin-cond_qdot}. \\
    In artificial architectures, the microscopic properties such as energy spectrum, magnetic moment and the wavefunction spatial distribution, can be varied via gate voltage or magnetic field, without altering the material's structure or chemical composition, which is the case with real materials. As a consequence, in real materials tuning a single parameter, affect the other microscopic parameters and the Hamiltonian of system, which in turn make the two quantum phase transitions of the system, often belong to two different but related materials. The situation is different with quantum dots, where parameters can be controlled continuously and nearly independently, which allows the access to the critical features and phase diagram. \\
    Quantum dots were used, in an experiment by Oreg and Goldhaber-Gordon, to observe the two-channel Kondo state, which describes a quantum phase transition between non-Fermi and Fermi liquids in metals~\cite{oreg2003}.
    \item \textbf{Two-dimensional electron systems:}\\
    Quantum phase transitions can be realized experimentally in systems where the electrons are free to move in a plane but their motion perpendicular to the plane is quantized in a confined using a potential well. This type of systems is called: two-dimensional electron systems, and they were used for the metal-insulator transitions in perpendicular and zero magnetic field, as well as transitions in the Wigner crystal~\cite{wigcrys1988, 2d1992}.
    \item \textbf{The compound $\text{LiHoF}_{\boldsymbol{4}}$:}\\
    The Ising model is used to describe the low-lying magnetic excitations of the insulator $\text{LiHoF}_4$. Bitko, Rosenbaum, and Aeppli studied experimentally this compound by placing the material in a transverse magnetic field to the magnetic axis~\cite{bitko1996}. The transverse magnetic field creates quantum tunneling between the states of the ion Ho, which can destroy the long-range magnetic order, if applied sufficiently strong. Indeed, such a quantum phase transition was successfully observed, where the ferromagnetic moment was observed to vanish at the critical point.
    \item \textbf{Ultracold atoms:  }\\
   Ultracold atoms in optical lattices are a versatile platform to study the genuine properties of quantum many-body systems~\cite{ultracoldbook}. The superfluid-mott insulator quantum phase transition have been realized experimentally using ultracold atoms in optical lattices. The experiment done by Greiner, Bloch and collaborators consist of cooling ${}^{87}$Rb atoms to low temperatures so that their quantum statistics is dominant. Since ${}^{87}$Rb atoms are bosons, they can condense in a superfluid state via Bose-Condensation. To drive the atoms to insulator phase, Greiner and his collaborator used an optical lattice to apply a periodic potential on the atoms. This forces the system to localize in the minima of the periodic potential, which leads to a quantum phase transition to an insulating state~\cite{bloch2002}.
\end{itemize}
\chapter{Classical \& Quantum Information Theory \label{chap2}}
\section{What is information?}
Classical information theory aims at studying the problem of sending classical information, which can take several forms: letters in alphabet, speech and strings of bits, over communication channels ruled by the laws of classical physics. In 1948, Claude Shannon developed his revolutionary mathematical theory of communication, where he defines communication as every procedure by which one mind may effects another~\cite{shannon}. Shannon's theory sets three levels of communications problems:
\begin{enumerate}
    \item \textbf{The technical problem;}\\
    this level deals with the engineering details of the communication system. In particular, how accurate is the transfer of information between the sender and receiver.
    \item \textbf{The semantic problem;}\\
    after sending the information, how precisely do the transmitted message deliver the desired meaning? The semantic problem is tightly connected with the interpretation of the message by the receiver, as compared with the intended meaning by the sender. 
    \item \textbf{The effectiveness problem;}\\
    the essence of communication is to affect the physical conduct of the receiving party. Therefore, the effectiveness problems are dealing with the success of the desired interpretation of the sent message in affecting the physical act of the receiver.
\end{enumerate}
So far we have not given a clear definition of the key player of the theory of communication, that is information. The word information relates, according to Shannon, not so much to what do we say, but instead to what we could say. This implies that information is a measure of our freedom of choice when we select a message. Although, information must not be confused with meaning, as we can have two messages, one full of meaning and the other one is pure nonsense, that are equivalent from an information point of view.

\begin{figure}[t!]
    \begin{tikzpicture}
        \node[thick,rectangle,draw,minimum width = 2cm, minimum height = 2cm] (r1) at (0,0) {};
        \node[thick,rectangle,draw,minimum width = 2cm, minimum height = 2cm] (r2) at (3,0) {};
        \draw[->, ultra thick] (r1) -- (r2);
        \node[thick,rectangle,draw,minimum width = 1cm, minimum height = 1cm] (r0) at (6.5,0) {};
        \draw[->, ultra thick] (r2) -- (r0);
        \node[thick,rectangle,draw,minimum width = 2cm, minimum height = 2cm] (r_b) at (6.5,-4) {};
        \draw[->, ultra thick] (r_b) -- (r0);
        \node[thick,rectangle,draw,minimum width = 2cm, minimum height = 2cm] (r3) at (10,0) {};
        \draw[->, ultra thick] (r0) -- (r3);        
        \node[thick,rectangle,draw,minimum width = 2cm, minimum height = 2cm] (r4) at (13,0) {};
        \draw[->, ultra thick] (r3) -- (r4);
        \node[above] (r1) at (0,1.5) {Information};
        \node[above] (r1) at (0,1.0) {source};
        \node[above] (r2) at (3,1.0) {Transmitter};
        \node[above] (r1) at (5,-0.7) {Signal};
        \node[above] (r1) at (8.,-0.7) {Received};
        \node[above] (r1) at (8 ,-1.3) {Signal};
        \node[above] (r3) at (10 ,1.5) {Receiver};
        \node[above] (r4) at (13 ,1.5) {Destination};
        \node[above] (r4) at (11.5,-1.7) {Message};
        \node[above] (r4) at (1.5,-1.7) {Message};
        \node[above] (r4) at (6.5,-5.7) {Noise source};
    \end{tikzpicture}
    \caption{Sketch of a communication system.}
    \label{comm_sys}
\end{figure}
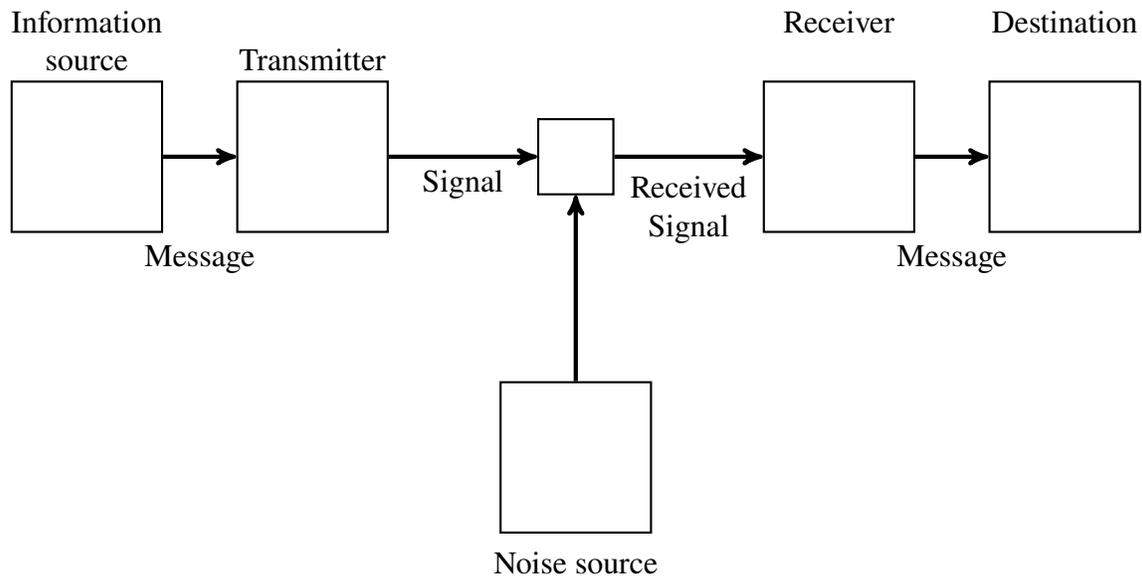

\noindent Classically, information processing is done via a communication system. Figure~\eqref{comm_sys}, depicts a sketch of the pathway of information, carried in a message, through a communication system. Out of a set of possible messages, the information source selects a desired message, e.g. words. The transmitter transforms the message into a signal, and send it across the communication channel to the receiver. The latter acts as an inverse transmitter, which reverse the signal back to a message and delivers it to the final destination. Examples of this process are encountered in our daily life. For example, in oral speech, the brain act as the information source. The vocal system produces the signal by varying air pressure and the communication channel is the air. Information processing is never perfect and unwanted noise, in the form of distortions of sound as in telephony, static in radio or errors in a picture,  can change the sent message. Fortunately, error correcting codes exist in order to mitigate the undesirable effect of noise in communication protocols.

\section{Shannon information}
The amount of information is measured by the logarithm, to the base 2, of the number of available choices. The case when there are only two choices has an information content of unity, as $\log_2(2)=1$. The unit of information is the ``bit'', which is a concatenation of ``binary digit'', proposed by John W. Turkey~\cite{shannon}. The use of the logarithm function as a measure of the amount of information seems even more natural as we consider more complex examples. In a system of four on-off switches, what is the amount of information? The natural answer is four bits, given the fact that in a single switch the information is unity. The number of possibilities in a system of four switches is $2^4=16$. The switches can be in 16 different configurations, which are: 0000, 0001, 0010, 0100, 0101, 0110, 1000, 1100, 1010, 1001, 0011, 1110, 1101, 1011, 0111 and 1111. The first and last configuration corresponds, respectively, to switches all on and off. The logarithmic measure associates 4 bits of information to this case, as $\log_2 16 = 4$. Doubling the amount of switches, squares the number of possible messages. Therefore, intuitively doubles the amount of information when measured by the logarithm. 

\noindent A general aim of information theory is to analyze communication systems, when the information source selects elementary sequences of symbols in order to form the message. In this situation, the output follows a stochastic process, where the symbols are chosen according to a probability distribution, in which the probabilities are not independent. 
In real life, humans form speech in ways that make the constituent words, probabilistically, tightly connected. For example after the three words ``in the event'', saying ``that'' is more probable than saying ``dog''. This class of stochastic process, where an event is dependent on previous events is called ``Markovian process''. To measure the amount of information in Markovian processes, Shannon borrowed the concept entropy from thermodynamics. Entropy is related to the degree of randomness in a physical system and its connection to the concept of information is intuitive, when seen in the following way: information is associated with the amount of freedom of choice in constructing messages. A situation where a system is highly organized, imply few choices. Therefore the entropy and information is low.     

\begin{figure}[t!]
    \centering
    \includegraphics[scale=0.7]{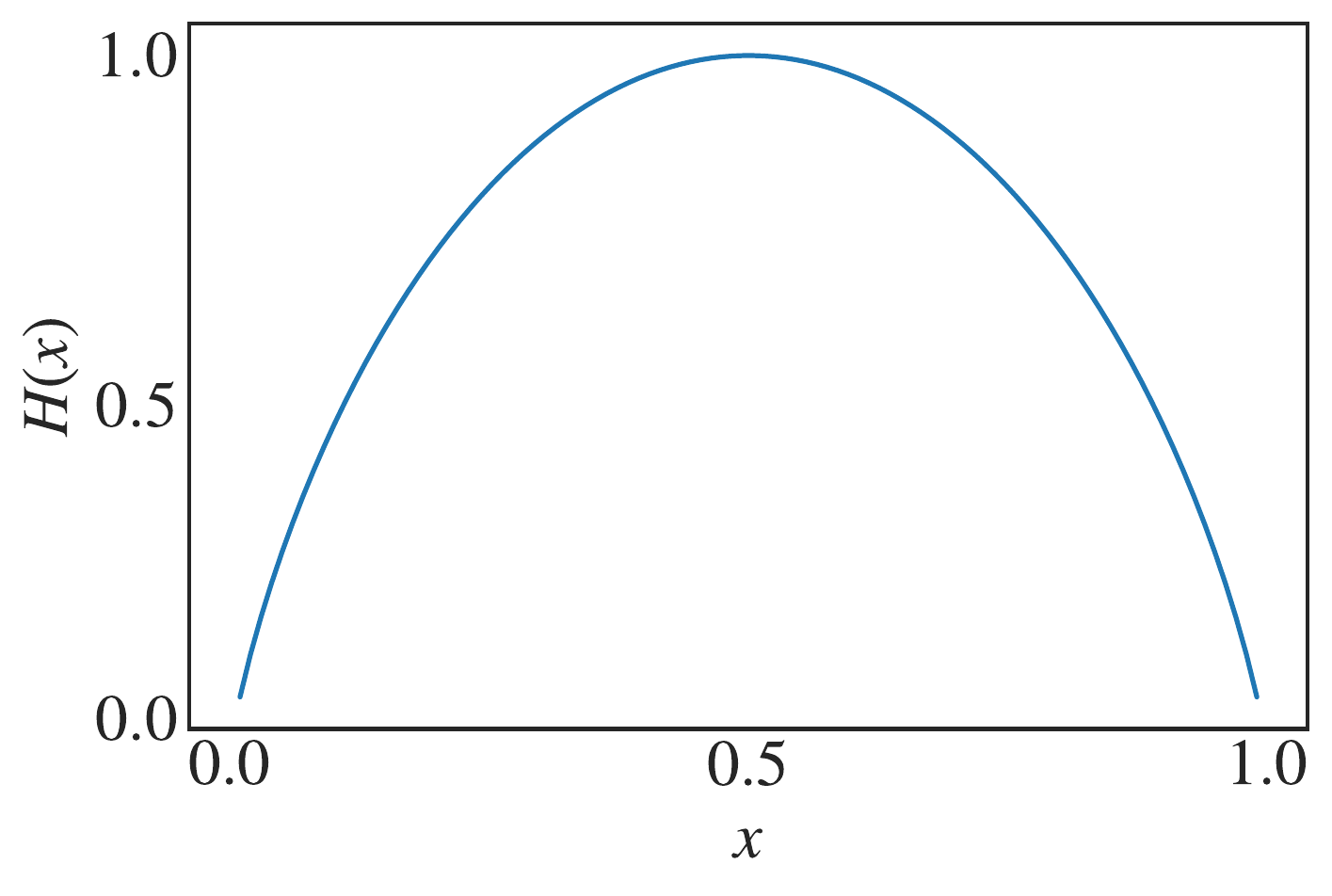}
    \caption{The Shannon entropy~\eqref{shann_entro}, for the case of a biased coin toss.}
    \label{coin_toss}
\end{figure}

\noindent For a set of $n$ independent messages, whose corresponding probabilities of choice are $p_1, p_2, \dots, p_n$. The entropy-like expression to measure the amount of information content is called ``the Shannon entropy'', it is given by
\begin{equation}
    H(n)=-\sum_i p_i(n) \log_2 p_i(n).
    \label{shann_entro}
\end{equation}
It satisfy the following properties
\begin{enumerate}
    \item $H(n)$ is monotonically decreasing in $n$. Gaining knowledge on a probable event, decreases the information on an observed event.
    \item $H(n)\geq 0$, information is a non-negative quantity. This is the motivation for introducing the negative sign in~\eqref{shann_entro}.
    \item An event occurring with unit probability, delivers zero information. i.e. $H(1)=0$.
    \item $H(n_1,n_2)=H(n_1)+H(n_2)$, if the events $n_1$ and $n_2$ are independent.
\end{enumerate}
Figure~\eqref{coin_toss} represents an application of Shannon entropy~\eqref{shann_entro}, for the case of a biased coin toss. The probability for heads is $p_h=x$ and for tails is $p_t=1-x$, so that the Shannon entropy~\eqref{shann_entro}, reduces to $H(x)=-x\log_2 x -(1-x)\log_2(1-x)$. From Figure~\eqref{coin_toss}, we see that $H(x)$ is maximum when the two outcomes are equally probable, i.e. $p_h=p_t=0.5$. As soon as one outcome becomes more probable, that is $p_h>p_t$, the value of $H(x)$ decreases and reaches zero when one outcome have unit probability. 

\noindent The observed outcome in the previous example can be generalized for situations with multiple events. More choices imply more information, which tend to increase with the homogeneity of the probability distribution, that is at equal probabilities. There is more information if we can select freely in a set of a hundred words than in set of twenty words.
\section{Quantum bit}
Shannon's communication theory invented the future.  All modern day communication systems, from optical, underwater to interplanetary, are based on his mathematical theory of communication. However, the laws of classical physics, offer very limited features in order to solve today's growing complex problems, such as simulation of physical systems and cryptography. Additionally, we are witnessing a fast reduction of the component size of technological devices, to an order where quantum features are dominant, e.g. transistors. The superposition principle and entanglement are the main features of quantum theory, with no classical counterpart, are used in order to perform tasks that are difficult or even impossible using  classical protocols, such as quantum teleportation, Shor's factorization algorithm and quantum cryptography~\cite{telepo, shor, crypto}.

\noindent Quantum information theory seeks to exploit the rich features of quantum systems in designing new information processing protocols. The field emerged in mid-1980, since then various remarkable advances have been made in order to design quantum devices, such as: D-Wave systems~\cite{nielsen2002quantum}. The physical carrier of classical information is the bit, taking two values: either 0 or 1. In contrast, the indivisible unit of quantum information is the quantum bit or qubit, which can be represented mathematically as a vector in a complex Hilbert space, with two mutually orthogonal state $\ket{0}$ and $\ket{1}$ in the following way
\begin{equation}
    \ket{\psi}=\alpha \ket{0} + \beta \ket{1},
    \label{qubit}
\end{equation}
where $(\alpha, \beta) \in \mathbb{C}$, satisfying the normalization condition $|\alpha|^2 + |\beta|^2 =1$. The linear form of the qubit~\eqref{qubit}, allows it to take infinite possible superpositions, which is a feature due to the superposition principle. This implies that the amount of information that can be encoded into a qubit, in theory, is \textit{infinite}. However, the amount of information extracted from a qubit, via single measurement, is only one bit of classical information associated with the probability of the measured state. Another property that stems from the linearity of quantum mechanics is the no-cloning theorem, which forbids the creation of independent copies of an unknown qubit. This makes qubits very robust for quantum communication protocols as they prevent the creation of copies of transmitted quantum cryptographic key~\cite{quantum_comm_book}.
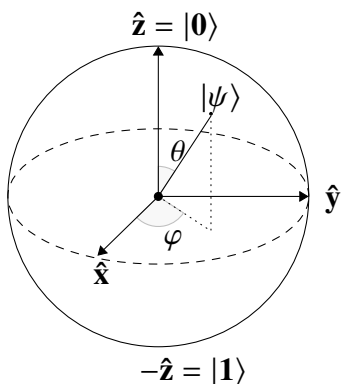
\begin{figure}[t!]
\centering
    \begin{tikzpicture}[line cap=round, line join=round, >=Triangle]
      \clip(-2.19,-2.49) rectangle (2.66,2.58);
      \draw [shift={(0,0)}, lightgray, fill, fill opacity=0.1] (0,0) -- (56.7:0.4) arc (56.7:90.:0.4) -- cycle;
      \draw [shift={(0,0)}, lightgray, fill, fill opacity=0.1] (0,0) -- (-135.7:0.4) arc (-135.7:-33.2:0.4) -- cycle;
      \draw(0,0) circle (2cm);
      \draw [rotate around={0.:(0.,0.)},dash pattern=on 3pt off 3pt] (0,0) ellipse (2cm and 0.9cm);
      \draw (0,0)-- (0.70,1.07);
      \draw [->] (0,0) -- (0,2);
      \draw [->] (0,0) -- (-0.81,-0.79);
      \draw [->] (0,0) -- (2,0);
      \draw [dotted] (0.7,1)-- (0.7,-0.46);
      \draw [dotted] (0,0)-- (0.7,-0.46);
      \draw (-0.08,-0.3) node[anchor=north west] {$\varphi$};
      \draw (0.01,0.9) node[anchor=north west] {$\theta$};
      \draw (-1.01,-0.72) node[anchor=north west] {$\mathbf {\hat{x}}$};
      \draw (2.07,0.3) node[anchor=north west] {$\mathbf {\hat{y}}$};
      \draw (-0.5,2.6) node[anchor=north west] {$\mathbf {\hat{z}=|0\rangle}$};
      \draw (-0.4,-2) node[anchor=north west] {$-\mathbf {\hat{z}=|1\rangle}$};
      \draw (0.4,1.65) node[anchor=north west] {$|\psi\rangle$};
      \scriptsize
      \draw [fill] (0,0) circle (1.5pt);
      \draw [fill] (0.7,1.1) circle (0.5pt);
    \end{tikzpicture}
\caption{Bloch Sphere of a two-dimensional quantum system.}
\label{bloch_sphere}
\end{figure}
In general, the pure state of a qubit can be written as
\begin{equation}
    \ket{\psi(\theta,\varphi)}=e^{-i^\varphi/2}\cos(\theta/2) \ket{0}+e^{i^\varphi/2}\sin(\theta/2)\ket{1},
    \label{general_qubit}
\end{equation}
where $\theta$ and $\varphi$ are real numbers, satisfying $0\leq \theta \leq \pi$ and $0 \leq \varphi \leq 2\pi$. The qubit~\eqref{general_qubit} can be represented geometrically in the Bloch sphere as shown in Figure~\eqref{bloch_sphere}. The classical bit is limiting case of the qubit~\eqref{general_qubit}, where we know for sure the vector is always either up ($\ket{0}$) or down ($\ket{1}$).

\noindent For mixed states, the statistical properties of quantum systems are described using the density matrix $``\rho$'', which is a self-adjoint positive semi-definite operator with unit trace. The density matrix can be decomposed in terms of the identity $\mathbb{I}$, and the Pauli matrices:
\begin{equation}
    \sigma^x=\begin{pmatrix}
     1&0\\
     0&-1
    \end{pmatrix},%
    \sigma^y=\begin{pmatrix}
     0&-i\\
     i&0
    \end{pmatrix},%
    \sigma^z=\begin{pmatrix}
     1&0\\
     0&1
    \end{pmatrix},
    \label{qubit_pauli_matrix}
\end{equation}
as
\begin{equation}
    \rho = \frac{1}{2} (\mathbb{I}+\vec{n}.\vec{\sigma}),
\end{equation}
where $\vec{n}$ is the Bloch vector, and $\vec{\sigma}=(\sigma^x,\sigma^y,\sigma^z)$ is a vector of Pauli matrices ~\eqref{qubit_pauli_matrix}. The positivity condition of density matrices enforces $n\leq 1$. Equality is attained for pure states, where $\Tr(\rho^2)=\Tr(\rho)$, while the rest of the values are for mixed states. Geometrically, in the Bloch sphere, c.f. Figure~\eqref{bloch_sphere}, the surface of the sphere corresponds to all pure states of the qubit, while all mixed states lie within the sphere. The center of the sphere is described by a completely mixed state: $\rho=\frac{1}{2}\mathbb{I}$.

\begin{figure}[t!]
    \centering
    \includegraphics[scale=0.7]{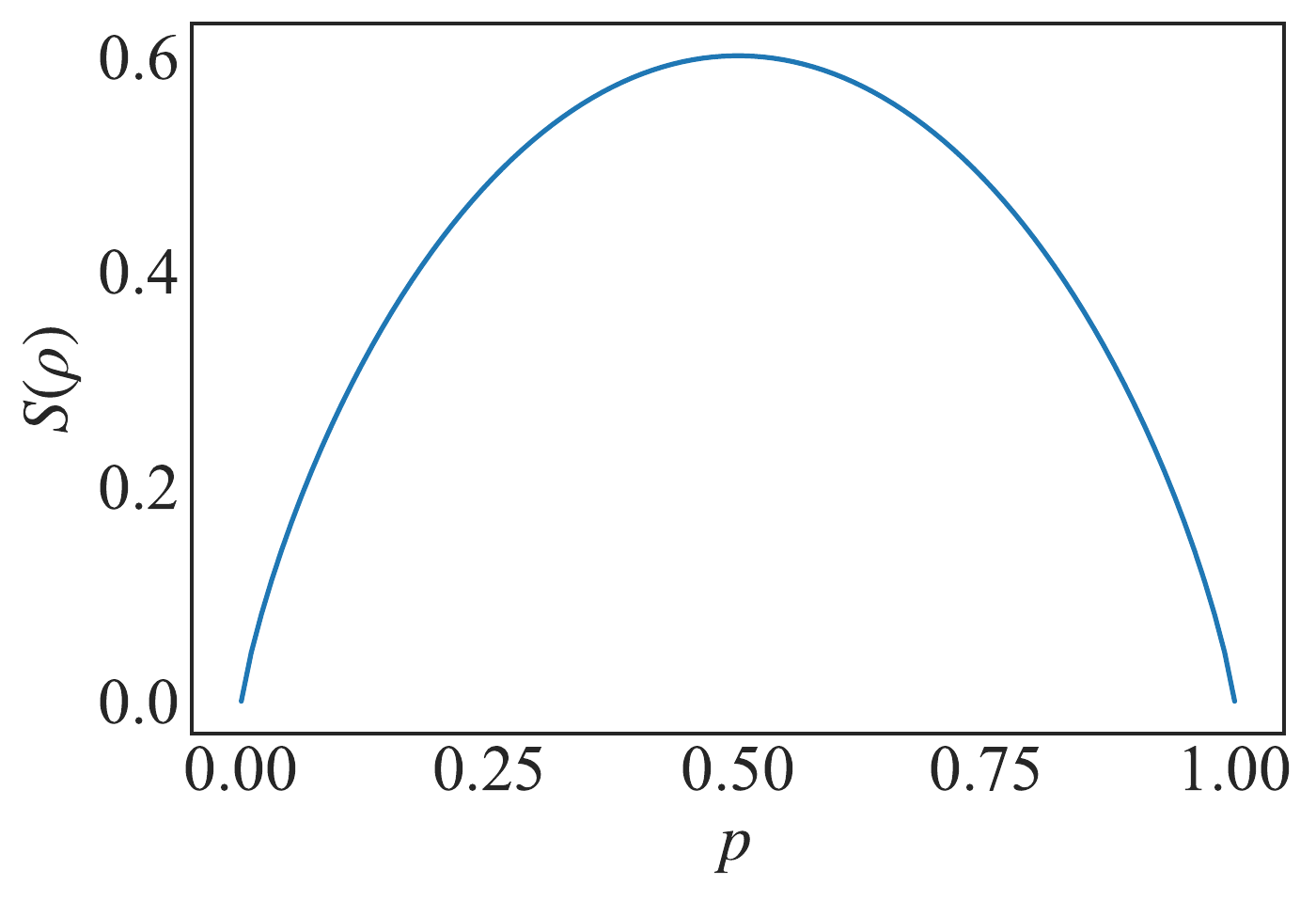}
    \caption{The Von Neumann entropy $S(\rho)$~\eqref{entropy_vn}, with respect to the mixing parameter $p$ for the density operator ~\eqref{werner_state1}.}
    \label{werner_vnentropy}
\end{figure}

\noindent The amount of information in a probability distribution is measure by the Shannon entropy~\eqref{shann_entro}. For quantum systems, the density operator replaces the probability distribution. The Von Neumann entropy generalizes the concept of the Shannon entropy~\eqref{shann_entro}, to quantum states and can be written as
\begin{equation}
    S(\rho)=-\Tr \rho \log_2(\rho).
    \label{entropy_vn}
\end{equation}
If the $\lambda_i$ are the eigenvalues of the density matrix $\rho$, the Von Neumann entropy~\eqref{entropy_vn}, can be simplified into
\begin{equation}
    S(\rho)=- \sum_i\lambda_i \log_2(\lambda_i).
    \label{eigen_entropy_vn}
\end{equation}
The Von Neumann entropy~\eqref{entropy_vn} is zero for known pure states, similarly with the Shannon entropy~\eqref{shann_entro}, of events occurring with unit probability. For a completely mixed density operator, living in a $d$-dimensional Hilbert space, the entropy reduces to $\log_2(d)$. Some other mathematical properties of the Von Neumann entropy~\eqref{entropy_vn}, are given next
\begin{enumerate}
    \item \textbf{Concavity;}\\
    for a collection of positive probabilities, which sum up to unity and a density matrix $\rho_i$, $S(\rho)$ is concave:
        \begin{equation}
            S \left( \sum_{i=1}^k p_i\rho_i \right) \geq \sum_{i=1}^k S(\rho_i).
        \end{equation}
    \item \textbf{Additivity;}\\
    given two independent systems $A$ and $B$ described, respectively, by two density matrices $\rho_A$ and $\rho_B$, the Von Neumann entropy $S(\rho)$ is additive:
        \begin{equation}
            S(\rho_A \otimes \rho_B) = S(\rho_A)+S(\rho_B).
        \end{equation}
    \item \textbf{Strong subadditivity;}\\
    for any three systems $A$, $B$ and $C$, strong subadditivity implies that
        \begin{equation}
            S(\rho_{ABC})+S(\rho_B) \leq S(\rho_{AB}) + S(\rho_{BC}),
        \end{equation}
    which automatically insures the subadditivity of $S(\rho)$, by
        \begin{equation}
            S(\rho_{AC}) \leq S(\rho_{A}) + S(\rho_{C}),
        \end{equation}
    where equality is reached for an uncorrelated system.
\end{enumerate}
The Von Neumnann entropy~\eqref{entropy_vn}, reduces to the Shannon entropy~\eqref{shann_entro}, when the density operator $\rho$ is written in terms of orthogonal states. Consider $\rho=\sum_i p_i \ket{\psi_i}\bra{\psi_i}$, the Von Neumnann entropy is
\begin{align}
    S(\rho)&= -\Tr \rho \log_2\rho, \nonumber \\
    &=-\Tr \left[\sum_i p_i \ket{\psi_i}\bra{\psi_i} \log_2 \left( \sum_j p_j \ket{\psi_j}\bra{\psi_j} \right) \right],\nonumber \\
    &=- \sum_k \bra{\psi_k} \sum_i p_i \ket{\psi_i}\bra{\psi_i} \left( \sum_j p_j \ket{\psi_j}\bra{\psi_j} \right) \ket{\psi_k}, \nonumber \\
    &=-\sum_{i,j,k} \delta_{k,j} p_j \delta_{j,i} \log_2 p_i \delta_{ik}, \nonumber \\
    &=-\sum_i p_i \log_2 p_i, \nonumber \\
    &=H(p_i).
\end{align}
However, for non-orthogonal quantum states, the Von Neumnann entropy~\eqref{entropy_vn}, and the Shannon entropy~\eqref{shann_entro}, are not equivalent. This can be seen by considering the following example, where the density operator is written as
\begin{equation}
    \rho=p \ket{0}\bra{0}+\frac{1-p}{2}(\ket{0}+\ket{1})(\bra{0}+\bra{1}),
    \label{werner_state1}
\end{equation}
Figure~\eqref{werner_vnentropy} shows the behaviour of the Von Neumnann entropy~\eqref{entropy_vn}, with respect to the mixing parameter $p$ for the density operator~\eqref{werner_state1}. The behaviour is similar to that of the Shannon entropy reported in Figure~\eqref{coin_toss}. For $p=0$ and $p=1$, the density operator ~\eqref{werner_state1}, reduces to a pure state; thus, the entropy is zero. Switching on the mixing parameter $p$, increases the indistinguishablity of the quantum state, therefore, raises the entropy which reaches its maximum value of $0.6$ bits at equal mixture, i.e. $p=0.5$. In contrast, we see in Figure~\eqref{coin_toss} that the Shannon entropy~\eqref{shann_entro}, reaches 1 bit at equal probability. Thus, the cost of transmitting a quantum symbol is 0.6 bits, while the price is higher for transmitting a classical bit of information. Generally, we have
\begin{equation}
    S(\rho)\leq H(p_i).
\end{equation}
Equality holds for quantum systems written in terms of orthogonal states. The Von Neumnann entropy~\eqref{entropy_vn}, is not basis dependent, while it is the opposite for the Shannon entropy~\eqref{shann_entro}.

\section{Entanglement}
The extent of the features of quantum systems grows richer, when we consider composite systems. The most striking property of composite quantum systems is the entanglement, which has no classical counterpart~\cite{EPR}. This prevented physicists from providing a definition that encompasses the nature of the entanglement outside the usual mathematical way.  For two parties, Alice ($A$) and Bob ($B$), the total quantum state can be written as
\begin{equation}
    \ket{\psi_{AB}}=\ket{a} \otimes \ket{b},
    \label{psi}
\end{equation}
where the states $\ket{a}$ and $\ket{b}$ belong, respectively, to the local Hilbert spaces $\mathcal{H}_A$ and $\mathcal{H}_B$. States of the form of~\eqref{psi}, are called separable or product states. Quantum mechanics allow states to be in another possibility, through superposition, which does not always yield product states:
\begin{equation}
    \ket{\psi_{AB}^{'}}=\frac{1}{N}(\ket{a_1} \otimes \ket{b_1}+\ket{a_2} \otimes \ket{b_2}),
    \label{psi_entangled}
\end{equation}
where $N$ is a normalization constant, derived through the condition $\braket{\psi|\psi}=1$. If the state $\ket{\psi_{AB}^{'}}$~\eqref{psi_entangled}, cannot be written in a product form, i.e. $\ket{\psi_{AB}^{'}} \neq \ket{a} \otimes \ket{b} $. The state is said to be \textit{entangled}. A simple example of an entangled state of two qubits is
\begin{equation}
    \ket{\psi^+}=\frac{1}{\sqrt{2}} (\ket{0}_A \ket{1}_B+\ket{1}_A\ket{0}_B).
\end{equation}
If Alice measures her qubit and finds the state $\ket{0}$, the state of Bob's qubit collapses to the state $\ket{1}$. As a consequence, the parties of an entangled state cannot be described individually and independently from one another. This feature makes the entanglement a robust resource for secure communications and efficient cryptography protocols. Indeed, in 1991 Ekert presented the first quantum key distribution protocol~\cite{ekert91},  that is based on entanglement, allowing two parties, Alice and Bob, to communicate in a completely secure way. Classical cryptography protocols are based on the conjecture that a large number is hard to factorize, whereas Ekert's protocol is provably secure. Another application of entanglement is quantum dense coding, that is a protocol developed by Bennett and Wiesner in 1992, which allows two entangled parties to communicate two bits of classical information by sending only a single qubit~\cite{weisner92}. Bennett showed, in 1993, that unknown quantum states can be perfectly communicated through quantum teleportation~\cite{telepo}, and in 1997 the protocol was verified experimentally by two research groups, led by Anton Zeillinger and Sandu Popescu~\cite{exp_telep, exp_telep_popescu}.

\noindent The success of entanglement-based quantum protocols is highly connected with the state shared between the parties. In quantum dense coding, teleportation and cryptography, Alice and Bob need to share entangled singlet states. Quantum systems are, in general, represented by mixed states and pure quantum state are not always singlets. Therefore, a general resource theory for mixed quantum states is needed in order to assess their capability in succeeding quantum protocols. This is done by quantifying the amount of entanglement in quantum mixed states. A prominent measure is the distillable entanglement, which is defined as the maximal number of singlets that can be obtained per copy of a given mixed state via local operations and classical communication (LOCC), if the number of copies goes to infinity. LOCC are a set of operations used by two parties, where Alice performs a local operation on her part of the system and communicates the result via classical means, e.g. telephone. LOCC operations are useful for state preparation, state discrimination, and entanglement transformations. The distillable entanglement is hard to evaluate and exact expressions are only available for few simple systems, which pushed toward the development of other entanglement measures. An entanglement monotone $E(\rho)$, is a measure satisfying the following properties~\cite{rev_entanglement}:
\begin{enumerate}
    \item $E(\rho)$ does not increase under LOCC.
    \item $E(\rho)\geq0$, the equality holds when $(\rho)$ is separable.
    \item $E(\rho)$ is invariant under local unitary operations.
    \item $E(\rho)$ does not increase under different preparation of $\rho$, i.e. $\sum_i p_i E(\rho_i) \geq E(\sum_i p_i \rho_i)$.
\end{enumerate}
For a pure bipartite state $\ket{\psi_{AB}}$, the entanglement can be quantified by the von Neumann entropy of the reduced density operator $\rho_A=\Tr_B(\ket{\psi_{AB}} \bra{\psi_{AB}})$:
\begin{equation}
    E(\ket{\psi_{AB}})=S(\rho_A)=-\sum_i \lambda_i \log_2 \lambda_i,
    \label{dist_ent}
\end{equation}
where $\lambda_i$ are the eigenvalues of $\rho_A$. The distillable entanglement reduces to the von Neumann entropy~\eqref{dist_ent}, for all pure states.

\noindent The quantification of entanglement in mixed states $\rho_{AB}$, can be done via convex roof measures and distance based measures; in this thesis, we focus on the former. Following the convex roof construction, any measure of the entanglement $E(\rho)$ for pure states can be extended to mixed states via
\begin{equation}
    E(\rho)=\inf_{p_i,\ket{\psi_i}} \sum_i p_i  E(\ket{\psi_i}),
    \label{ent_measure}
\end{equation}
where the infimum is taken over all possible decompositions $\{ p_i, \ket{\psi_i}\}$ of the density operator $\rho$. The most popular measure of entanglement for bipartite quantum systems is the entanglement of formation. For pure states, it reduces to the von Neumann entropy~\eqref{dist_ent} of the subsystems, and for mixed states Wootters solved the infimum in~\eqref{ent_measure}, for the case of systems composed of two qubits~\cite{wootters_concurrence}. The formula developped by Wooters can be written as
\begin{equation}
    E_f(\rho)=h \left( \frac{1}{2}+\frac{1}{2}\sqrt{1-C^2(\rho)} \right),
    \label{entanglement_formation}
\end{equation}
where $h(x)=-x\log_2(x)-(1-x)\log_2(1-x)$ denotes the binary entropy and $C(\rho)$ is the conccurence, given by
\begin{equation}
    C(\rho)=\max\{0,\lambda_1-\lambda_2-\lambda_3-\lambda_4\},
    \label{concurrence}
\end{equation}
where the $\lambda_i$'s are the square roots of the eigenvalues of the matrix $\rho\Tilde{\rho}$ in decreasing order, with $\Tilde{\rho}$ being a transformed matrix of $\rho$, i.e. $\Tilde{\rho}=(\sigma^y \otimes \sigma^y) \rho^* (\sigma^y \otimes \sigma^y)$, $\sigma^y$ is the second Pauli matrix ~\eqref{qubit_pauli_matrix}. Throughout this dissertation, the concurrence~\eqref{concurrence}, will be used as the main measure of entanglement in quantum spin systems.
\section{Quantum discord}
Entanglement quantifies the class of quantum correlations that can be conceptualized under the separability paradigm. For general forms of quantum correlations the situation is difficult and not clear. Extending the notion of distilable entanglement, under LOCC, to general quantum correlations is meaningless, as LOCC creates an arbitrary amount of quantum correlations. This implies that a measure of distilable quantum correlations is infinite for all quantum states. During the past two decades, several approaches have been developed in order to measure and prove the existence of general forms of quantum correlations beyond entanglement. Quantum discord is the first measure among such approaches, that was discovered around 2000, which is based on the quantization of concepts from classical information theory~\cite{discord1,discord2, discord3}. 

\noindent Quantum discord is derived by noticing that the mutual information $I(X:Y)$, between two random variables $X$ and $Y$ can be expressed in two different ways. The mutual information is given by
\begin{equation}
    I(X:Y)=H(X)+H(Y)-H(X,Y),
    \label{mutual_info}
\end{equation}
where $H(X)$ is the Shannon entropy~\eqref{shann_entro}, and $H(X,Y)$ is the Shannon entropy for the joint probability distribution $p_{X,Y}$. The mutual information~\eqref{mutual_info}, is equivalent to
\begin{equation}
    J(X:Y)=H(X)-H(X|Y).
    \label{mutual_info2}
\end{equation}
Here, $H(X|Y)$ is the conditional entropy and can be written as
\begin{equation}
    H(X|Y)=\sum_{y \in Y} p_y H(X|Y=y)=H(X,Y)-H(Y).
\end{equation}
The equality of $I(X:Y)$~\eqref{mutual_info}, and $J(X:Y)$~\eqref{mutual_info2}, for classical probability distributions stems from Bayes' rule: $p_{x|y}=p_{xy}/p_y$, which can be used to prove $J(X:Y)$~\eqref{mutual_info2}. However, this does not hold when quantum theory is applied. For a quantum system $\rho_{AB}$, the quantum mutual information is given by
\begin{equation}
    I_{AB}=S(\rho_A)+S(\rho_B)-S(\rho_{AB}),
    \label{quantum_I}
\end{equation}
where $S(.)$ is the von Neumann entropy and the reduced density matrix $\rho_A=\Tr_B(\rho_{AB})$ ($\rho_B=\Tr_A(\rho_{AB}))$. The quantum mutual information~\eqref{quantum_I}, is the natural and straightforward quantization of the classical mutual information ~\eqref{mutual_info}. In contrast, writing the quantum version of $J(X:Y)$~\eqref{mutual_info2}, is not trivial. Replacing the Shannon entropy with the von Neumann entropy in this case, leads to a quantity that can be either positive or negative. The quantum conditional entropy is argued to be
\begin{equation}
    S_{A|B}= \min_{\{ \prod_k^B \} \in \mathcal{M}^B} \sum p_k S(\rho_{A|k}),
    \label{cnd_entropy}
\end{equation}
where the minimization is taken over all quantum measurements, $\{ \prod_k^B \}$, performed on the subsystem $B$ and $\mathcal{M}^B$ is the set of such measurements. $\{p_k, \rho_{A|k}\}$ is the post-measurement ensemble that is formed at Alice's side, and
$\rho_{A|k}= \Tr_B \left[ \prod_k^B \rho_{AB} \right]/p_k$  with $p_k=\Tr_{AB} \left[ \prod_k^B \rho_{AB}\right]$. The quantum version of $J(X:Y)$ ~\eqref{mutual_info2}, can now be written as
\begin{equation}
    J_{A|B}=S(\rho_A)-S(A|B).
    \label{classical_corr}
\end{equation}
It is argued that $I_{AB}$~\eqref{quantum_I}, and $J_{A|B}$~\eqref{classical_corr}, quantify, respectively, total and classical correlations in a bipartite quantum system $\rho_{AB}$. Therefore, their difference is proposed as a measure of \textit{all} quantum correlations. Thus, quantum discord is:
\begin{equation}
    D(\rho_{AB})=\min_{ \{ \prod_i^B \}} \left[ I_{AB}-J_{A|B} \right].
    \label{quantum_discord}
\end{equation}
Here the minimization is over all measurements $\{ \prod_i^B \}$, in order to have a measurement-independent quantity. Similarly with entanglement measures, quantum discord satisfies a number of properties
\begin{enumerate}
    \item $D(\rho_{AB}) \geq 0$, since in general $I_{AB}\geq J_{A|B}$. Equality is reached for classical systems, where a local measurement on subsystem $B$ does not disturb the system. In contrast, states with zero entanglement is not always a sign of classical system.
    \item Quantum discord is asymmetric, i.e. $D\left(\rho_{AB}^{ \{ \prod_i^B \} }\right) \neq D\left(\rho_{AB}^{ \{ \prod_i^A \} }\right) $, which is due to the conditional entropy~\eqref{cnd_entropy}, being non-symmetric.
    \item Similarly to entanglement measures, quantum discord~\eqref{quantum_discord}, is invariant under local unitary transformations.
    \item For pure bipartite states, quantum discord~\eqref{quantum_discord}, reduces to the entanglement measured via the von Neuman entropy~\eqref{entropy_vn}.
\end{enumerate}
We illustrate the difference between quantum discord~\eqref{quantum_discord}, and the entanglement for mixed states, by considering the example of a class of Werner states, given by
\begin{equation}
    \rho_W= \frac{1-p}{4} \mathbb{I}_4 + \frac{p}{2} \left[ (\ket{01}-\ket{10})(\bra{01}-\bra{10}) \right],
    \label{werner_state}
\end{equation}
\begin{figure}[H]
    \centering
    \includegraphics[scale=0.7]{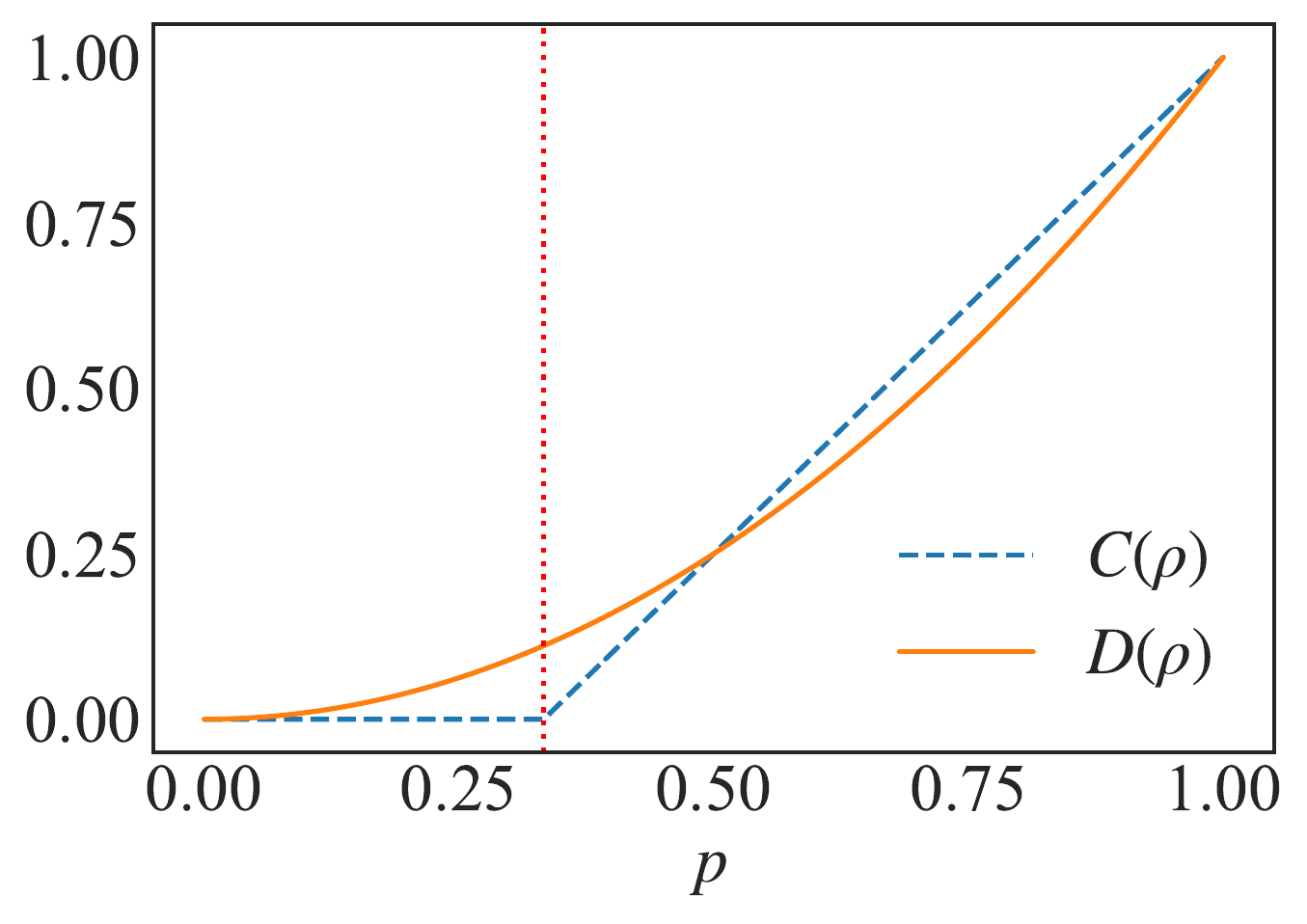}
    \caption{Quantum discord~\eqref{quantum_discord}, and the concurrence~\eqref{concurrence}, for the Werner state~\eqref{werner_state}, with respect to the relative weight $p$.}
    \label{plot_werner_qdcnc}
\end{figure}
\noindent where $p$ is the relative weight. The Werner state~\eqref{werner_state}, is separable for $p\leq1/3$ and entangled for $p>1/3$. This is confirmed in Figure~\eqref{plot_werner_qdcnc}, which depicts the behaviour of quantum discord~\eqref{quantum_discord}, and the concurrence~\eqref{concurrence}, in the Werner state~\eqref{werner_state}, with respect to the relative weight $p$. In contrast, we see that quantum discord is non-zero in the whole range of $p$. This simple example shows the resilience of quantum discord in capturing general forms of quantum correlations.

\noindent Despite the efficiency of quantum discord in capturing general forms of quantum correlations, the usefulness of states with non-zero quantum discord is debated~\cite{barry_discord}. Moreover, a quantum resource theory for separable states with non-zero quantum discord is still lacking. There are a few quantum information processing tasks, where it is claimed that they require states with non-zero quantum discord and without having any amount of entanglement, such as quantum state merging protocol~\cite{quantum_state_merg}, identification of quantum channels~\cite{quantum_channel} and quantum metrology~\cite{metrology1,metrology2}. Quantum discord has been shown to play a role in the efficiency of quantum refrigerator~\cite{discord_refr} and in studying photosynthesis in biological systems~\cite{bio_discord}.

\section{Quantum coherence}
Coherence constitutes the principal building block of today's physical theories and underlies several phenomena, such as interference and optical coherence in the theory of electromagnetism, and the quantization of energy in quantum mechanics~\cite{quant_coherence}. Coherence plays a central role in the field of quantum information theory, as it is the basic ingredient for creating quantum interference and entanglement, which are the mandatory resources for performing quantum information tasks. The exploration of single particle coherence, enabled the realization of several devices, which are classified under the ``Quantum technologies 1.0'', such as the laser. The emergence of quantum information science introduced revolutionary and new protocols, for using quantum phenomena as resources to perform tasks that are not possible using classical laws. This resource-driven point of view, paved the way for constructing the resource theory of entanglement, and motivated the development of a mathematical framework that encompasses a wider range of quantum features. Quantum technologies 2.0 comprise applications where quantum coherence is the primary resource, this includes quantum-enhanced metrology, quantum thermal machines and communication protocols~\cite{coherence_review}.

\noindent Coherence is naturally a basis dependent quantity. Therefore, the first step in constructing the resource theory for quantum coherence is by choosing a reference basis. Given a finite $d$-dimensional Hilbert space $\mathcal{H}$, with reference orthonormal basis $\{ \ket{i} \}_{i=0,\dots,d-1} $. All incoherent density operators $\rho$, can be written in the following form
\begin{equation}
    \rho = \sum_{i=0}^{d-1} p_i \ket{i}\bra{i}.
    \label{incoherent_rho}
\end{equation}
Here, $p_i$'s are probabilities and $\rho \subset \mathcal{I}$, where $\mathcal{I} \subset \mathcal{B(H)}$ denotes the set of incoherent states that belong to $\mathcal{B(H)}$: the set of all bounded trace operators on $\mathcal{H}$. Incoherent states~\eqref{incoherent_rho}, are diagonal in the basis $\ket{i}$, which implies that they are accessible without any cost.

\noindent For a single qubit, the reference basis can be the eigenstates of the Pauli matrix $\sigma^z$, i.e. the computational basis: $\{ \ket{0}, \ket{1}\}$. Therefore, any density matrix $\rho$ with non-zero diagonal elements is outside the set $\mathcal{I}$ and hence has non-zero resource content. In this case of multipartite systems, the reference basis is constructed by the tensor product of the local reference basis states of each subsystem. For example, in the case of systems composed of $N$-qubits, the set of incoherent states is $\mathcal{I}=\{ \ket{0}, \ket{1}\}^{\otimes N}$.

\noindent Similarly with entanglement and quantum discord, quantifying quantum coherence follows certain postulates. A proper measure of coherence $C(\rho)$ needs to satisfy:
\begin{enumerate}
    \item $C(\rho) \geq 0$, with equality if and only if $\rho$ is incoherent.
    \item $C(\rho)$ is a coherence monotone, which means that it does not increase under incoherent operations $\Lambda$, i.e. $C(\Lambda[\rho]) \leq C(\rho) $. 
    \item Strong monotonicity, that is  $C(\rho)$ does not increase on average under selective incoherent operations, i.e. $\sum_i p_i C(\sigma_i) \leq C(\rho)$, where $p_i=\Tr \left[ K_i \rho K_i^{\dagger} \right]$ represents probabilities, $\sigma_i=K_i \rho K_i^{\dagger}/p_i$ are the post-measurement states and Kraus operators $K_i$.  
    \item Convexity: $\sum_i p_i C(\rho_i) \geq C\left( \sum_i p_i \rho_i \right) $.
    \item For pure states, $C(\rho)$ reduces to the von Neumann entropy~\eqref{entropy_vn}.
    \item Additivity: for a separable state $\rho=\rho_A \otimes \rho_B$, the coherence $C(\rho)$ is additive, i.e. $C(\rho)=C(\rho_A)+C(\rho_B) $.
\end{enumerate}
A meaningful measure of quantum coherence $C(\rho)$ is one that satisfies, at least, postulates 1 and 2. Measures of quantum coherence can be classified into two categories: distance-based and convex roof quantifiers. In general, distance-based quantum coherence $C_D(\rho)$ is defined as
\begin{equation}
    C_D(\rho)=\inf_{\sigma \in \mathcal{I}} D(\rho,\sigma),
    \label{distance_coherence}
\end{equation}
where $D(\rho,\sigma)$ measures the distance between the state $\rho$ and its closest incoherent state $\sigma$, whereas infimum is taken over all incoherent states in $\mathcal{I}$. 
A suitable candidate for $D(\rho,\sigma)$, is the quantum relative entropy. It is given by
\begin{equation}
    S(\rho||\sigma)=\Tr \left[ \rho \log_2 \rho \right] - \Tr \left[ \rho \log_2 \sigma \right].
\end{equation}
Therefore, the corresponding measure of quantum coherence is the relative entropy of coherence, written as
\begin{equation}
    C_r(\rho)=\min_{\sigma \in \mathcal{I}}S(\rho||\sigma),
    \label{relative_coherence}
\end{equation}
which satisfies the positivity, monotonicity and strong monotonicity postulates. In the same fashion with entanglement, the convex roof quantum coherence measure for mixed states can be written as
\begin{equation}
    C(\rho)=\inf_{\{ p_i,\ket{\psi_i}\} } \sum_i p_i  C(\ket{\psi_i}),
\end{equation}
where the infinimum is taken over all possible decompositions $\{ p_i, \ket{\psi_i}\}$ of the density operator $\rho$. When $\rho=\sum_{ij} \rho_{ij} \ket{i}\bra{j} $, the coherence of formation is equal to entanglement of formation~\eqref{entanglement_formation}, of the maximally correlated state $\rho_{mc}=\sum_{ij} \rho_{ij} \ket{ii}\bra{jj}$. Thus, we have
\begin{equation}
    C_f(\rho)=E_f(\rho_{mc}).
    \label{coh_formation}
\end{equation}
Through the formula for the entanglement of formation of two-qubits systems~\eqref{entanglement_formation}, we can write the coherence of formation~\eqref{coh_formation}, for any single-qubit states as
\begin{equation}
    C_f(\rho)=h \left( \frac{1+\sqrt{1-4|\rho_{01}|^2} }{2} \right),
\end{equation}
where $h(x)=-x\log_2(x)-(1-x)\log_2(1-x)$ is the binary Shannon entropy.

\noindent In the remainder of this dissertation, we will be interested in studying a special figure of merit of quantum coherence, that combines geometric and entropic properties. To this end, we introduce the quantum Jensen-Shannon divergence~\cite{QJSD, coherencePRL}
\begin{equation}
    \mathcal{J}(\rho,\sigma) = S\Big(\frac{\rho+\sigma}{2}\Big)-\frac{S(\rho)}{2}-\frac{S(\sigma)}{2},
    \label{QJSD}
\end{equation}
where $S(.)$ is the von Neumann entropy~\eqref{entropy_vn}. The quantum Jensen-Shannon divergence~\eqref{QJSD}, is known to be a distance measure bounded between $0\leq \mathcal{J} \leq 1$. Additionally, it is well defined irrespective of the support of $\rho$ and $\sigma$. Then, the quantum coherence is defined as
\begin{equation}
    QC(\rho)=\sqrt{ \mathcal{J}(\rho,\sigma) }.
    \label{sqrt_qjsd}
\end{equation}
The measure~\eqref{sqrt_qjsd}, will be the main figure of merit for quantum coherence in the next chapter.
\chapter{Quantum Correlations in Spin Systems\label{chap3} }
In this chapter we apply the quantum correlation approach to quantum spin systems, in order to study their quantum critical features. In particular, we will analyze the critical phenomena present in the $XY$ model via the entanglement, quantum discord and quantum coherence~\cite{Zakaria2019}.
\section{The $XX$ model \label{xymodel}}
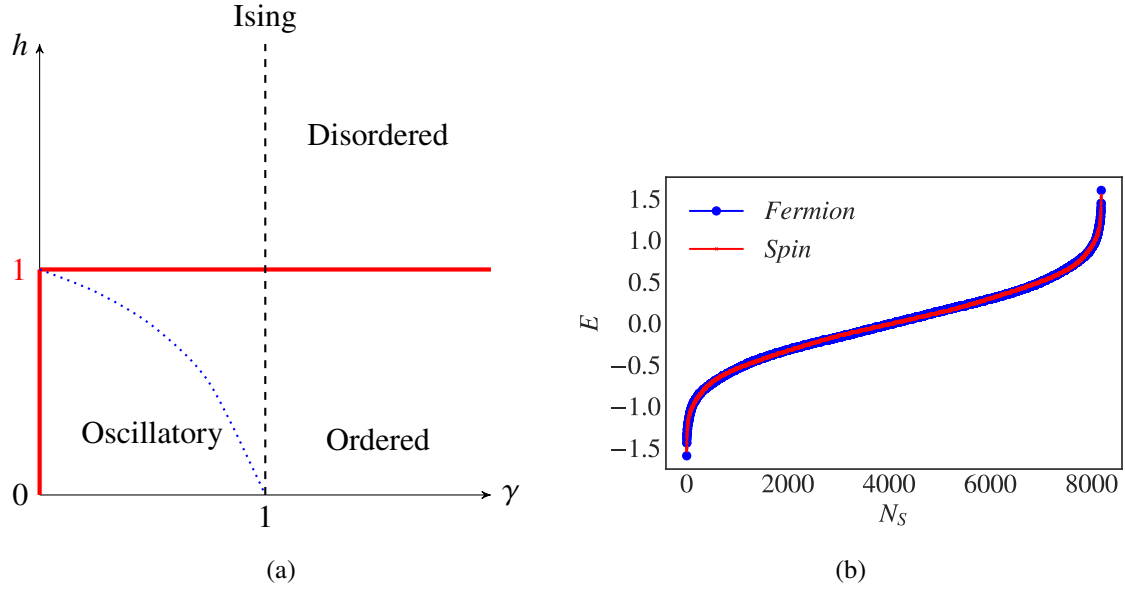
\begin{figure}[t!]
        \centering
        \begin{subfigure}[t]{0.49\textwidth}
    	\begin{tikzpicture}[scale=1.5]
        \draw[-] node[left]{$0$} (0,0)--(2,0) node[below]{$1$};
        \draw[->] (2,0)--(4,0) node[right]{$\gamma$};
        \draw[-,ultra thick,red] (0,0)--(0,2) node[left]{$1$};
        \draw[->] (0,2)--(0,4) node[left]{$h$};
        \draw[ultra thick, red]  (0,2)--(4,2);
        \draw[dashed, thick]  (2,0)--(2,4)node[above]{Ising};
        \node[above] (A) at (1,0.3) {Oscillatory};
        \node[above] (A) at (3,0.3) {Ordered};
        \node[above] (A) at (3,3) {Disordered};
        \draw [dotted,blue,thick] plot [smooth] coordinates {(0,2) (0.5,1.8) (1,1.5) (1.5,1)  (2,0)};
    	\end{tikzpicture}
    	\caption{}
    	\label{PH_XY}
        \end{subfigure}  
        \begin{subfigure}[t]{0.49\textwidth}
            \includegraphics[scale=0.5]{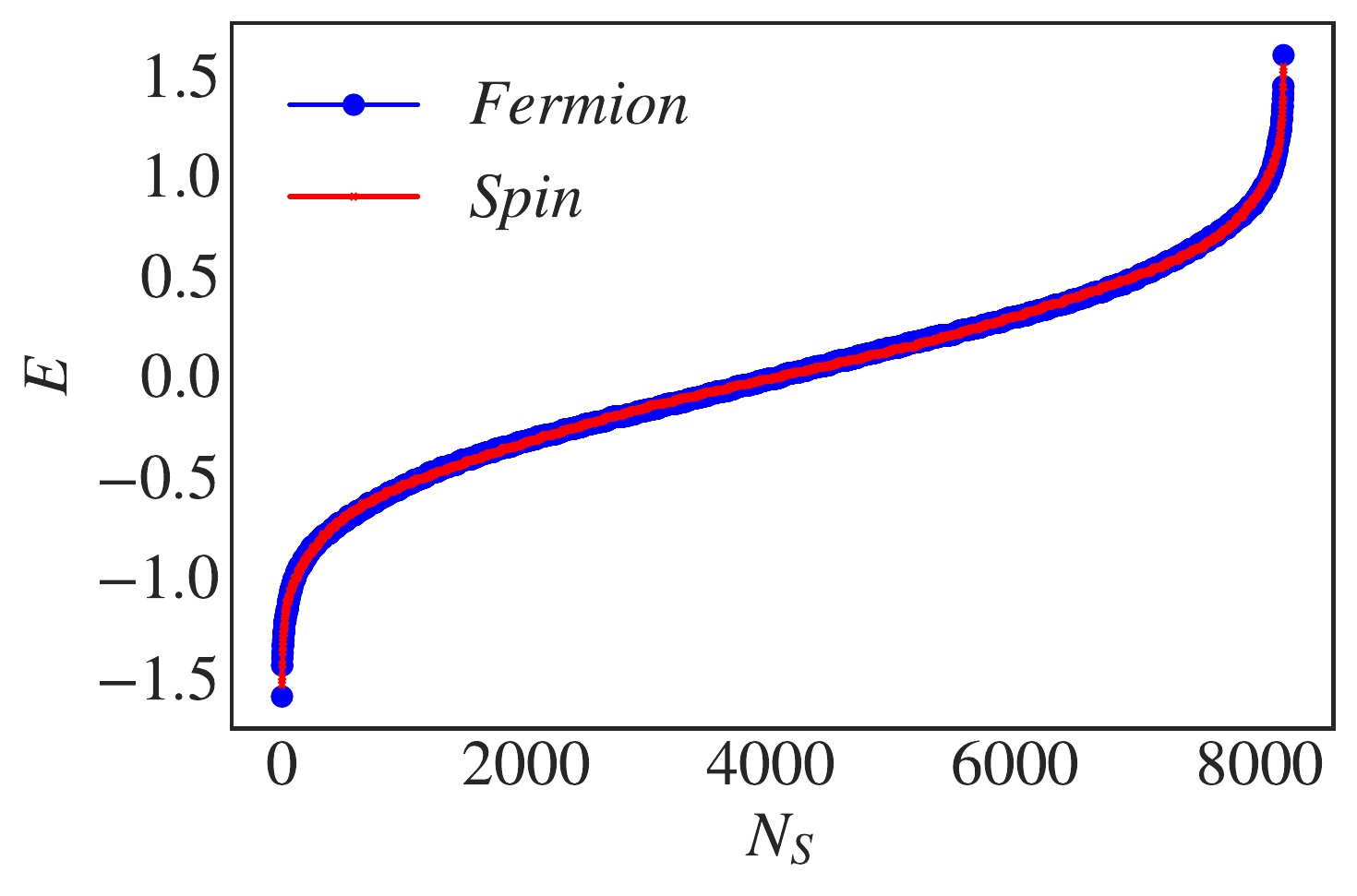}
            \caption{}
            \label{fig_jw_quspin}
        \end{subfigure}
        \caption{Phase diagram of the $XY$ model~\eqref{eq1}, in the $\gamma$-$h$ plane, and (b) the Spectrum of the $XX$ model~\eqref{eq1}, in the spin and fermion representation for $N=14$.}
    \end{figure}

The Hamiltonian of the $XY$ model describes a set of localized spin-$\frac{1}{2}$ particles interacting with nearest-neighbors exchanging coupling in a one-dimensional chain, subject to an external magnetic field $h$. It is given by 
    \begin{equation}
        \mathcal{H}_{XY}=J\sum_{i=1}^N \big((1+\gamma)S_i^xS_{i+1}^x+(1-\gamma)S_i^yS_{i+1}^y\big)-h\sum_{i=1}^NS_i^z,
        \label{eq1}
    \end{equation}
where $S_i$ is the spin-$\frac{1}{2}$ operator on site $i$, $J$ is the interaction energy between the spins and $\gamma$ is the anisotropy parameter. A positive (negative) exchange coupling $J$ favors anti-ferromagnetic (ferromagnetic) ordering of the spins. The $(\gamma-h)$ phase diagram at zero-temperature of this model is shown in Figure~\eqref{PH_XY}, where we see a critical line for $h=1$ and for $\gamma=0$ and $h<1$, shown in red. When $\gamma=1$ the model corresponds to the quantum Ising chain ~\eqref{eq_quantum_ising}. An additional critical phenomenon occurs in the $XY$ model~\eqref{eq1}, on the line $\gamma^2+h^2=1$, where the ground state is completely factorized in terms of product of single spin states. For simplicity, in the remainder of this chapter we work in the isotropic limit ($XX$ chain), i.e. $\gamma=0$.

\noindent The Hamiltonian~\eqref{eq1}, belongs to the class of exactly solvable models~\cite{franchini2017, baxterbook}. To solve the Hamiltonian~\eqref{eq1}, we first map each spin operator to a fermion operator, using the Jordan-Wigner transformation~\cite{jordan-wigner}. For simplicity, we introduce the spin ladder operators
    \begin{equation}
        S^{\pm}=S^x\pm i S^y 
        \quad\text{ with}\quad
        S^+=\begin{pmatrix}
        0&1\\0&0
        \end{pmatrix},
        \quad \quad
        S^-=\begin{pmatrix}
        0&0\\1&0
        \end{pmatrix}.
        \label{eq2}
    \end{equation}
Jordan and Wigner showed that the spin ladder operators can be represented exactly by the fermion operators ($c_i$'s) with the following mapping
    \begin{equation}
        S_i^+=c_i^+\exp{\left( \displaystyle{-j\pi\sum_{l=1}^{i-1}c_l^+c_l} \right) },
        \quad \quad
        S_i^-=\exp{\left(\displaystyle{j\pi\sum_{l=1}^{i-1}c_l^+c_l}\right)}c_i,
        \quad \quad
        S_i^z=c_i^+c_i-\frac{1}{2},
        \label{eq3}
    \end{equation}
where $j$ is the imaginary unit. Rewriting the Hamiltonian~\eqref{eq1}, in terms of the ladder operators~\eqref{eq2}, and by applying the transformation~\eqref{eq3}, we can cast the Hamiltonian~\eqref{eq1}, into the following form
    \begin{equation}
        \mathcal{H}_{XX}=J\sum_{i=1}^N\left(c_i^+c_{i+1}+c_{i+1}^+c_i\right)-h\sum_{i=1}^N\left(c_i^+c_i-\frac{1}{2} \right).
        \label{hxx}
    \end{equation}
The next step toward diagonalizing the $XX$ model~\eqref{eq1}, is making use of its translation invariance. This symmetry property enable us to write the fermion operators, $c_i$ and $c_i^{\dagger}$, in terms of the Fourier transform as
    \begin{equation}
        d_k=\frac{1}{\sqrt{N}} \sum_{i=1}^N \exp{(\displaystyle{-jik})}c_i,
        \quad\text{and}\quad
        d_k^+=\frac{1}{\sqrt{N}}\sum_{i=1}^N \exp{(\displaystyle{jik})}c_i^+.
    \end{equation}
The final diagonalized form of the Hamiltonian~\eqref{hxx}, written in terms of the
spinless fermion creation and annihilation operators is

    \begin{equation}
    \mathcal{H}_{XX}=\sum_k \epsilon(k)d_k^+d_k,
    \quad\text{with}\quad
    \epsilon(k)=J\cos(k)-h.
    \label{diag_xx}
    \end{equation}
Without loss of generality, the coupling constant $J$ will be equal to unity in the remaining of this chapter. The Jordan-Wigner transformation can be verified, numerically, using ``QuSpin'' a Python package for dynamics and exact diagonalisation of quantum many-body systems~\cite{quspin1, quspin2}.

\noindent In Figure~\eqref{fig_jw_quspin} we show for a chain of $N=14$ spins, $J=1$ and $h=1$ with periodic boundary conditions the energy spectrum $E$ of the $XX$ chain~\eqref{eq1}, with respect to the number of states $N_S$. The red solid line corresponds to the eigenenergies of the usual spin representation of the $XX$ model~\eqref{eq1}, while the blue crosses are the eigenvalues of the fermionic Hamiltonian~\eqref{diag_xx}, obtained by the Jordan-Wigner mapping. We see that for each spin state there is a corresponding fermion state, which means that the two representations match exactly, as predicted.
\section{Quantum correlations}
We are interested in studying the quantum correlations present in an infinite-chain of spin-$\frac{1}{2}$, described via the isotropic $XX$ model~\eqref{eq1}. For this aim, we evaluate the reduced density matrix of two spins $i$ and $j$ distant by some lattice spacing $m$, where $i=j+m$. Using Pauli basis expansion~\cite{pauli_basis_exp}, we have
    \begin{equation}
        \rho_{i,i+m}=\frac{1}{4}\sum_{\alpha,\beta=0}^3 p_{\alpha \beta}  \sigma_i^{\alpha}\otimes\sigma_{i+m}^{\beta},
    \end{equation}
with $p_{\alpha \beta}=\langle\sigma_i^{\alpha}\sigma_{i+m}^{\beta}\rangle$ is the two-site correlation function, and $\sigma_i^\alpha$ are the Pauli matrices, $\alpha=0,1,2,3$. In terms of Pauli ladder operators $\sigma^{\pm}=\frac{1}{2}(\sigma^x\pm i\sigma^y)$, we can write in the computational basis $\{\ket{\uparrow\uparrow},\ket{\uparrow\downarrow},\ket{\downarrow\uparrow},\ket{\downarrow\downarrow}\}$, the density matrix of two spins as follows
    \begin{equation}
        \rho_{i,i+m}=
        \begin{pmatrix}
        \langle P_i^{\uparrow}P_{i+m}^{\uparrow} \rangle&\langle P_i^{\uparrow}\sigma_{i+m}^-\rangle&\langle \sigma_i^-P_{i+m}^{\uparrow}\rangle&\langle \sigma_i^-\sigma_{i+m}^-\rangle\\
        \langle P_i^{\uparrow}\sigma_{i+m}^+\rangle&\langle P_i^{\uparrow}P_{i+m}^{\downarrow} \rangle&\langle \sigma_i^-\sigma_{i+m}^+\rangle&\langle \sigma_i^-P_{i+m}^{\downarrow}\rangle\\
        \langle \sigma_i^+P_{i+m}^{\uparrow}\rangle&\langle \sigma_i^+\sigma_{i+m}^-\rangle&\langle P_i^{\downarrow}P_{i+m}^{\uparrow} \rangle&\langle P_i^{\uparrow}\sigma_{i+m}^{-}\rangle\\
        \langle \sigma_i^+\sigma_{i+m}^+\rangle&\langle \sigma_i^+P_{i+m}^{\downarrow}\rangle&\langle P_i^{\uparrow}\sigma_{i+m}^{+}\rangle&\langle P_i^{\downarrow}P_{i+m}^{\downarrow} \rangle
        \end{pmatrix},
        \label{eq5}
    \end{equation}
where $P^{\uparrow}=\frac{1}{2}(1+\sigma^z)$ and $P^{\downarrow}=\frac{1}{2}(1-\sigma^z)$. The brackets are average values taken in the ground state of the isotropic $XX$ model~\eqref{eq1}.

\noindent The Hamiltonian~\eqref{eq1} satisfies several symmetries that reduce the number of non-zero elements of the density matrix~\eqref{eq5}. For instance, translation invariance means that $\rho_{i,i+m}=\rho_{i+m,i}$, therefore we have $p_{\alpha\beta}=p_{\beta\alpha}$. The global phase-flip symmetry implies that the commutator $\left[\sigma_i^z\sigma_{i+m}^z,\rho_{i,i+m}\right]=0$, then the coefficients $p_{01}=p_{10}$, $p_{02}=p_{20}$, $p_{13}=p_{31}$ and $p_{23}=p_{32}$ must be zero. Furthermore, the $z$-component of the total magnetization commutes with the Hamiltonian~\eqref{eq1}, $\left[ \sum_{i} \sigma_i^z,\mathcal{H_{XX}} \right]=0$ which implies that a coherent superposition of basis states $\ket{\uparrow\uparrow}$ and $\ket{\downarrow\downarrow}$ cannot be possible for any two spins. Then the coefficient $p_{30}$ and $p_{03}$ corresponding to the matrix elements $\ket{\uparrow\uparrow}\bra{\downarrow\downarrow}$ and $\ket{\downarrow\downarrow}\bra{\uparrow\uparrow}$ are zero. The density matrix~\eqref{eq5}, reduces to 

    \begin{equation}
    \rho_{i,i+m}=
        \begin{pmatrix}
        X_{i,i+m}^+&0&0&0\\
        0&Y_{i,i+m}^+&Z_{i,i+m}^*&0\\
        0&Z_{i,i+m}&Y_{i,i+m}^-&0\\
        0&0&0&X_{i,i+m}^-
        \end{pmatrix}.
        \label{eq6_mat}
    \end{equation}

\noindent In terms of the occupation number operator $n_i=c_i^+c_i$, the elements of the density matrix~\eqref{eq6_mat}, are given by
    \begin{equation}
        \begin{aligned}
        X_{i,i+m}^+&=\langle n_in_{i+m} \rangle,
        \quad\text{}\quad
        X_{i,i+m}^-=1- \langle n_i \rangle - \langle n_{i+m} \rangle+\langle n_in_{i+m} \rangle, \\
        \quad\text{}\quad
        Y_{i,i+m}^+&=\langle n_i \rangle-\langle n_i n_{i+m}\rangle,
        \quad\text{}\quad
        Y_{i,i+m}^-=\langle n_{i+m} \rangle-\langle n_{i+m}n_i\rangle, \\
        Z_{i,i+m}&=\langle c_i^+\Big(\prod_{k=i}^{i+m-1}\{1-2c_k^+c_k\}\Big)c_{i+m} \rangle.
        \end{aligned}
        \label{eq7}
    \end{equation}
We investigate various contributions of correlations at long distances of $m$ in the $XX$ model~\eqref{eq1}. We study the correlations between second nearest $(2N)$ $m=2$, third nearest $(3N)$ $m=3$ and the fourth neighbor $(4N)$ $m=4$. Then, by using Wick's theorem~\cite{wick1950} the set of equations~\eqref{eq7}, reduce to
\begin{equation}
    \begin{aligned}
        X_{i,i+m}^+&=f_0^2-f_m^2,
        \quad\text{}\quad
        X_{i,i+m}^-=1-2f_0+f_0^2-f_m^2,\\
        Y_{i,i+m}^+&=Y_{i,i+m}^-=f_0-f_0^2+f_m^2,
    \label{eq_diag}
    \end{aligned}
\end{equation}
\noindent while for the element $Z_{i,i+m}$ we have
\begin{equation}
    \begin{aligned}
    \quad\text{for $m=2$:}\quad
    Z_{i,i+2}&=f_2-2f_0f_2+2f_1^2,\\
    \quad\text{for $m=3$:}\quad
    Z_{i,i+3}&=4(f_1^3-2f_0f_1f_2+f_2^2f_1+f_0^2f_3-f_1^2f_3+f_1f_2-f_0f_3)+f_3, \\
    \quad\text{for $m=4$:}\quad
    Z_{i,i+4}&=8(f_1^4-3f_0f_1^2f_2+2f_1^2f_2^2+2f_0^2f_1f_3+f_0^2f_2^2-f_2^4-2f_0f_1f_2f_3\\&+2f_1f_2^2f_3-2f_1^3f_3+f_1^2f_3^2-f_0f_2f_3^2-f_0^3f_4+2f_0f_1^2f_4-2f_1^2f_2f_4\\&+f_0f_2^2f_4)+4(3f_1^2f_2-2f_0f_2^2-4f_0f_1f_3+2f_1f_2f_3+3f_0^2f_4-2f_1^2f_4\\&+f_2f_3^2-f_2^2f_4)+2(2f_1f_3-3f_0f_4+f_2^2)+f_4.
    \label{eqz}
    \end{aligned}
\end{equation}
\noindent For a non-negative integer number $m$, the function $f_m$ is given by
    \begin{equation}
        f_m=\frac{1}{\pi}\int_{0}^{\pi}\cos(km)g(k)dk,
        \label{eq11}
    \end{equation}
\noindent where $g(k)=\frac{1}{1+\displaystyle{e^{\beta \epsilon(k)}}}$ is the Fermi-Dirac distribution, $\beta=\frac{1}{k_B T}$ and the Boltzmann constant is taken equal to unity.
\subsection{Entanglement}

    \begin{figure}[t!]
     \subfloat[$T=0$\label{fig4.2.1a}]{%
       \includegraphics[width=0.5\textwidth]{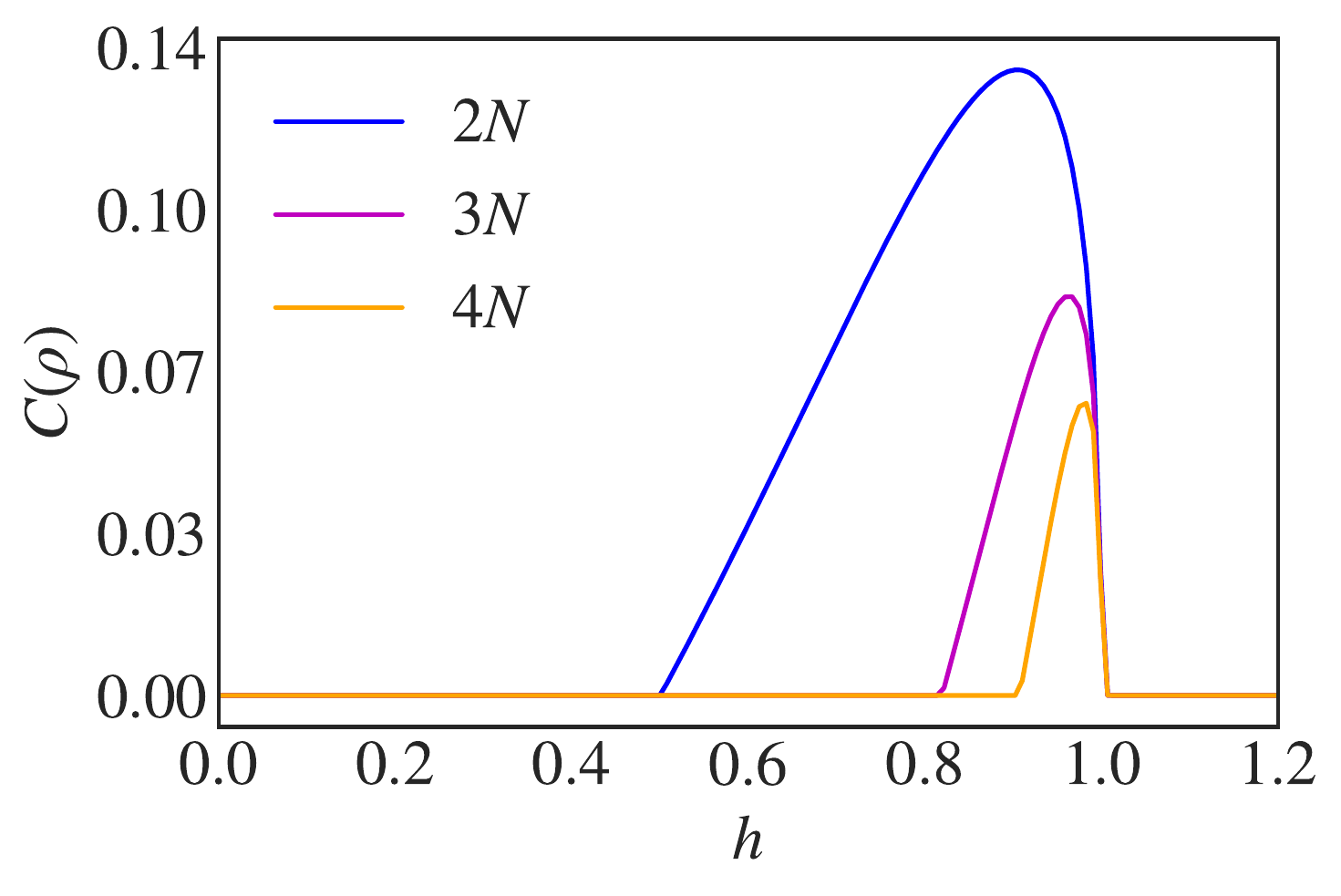}
       }
     \hfill
     \subfloat[$h=0.95$\label{fig4.2.1b}]{%
       \includegraphics[width=0.5\textwidth]{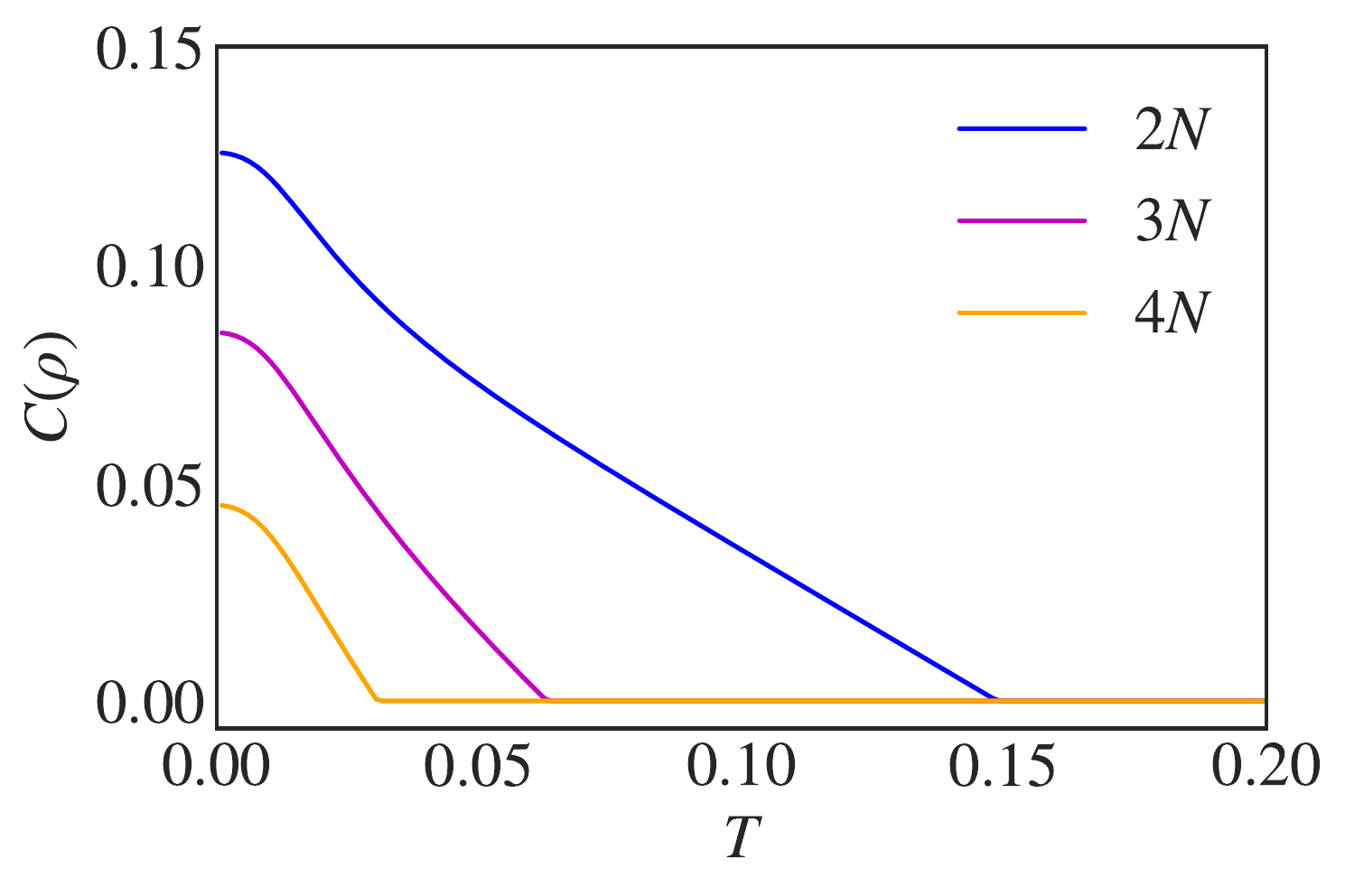}
       }
     \caption{The concurrence~\eqref{cnc_dm}, for the $2N$, $3N$ and $4N$ spin pairs. In (a) $T=0$ with respect to the magnetic field $h$ and (b) $h=0.95$ with respect to the temperature $T$.}
     \label{fig4.2.1}
   \end{figure}
We evaluate the amount and the behaviour of the entanglement in the two site density matrix $\rho_{i,i+m}$~\eqref{eq6_mat}, using the concurrence~\eqref{concurrence}. In this case, the concurrence takes the following form
    \begin{equation}
        C(\rho_{i,i+m})=\max\Big\{0,2\Big(|Z_{i,i+m}|-\sqrt{X_{i,i+m}^+X_{i,i+m}^-}\Big)\Big\},
        \label{cnc_dm}
    \end{equation}
where $Z_{i,i+m},X_{i,i+m}^+$ and $X_{i,i+m}^-$ are given by the set of equations~\eqref{eq_diag} and~\eqref{eqz}. 
Figure~\eqref{fig4.2.1a} shows the behaviour of the concurrence~\eqref{cnc_dm}, at zero temperature with respect to the magnetic field $h$, for second-, third- and fourth-nearest neighbours ($2N$, $3N$ and $4N$). None of the spins are entangled at $h=0$, and as $h$ is increased we see the creation of the entanglement at the critical field $h_c^E$ of $\frac{1}{2}$ for $m=2$, $\frac{4}{5}$ for $m=3$, and $\frac{9}{10}$ for $m=4$~\cite{fumani2018, nishimori2005_ent}. Further increase of $h$ beyond $h_c^E$ enhances the entanglement, between all the pairs until its saturation close to the quantum critical point $h_c \sim 1$, where the entanglement drops quickly and dies out.  In Figure~\eqref{fig4.2.1b}, we show the behaviour of the concurrence~\eqref{cnc_dm}, with respect to the temperature $T$, for $h=0.95$. For all the pairs, we see the entanglement start from a maximum value and drops quickly to zero with the temperature. This shows that entanglement is not resilient against thermal fluctuations. 

\begin{figure}[t!]
        \subfloat[$2N$\label{fig4.2.1_a}]{%
       \includegraphics[width=0.33\textwidth]{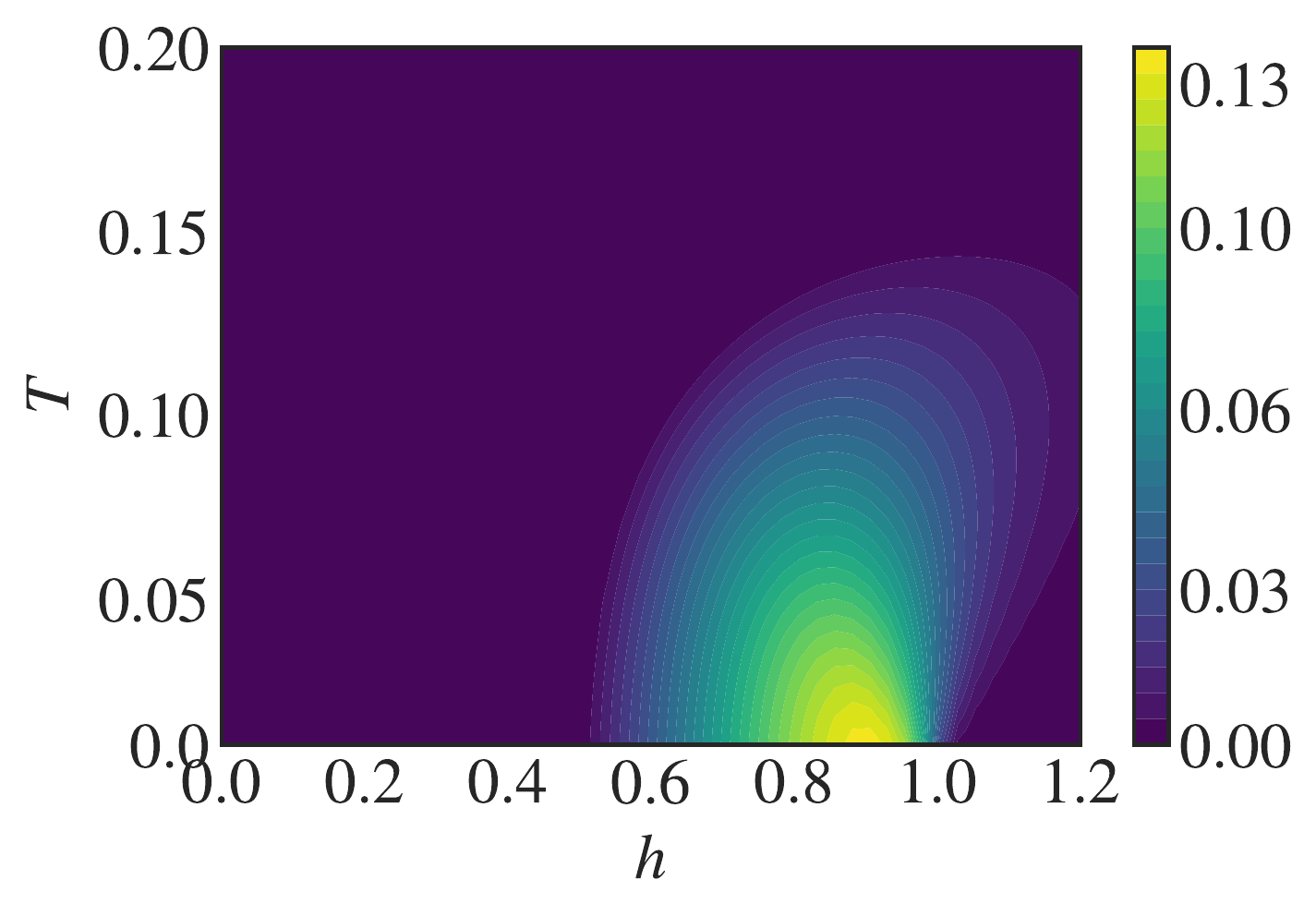}
       }%
     \subfloat[$3N$\label{fig4.2.1_b}]{%
       \includegraphics[width=0.33\textwidth]{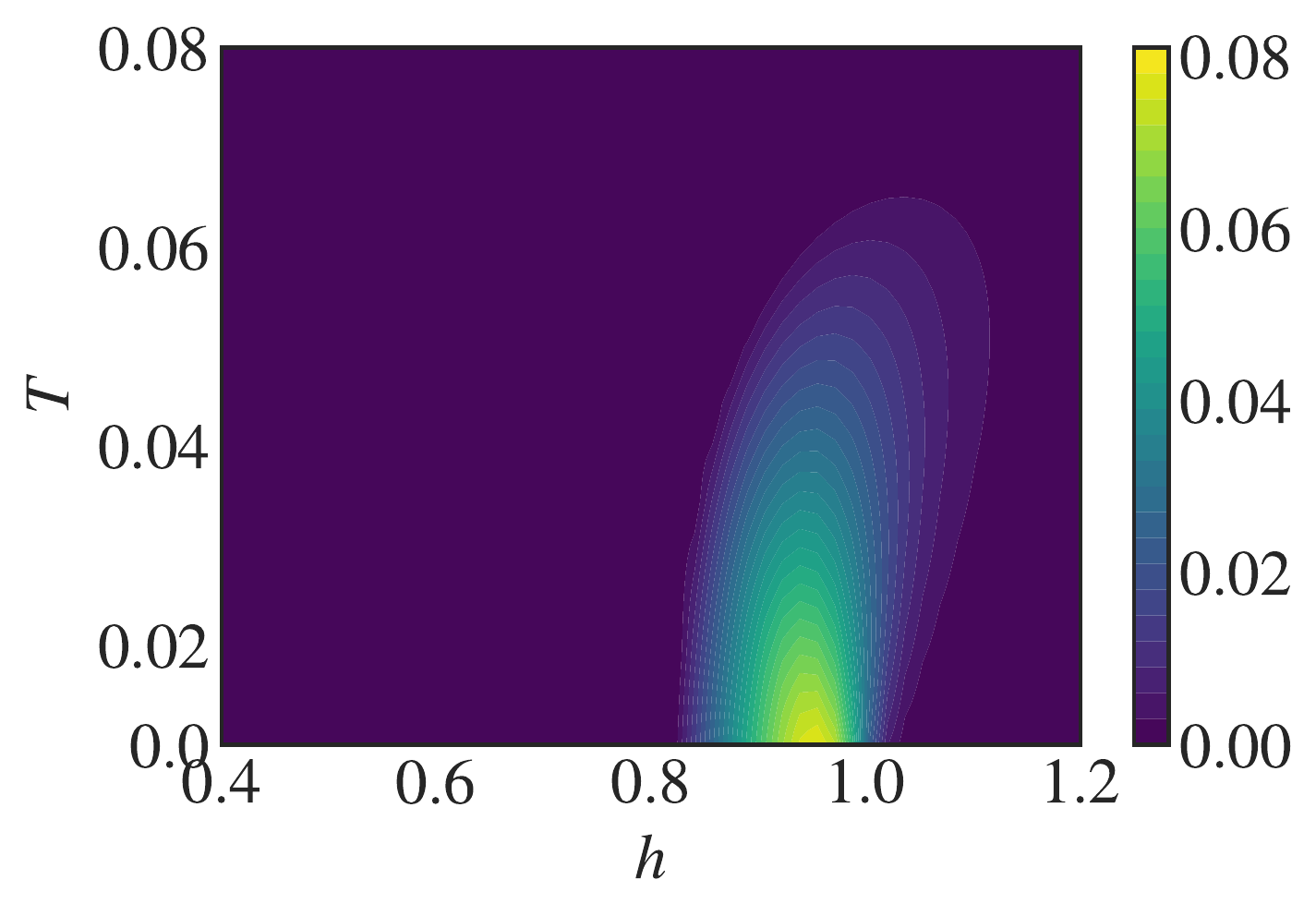}
       }%
     \subfloat[$4N$\label{ffig4.2.1_c}]{%
       \includegraphics[width=0.33\textwidth]{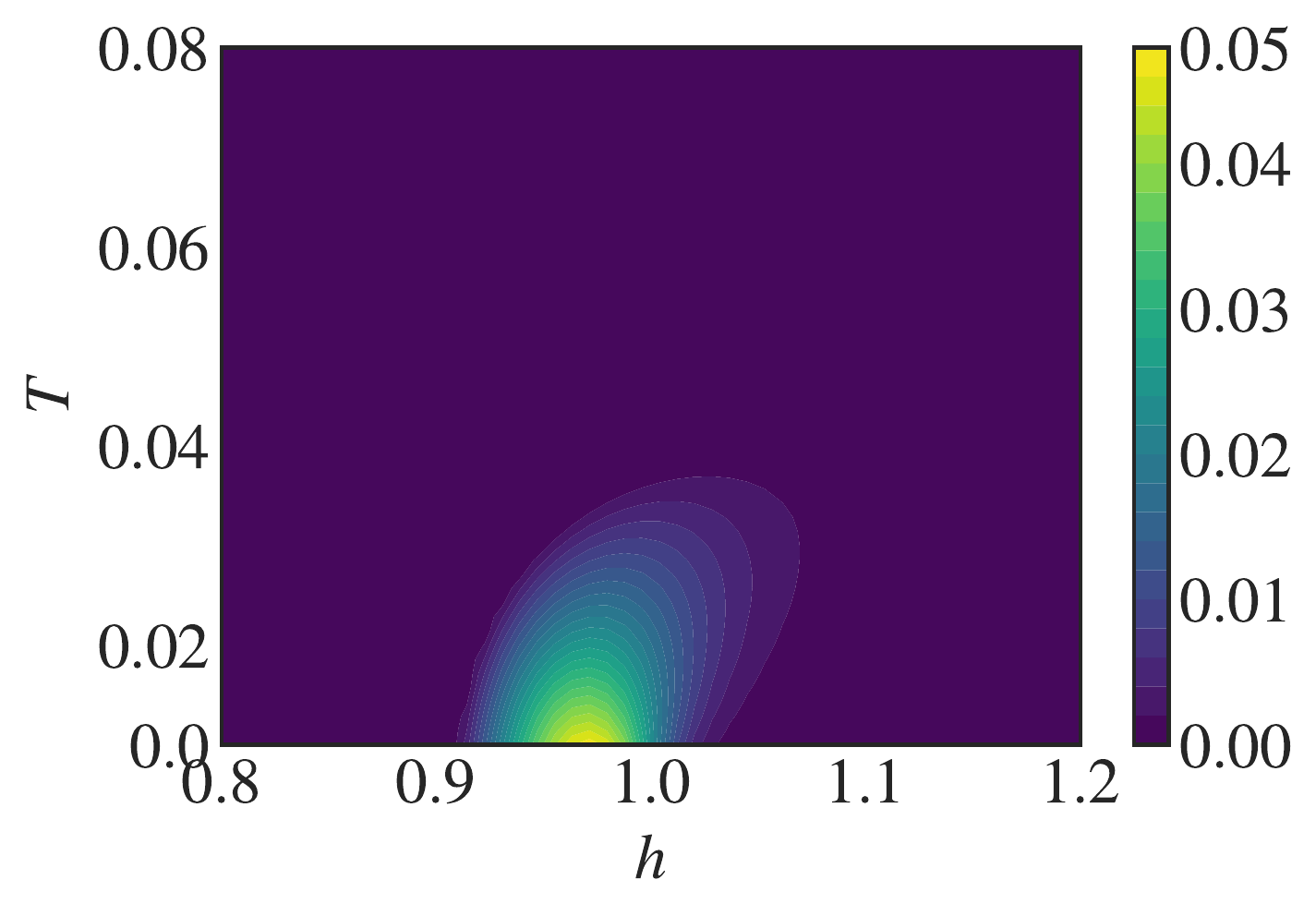}
       }
     \caption{$T-h$ phase diagram of the concurrence~\eqref{cnc_dm}, for (a) $2N$, (b) $3N$, (c) $4N$, spin pairs.}
     \label{fig4.1.1_}
   \end{figure}

\noindent The reported behaviours of the concurrence~\eqref{cnc_dm}, can be seen more clearly with the $T-h$ phase diagram depicted in Figure~\eqref{fig4.1.1_}, where we show density plots of the concurrence~\eqref{cnc_dm}, for all the spins pairs. We see that the entangled region is small in the $T-h$ plane, and decreases in size as the distance between the spins $m$ is increased. Furthermore, the entanglement is maximum around the quantum critical point. This is due to the fact that at a quantum phase transition, fluctuations occur at all length and time scales, i.e. the divergence of the correlations length $\xi$~\cite{Sachdev}.
\subsection{Quantum discord and classical correlations }
Quantum discord~\eqref{quantum_discord}, is difficult to solve due to the minimization procedure of the conditional entropy. Furthermore, obtaining an analytical expression for the classical correlations~\eqref{classical_corr}, is not an easy task for general states. However, for a special class of density matrices, called X-states, an explicit expression is available~\cite{discord_X_state}. An X-state is described by the following density matrix
\begin{equation}
    \rho_{AB}=
        \begin{pmatrix}
         \rho_{11}&0&0&\rho_{14}\\
         0&\rho_{22}&\rho_{23}&0\\
         0&\rho_{32}&\rho_{33}&0\\
         \rho_{41}&0&0&\rho_{44}
        \end{pmatrix}.
    \end{equation}.
We evaluate classical correlations ($CC$), and quantum discord ($QD$) as follows
\begin{equation}
            \begin{aligned}
            CC(\rho_{AB})=\max\{CC_1,CC_2\},\\
            QD(\rho_{AB})=\min\{QD_1,QD_2\},
        \end{aligned}
\end{equation}
where 
\begin{equation}
            \begin{aligned}
            CC_j&=H(\rho_{11}+\rho_{22})-D_j,\\
            QD_j&=H(\rho_{11}+\rho_{33})+\sum_{k=1}^4 \lambda_k \log_2(\lambda_k)+D_j,
        \end{aligned}
\end{equation}
and
\begin{equation}
            \begin{aligned}
            D_1&=H(w), D_2=-\sum_{j=1}^4 \rho_{jj} \log_2(\rho_{jj})-H(\rho_{11}+\rho_{33}),\\
            w&=\displaystyle{\frac{1+\sqrt{[1-2(\rho_{33}+\rho_{44})]^2+4(|\rho_{14}|+|\rho_{23}|)^2}}{2}},
        \end{aligned}
\end{equation}
$H(x)=-x\log_2(x)-(1-x)\log_2(1-x)$ being the binary Shannon entropy and $\lambda_k$ are the eigenvalues of the matrix $\rho_{AB}$. Then, for the two sites density matrix~\eqref{eq6_mat}, the analytical expression for $CC(\rho_{i,i+m})$ and $QD(\rho_{i,i+m})$ are given by
\begin{equation}
            \begin{aligned}
            CC_1(i,i+m)&=H(X_{i,i+m}^++Y_{i,i+m}^+)-H\Bigg(\frac{1+\sqrt{[1-2(Y_{i,i+m}^-+X_{i,i+m}^-)]^2+4|Z_{i,i+m}|^2}}{2}\Bigg),\\
            CC_2(i,i+m)&=H(X_{i,i+m}^++Y_{i,i+m}^+)+\{X_{i,i+m}^+\log_2(X_{i,i+m}^+)+Y_{i,i+m}^+\log_2(Y_{i,i+m}^+)\\&+Y_{i,i+m}^-\log_2(Y_{i,i+m}^-)+X_{i,i+m}^-\log_2(X_{i,i+m}^-)\}+H(X_{i,i+m}^++Y_{i,i+m}^-),
        \end{aligned}
\label{dm_cc}
\end{equation}
\begin{equation}
            \begin{aligned}
            QD_1(i,i+m)&=H(X_{i,i+m}^++Y_{i,i+m}^-)
            +\{X_{i,i+m}^+\log_2{X_{i,i+m}^+}+X_{i,i+m}^-\log_2{X_{i,i+m}^-}\\&+(Y_{i,i+m}^+-|Z_{i,i+m}|)\log_2{(Y_{i,i+m}^+-|Z_{i,i+m}|)}+(Y_{i,i+m}^-+|Z_{i,i+m}|)\log_2{(Y_{i,i+m}^-+|Z_{i,i+m}|)}\}\\&
            +H\Bigg(\frac{1+\sqrt{[1-2(Y_{i,i+m}^-+X_{i,i+m}^-)]^2+4|Z_{i,i+m}|^2}}{2}\Bigg),\\
            QD_2(i,i+m)&=H(X_{i,i+m}^++Y_{i,i+m}^-)
            +\{X_{i,i+m}^+\log_2{X_{i,i+m}^+}+X_{i,i+m}^-\log_2{X_{i,i+m}^-}\\&+(Y_{i,i+m}^+-|Z_{i,i+m}|)\log_2{(Y_{i,i+m}^+-|Z_{i,i+m}|)}+(Y_{i,i+m}^-+|Z_{i,i+m}|)\log_2{(Y_{i,i+m}^-+|Z_{i,i+m}|)}\}\\&
            -\{X_{i,i+m}^+\log_2(X_{i,i+m}^+)+Y_{i,i+m}^+\log_2(Y_{i,i+m}^+)+Y_{i,i+m}^-\log_2(Y_{i,i+m}^-)+X_{i,i+m}^-\log_2(X_{i,i+m}^-)\}\\&
            -H(X_{i,i+m}^++Y_{i,i+m}^-),
        \end{aligned}
\label{dm_qd}
\end{equation}
where $X_{i,i+m}^+$,$X_{i,i+m}^-$,$Y_{i,i+m}^+$,$Y_{i,i+m}^-$ and $Z_{i,i+m}$ are given by the set of equations~\eqref{eq_diag} and~\eqref{eqz}.
\begin{figure}[t!]
     \subfloat[$T=0$\label{fig4.2.2a}]{%
       \includegraphics[width=0.5\textwidth]{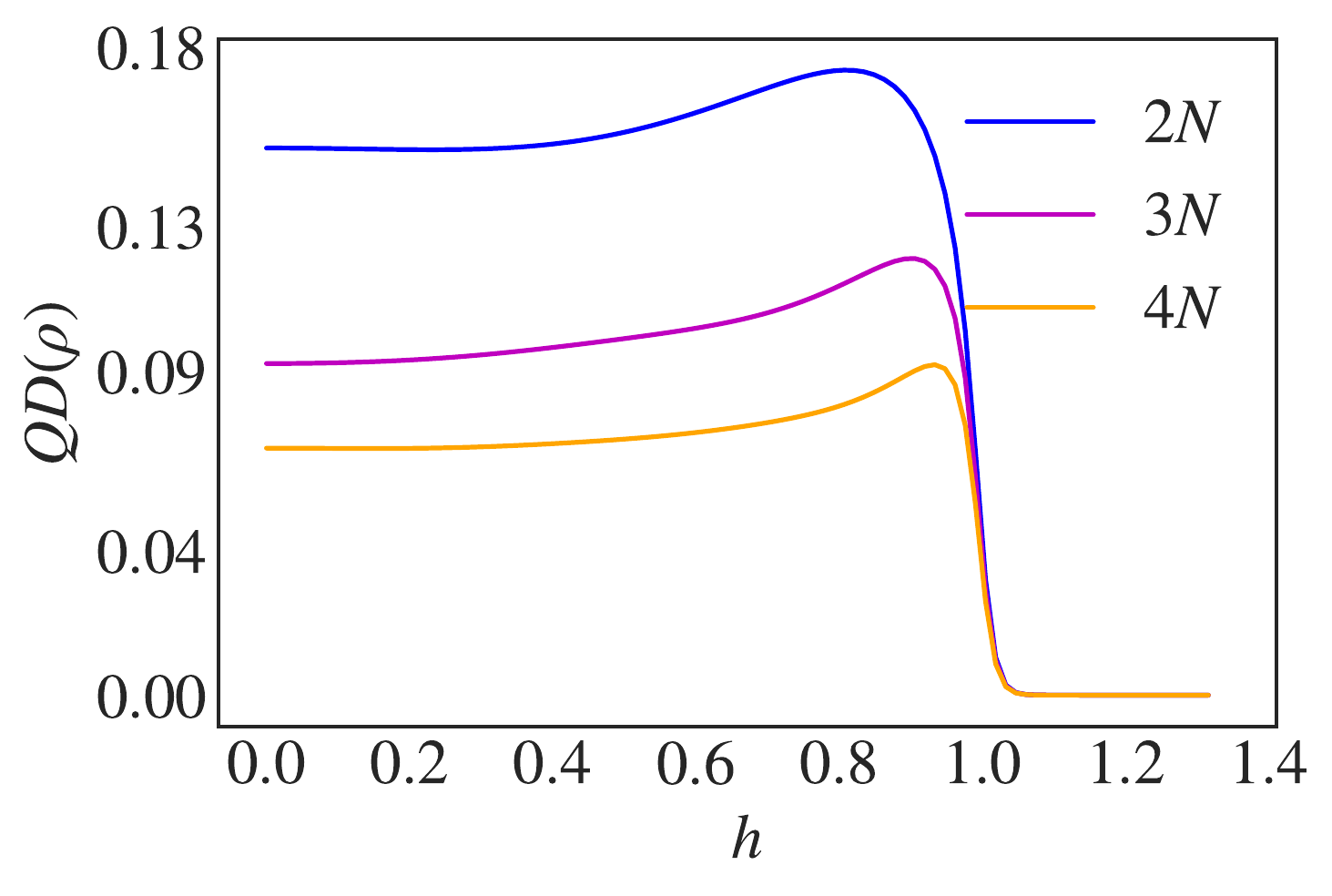}
       }
     \hfill
     \subfloat[$h=0.0$\label{fig4.2.2b}]{%
       \includegraphics[width=0.5\textwidth]{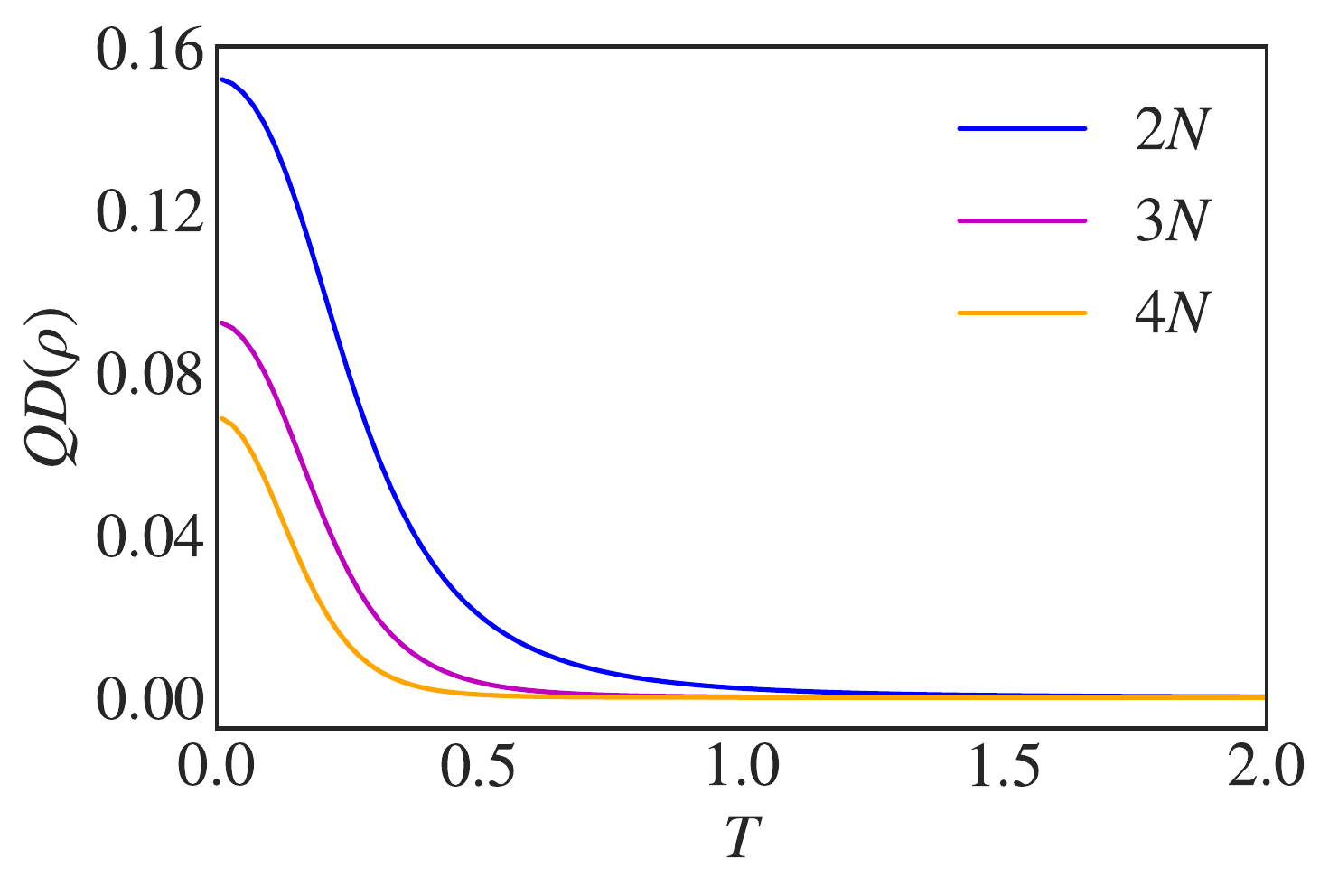}
       }
     \caption{Quantum discord~\eqref{dm_qd}, for the $2N$, $3N$ and $4N$ spin pairs. In (a) $T=0$ with respect to the magnetic field $h$ and (b) $h=0$ with respect to the temperature $T$. }
     \label{fig4.2.1}
\end{figure}
\noindent Figure~\eqref{fig4.2.2a} shows the behaviour of quantum discord~\eqref{dm_qd}, at zero temperature with respect to the magnetic field $h$, for second-, third- and fourth-nearest neighbours. Unlike the concurrence, quantum discord is non-zero for $h<1$, and saturates around the quantum critical $h_c \sim 1$ where it tends to zero \textit{asymptotically}. In general, quantum discord originates from the coherence that arises from quantum superposition, which exists in the subsystems of a quantum system and persists even if the system is in a product state, while the amount of entanglement in a separable state is always zero. 

\noindent The effect of temperature on quantum discord~\eqref{dm_qd}, at zero magnetic field is reported in Figure~\eqref{fig4.2.2b}. We see a similar behavior to that of the entanglement, c.f. Figure~\eqref{fig4.2.1b}. The difference is that quantum discord is more robust to thermal fluctuations, which do not kill the quantumness of system, as quantum discord stay asymptotically around 0, even for high values of $T$. Only decoherence makes quantum discord vanish.

   \begin{figure}[b!]
        \subfloat[$2N$\label{fig4.2.2_a}]{%
       \includegraphics[width=0.33\textwidth]{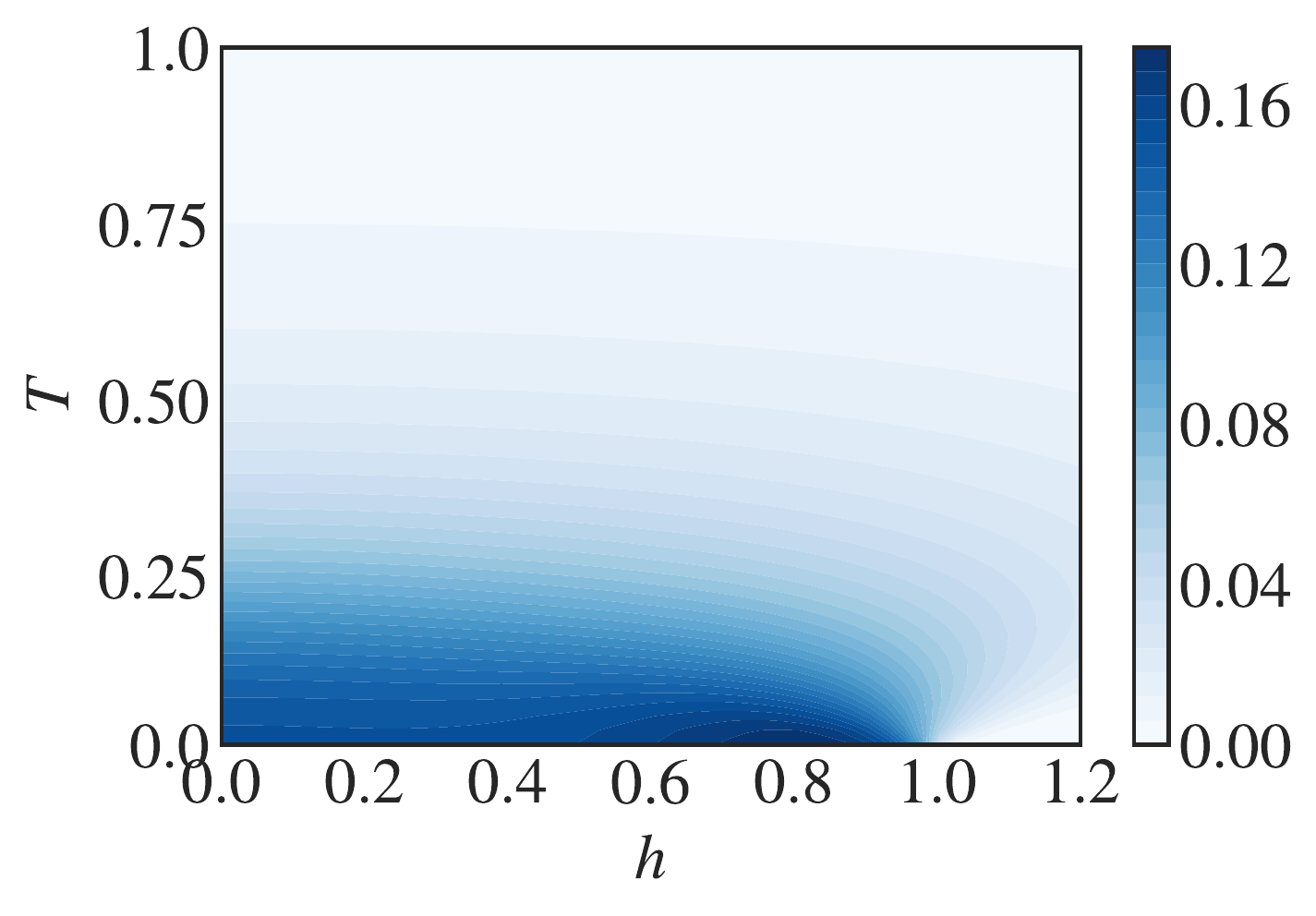}
       }%
     \subfloat[$3N$\label{fig4.2.2_b}]{%
       \includegraphics[width=0.33\textwidth]{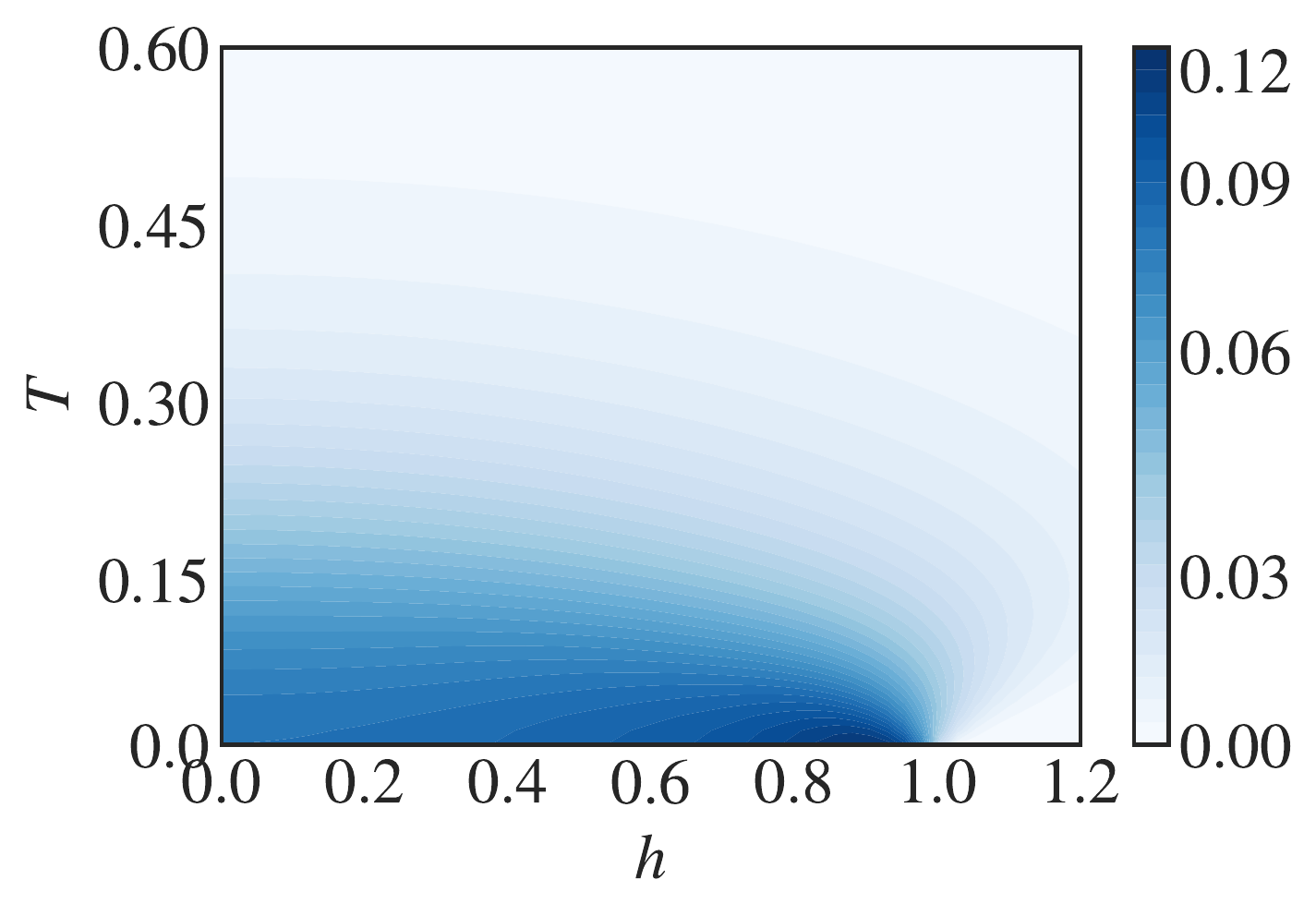}
       }%
     \subfloat[$4N$\label{fig4.2.2_c}]{%
       \includegraphics[width=0.33\textwidth]{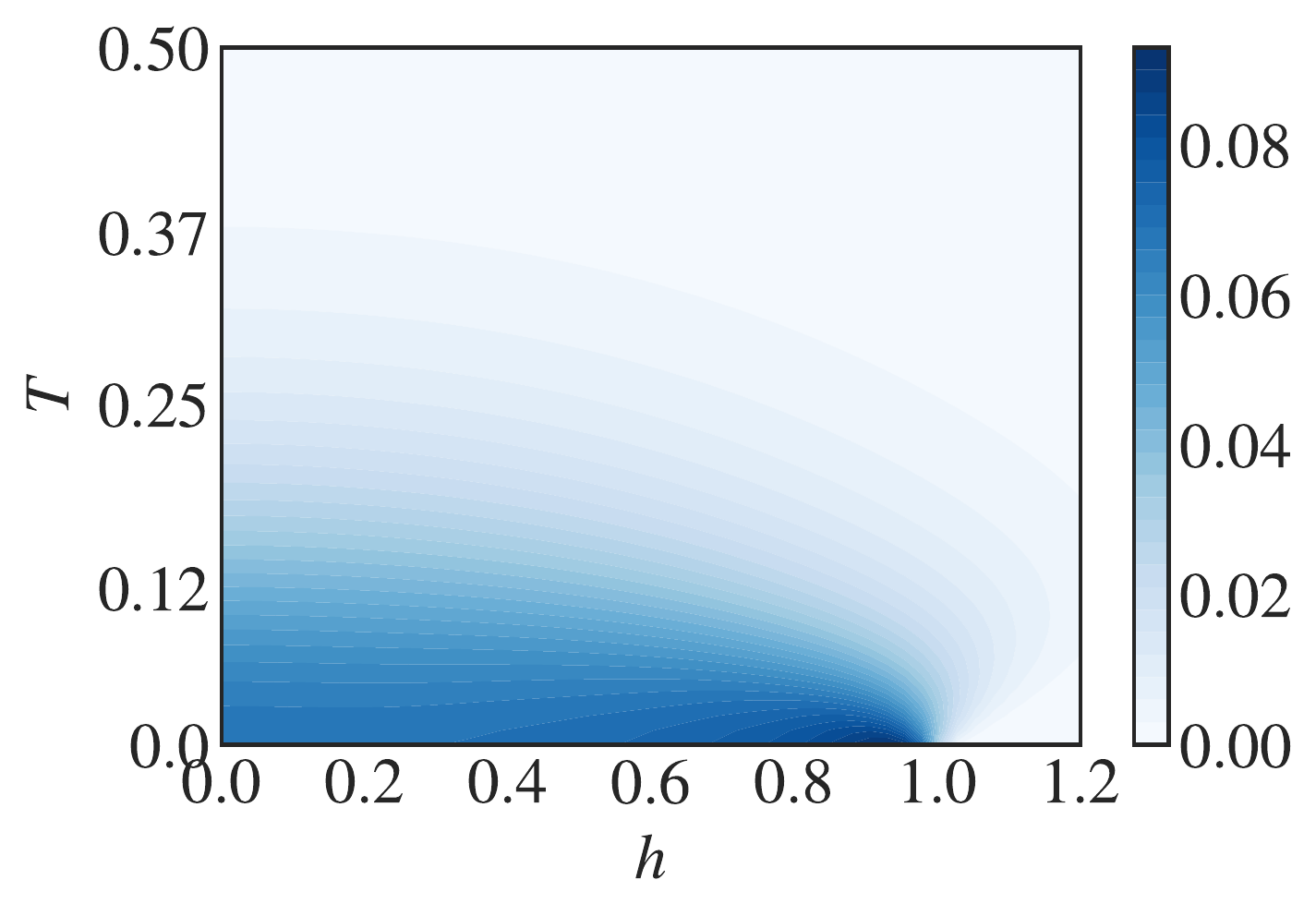}
       }
     \caption{$T-h$ phase diagram of quantum discord ~\eqref{dm_qd}, for (a) $2N$, (b) $3N$, (c) $4N$, spin pairs.}
     \label{fig4.1.2_}
   \end{figure}
\noindent The $T-h$ phase diagram of quantum discord~\eqref{quantum_discord}, is reported in Figure~\eqref{fig4.1.2_}. For all the spin pairs, we see that quantum discord persists for all values of temperature and magnetic field. Only its magnitude reduces as the distance between the spins $m$ increases. 

\begin{figure}[t!]
        \subfloat[$T=0$\label{fig4.2.2__a}]{%
       \includegraphics[width=0.49\textwidth]{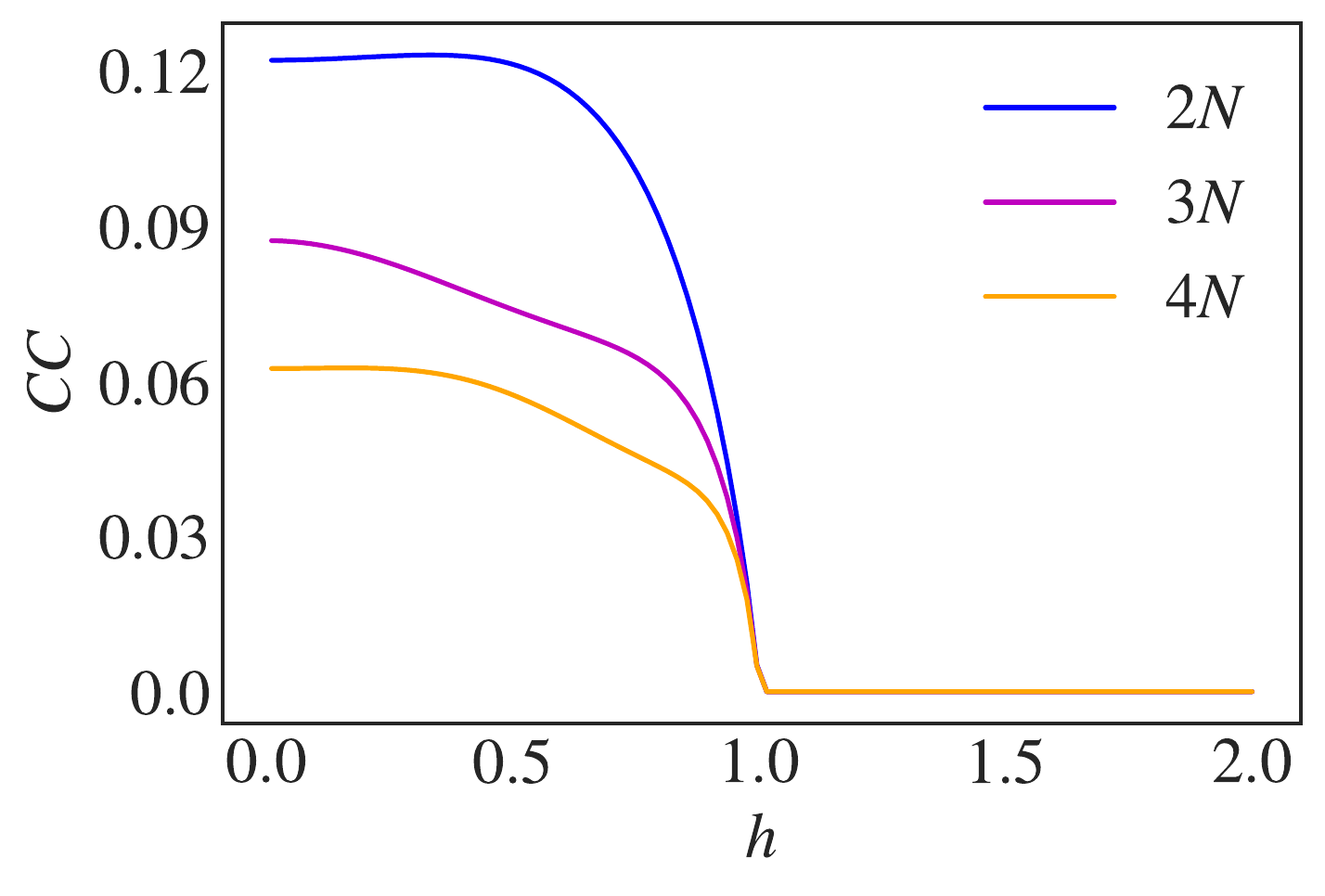}
       }%
     \subfloat[$h=0$\label{fig4.2.2__b}]{%
       \includegraphics[width=0.49\textwidth]{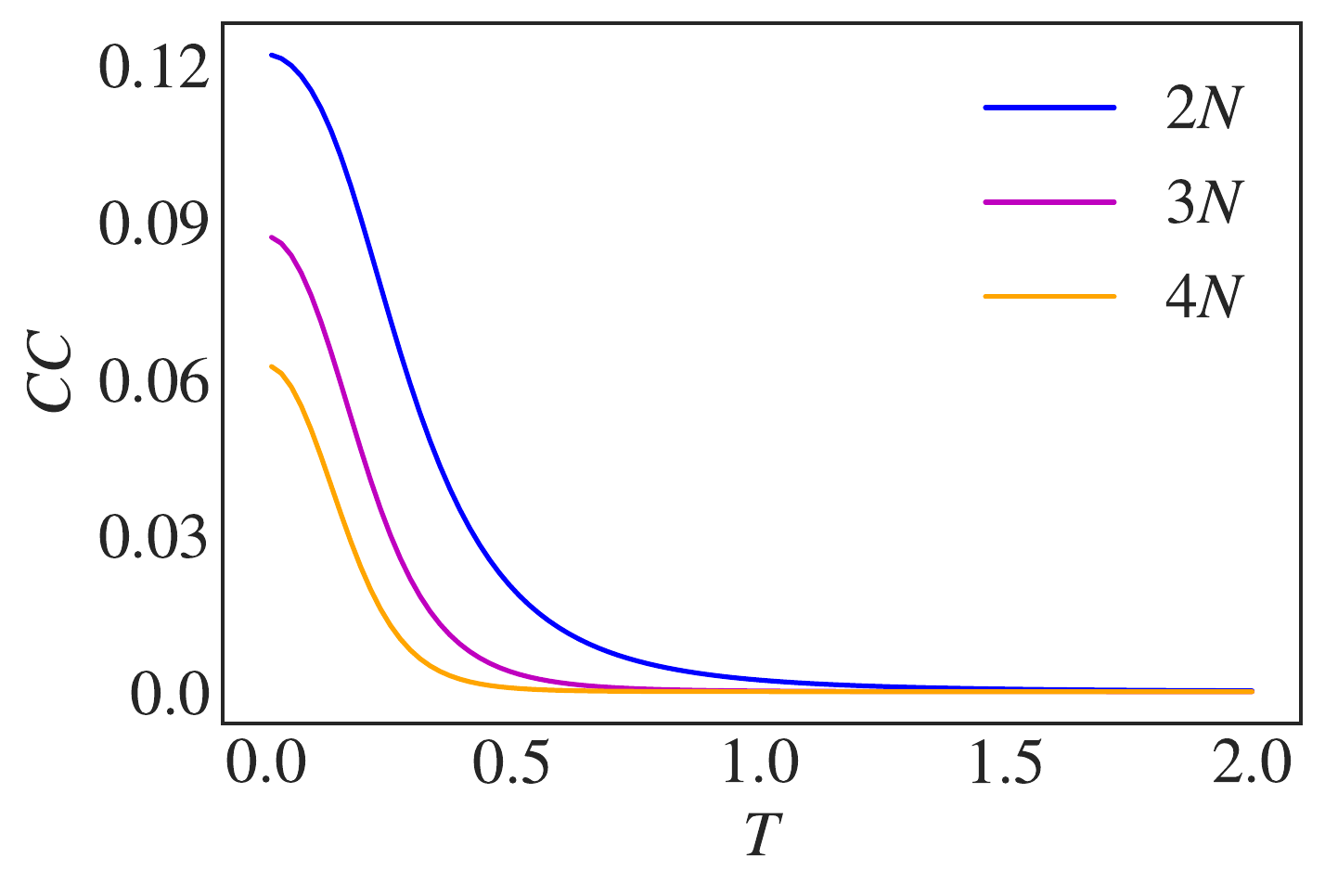}
       }
     \caption{Classical correlation~\eqref{dm_cc}, for all the spins pairs. In (a) $T=0$ with respect to the magnetic field, and (b) $h=0$ with respect to the temperature $T$.}
     \label{fig4.1.2__}
\end{figure}

\noindent The behaviour of classical correlations~\eqref{dm_cc}, with respect to the magnetic field and the temperature is presented, respectively, in Figure~\eqref{fig4.2.2__a} and Figure~\eqref{fig4.2.2__b}. We report a similar behaviour to quantum discord, c.f. Figure~\eqref{fig4.2.1}. The difference arises only around the quantum critical point $h_c \sim 1$, where quantum discord saturates while classical correlations continue to decrease in this point. Furthermore, we report the dominance of quantum correlations over classical correlations, while in the transverse Ising model the inverse phenomenon takes place~\cite{Maziero2010}.

\subsection{Quantum Coherence}
We evaluate the quantum Jensen-Shannon divergence~\eqref{QJSD}, for the two-site density matrix~\eqref{eq6_mat}. In terms of $Y_{i,i+m}^+$,$Y_{i,i+m}^-$ and $Z_{i,i+m}$ given by the set of equations~\eqref{eq_diag} and~\eqref{eqz}, the analytical expression of quantum coherence in the $XX$ model~\eqref{eq1}, is given by

\begin{equation}
    \begin{aligned}
            \mathcal{QC}(\rho_{i,i+m})&=\Bigg[\Big\{
            -\Big(Y_{i,i+m}^+-\frac{|Z_{i,i+m}|}{2}\Big)\log_2{\Big(Y_{i,i+m}^+-\frac{|Z_{i,i+m}|}{2}\Big)}-\Big( Y_{i,i+m}^-+\frac{|Z_{i,i+m}|}{2}\Big)\\&\log_2{\Big(Y_{i,i+m}^-+\frac{|Z_{i,i+m}|}{2}\Big)}\Big\}-\frac{1}{2}\Big\{-(Y_{i,i+m}^+-|Z_{i,i+m}|)\log_2{(Y_{i,i+m}^+-|Z_{i,i+m}|)}\\&-(Y_{i,i+m}^-+|Z_{i,i+m}|)\log_2{(Y_{i,i+m}^-+|Z_{i,i+m}|)} \Big\}-\frac{1}{2}\Big\{-Y_{i,i+m}^+\log_2{Y_{i,i+m}^+}\\& -Y_{i,i+m}^-\log_2{Y_{i,i+m}^-}\Big\}\Bigg]^{\displaystyle{{\frac{1}{2}}}}.
\end{aligned}
\label{QJSD_qc}
\end{equation}

\begin{figure}[t!]
     \subfloat[$T=0$\label{fig4.2.3a}]{%
       \includegraphics[width=0.5\textwidth]{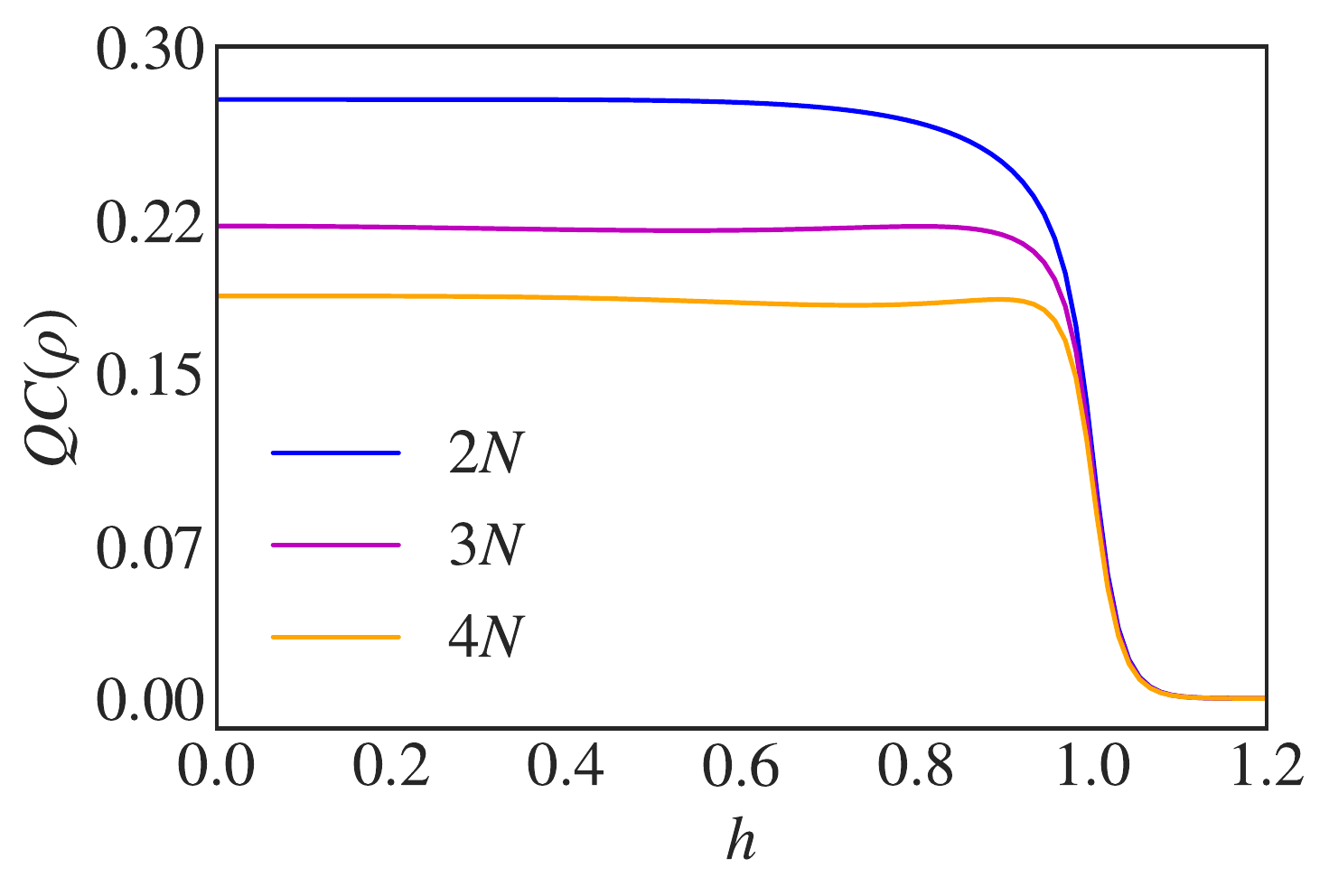}
       }
     \hfill
     \subfloat[$h=0.0$\label{fig4.2.3b}]{%
       \includegraphics[width=0.5\textwidth]{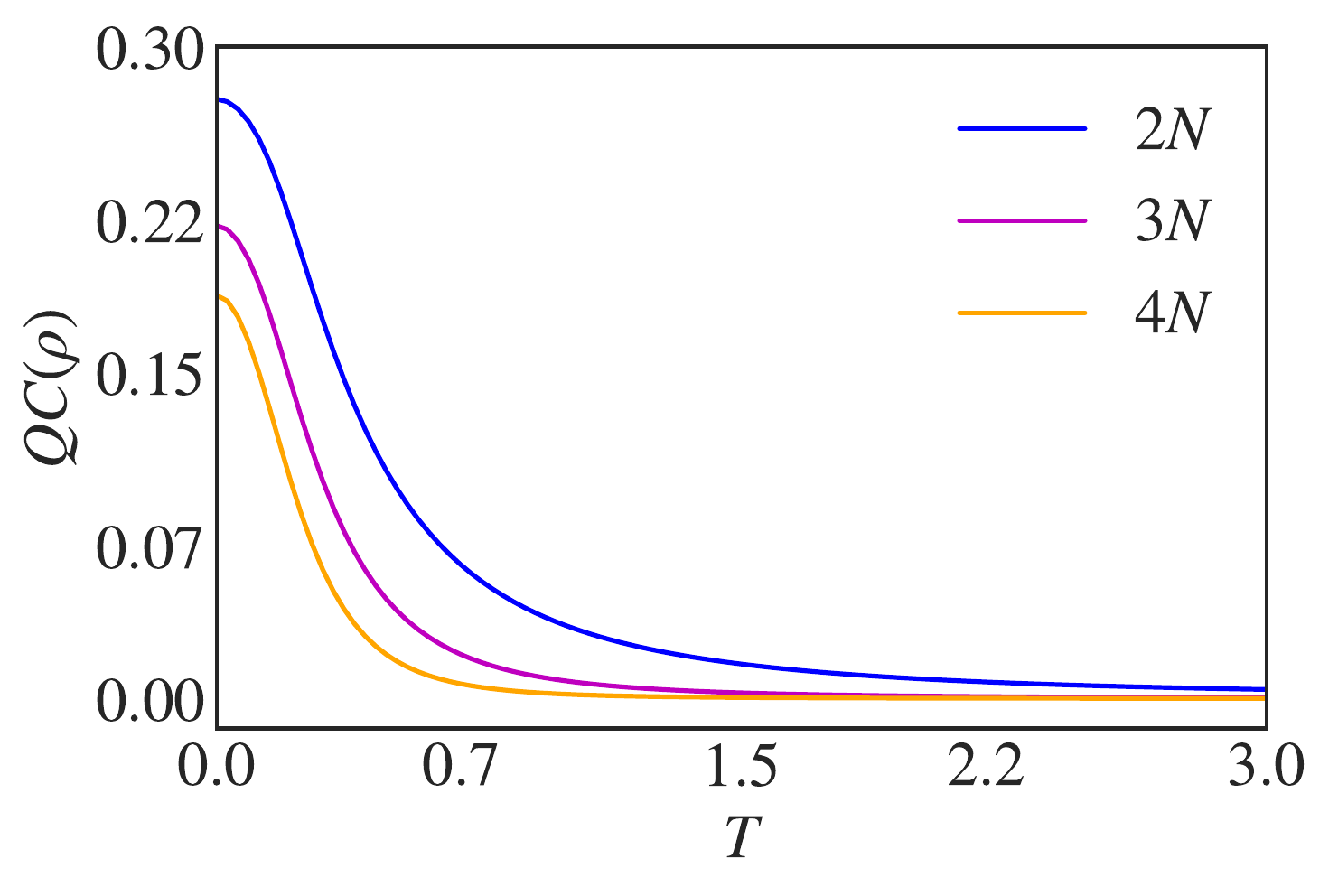}
       }
     \caption{Quantum coherence~\eqref{QJSD_qc}, for all the spins pairs. In (a) $T=0$ with respect to the magnetic field, and (b) $h=0$ with respect to the temperature $T$.}
     \label{fig4.2.3}
\end{figure}

\begin{figure}[t!]
        \subfloat[$2N$\label{fig4.2.2_a}]{%
       \includegraphics[width=0.33\textwidth]{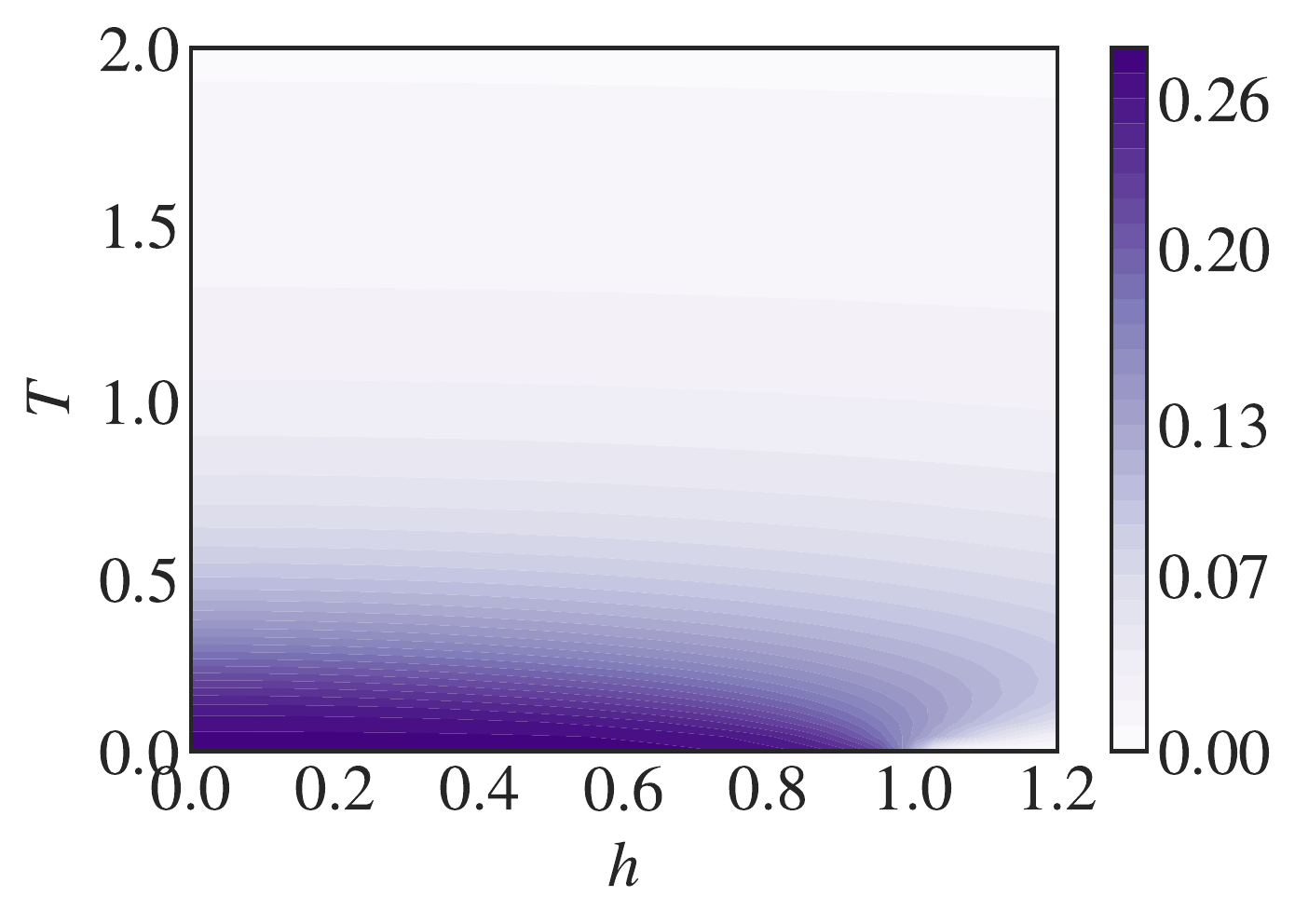}
       }%
     \subfloat[$3N$\label{fig4.2.2_b}]{%
       \includegraphics[width=0.33\textwidth]{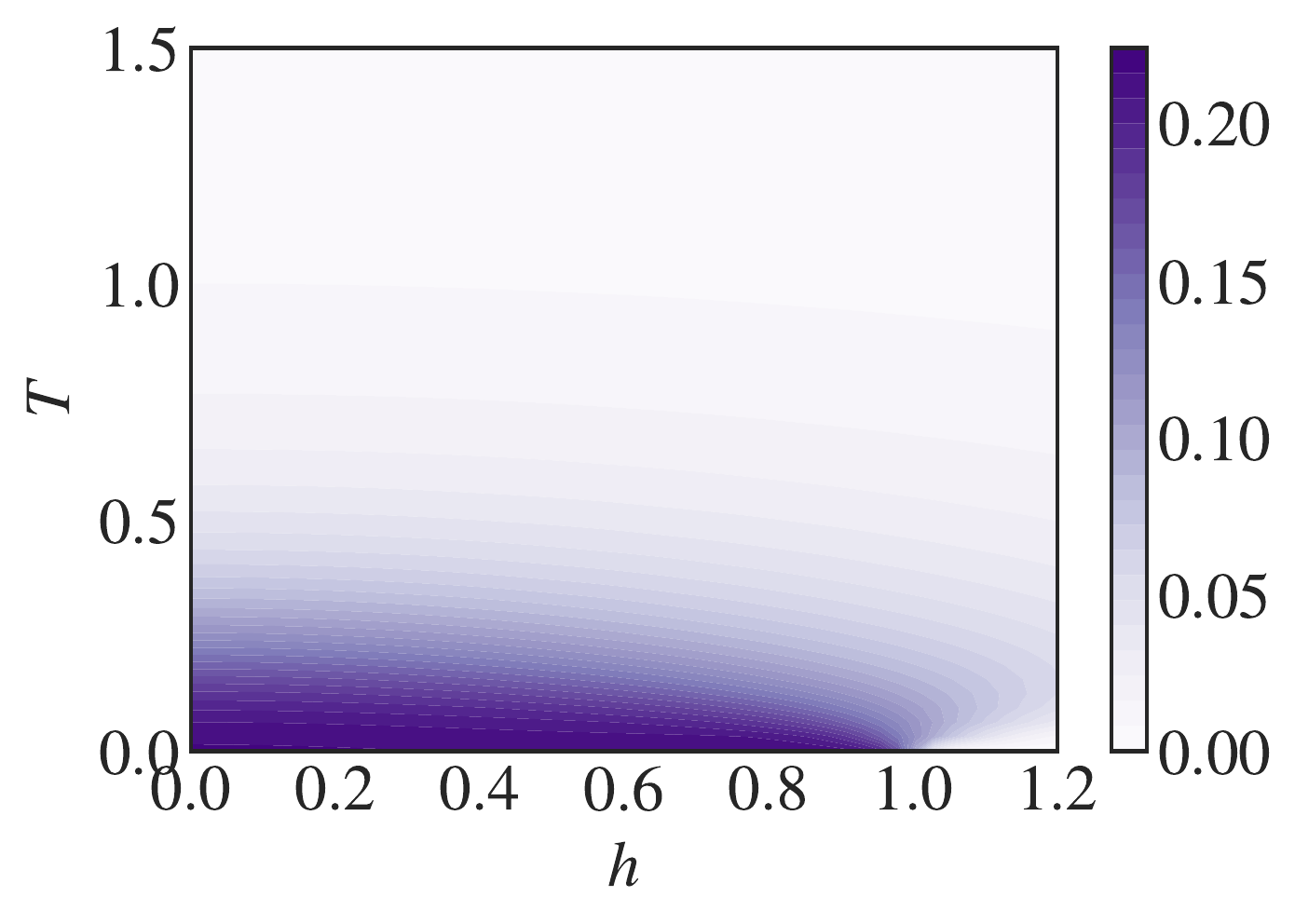}
       }%
     \subfloat[$4N$\label{ffig4.2.2_c}]{%
       \includegraphics[width=0.33\textwidth]{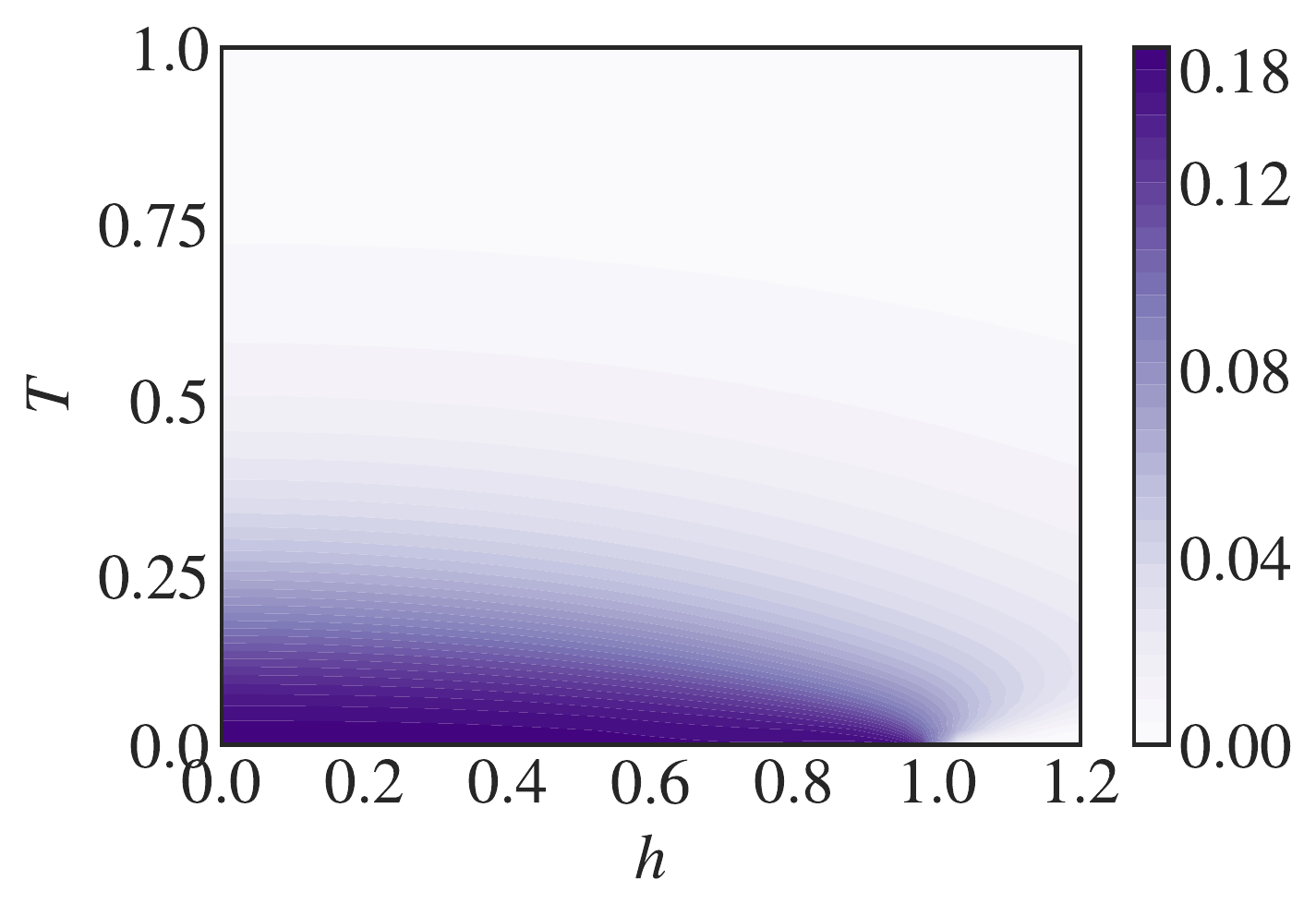}
       }
     \caption{$T-h$ phase diagram of quantum coherence~\eqref{QJSD_qc}, for (a) $2N$, (b) $3N$, (c) $4N$, spin pairs.}
     \label{fig4.2.3_}
   \end{figure}

\noindent The behaviour of quantum coherence~\eqref{QJSD_qc}, with respect to the magnetic field (temperature) at $T=0$ $(h=0)$ is shown in Figure~\eqref{fig4.2.3a} (Figure~\eqref{fig4.2.3b}), which is very similar to quantum discord. This is due to that fact that coherence stems from individual sites (local coherence) or may be distributed along sites (intrinsic coherence)~\cite{coherencePRL}  and, one can show that due to spin-flip symmetry of the $XX$ chain~\eqref{eq1}, the one site density matrix is diagonal which means that the local coherence is zero for the $XY$ model. Hence, the total coherence observed in the system will entirely originate from the correlations between the two spins, and since quantum discord represents total quantum correlations, the behavior of quantum coherence will resemble  that of quantum discord. It is worthwhile noting that in some models (e.g. Heisenberg spin models with Dzyaloshinsky-Moriya interactions~\cite{coherence_india1,coherence_india2}) the inverse behavior is observed, and the behavior of quantum coherence is reminiscent of entanglement not quantum discord. Moreover, it is argued that quantum discord can be understood from the discrepancy between the relative quantum coherence for the total system and that of the subsystem chosen for the calculation of quantum discord. We confirm the similarity of quantum coherence and quantum discord by the $T-h$ phase diagram in Figure~\eqref{fig4.2.3_}.
\section{Quantum phase transitions}
\begin{figure}[t!]
     \subfloat[Concurrence\label{fig4.3a}]{%
       \includegraphics[width=0.33\textwidth]{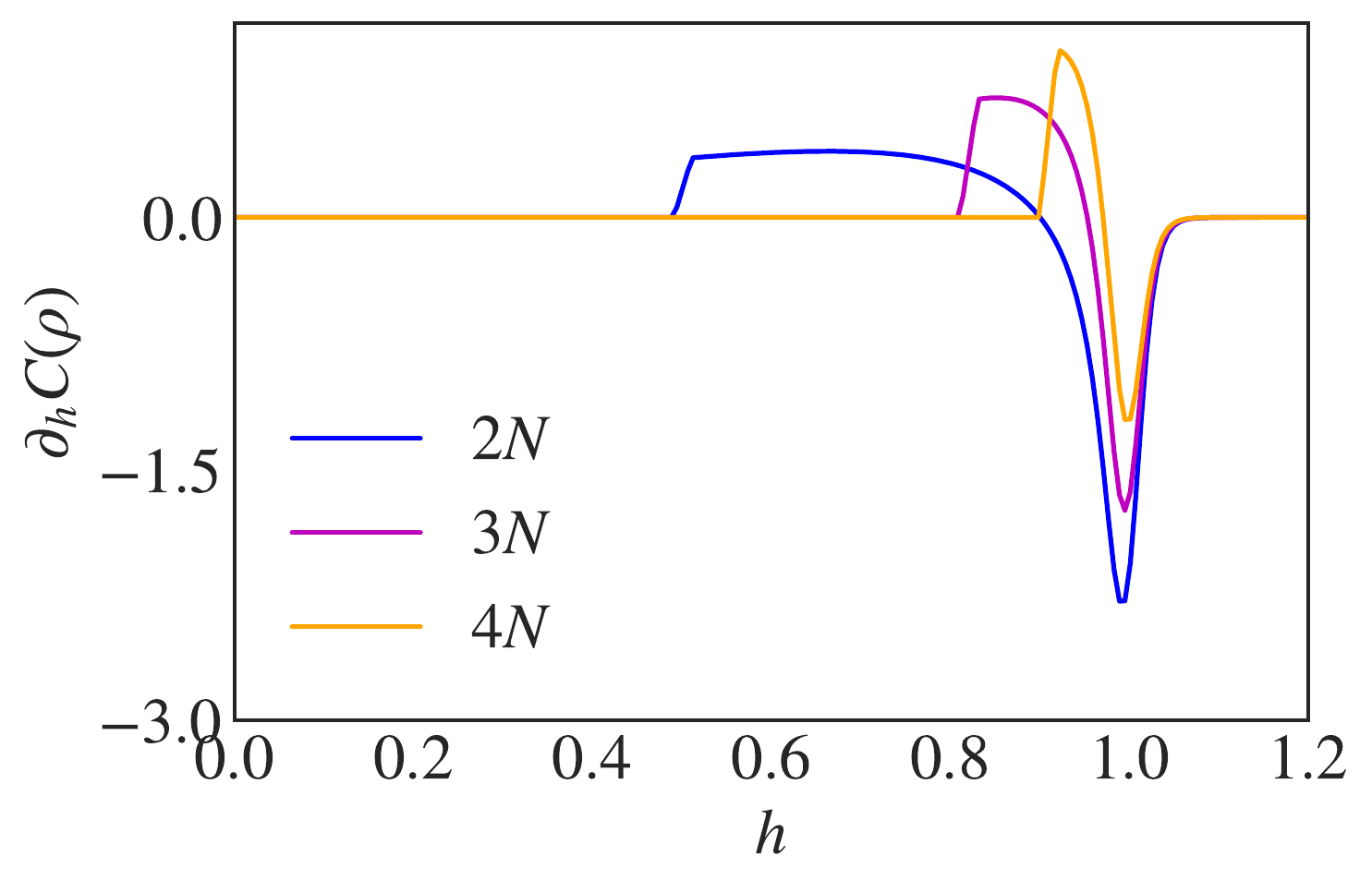}
       }%
     \subfloat[Quantum discord\label{fig4.3b}]{%
       \includegraphics[width=0.33\textwidth]{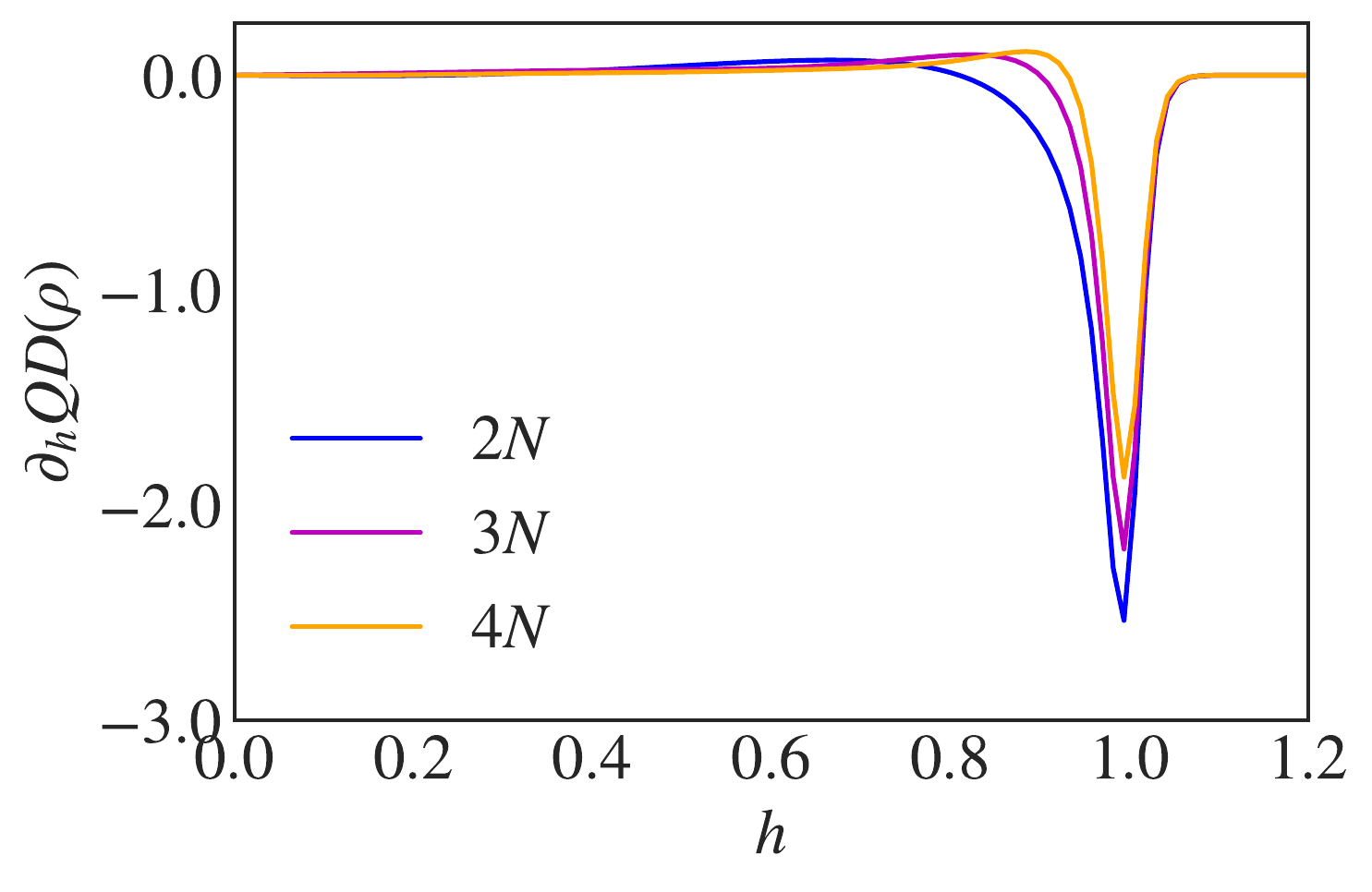}
       }%
     \subfloat[Quantum coherence\label{ffig4.3c}]{%
       \includegraphics[width=0.33\textwidth]{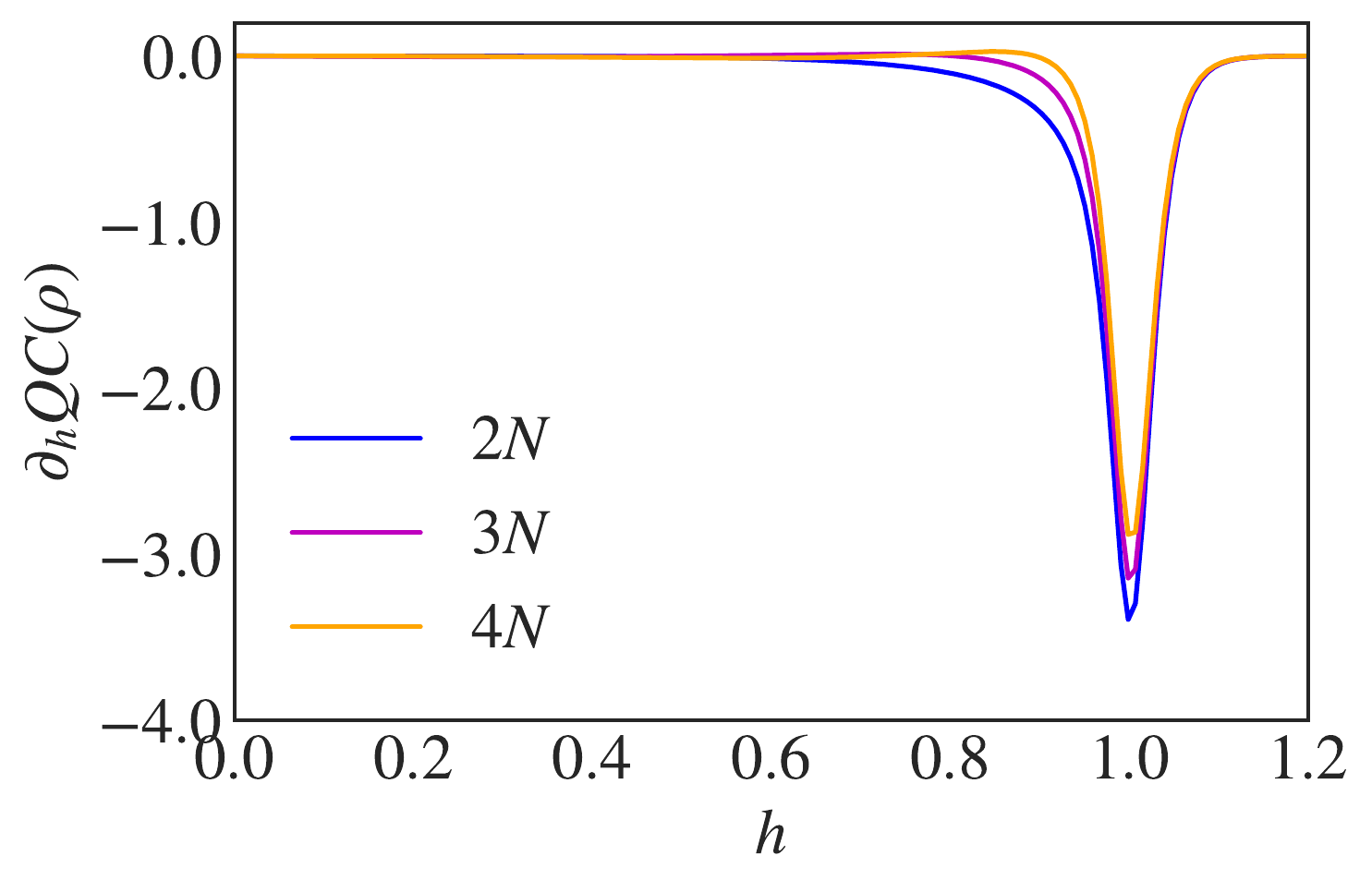}
       }
     \caption{First derivative of (a) the concurrence~\eqref{cnc_dm} , (b) quantum discord~\eqref{dm_cc}, and (c) quantum coherence~\eqref{QJSD_qc}, with respect to the magnetic field $h $ at zero temperature for the $2N$, $3N$ and $4N$ spin pairs.}
     \label{fig4.3}
\end{figure}
The $XX$ model undergoes, in the thermodynamic limit, a second-order quantum phase transition at the critical field $h_c=1$, where the system transits from Mott-insulator phase to superfluid phase or vice versa. The ground state of the Mott-insulator phase is a separable state while the ground state of superfluid phase is argued to be an entangled state~\cite{zhang2012classification}. Moreover, quantum phase transition are followed by a specific change in the nature of quantum correlations in the ground state of the system. This is confirmed by the results of the previous sections, where we witnessed the saturation of entanglement, quantum discord and quantum coherence at the vicinity of the quantum critical point and their quick drop to zero, as the system crosses to the other phase.

\noindent Second-order quantum phase transitions appear as singularities in the ground state of the quantum system, which manifest itself as divergences in the derivatives of thermodynamical quantities, i.e. free energy. The situation is similar for quantum correlation measures, taking the first derivative of the concurrence~\eqref{cnc_dm}, quantum discord~\eqref{dm_qd}, and quantum coherence~\eqref{QJSD_qc}, reveal the presence of the quantum critical point. This is shown in Figure~\eqref{fig4.3}, where we plot the first derivative of the studied figures of merit, for all the spins pairs, and we see that all the measures diverge at the quantum critical point $h_c=1$. Thus, quantum correlations measures can successfully reveal the critical properties of the isotropic $XY$ model.
\section{Summary}
\noindent Through this chapter, we presented a comparative study between various types of pairwise correlations that may emerge in a quantum spin system, such as: entanglement, quantum discord, classical correlations and quantum coherence. The analytical expressions of the pairwise quantities were provided using the Jordan–Wigner transformation, which we verified numerically, and the ``Pauli basis expansion''. Then, we showed the dominance of quantum correlations over their classical counterpart, the potential use of quantum discord in quantum information processing being robust against the temperature, and the role of quantum entanglement, quantum discord and quantum coherence in detecting the second-order quantum phase transition in the $XX$ model. Finally, we focused on quantum coherence, an essential feature of quantum mechanics which we proved to be a long range quantity that outperforms entanglement and quantum discord in the robustness to thermal perturbations. The present study shows that quantum correlations measures are very promising candidates in revealing the salient features of quantum-many body systems. In the next part, we analyze another approach to represent and detect quantum critical phenomena in quantum phase space.

\part{Quantum Phase Transitions in Phase Space}

\chapter{Phase Space Formulation of Quantum Mechanics \label{chap4}}
Quantum theory proved to be very successful in describing the features of nano-scale systems, e.g. determination of energy levels of atoms and calculation of transition probabilities. However, a unified interpretation of quantum mechanics is still under debate, due to the existence of many different formulations~\cite{quantum_formulation}. Heisenberg's, Schr\"odinger's and Feynmann's formulations, to cite a few. Additionally, quantum theory can be represented using the density matrix, the second quantization and the variational approach. Finally, the pilot wave, the Hamilton-Jacobi and the Wigner formulation are also valid reformulations of quantum mechanics. Each of these formulations has its own advantages and most of them are equivalent. Nevertheless, the different formulations of quantum theory are often based on distinct views of reality.

\noindent In this chapter, we focus on a special formulation of quantum theory, which is based on the Wigner function~\cite{wigner_review_1}. The approach is different in character, as it is a formulation of quantum theory in phase space. As a consequence, it is not equivalent to most of the other formulations of quantum mechanics. The Wigner formalism, being accessible experimentally, has received prominent attention in the field of quantum optics and quantum information theory, in order to explore the features of quantum systems. We start from the formulation of the Wigner function for infinite-dimensional systems and we discuss its features and its experimental implementations. The Wigner function can take negative values, as a consequence it is a quasi-probability distribution. Therefore, one of the goals of this chapter is to discuss the interpretation of negative probabilities and their interplay with non-classicality. Additionally, we focus also on the extension of the Wigner formalism to finite-dimensional quantum systems, its features and its experimental implementations~\cite{wigner_review_2}.
\section{The Wigner function}
The phase space formulation of quantum mechanics dates back to 1932, when Eugene Wigner developed the inverse of the Weyl transformation, which takes a Hamiltonian in phase space and maps it into a quantum mechanical operator.  The Wigner function does the opposite, as it takes a quantum mechanical operator and represents it in phase space~\cite{wigner1932}. To complete the statistical aspect of quantum theory, Moyal and Gronewold pushed the development of the Wigner function further in the 40's, by describing the evolution of quantum systems in phase space, and relating the Wigner function to the expectation value of any quantum operator~\cite{moyal_1949,GROENEWOLD}.

\noindent According to the laws of classical physics, the state of a system can be described by a point in a $6N$ dimensional phase space, where the axes are the position $x$ and momentum $p$. In classical physics, the position and momentum commute, hence we can determine the position and momentum of a particle simultaneously. In contrast, the uncertainty principle in quantum mechanics forbids the measurement of $x$ and $p$ at the same time. Therefore, the position $P(x)$ and momentum $Q(x)$ follow probability densities

\begin{align}
    P(x)&=| \psi (x)|^2,\\
    Q(k)&=|\phi (k)|^2,
\end{align}
where $\psi(x)$ and $\phi(k)$ represent wave-functions and are related via the Fourier transform:
\begin{equation}
    \phi(k)=\frac{1}{\sqrt{2\pi}} \int_{-\infty}^{+\infty} \psi(x)e^{-ikx} dx.
\end{equation}
To adequately describe quantum mechanics in phase space, it is desirable to have a single function with both probabilities $P(x)$ and $Q(k)$. The Wigner function is constructed to achieve this task, and can be written as
\begin{equation}
    W(x,p)=\frac{1}{2\pi \hbar} \int_{-\infty}^{+\infty}  \psi^{*}\left(x-\frac{y}{2}\right)\psi\left(x+\frac{y}{2}\right)  e^{-\frac{py}{\hbar}} d_y.
    \label{cont_wf}
\end{equation}
In general, it can be constructed for any operator $A$ via the expectation value of a kernel, $\Delta(\alpha)$, such that
\begin{equation}
    W_A (\alpha)=\Tr \left[ A \Delta(\alpha) \right],
    \label{gen_wf}
\end{equation}
where $\alpha=\frac{(x+ip)}{\sqrt{2}}$. The kernel $\Delta(\alpha)$, can be written in terms of a parity and displacement operator as
\begin{equation}
    \Delta(\alpha)=D(\alpha) \Pi D^{\dagger}(\alpha),
    \label{kernel}
\end{equation}
where the parity operator $\Pi$ is written as
\begin{equation}
    \Pi=2e^{i\pi a^{\dagger}a},
\end{equation}
and the displacement operator $D(\alpha)$, is given by
\begin{equation}
    D(\alpha) = \exp(\alpha a^{\dagger} - \alpha^* a)=e^{-\frac{1}{2} |\alpha|^2} e^{\alpha a^{\dagger}} e^{-\alpha^* a}
    =e^{\frac{1}{2} |\alpha|^2} e^{-\alpha^* a} e^{\alpha a^{\dagger}}. 
    \label{displacement}
\end{equation}
Here, $a^{\dagger}$ and $a$ are creation and annihilation operators, respectively. The displacement operator~\eqref{displacement} displaces any arbitrary state in phase space. For example, a coherent state $\ket{\alpha}$ can be defined as the displacement of the vacuum state $\ket{0}$~\cite{glauber1963,review_coh_state1990}
\begin{equation}
    D(\alpha) \ket{0}=\ket{\alpha}.
\end{equation}
Any construction of the Wigner function needs to satisfy five criteria, called the Stratonovich-Weyl correspondence~\cite{GWF2016}. They are given as follows
\begin{enumerate}
	\item \textbf{Linearity;} \\ we can fully reconstruct any operator $A$ from $W_A (\alpha)$ and vice versa, via the mapping $W_A (\alpha)=\text{Tr} \left( A \Delta(\alpha) \right)$ and $A=\int_{\alpha} W_A (\alpha) \hat{\Delta}(\alpha) d\alpha$.
	\item \textbf{Reality;} \\ $W_A (\alpha)$ is always real and normalized to unity.
	\item \textbf{Covariance;} \\ if $A$ is invariant under global unitary operations then so is $W_{A}(\alpha)$.
	\item \textbf{Traciality;} \\ the overlap between states, defined by the definite integral $\int_{\alpha} W_{A} W_{A^{\prime}} d\alpha=\text{Tr} \left(A A^{\prime}\right)$, exists and is considered a unique property of the Wigner function.
	\item \textbf{Standardization;} \\ $W_A (\alpha)$ is “standardized” so that the definite integral over all space $\int_{\alpha} W_A (\alpha) d\alpha = \Tr [A]$ exists and $\int_{\alpha} \Pi(\alpha) d\alpha=\mathbb{I}$.
\end{enumerate}
In the following, the operator $A$ will be replaced by the density matrix $\rho$. We can immediately see that the Wigner function contains the same amount of information as the density operator, since it is possible to map the two functions into each other.
\section{Negativity and non-classicality}
\begin{figure}[t!]
    \subfloat[\label{ex_fock}]{%
       \includegraphics[width=0.49\textwidth]{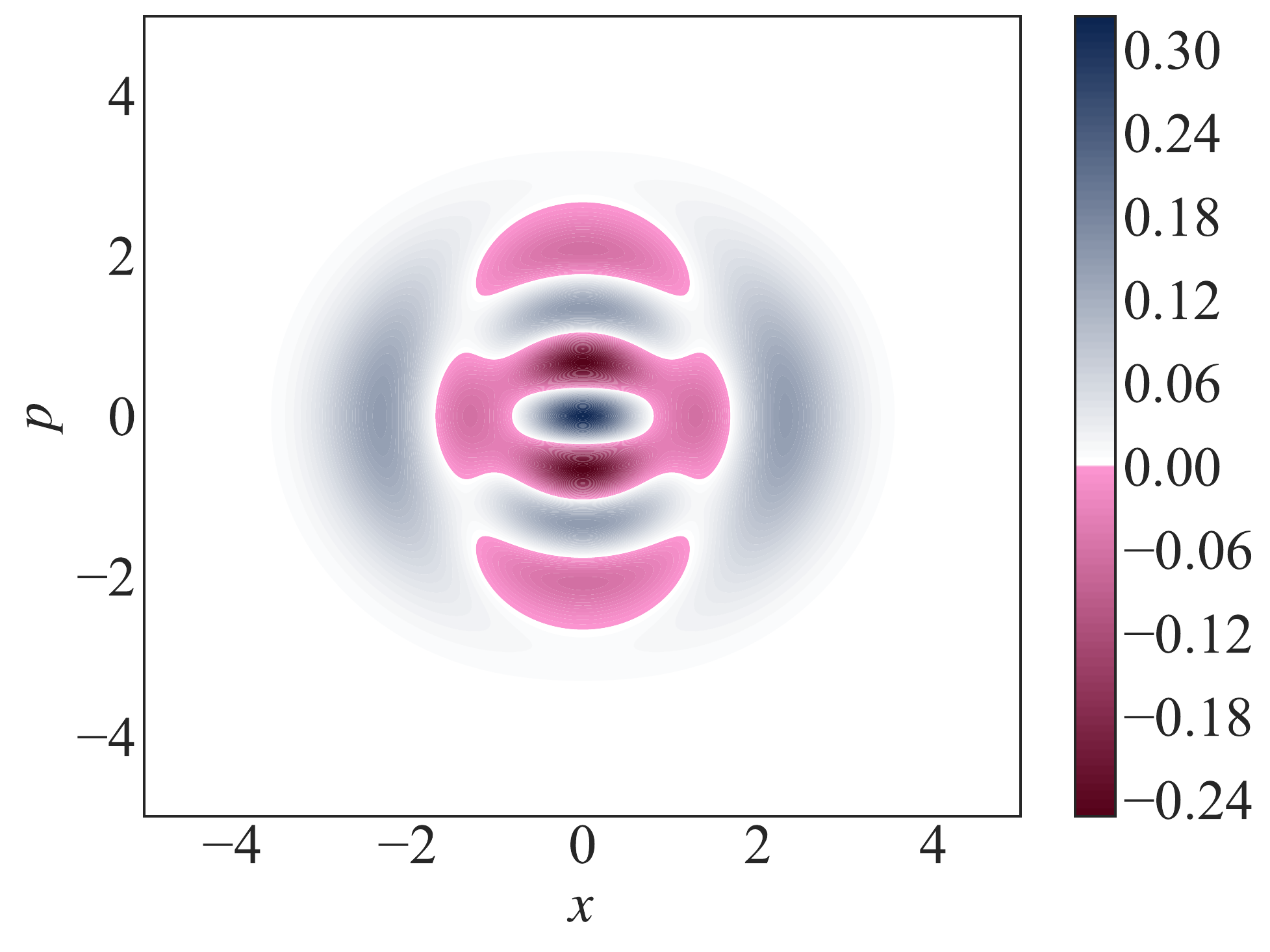}
       }%
     \subfloat[\label{ex_cat}]{%
       \includegraphics[width=0.49\textwidth]{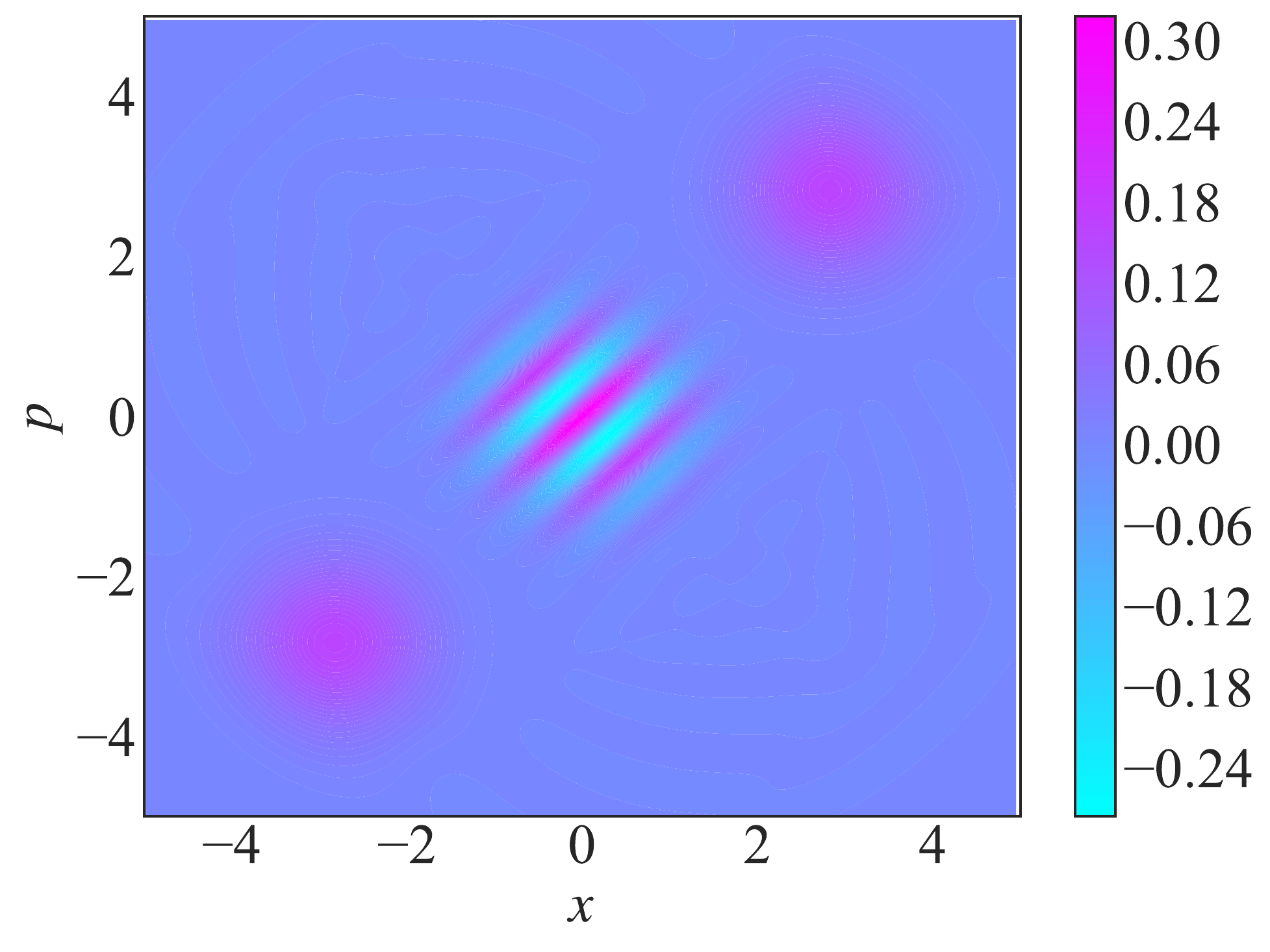}
       }
     \caption{Wigner function for (a) the Fock states~\eqref{fock}, and (b) the Cat states~\eqref{cat}.}
	\label{ex_wigner}
   \end{figure}

The traciality property of the Wigner function cannot hold unless the Wigner function takes negative values. As a consequence, the Wigner function is not a probability distribution and
can not describe a physical quantity. Instead, it is a quasi-probability distribution that can be interpreted, when integrating either over $x$ or $p$ in phase space. Figure~\eqref{ex_wigner} represents the Wigner function for (a) Fock states, and (b) cat states, given by

\begin{align}
    \ket{\psi}_{\text{Fock}}&=\frac{1}{\sqrt{2}} \left( \ket{0}_N+\ket{1}_N \right), \label{fock} \\
    \ket{\psi}_{\text{Cat}}&=\frac{1}{\sqrt{2}} \left( \ket{\beta} + \ket{-\beta} \right), \label{cat}
\end{align}
where $N=10$ is the dimension of the Fock state, and $\ket{\beta}=e^{-\frac{1}{2}|\beta|^2}\sum_{n=0}^{+\infty} \frac{(-\beta)^n}{\sqrt{n!}} \ket{n}$ is a coherent state. We see for both examples that the Wigner function is negative in regions of the phase space. What does this mean?
Since the early days of quantum mechanics, negative probabilities were under debate. In 1942, Dirac discussed them in the framework of negative energies~\cite{dirac1942}. Groenewold argued that negative probabilities are necessary in order to describe quantum systems in phase space~\cite{GROENEWOLD}. In the 1980s, Feynmann provided a simple argument for negative probabilities, where he considered them as a step in a total sum~\cite{feynamn_neg_prob}. Therefore, he concluded that the negativity should not be studied in isolation. 

\begin{figure}[t!]
    \subfloat[\label{neg_werner2}]{%
       \includegraphics[width=0.49\textwidth]{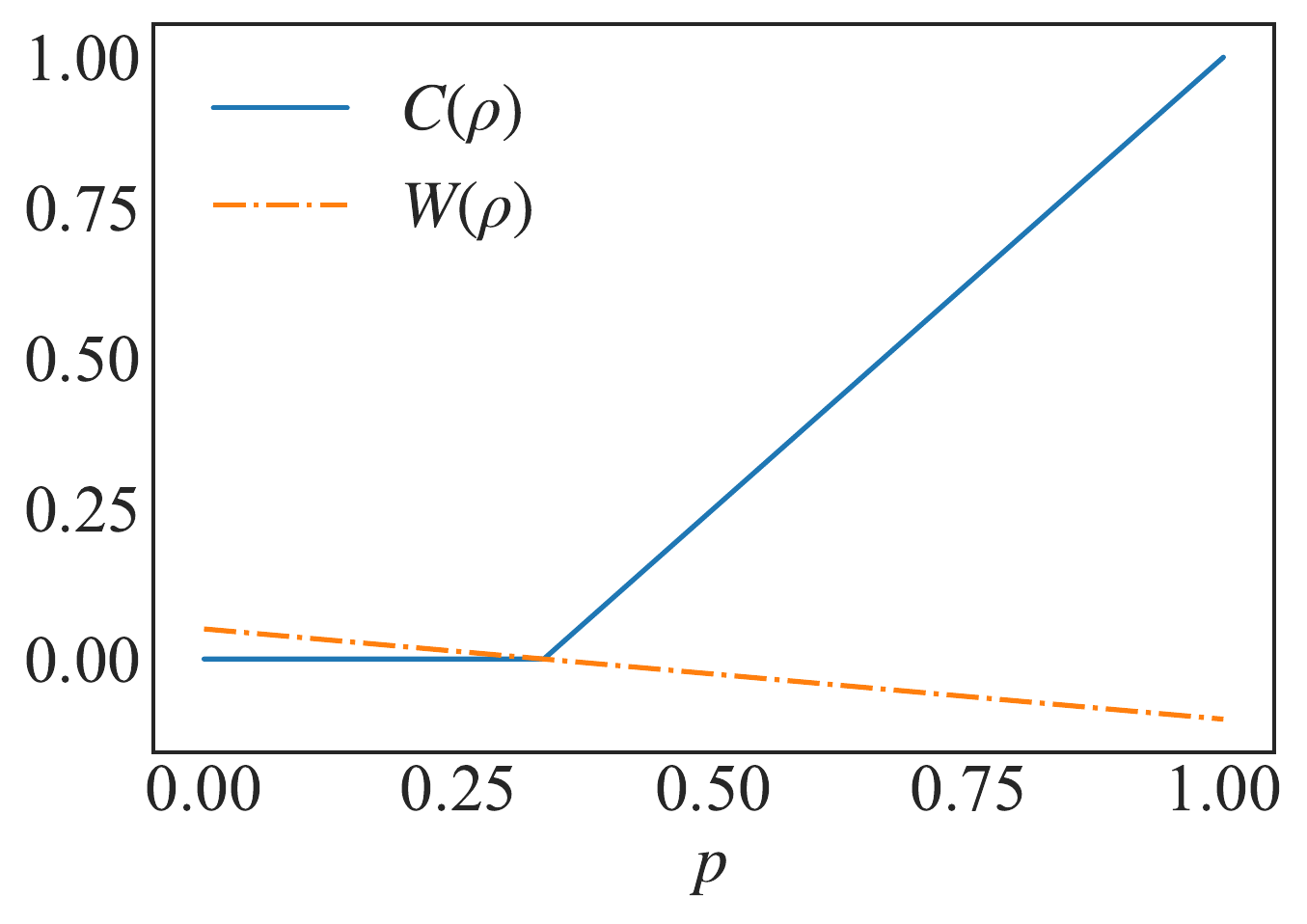}
       }%
     \subfloat[\label{squeezed}]{%
       \includegraphics[width=0.49\textwidth]{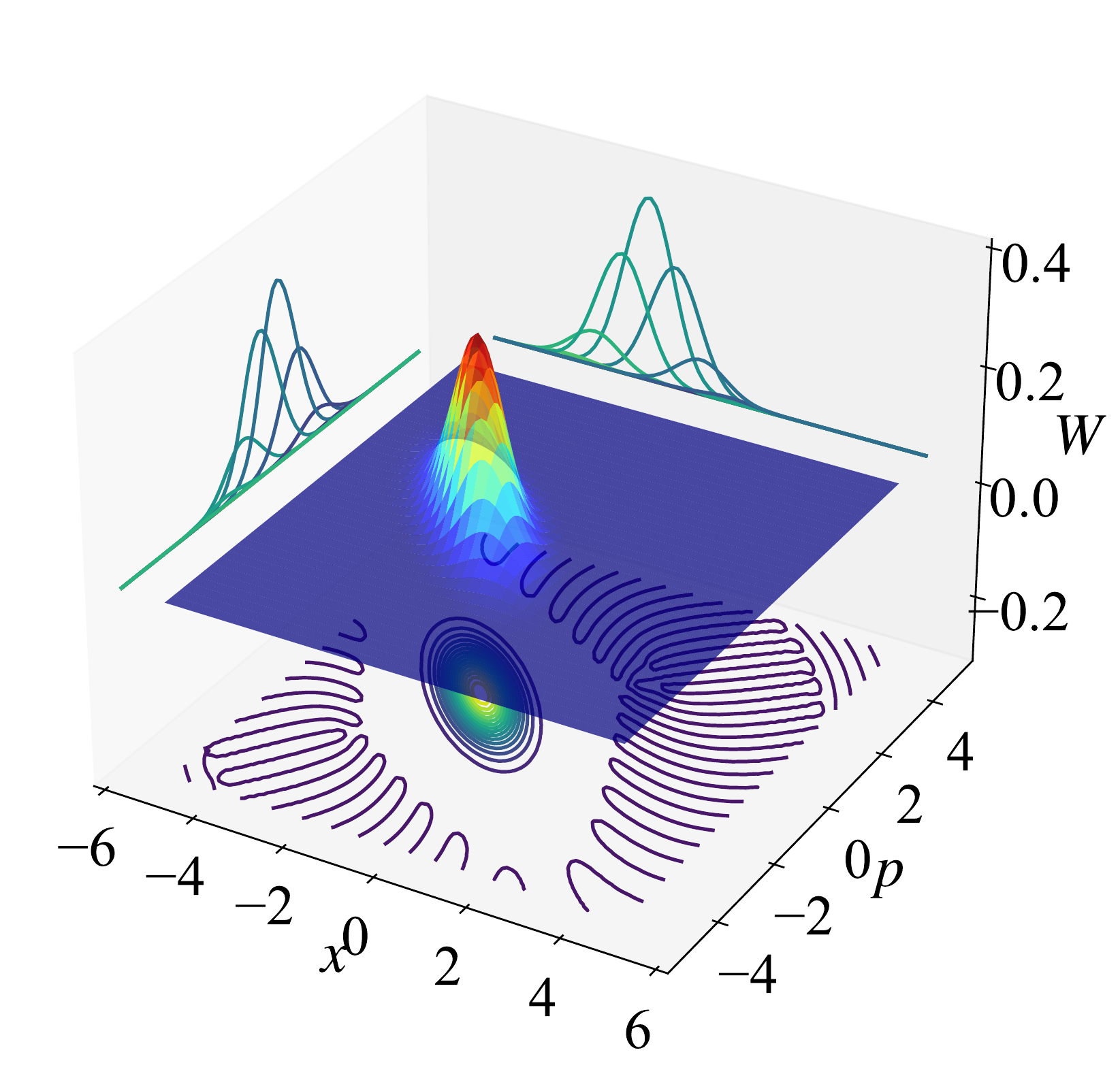}
       }
     \caption{(a) The Wigner function~\eqref{cont_wf} and the concurrence~\eqref{concurrence} in a Werner state~\eqref{werner_state1}, with respect to the mixing parameter $p$, and (b) The Wigner function~\eqref{cont_wf}, for two mode squeezed states.}
	\label{ex_werner_wigner}
   \end{figure}

\noindent The negativity of the Wigner function is generally accepted that it is arising due to non-classical features of quantum systems, such as quantum correlations or quantum interference. A system consisting of two coherent states in superposition, will have positive and negative oscillations which are due to the coherence of two states, cf. Figure~\eqref{ex_fock} and~\eqref{ex_cat}. These oscillations are precursor of quantum correlations. This motivated the proposition of measuring quantum correlations with the negative volume of the Wigner function~\cite{negativity2018}, given by
\begin{equation}
    N(W_{\rho}(\alpha))=\frac{1}{2} \int_{\alpha} |W_{\rho}(\alpha)| d\alpha -1.
    \label{neg_wig}
\end{equation}
The versatility of the Wigner function in revealing quantum correlations can be seen by considering the example of Werner states~\eqref{werner_state1}. Figure~\eqref{neg_werner2} shows the Wigner function and the concurrence~\eqref{concurrence} in the case of the Werner state, with respect to the mixing parameter $p$. We notice that as soon as the concurrence is non-zero at $p\!>\!\frac{1}{3}$, the Wigner function takes negative values, indicating the presence of quantum correlations in the form of entanglement. 

\noindent However, the negativity of the Wigner function is not capable of capturing all forms of quantum correlations, as a consequence of Hudson's theorem, which relates the negativity to the non-Gaussianity of the state~\cite{hudson}. For example, squeezed states are written in the form of squeezed Gaussian states, which have no negative Wigner function. Hence, negativity cannot capture the quantum correlations in this system. Figure~\eqref{squeezed} shows the three dimensional plot of the Wigner function for squeezed states, which is positive for all values of $x$ and $p$.
Therefore, the negativity~\eqref{neg_wig} should be used carefully when measuring the quantumness of the system. The success of the negativity in revealing non-classical behavior is only limited to some cases, where quantum correlations arise in a particular form. Hence, a consensus on the use of negativity to reveal quantum correlations is still lacking.
\section{Measuring quasi-probabilities}
The Wigner function has several applications in statistical mechanics, quantum chemistry, classical optics and signal processing~\cite{ferry_wigner_book}. However, the use of quasi-probability distribution flourished in the field of quantum optics with the construction of quantum states via quantum tomography and the measurement of the spectral phase of ultra-short laser pulses by frequency-resolved optical gating methods~\cite{frog}.  There are many approaches to generate the Wigner function experimentally, in the following we discuss two of them. The first consists of generating the density matrix $\rho$ to produce the Wigner function, while the second approach takes direct measurement of phase space points, without the need to generate the density operator. 
\begin{enumerate}
    \item \textbf{Creation of the density operator~\cite{direct_mesurement_wigner}.}\\
    This procedure consists of making use of cyclic permutation within the trace operator to simplify the measurement of quantum systems in phase space into parity measurement of a displaced state
\begin{align}
    W_{\rho}(\alpha)&=\Tr \left[ \rho \Delta(\alpha) \right]=\Tr \left[ \rho D(\alpha) \Pi D^{\dagger}(\alpha)  \right], \nonumber \\
    &=\Tr \left[ D^{\dagger}(\alpha)  \rho D(\alpha) \Pi     \right]=\Tr \left[ \rho(-\alpha) \Pi \right],
\end{align}
where $D(-\alpha)=D^{\dagger}(\alpha)$, and $\rho(\alpha)=D(\alpha) \rho D^{\dagger}(\alpha)$ is the rotated density matrix. Therefore, by creating, experimentally, the quantum state and displacing it by $\alpha$ we can measure the Wigner function by applying the parity operator using classical computation of diagonal elements or by coupling the state to a two-level system, where the two levels act as the $\pm 1$ parity values.
    \item \textbf{Direct measurement of phase space points~\cite{direct_mesurement_weyl1, direct_mesurement_weyl2}.}\\
    In this method, the qubit is measured in the $x$ and $y$ Pauli basis, instead of taking parity measurements. First, we couple the state $\rho$ with another qubit in the state $\ket{+}=(\ket{0}+\ket{1})/\sqrt{2}$, such that
    \begin{equation}
        \rho_{tot}(0)=\ket{+}\bra{+} \otimes \rho.
    \end{equation}
    Then, a unitary transformation is performed of the form
    $\rho_{tot}(\tilde{\alpha})=R(\tilde{\alpha})\rho_{tot}(0)R^{\dagger}(\tilde{\alpha})$, where
    \begin{equation}
        R(\tilde{\alpha})=\exp\left( \frac{1}{2} \sigma^z \otimes \left[ a^{\dagger} \tilde{\alpha}-a\tilde{\alpha}^* \right] \right).
    \end{equation}
    After the preparation of the state in the desired phase point $\alpha$. We generate the characteristic function, $\chi_{\rho}(\tilde{\alpha})$, by measuring the qubit in the $x$ and $y$ basis, which gives
    \begin{equation}
        \chi_{\rho}(\tilde{\alpha})=\langle \sigma^x \rangle +i\langle \sigma^y \rangle,
    \end{equation}
    where $\langle \sigma^x \rangle=\Tr \left[ \sigma^x \rho_{tot}(\tilde{\alpha}) \right]$ and $\langle \sigma^y \rangle=\Tr \left[ \sigma^y \rho_{tot}(\tilde{\alpha}) \right]$. Then by performing the Fourier transform of the characteristic function $\chi_{\rho}(\tilde{\alpha})$, we can generate the quasi-probability distribution.
\end{enumerate}
\section{Qubits in phase space}
The focus of this section is on representing finite dimensional quantum systems in phase space. Several formulations are available in the literature~\cite{wigner_review_2}, here we study two particular representations for two-level qubits: Wootters'~\cite{DWFbook} and Stratonovich ~\cite{GWF2016,GWF_entanglement_PRA}. The latter considers the qubit in the Bloch sphere, while the former represents qubit states in a toroidal lattice.
\subsection{The Stratonovich kernel}
A general extension of the Wigner function~\eqref{gen_wf} to finite-dimensional systems requires the construction of a kernel $\Delta(\Omega)$ that reflects the symmetries of the system at hand. For a qubit, Tilma \textit{et al}~\cite{GWF2016} argued that the parity operator $\Pi$ has analogous properties to the Pauli matrix $\sigma^z$ which rotates the qubit by $\pi$ about the $z$-axis of the Bloch sphere in the Pauli representation, while the displacement operator $D$ is equivalent to the SU(2) rotation operator, given by
\begin{equation}
    U(\theta,\varphi,\phi)=e^{i\sigma^z\varphi}e^{i\sigma^y\theta}e^{i\sigma^z\phi}.
\end{equation}
$U(\theta,\varphi,\phi)$ displaces the two level quantum state along the surface of the Bloch sphere. This line of thought leads to the following kernel for a qubit
\begin{equation}
\Delta(\theta,\varphi)=\left[U(\theta,\varphi,\phi) \Pi U^{\dagger}(\theta,\varphi,\phi)  \right],
\label{startonovich_kernel}
\end{equation}
where $\theta\!\in\![0,\frac{\pi}{2}]$ and $\varphi\!\in\![0,2\pi]$ parameterize the representation in phase space and $\Pi\!=\! \frac{1}{2} \left( \mathbb{I}\!-\!\sqrt{3} \hat{\sigma}_z \right)$ is a Hermitian operator. Due to the commutation of $\sigma^z$ with $\Pi$, $\phi$ does not contribute to the function. The generalization to a composite system of qubits is straightforward by performing the tensor product
\begin{equation}
	\Delta(\theta,\varphi)=\bigotimes_i^N U \Pi U^{\dagger}.
\end{equation}
The choice of the kernel  $\Delta(\Omega)$ and the set of parameters $\Omega$ is not unique to define the Wigner function. The advantage of the kernel~\eqref{startonovich_kernel} is that it can be applied to construct the Wigner function for any arbitrary quantum system, represented by the density matrix $\rho$. For $d-$dimensional quantum systems, we can use the generalized Gell-Mann matrices $\Lambda_i$ in order to construct a natural extension of the kernel~\eqref{startonovich_kernel}~\cite{gell-mann}
\begin{equation}
    \Delta^d(\Omega)=\frac{1}{d}U(\Omega)\Pi^d U^{\dagger}(\Omega),
\end{equation}
where the parity operator is written as
\begin{equation}
    \Pi^d=\mathbb{I}_{dxd}-\mathcal{N}(d) \Lambda_{d^2 -1}.
\end{equation}
Here, $\mathcal{N}(d)=\sqrt{\frac{d(d+1)(d-1)}{2}}$ is a normalization constant, and $\Lambda_{d^2 -1}$ is a $d\! \times \!d$ diagonal matrix, with entries: $\sqrt{2/d(d-1}$ except  $\left(\Lambda_{d^2 -1}\right)_{d,d}=-\sqrt{2(d-1)/d}$.

\begin{figure}[t!]
    \subfloat[One qubit\label{neg_werner}]{%
       \includegraphics[width=0.33\textwidth]{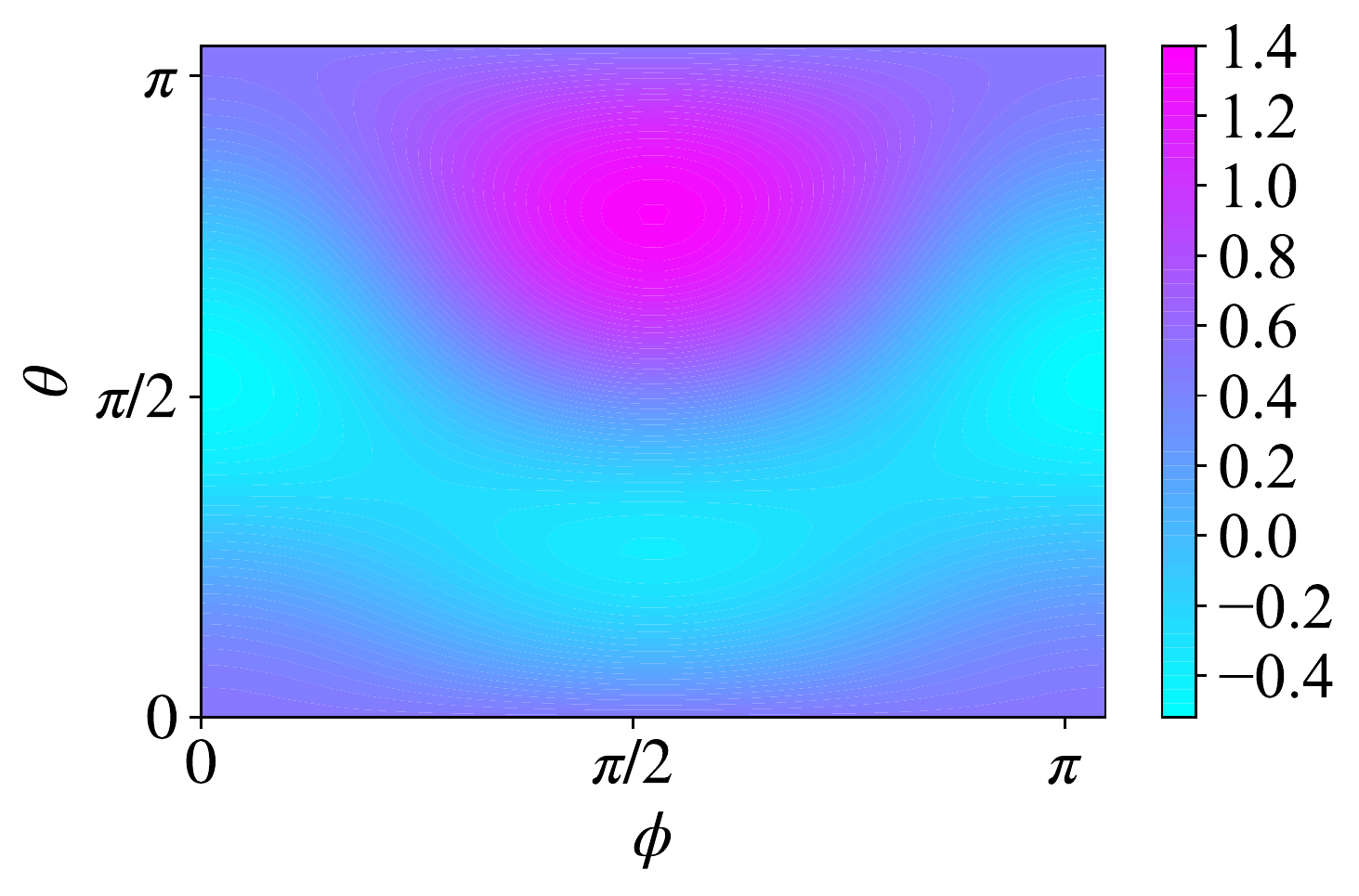}
       }%
     \subfloat[Bell state\label{squeezed}]{%
       \includegraphics[width=0.33\textwidth]{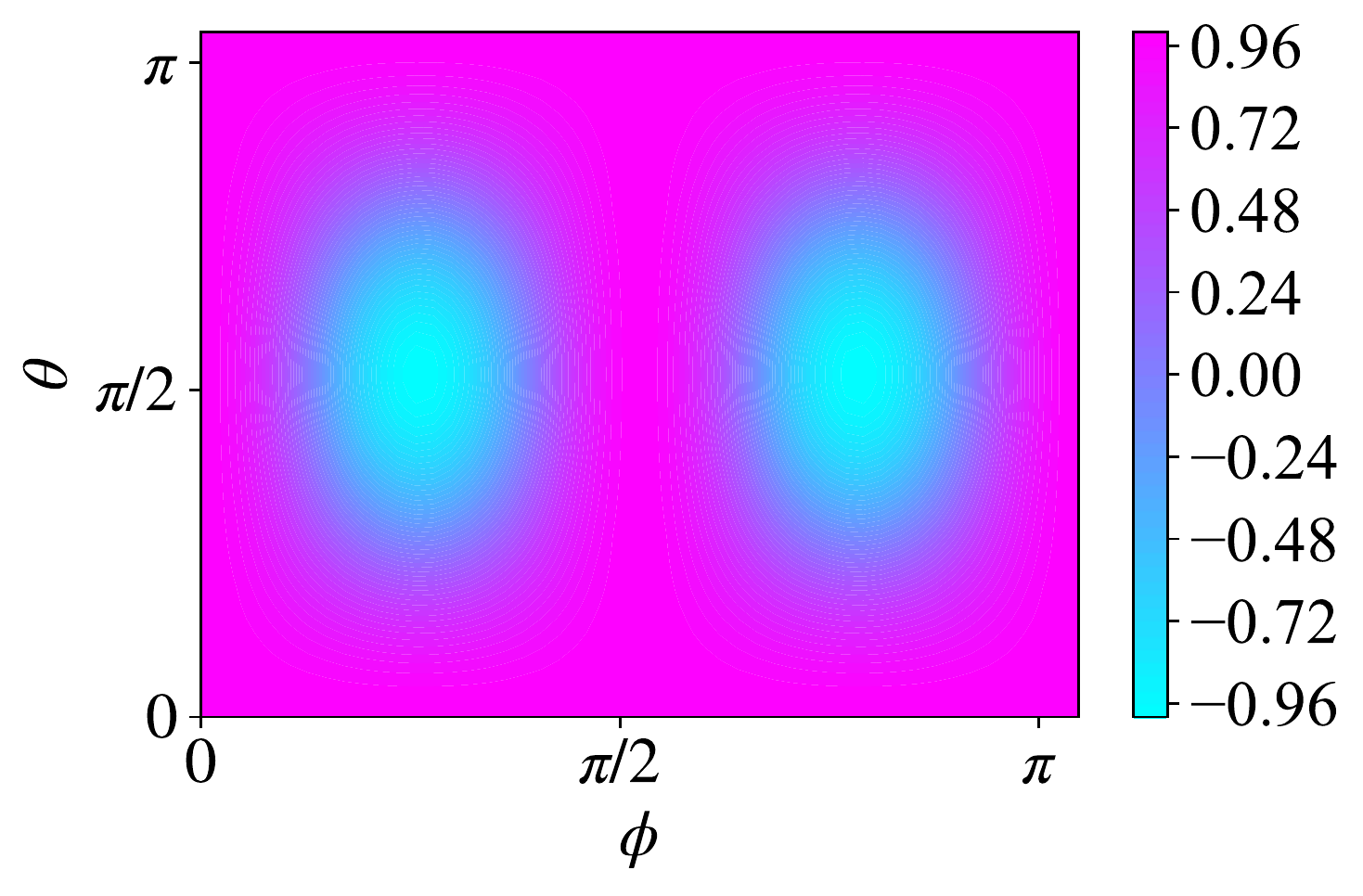}
       }%
     \subfloat[GHZ state\label{squeezed}]{%
       \includegraphics[width=0.33\textwidth]{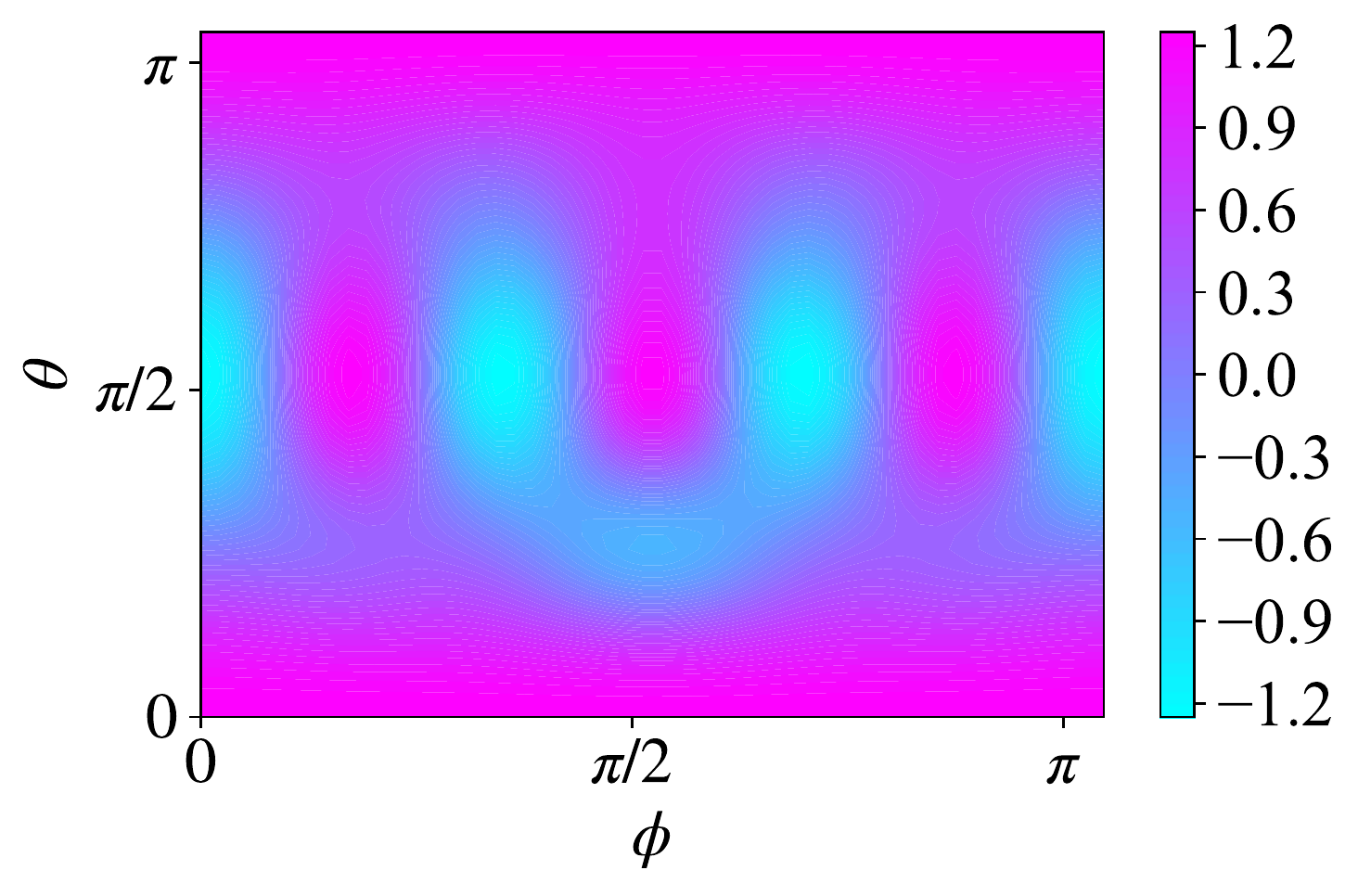}
       }
     \caption{The Wigner function using the Stratonovich kernel~\eqref{startonovich_kernel}, for (a) one qubit, (b) Bell state, and (c) GHZ state.}
	\label{ex_stratonovich}
\end{figure}

\noindent We represent in Figure~\eqref{ex_stratonovich} density plots of the Wigner function using the Stratonovich kernel~\eqref{startonovich_kernel}, for three quantum systems
\begin{align}
    \ket{\psi}&=\frac{1}{\sqrt{2}}\left( \ket{0} + \ket{1} \right) && \text{one qubit},\\
    \ket{\Phi}&=\frac{1}{\sqrt{2}} \left( \ket{00} + \ket{11} \right) && \text{Bell state},\\
    \ket{\Psi}&=\frac{1}{\sqrt{2}} \left( \ket{000} + \ket{111} \right) && \text{GHZ state}.
\end{align}

\noindent In the three examples, the Wigner function takes both negative and positive values in a similar fashion to its continuous counterpart. Negativity in these cases, can be interpreted as a sign of superposition. Symmetry also plays an important role in the negativity, as any pure state of a qubit can be written as the superposition of two antipodal pure states. This shows the versatility of the Stratonovich kernel in revealing the salient features of symmetric quantum states.  
\subsection{The Wootters' kernel}
Wootters' Winger function~\cite{wootters} can be seen as a generalization of Feynman's phase space construction for two-level systems~\cite{feynamn_neg_prob}. Feynman treated the case of qubits, while Wootters extended the construction to systems with prime powered dimension. Feynman phase space is constructed by writing the probability distributions as
\begin{align}
    p_{++}&= \frac{1}{2} \langle \mathbb{I} + \sigma^x + \sigma^y + \sigma^z \rangle, &&
    p_{+-}= \frac{1}{2} \langle \mathbb{I} - \sigma^x - \sigma^y + \sigma^z \rangle,\\
    p_{-+}&= \frac{1}{2} \langle \mathbb{I} + \sigma^x - \sigma^y - \sigma^z \rangle, &&
    p_{--}=\frac{1}{2} \langle \mathbb{I} - \sigma^x + \sigma^y - \sigma^z \rangle,
    \label{feynmann}
\end{align}
which represent the joint probability of finding the qubit aligned with the $\pm z$ and $\pm x$ axes, respectively. In contrast, Wootters phase space can be constructed by turning the set of equations~\eqref{feynmann}, into a phase point operator. Hence, the phase space is an $N\!\times\!N$ grid, labelled by a pair of coordinates $(x,p)$, each taking values from $0$ to $N-1$  and for each coordinate we define the usual addition and multiplication mod $N$. If the dimension of the system is $N\!=\!q^k$, with $q$ a prime and $k$ an integer greater than 1, the phase space is constructed by performing the $k$-fold Cartesian product of $q\!\times\!q$ phase spaces. Naturally, the simplest example one can consider is a system with two orthogonal states, i.e. a qubit  with $N\!=\!2$, whose discrete phase space consists of four points, while for a composite system of two qubits, i.e. $N\!=\!2^2$ the phase space is formed by 16 points, cf. Table~\eqref{tab1}. Each point in the phase space is described by the discrete phase point operator, $A(x_i,p_i)$. For a single qubit it is given by
\begin{equation}
A(x_1,p_1)=\frac{1}{2} \Big( \mathbb{I} + (-1)^{x_1} \sigma^z+(-1)^{p_1} \sigma^x+(-1)^{x_1+p_1} \sigma^y \Big),
\label{wootters_kernel}
\end{equation}

\begin{table}[t!]
	\begin{subtable}{.3\linewidth}
		\centering
		\captionsetup{justification=centering}
		\caption{One qubit}
		\tikzmark{t}\\
		\tikzmark{l}
		\begin{tabular}{c   c     c     c}
			& & $0$ & $1$\\[-8pt]
			& \multicolumn{1}{@{}l}{\tikzmark{x}}\\
			0 & & .&.\\
			1 & & . &.\\
		\end{tabular}
		\tikzmark{r}\\
		\tikzmark{b}
		\tikz[overlay,remember picture] \draw[-triangle 45] (x-|l) -- (x-|r) node[right] {$p_1$};
		\tikz[overlay,remember picture] \draw[-triangle 45] (t-|x) -- (b-|x) node[below] {$x_1$};
		\label{tab1a}
	\end{subtable}%
	\hfill
	\begin{subtable}{.6\linewidth}
		\centering
		\captionsetup{justification=centering}
		\caption{Two qubits}
		\tikzmark{t}\\
		\tikzmark{l}
		\begin{tabular}{c   c     c     c c c}
			& & $00$ & $01$&10&11\\[-8pt]
			& \multicolumn{1}{@{}l}{\tikzmark{x}}\\
			00 & & .&.&.&.\\
			01 & & . &.&.&.\\
			10&&.&.&.&.\\
			11&&.&.&.&.
		\end{tabular}\tikzmark{r}\\
		\tikzmark{b}
		\tikz[overlay,remember picture] \draw[-triangle 45] (x-|l) -- (x-|r) node[right] {$(p_1,p_2)$};
		\tikz[overlay,remember picture] \draw[-triangle 45] (t-|x) -- (b-|x) node[below] {$(x_1,x_2)$};
		\label{tab1b}
		\end{subtable}
	\vskip0.5cm
	\caption{Discrete phase space for (a) one qubit and (b) two qubits.}
\label{tab1} 
\end{table}
\noindent where $\sigma^i$ ($i=x,y,z$) are the usual Pauli operators. For composite systems the phase point operators are constructed from the tensor product of the phase point operators of the corresponding subsystems, i.e.  $A(x_1\dots x_k,p_1\dots p_k)\!=\!A(x_1,p_1)\otimes A(x_2,p_2) \otimes \dots \otimes A(x_k,p_k)$. Since the $A(x_i,p_i)$'s form a complete orthogonal basis of the Hermitian $N\!\times\!N$ matrices, any density matrix can be decomposed as $\rho\!\!=\!\!\sum\limits_{(x_i,p_i)}W(x_i,p_i)A(x_i,p_i)$, where the real-valued coefficients
\begin{equation}
W(x_i,p_i)=\frac{1}{N} \text{Tr} (\rho A(x_i,p_i)),
\label{wooters_wigner}
\end{equation}
correspond to the discrete Wigner function and $N$ is the dimension of the overall system. We can relate Feynman's and Wootters' phase space, by noticing that $W(0,0)=p_{++}$, $W(0,1)=p_{+-}$, $W(1,0)=p_{-+}$, and $W(1,1)=p_{--}$. 

\noindent The Wootters kernel~\eqref{wootters_kernel} can be written in terms of the Stratonovich kernel. This can be achieved by introducing the displacement operator for Wootters' phase space
\begin{equation}
    D(x,p)=\exp{\left(\frac{1}{2}i\pi xp\right)} (\sigma^x)^p (\sigma^z)^x,
\end{equation}
while the generalized parity operator is
\begin{equation}
    A(0,0)=\frac{1}{2}( \mathbb{I} + \sigma^x + \sigma^y + \sigma^z).
\end{equation}
Therefore, the kernel in displaced parity form is
\begin{equation}
    A^{\prime}(x,p)=D(x,p)A(0,0)D^{\dagger}(x,p).
\end{equation}
The eigenvalues of $A(0,0)$ are $(1\pm \frac{\sqrt{3}}{2})$, which are the same for the parity operator of the Stratonovich kernel $\Pi$. Thus, we can diagonalize $A(0,0)$ as
\begin{equation}
    A(0,0)=U(\theta,\phi)\Pi U^{\dagger}(\theta,\phi),
\end{equation}
where $\theta=\arccos(\frac{1}{\sqrt{3}})$, and $\phi=-\frac{\pi}{4}$. Hence, we can write the Wootters kernel $A^{\prime}(x,p)$ in terms of the Stratonovich kernel as
\begin{equation}
    A^{\prime}(x,p)=\Delta \left(\theta+p\pi,\phi+ \frac{2x-p}{2} \pi \right).
    \label{map_wott_stra}
\end{equation}
The result~\eqref{map_wott_stra} implies that Wootters' Wigner function is a sub-quasi probability distribution of the Stratonovich Wigner function.
\section{Measuring qubits in phase space}
The experimental techniques to measure qubits in phase space are based upon their continuous counterpart. Re-creation of the density matrix was used in order to simulate an atomic cat state via Rydberg atoms with angular momentum $j=25$~\cite{rydb_exp_wigner}. The atom in this experiment is represented by a large angular momentum space, with Hilbert space dimension of 51. Alternatively, tomography techniques such as discrete representation of operators for spin systems (DROPS) can be used to measure multi-qubit states~\cite{drops1,drops2}. The disadvantage of these methods is that they require full tomography to construct the full density matrix, which becomes exponentially difficult as the number of qubits increases. Hence, the motivation behind developing techniques that do not require full tomography in constructing the density operator.

\noindent The displaced parity formalism can be used in order to measure the phase space of quantum states directly. The Wigner function in this case takes the form:
\begin{align}
    W_{\rho}(\theta,\phi)&=\Tr \left[ \rho \Delta(\theta,\phi) \right]=\Tr \left[ \rho U(\theta,\phi,\Phi) \Pi U^{\dagger}(\theta,\phi,\Phi)  \right], \nonumber\\
    &=\Tr \left[ \rho(\theta,\phi,\Phi) \Pi \right] = \langle \Pi \rangle_{(\theta,\phi,\Phi)}.
\end{align}
Here, $\rho(\theta,\phi,\Phi)=U^{\dagger}(\theta,\phi,\Phi) \rho U(\theta,\phi,\Phi) $. The measurement procedure is to first rotate the generated state $\rho$ to the desired location in phase space in order to create $\rho(\theta,\phi,\Phi)$. Then, apply the appropriate parity measurement. This protocol has been used in order to measure a two-level Caesium atom~\cite{caesium} and a state produced via the Nitrogen vacancy center in diamond~\cite{nitrogen}. In both experiments, they prepared the state in a superposition of the ground and excited state. Then, the state is rotated in ($\theta,\phi$) by applying pulses. The protective measurement $p_{m}(\theta,\phi)$ for $m=\pm \frac{1}{2}$, are given by
\begin{equation}
    p_{\pm \frac{1}{2}}(\theta,\phi)=\Tr \left[ \rho U(\theta,\phi) P_{\pm \frac{1}{2}} U^{\dagger}(\theta,\phi) \right],
\end{equation}
where $P_{\pm \frac{1}{2}}$ represent projectors of the eigenstates of$\sigma^z/2$ with eigenvalues $\pm \frac{1}{2}$. Therefore, the Wigner function reduces to
\begin{align}
    W_{\rho}(\theta,\phi)&=\sum_{m=-1/2}^{1/2} p_m(\theta,\phi)\Pi_m, \nonumber \\
    &=\frac{1}{2} \left[ 1+\sqrt{3} \left( p_{1/2}(\theta,\phi) - p_{1/2}(\theta,\phi) \right) \right].
\end{align}
Full tomography for one qubit is to some extent feasible, as it requires few measurements. In contrast, multi-qubit systems are difficult to measure due to the exponential increase of the Hilbert space. An interesting approach to represent multi-qubit states is by taking advantage of symmetric states. In many situations, symmetric states are the main requirement to use for computational advantage or metrology. Coherent states~\cite{glauber1963}, GHZ states~\cite{ghz}, Cat states~\cite{cat}, W states or Dicke states~\cite{dicke}, are all prominent examples of symmetric states. 
\chapter{Quantum Criticality in Phase Space \label{chap5} }
In this chapter, We apply Wootters' discrete Wigner function and the Stratonovich Wigner function, to detect quantum phase transitions in critical spin-$\tfrac{1}{2}$ systems. We develop a general formula relating the phase space techniques and the thermodynamical quantities of spin models, which we apply to single, bipartite and multipartite systems governed by the $XY$ and the $XXZ$ models. We will see that this approach allows us to introduce a novel way to represent, detect, and distinguish first-, second- and infinite-order quantum phase transitions. Furthermore, we will show that the factorization phenomenon of the $XY$ model is only directly detectable by quantities based on the square root of the bipartite reduced density matrix. Finally, we will establish that phase space techniques provide a simple, experimentally promising tool in the study of many-body systems and we discuss their relation with measures of quantum correlations and quantum coherence~\cite{mzaouali2019}.
\section{The $XY$ model}
\begin{figure}[t!]
    \centering
    \includegraphics[scale=0.7]{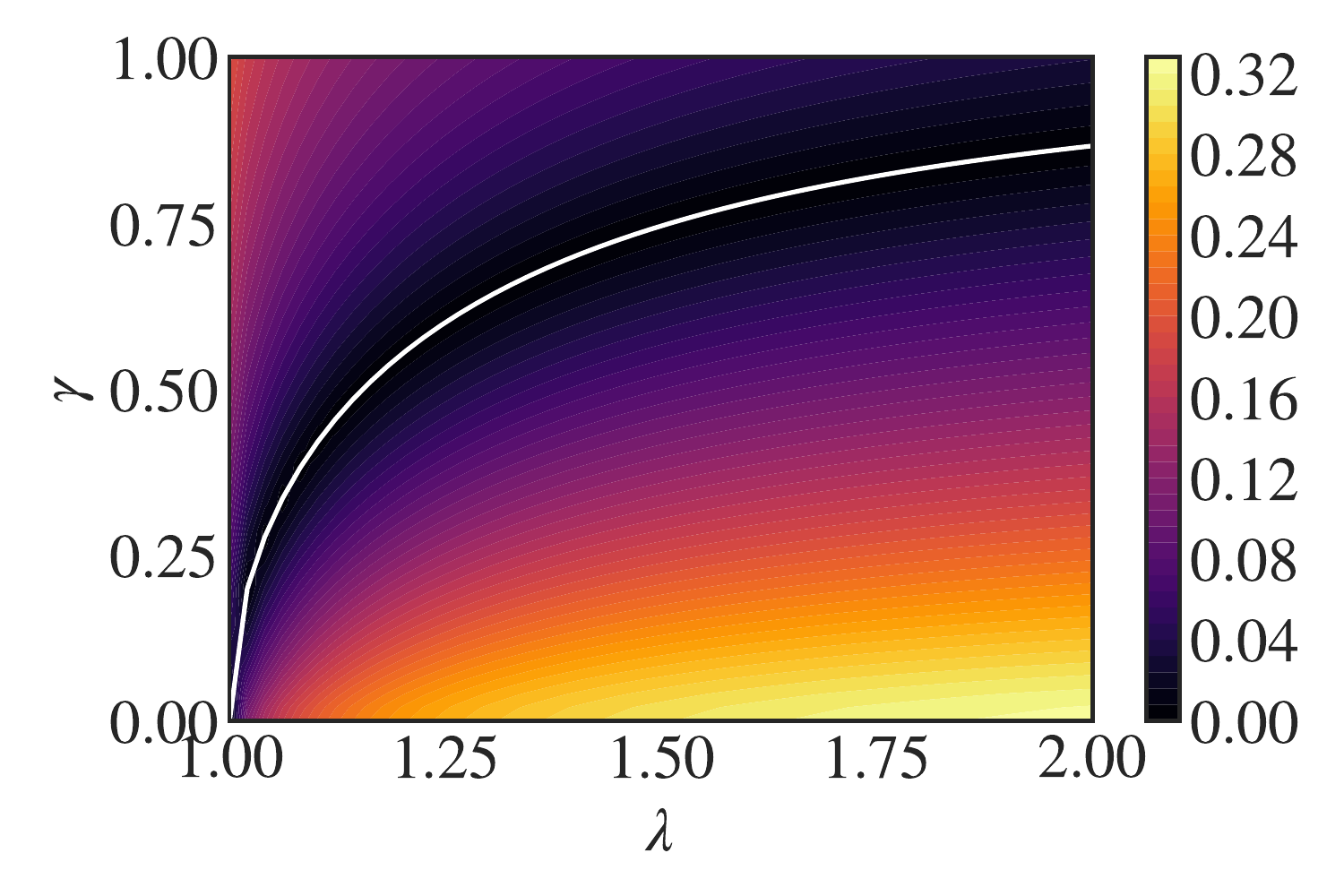}
    \caption{ $(\gamma - \lambda)$ phase diagram of the concurrence~\eqref{concurrence} in the $XY$ model~\eqref{xy}. The white curve represents the factorization line~\eqref{fact} of the ground state.}
    \label{fac_cnc_xy}
\end{figure}

The most natural candidate to investigate how phase space methods can explore criticality, are quantum spin chains, as they exhibit rich  quantum critical behavior. To test the validity of our idea, we start from a simple model of interacting spins: the $XY$ model, introduced in Section~\eqref{xymodel}:
\begin{equation}
	\mathcal{H}_{XY}\!=\!-\sum_{i=0}^{N-1} \left[ \frac{\lambda}{2} \bigg\{ \left(1\!+\!\gamma\right)\sigma_i^x\sigma_{i+1}^x+\left(1\!-\!\gamma\right)\sigma_i^y\sigma_{i+1}^y \bigg\}\!+\!\sigma_i^z \right],
	\label{xy}
\end{equation}
We have seen in the previous part that the $XY$ model~\eqref{xy} can be solved exactly and undergoes a second-order quantum phase transitions at $\lambda_c=1$, for $0\!\leq\!\gamma\!\leq\!1$. Moreover, this model exhibits a non-trivial factorization line, where the ground state of the model becomes completely factorized at 
\begin{equation}
	\lambda_f=\frac{1}{\sqrt{1\!-\!\gamma^2}}
	\label{fact}
\end{equation}
and is understood as an entanglement transition which is characterized by an energy level degeneracy~\cite{GiorgiPRB, CampbellPRA2013, GiampaoloPRA, AmicoPRB}. Figure~\eqref{fac_cnc_xy} shows the concurrence~\eqref{concurrence} of the $XY$ model~\eqref{xy} in the $(\gamma - \lambda)$ plane. The white curve is where the concurrence is zero, which indicates the separability of the ground state. In the following, the system of interest will be an infinite chain, and we calculate the reduced density matrix for a single-, two- and three-site system in order to evaluate the corresponding Wootters' discrete Wigner function (DWF).
\subsection{Single site}
\begin{figure}[t!]
   \subfloat[\label{ssDWF}]{%
       \includegraphics[width=0.49\textwidth]{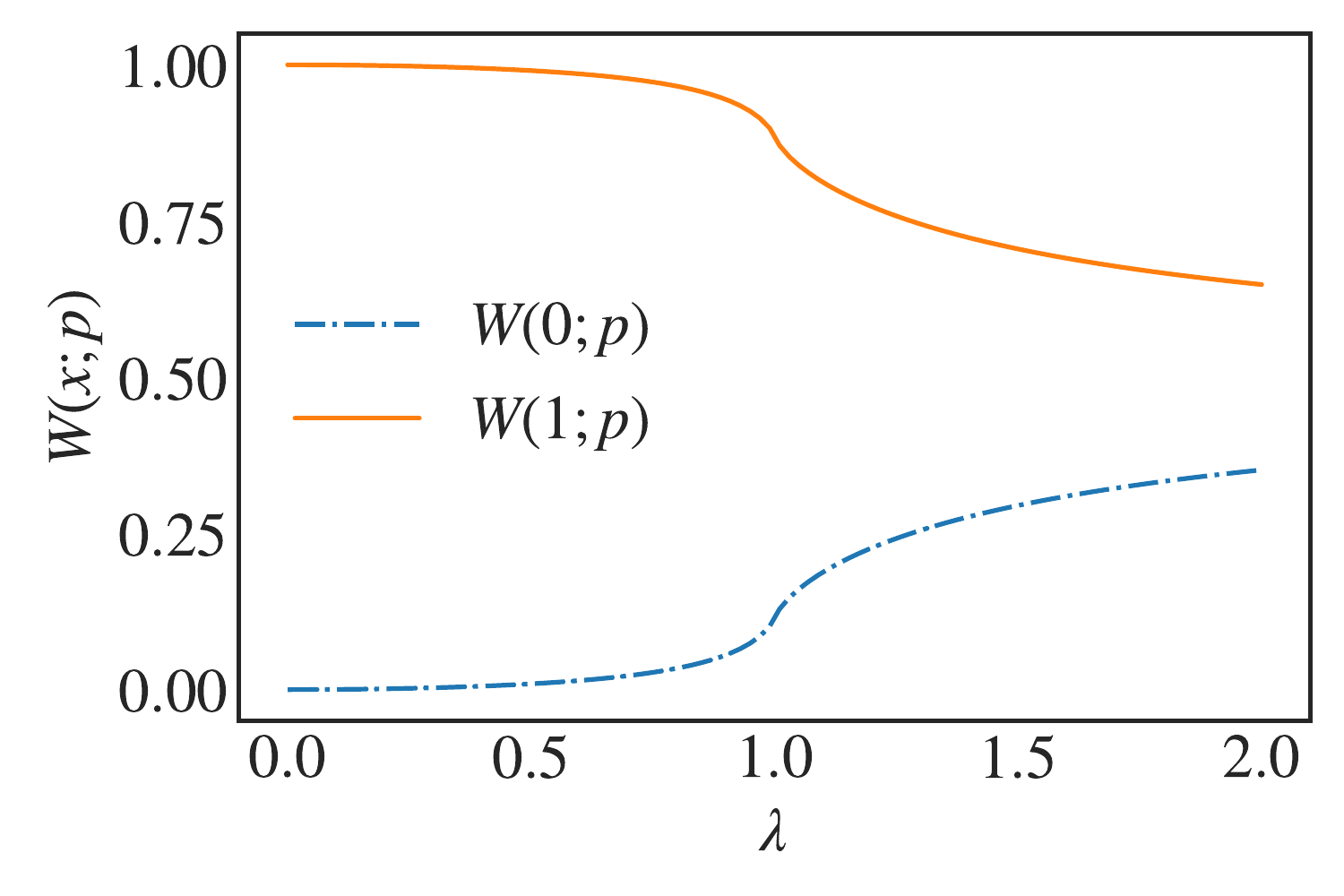}
       }%
     \subfloat[\label{der_ssDWF}]{%
       \includegraphics[width=0.49\textwidth]{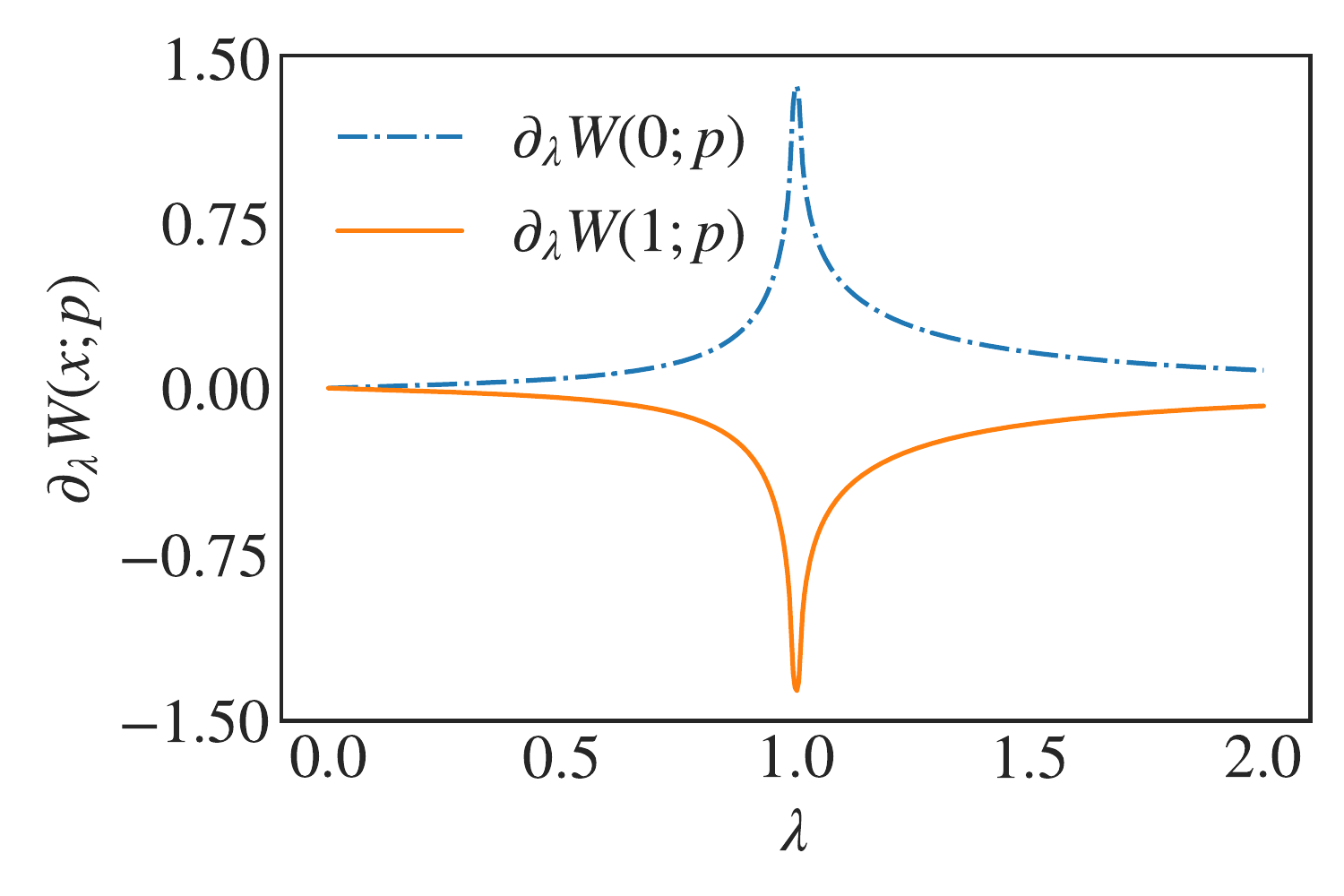}
       }\\
     \subfloat[\label{ssgwf}]{%
       \includegraphics[width=0.49\textwidth]{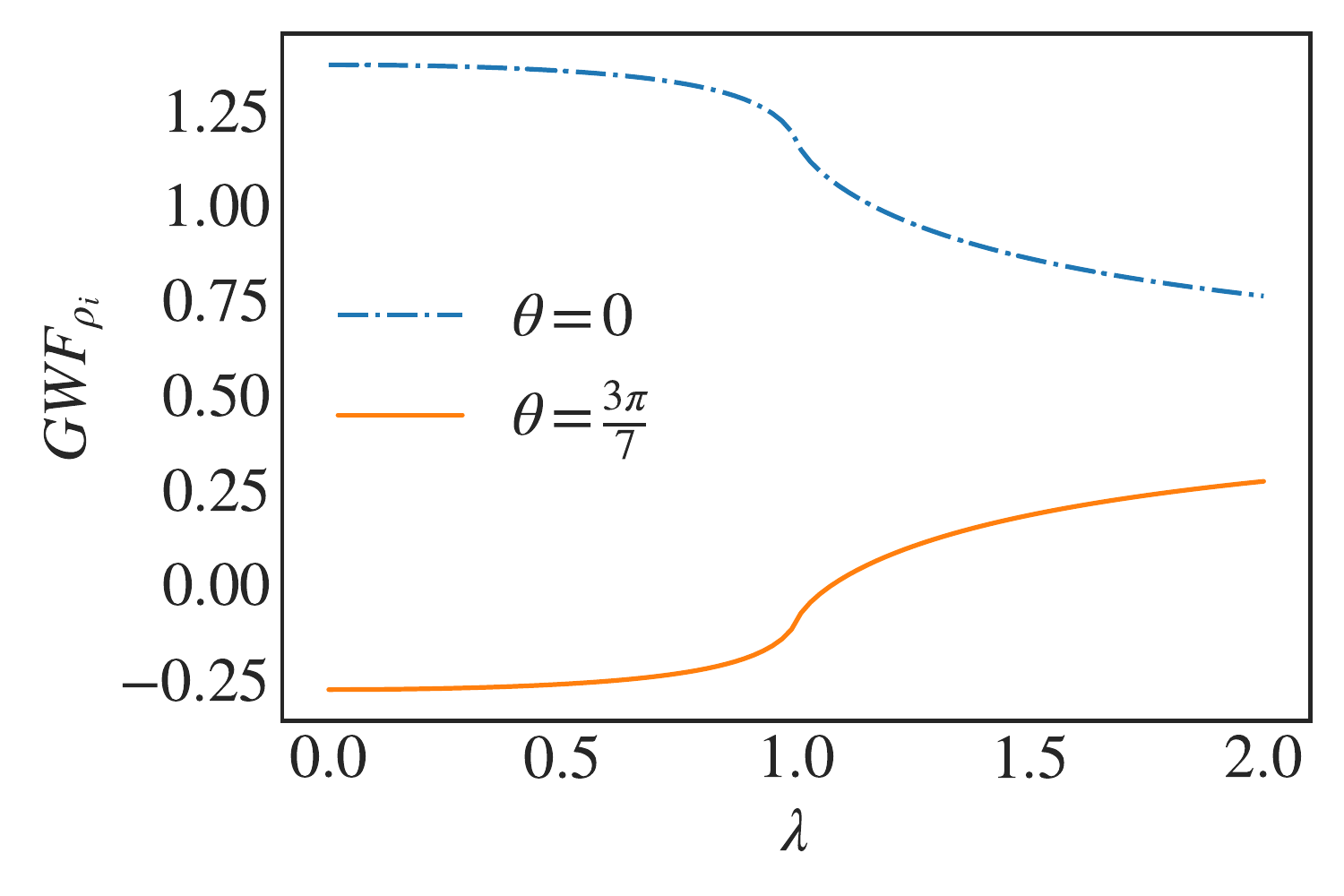}
       }%
     \subfloat[\label{der_ssgwf}]{%
       \includegraphics[width=0.49\textwidth]{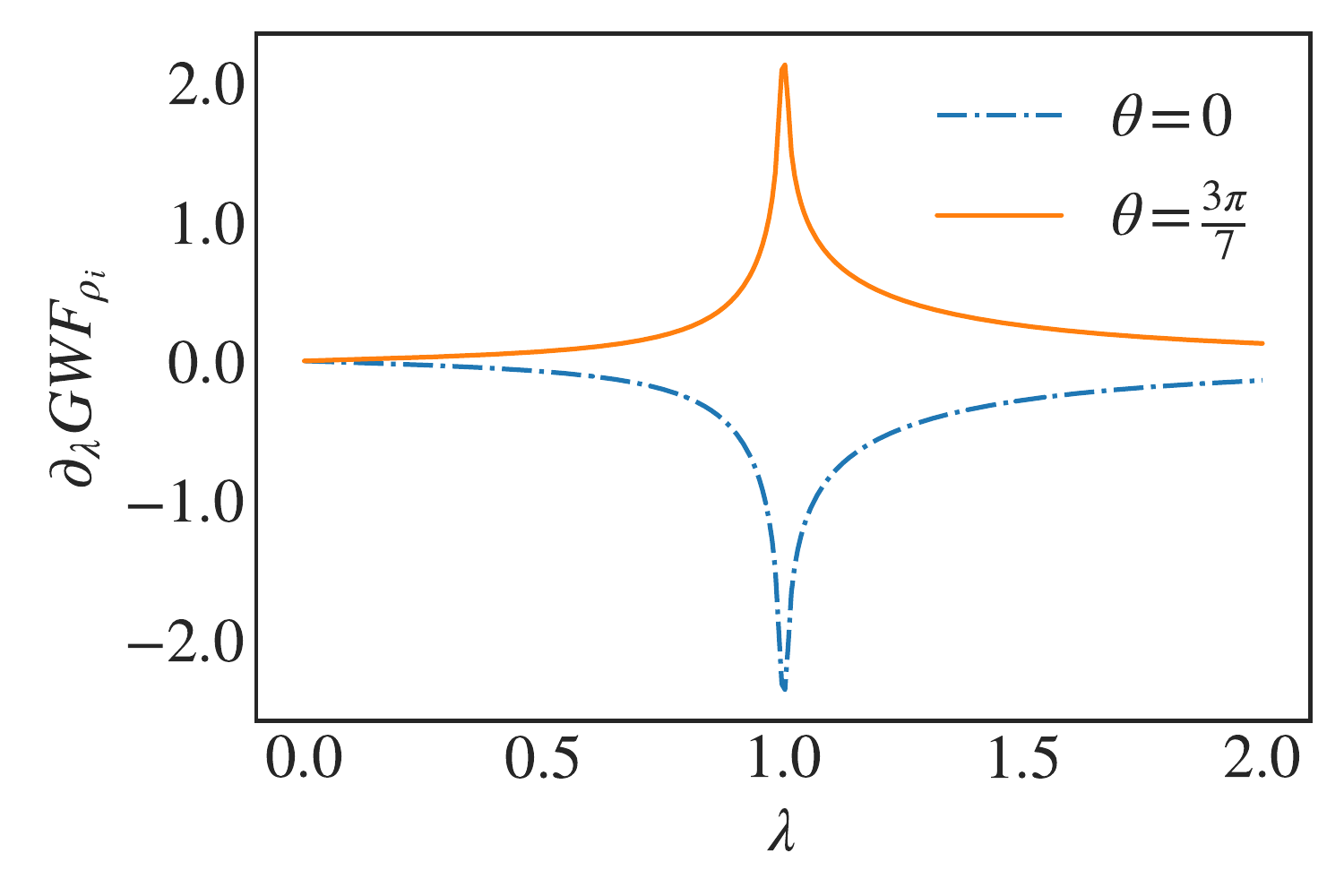}
       }
     \caption{(a) Discrete Wigner function and (b) its first derivative with respect to $\lambda$ for the single site $ XY $ model~\eqref{xy}. The two distinct behaviors of $W(x;p)$ and $\partial_{\lambda} W(x;p)$ correspond to the appropriate phase space points indicated in Table~\eqref{tab_ss1}. Panels (c) and (d) show the behavior of the maximum $(\theta=0)$, the minimum $\left(\theta \sim 3\pi/7\right)$ of the GWF (c) and their first derivative with respect to $\lambda$ (d), for the $XY$ model. In all plots $\gamma\!=\!0.5$.}
	\label{fig_ss_dwf}
\end{figure}

We start by considering a single site taken by performing the partial trace over all the other sites of an infinite chain. The reduced density matrix $\rho_i$ can be expressed as
	\begin{equation}
		\rho_i=\frac{1}{2} \sum_{\alpha=0}^{3} \langle \sigma^\alpha \rangle \sigma_i^\alpha,
		\label{sdm}
	\end{equation}
plugging~\eqref{sdm} in~\eqref{wooters_wigner} and taking into account the reality of the density matrix and the $\mathbb{Z}_2$ symmetry of quantum spin$-\frac{1}{2}$ chains, the DWF for one site takes the form
\begin{equation}
	W(x_1,p_1)=\frac{1}{2} \Big(
	 1+(-1)^{x_1}\langle \sigma^z \rangle \Big).
	 \label{DWF_single_site} 
\end{equation}
Due to the $\mathbb{Z}_2$ symmetry, the single site DWF only depends on $x_1$ and thus the DWF in this case consists of two distinct behaviors as depicted in Figures~\eqref{ssDWF},~\eqref{der_ssDWF} and Table~\eqref{tab_ss1}. Choosing $\gamma\!=\!0.5$ we see the concavity of the DWF changes after crossing the critical point, $\lambda_c\!\!=\!\!1$, which is further reflected by a divergence at $\lambda_c$ in the first derivative of the DWF with respect to $\lambda$. For this value of $\gamma$ the factorization point at $\lambda_f\!\sim\!1.1547$ and we find that the single site DWF shows no signatures of this phenomenon, which is to be expected since the reduced density matrix depends only on the magnetization and contains no information about correlations within the chain.

\noindent Turning our attention to the single site Stratonovich's generalized Wigner function (GWF), using the reduced density matrix~\eqref{sdm} and the Stratonovich kernel ~\eqref{startonovich_kernel} we find
\begin{equation}
	\text{GWF}_{\rho_i}(\theta)\!=\!\frac{1}{2} \left(1-\sqrt{3}\cos(2\theta)\langle \sigma_z \rangle \right). 
	\label{gwf_ss}
\end{equation}
\begin{table}[t!]
    		\centering
    		\tikzmark{t}\\
    		\tikzmark{l}
    		\begin{tabular}{c   c     c     c}
    			& & $0$ & $1$\\[-8pt]
    			& \multicolumn{1}{@{}l}{\tikzmark{x}}\\
    			0 & & \begingroup \color{blue!55} \textbf{\textemdash}\endgroup&\begingroup \color{blue!55} \textbf{\textemdash} \endgroup\\
    			1 & & \begingroup \color{orange!55} \textbf{-.-} \endgroup & \begingroup \color{orange!55} \textbf{-.-} \endgroup\\
    		\end{tabular}
    		\tikzmark{r}\\
    		\tikzmark{b}
    		\tikz[overlay,remember picture] \draw[-triangle 45] (x-|l) -- (x-|r) node[right] {$p_1$};
    		\tikz[overlay,remember picture] \draw[-triangle 45] (t-|x) -- (b-|x) node[below] {$x_1$};
    		\vskip0.5cm
    		\caption{Discrete phase space for the single site $XY$ model. Each symbol corresponds to a particular curve shown in Figure~\eqref{fig_ss_dwf}. }
    		\label{tab_ss1}
\end{table}
The single site GWF is insensitive to the angle $\varphi$ which is due to the $\mathbb{Z}_2$ symmetry. It is evident that the DWF~\eqref{DWF_single_site} corresponds to a particular choice of angle $\theta$ in the GWF~\eqref{gwf_ss}. In Figure~\eqref{ssgwf} we focus on two limits of the GWF~\eqref{gwf_ss} by fixing the value of the parameter $\theta$. For $\theta\!=\!0$ ($\theta\!\sim\!3\pi/7$) the GWF~\eqref{gwf_ss} is maximized (minimized), where we report an abrupt change in both limits after crossing the critical point $\lambda_c\!=\!1$ which is reflected in the first derivative of the GWF with respect to $\lambda$ (cf. Figure~\eqref{der_ssgwf}) by a divergence at the critical point $\lambda_c$. However, similarly to the analysis of the single site DWF, no sign of the factorization point manifests in the GWF. Due to the simple form of the single site density matrix, we find that both~\eqref{DWF_single_site} and~\eqref{gwf_ss} are directly related to the $\sigma_x$ coherence measures~\cite{CakmakPRB2014}. Furthermore, this is consistent with the fact that the presence of the interference terms in the original continuous Wigner function of a given quantum state indicates quantum coherence within the state.
\subsection{Two sites}
We extend our analysis to the case of a system of two sites $i$ and $j$ of an infinite quantum chain, with $i\!<\!j$ separated by some lattice spacing $m\!=\!j-i$. The reduced density matrix can be expressed as
\begin{equation}
\rho_{i,i+m}=\frac{1}{4}\sum_{\alpha,\beta=0}^3 p_{\alpha \beta}  \sigma_i^{\alpha}\otimes\sigma_{i+m}^{\beta},
\label{rho_ij}
\end{equation}
where $p_{\alpha \beta}=\langle\sigma_i^{\alpha}\sigma_{i+m}^{\beta}\rangle$ are the spin-spin correlation functions, ($\alpha,\beta)\!=\!0, \!1, \!2, \!3$. The two-site DWF can be concisely expressed as
\begin{equation}
\begin{split}
W_{\rho_{ij}}(x_1,x_2;p_1,p_2)=\frac{1}{16} \Big( 1+ \big[(-1)^{x_1}+(-1)^{x_2}\big]\langle \sigma^z \rangle\\+ (-1)^{p_1+p_2} \langle \sigma^x_i \sigma^x_{i+m} \rangle + (-1)^{x_1+x_2}\langle \sigma^z_i \sigma^z_{i+m} \rangle\\+ (-1)^{x_1+x_2+p_1+p_2} \langle \sigma^y_i \sigma^y_{i+m} \rangle \Big).
\end{split}
\label{eq4}
\end{equation}
On inspection it is evident that the DWF for a given choice of $(x_i,p_i)$ involves contributions from the various spin-spin correlation functions as well as the magnetization, which are central to spotlighting critical behavior. An advantage of the DWF~\eqref{eq4} is that it allows for a panoramic view of the properties of the system. In particular, evaluating the various DWF allows us to focus on contributions that are relevant in exhibiting the critical behavior. Since a given correlation measure will often depend on only specific spin-spin correlation functions, evaluating the DWF~\eqref{eq4} also allows a window into understanding the behavior of measures of quantum correlations across quantum phase transitions. A further advantage of the DWF~\eqref{eq4} is that any given DWF is experimentally accessible~\cite{expdwf2}. It is worth emphasizing that this expression is not specific to the models considered in this work, but applies to any Hamiltonian that is real and exhibits $\mathbb{Z}_2$ symmetry.

\begin{figure}[t]
    \subfloat[\label{tsDWF}]{%
       \includegraphics[width=0.49\textwidth]{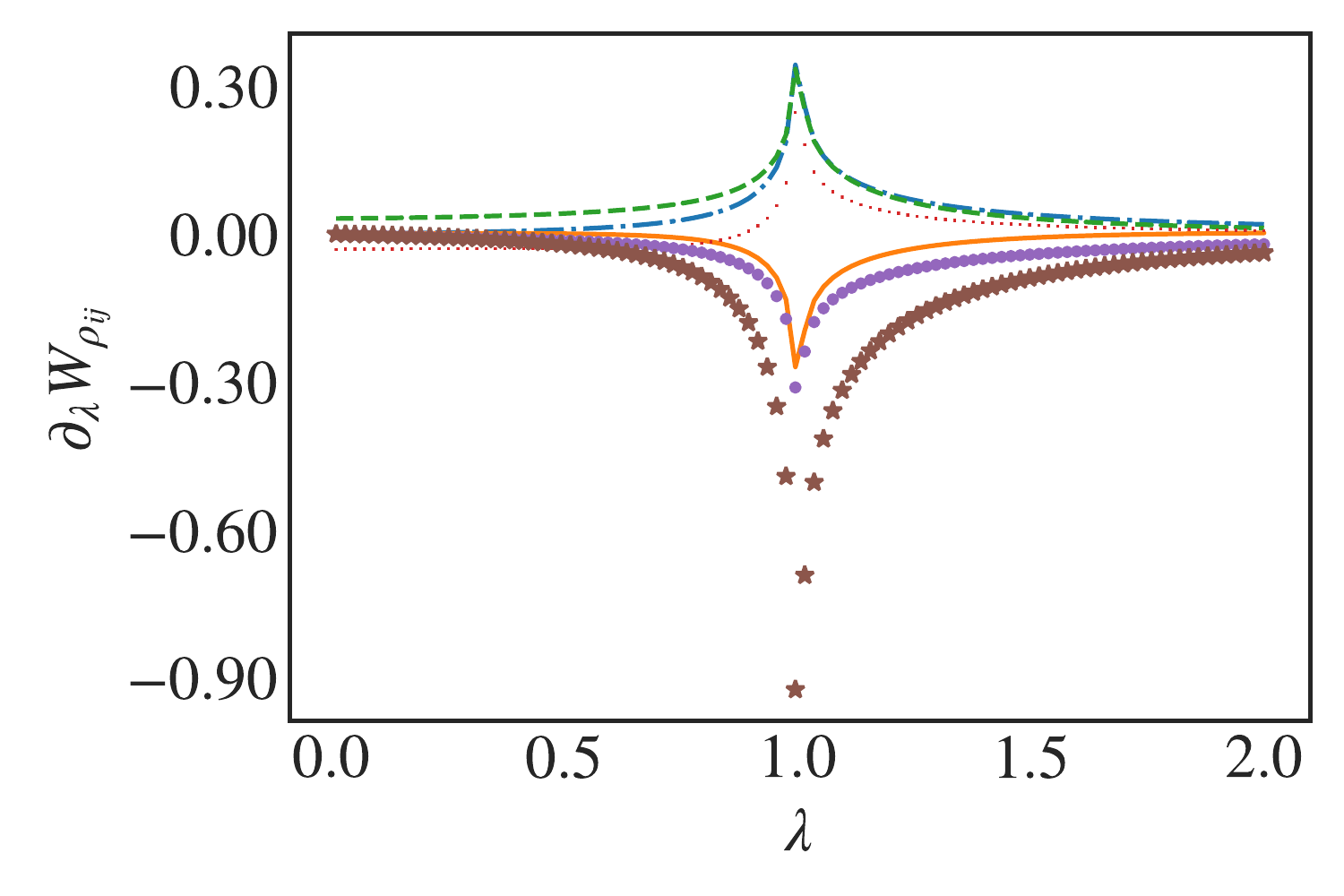}
       }%
     \subfloat[\label{sqtsDWF}]{%
       \includegraphics[width=0.49\textwidth]{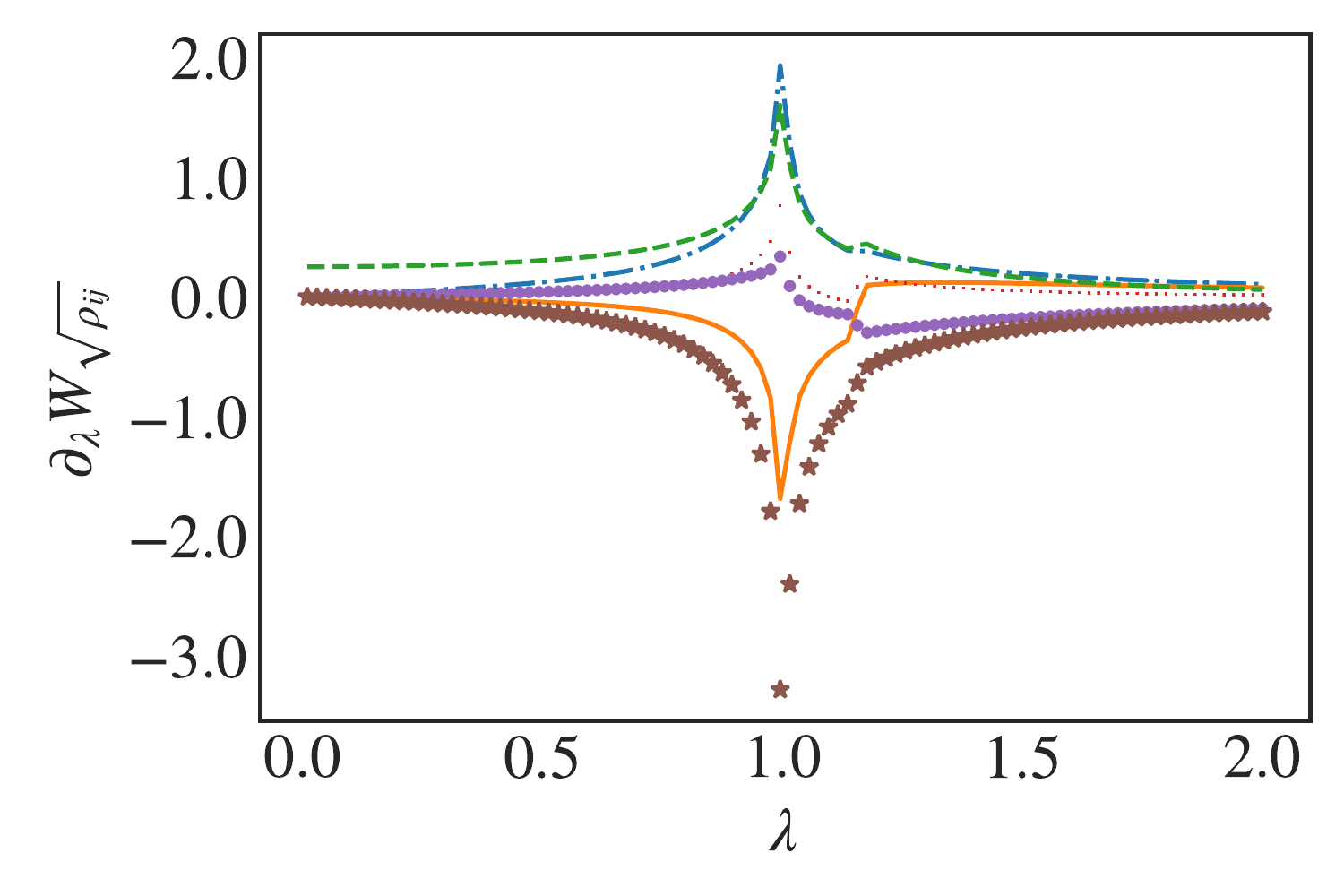}
       }\\
     \subfloat[\label{der_tsDWF}]{%
       \includegraphics[width=0.49\textwidth]{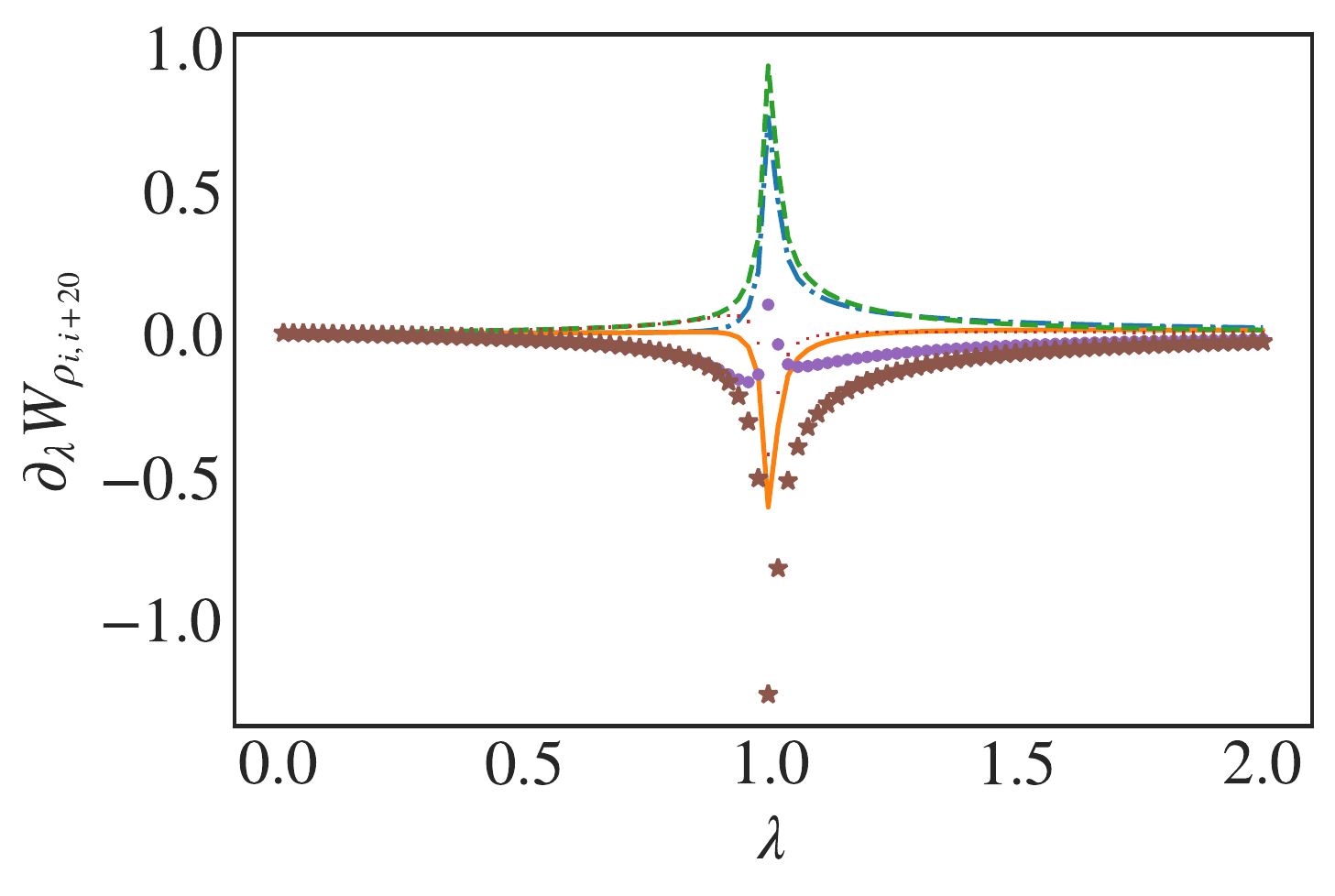}
       }%
     \subfloat[\label{der_sqssDWF}]{%
       \includegraphics[width=0.49\textwidth]{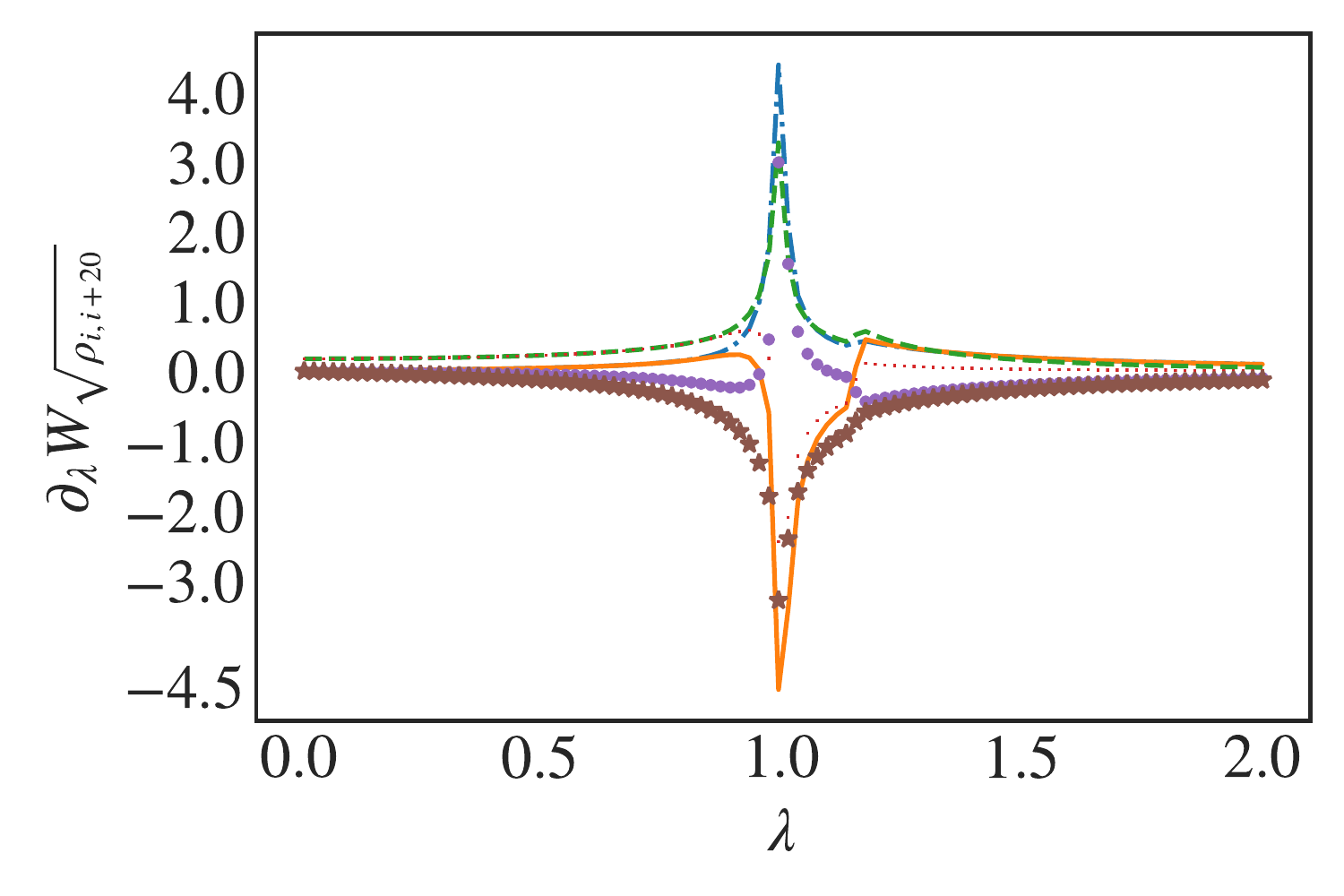}
       }
	\caption{First derivative with respect of $\lambda$ of the discrete Wigner function of (a) $W_{\rho_{ij}}$ and (b) $W_{\sqrt{\rho_{ij}}}$ for a pair of nearest neighbor spins for the $XY$ model. The six distinct behaviors correspond to the appropriate phase space points indicated in Table~\eqref{tabXY}. (c) and (d) show the same quantities for the long-range case of a pair of spins separated by 20 sites. In all panels $\gamma\!=\!0.5$}
	\label{fig_quantum phase transition_fp_xy}
\end{figure}

\noindent It is well known that various measures of bipartite quantum correlation accurately pinpoint the quantum phase transitions of this model~\cite{rozario, niel_osb, SarandyPRA2009, QPT2004, qptdiscord, Werlang2010, CampbellPRA2013, TonySciRep, qptcoherence, CakmakPRB2014}, therefore since the DWF is constructed from combinations of correlation functions that enter into the definition of such measures, it is not surprising that we find a qualitatively similar behavior. In line with these previous studies, Figure~\eqref{tsDWF} shows the first derivative of the DWFs for a pair of nearest neighbor spins. We see that there are six characteristic behaviors, c.f. Table~\eqref{tabXY}, and all of them clearly signal the quantum phase transition by showing a divergence in the first derivative at the critical point. Thus, as all discrete phase space points exhibit a qualitatively similar behavior, choosing to study any one in particular is sufficient to study the quantum phase transition.

\noindent It is interesting to note that, despite being dependent on all the relevant spin-spin correlation functions, there is no evidence of ground state factorization in the behavior of $\partial_{\lambda}W_{\rho_{ij}}$. In fact, as was shown for some coherence measures~\cite{CakmakPRB2014}, the factorization phenomenon is connected to an inherited discontinuity at the level of $\sqrt{\rho_{ij}}$ instead of $\rho_{ij}$. Remarkably we find a consistent behavior in the DWF by calculating $\partial_{\lambda} W_{\sqrt{\rho_{ij}}}$ in Figure~\eqref{sqtsDWF}. Now we find that the DWF develops a finite discontinuity at the factorization point for all six characteristic behaviors.

\begin{table}[t]
	\begin{subtable}{\linewidth}
		\centering
		\tikzmark{t}\\
		\tikzmark{l}
		\begin{tabular}{c   c     c     c c c}
			& & $00$ & $01$&10&11\\[-8pt]
			& \multicolumn{1}{@{}l}{\tikzmark{x}}\\
			00 & & \begingroup \color{blue!55} \textbf{-.-}\endgroup&\begingroup \color{orange} \textbf{\textemdash} \endgroup&\begingroup \color{orange} \textbf{\textemdash} \endgroup&\begingroup \color{blue!55} \textbf{-.-}\endgroup\\
			01 & & \begingroup \color{green} \textbf{ - - - }\endgroup&\begingroup \color{violet} \textbf{ \textellipsis\textellipsis }\endgroup&\begingroup \color{violet} \textbf{ \textellipsis\textellipsis }\endgroup&\begingroup \color{green} \textbf{ - - - }\endgroup \\
			10 & & \begingroup \color{green} \textbf{ - - - }\endgroup&\begingroup \color{violet} \textbf{ \textellipsis\textellipsis }\endgroup&\begingroup \color{violet} \textbf{ \textellipsis\textellipsis }\endgroup&\begingroup \color{green} \textbf{ - - - }\endgroup\\
			11 & & \begingroup \color{red} \huge $,$ \endgroup&\begingroup \color{Sepia} $\bigstar$ \endgroup&\begingroup \color{Sepia} $\bigstar$ \endgroup&  \begingroup \color{red} \huge $,$ \endgroup \\
		\end{tabular}\tikzmark{r}\\
		\tikzmark{b}
		\tikz[overlay,remember picture] \draw[-triangle 45] (x-|l) -- (x-|r) node[right] {$(p_1,p_2)$};
		\tikz[overlay,remember picture] \draw[-triangle 45] (t-|x) -- (b-|x) node[below] {$(x_1,x_2)$};
	\end{subtable}
	\vskip0.5cm
	\caption{Discrete phase space of the $ XY $ model. Each symbol corresponds to a particular curve shown in Figure~\eqref{fig_quantum phase transition_fp_xy}.}
	\label{tabXY}
\end{table}

\noindent Finally we examine the long range behavior of the DWF in the $XY$ model. Figure~\eqref{der_tsDWF} and~\eqref{der_sqssDWF} depict the first-order derivative of the DWF for a pair of spins $i$ and $j\!=\!i+m$ separated by $m\!=\!20$. For all phase space points, both first-order derivatives with respect to $\lambda$ of $W_{\rho_{i,i+20}}$ and $W_{\sqrt{\rho_{i,i+20}}}$ diverge at the critical point $\lambda_c\!=\!1$, revealing the quantum phase transition, while the discontinuity at the factorization point persists at long range only at the level of the first derivative of $W_{\sqrt{\rho_{i,i+20}}}$, which is consistent with the previous finding in the case of nearest neighbors.

\noindent As with the single site case, we extend the GWF formalism to the case of a system composed of two sites $i$ and $j$ with $i\!<\!j$, separated by some lattice spacing $m\!=\!j-i$. Plugging the reduced density matrix $\rho_{ij}$~\eqref{rho_ij} into the Stratonovich Wigner function~\eqref{gen_wf}, the two site GWF can be expressed as
\begin{align}
&\text{GWF}_{\rho_{ij}}(\theta_i,\varphi_i,\theta_j,\varphi_j)\!=\!\frac{1}{4} \Big[ 1-\sqrt{3} \left(\cos{2\theta_i} + \cos{2\theta_j} \right) \langle \sigma^z \rangle + \nonumber\\&
3\cos{2\varphi_i}\sin{2\theta_i}\cos{2\varphi_j}\sin{2\theta_j}  \langle \sigma^x_i\sigma^x_{j} \rangle + 3\sin{2\theta_i}\sin{2\theta_j} \nonumber \\&\sin{2\varphi_i}\sin{2\varphi_j} \langle \sigma^y_i\sigma^y_{j} \rangle +
3\cos{2\theta_i}\cos{2\theta_j} \langle \sigma^z_i\sigma^z_{j} \rangle \Big].
\label{gwf_ij}
\end{align}
In line with the DWF analysis, the GWF is written in terms of the various correlation functions with the additional dependence on the set of angles $(\theta_i,\varphi_i,\theta_j,\varphi_j)$. Again we see that~\eqref{gwf_ij} is a generalization of Wootters' DWF. 
\begin{figure}[t!]
	\centering
    \subfloat[\label{tsgwf}]{%
       \includegraphics[width=0.49\textwidth]{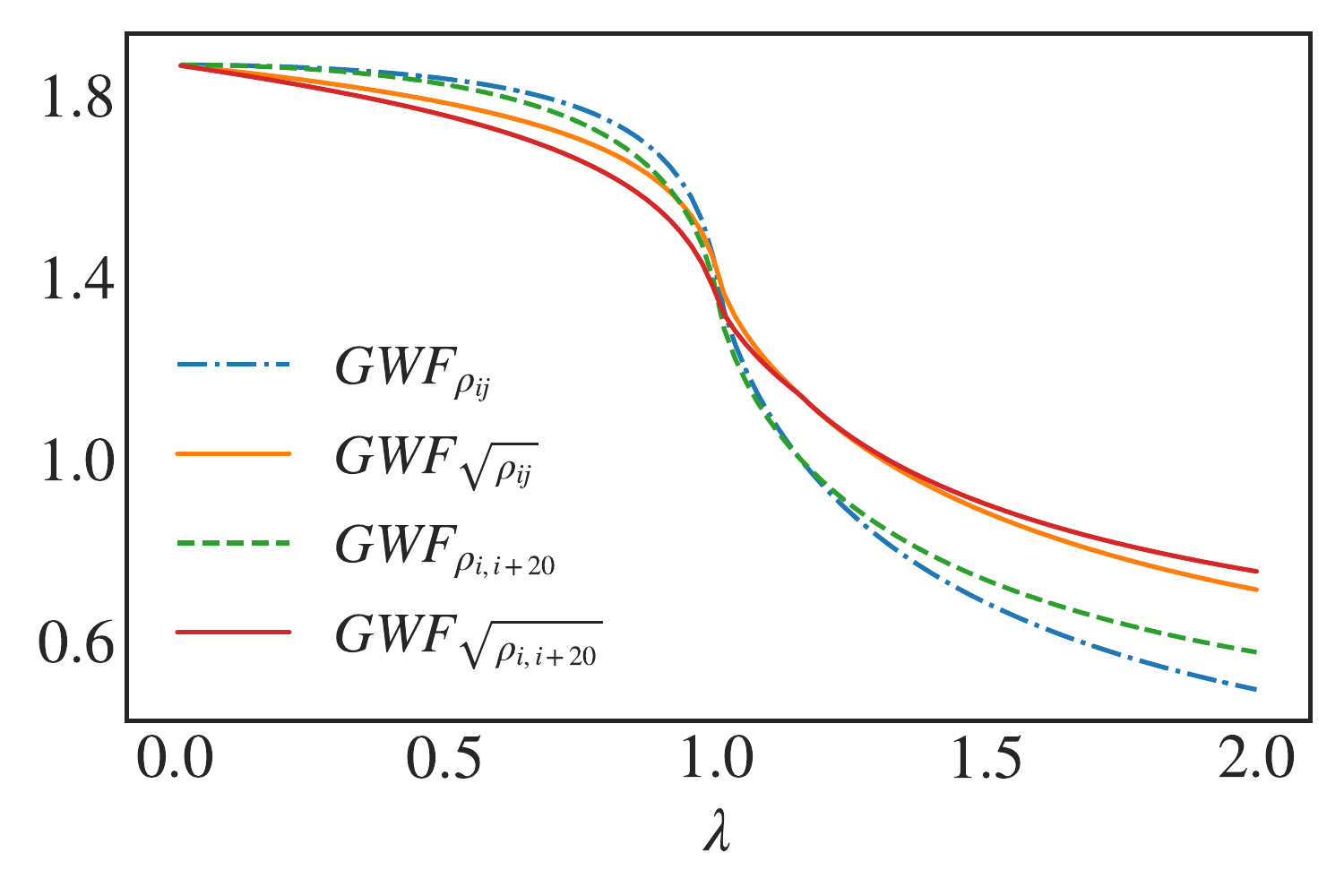}
       }%
     \subfloat[\label{sqtsgwf}]{%
       \includegraphics[width=0.49\textwidth]{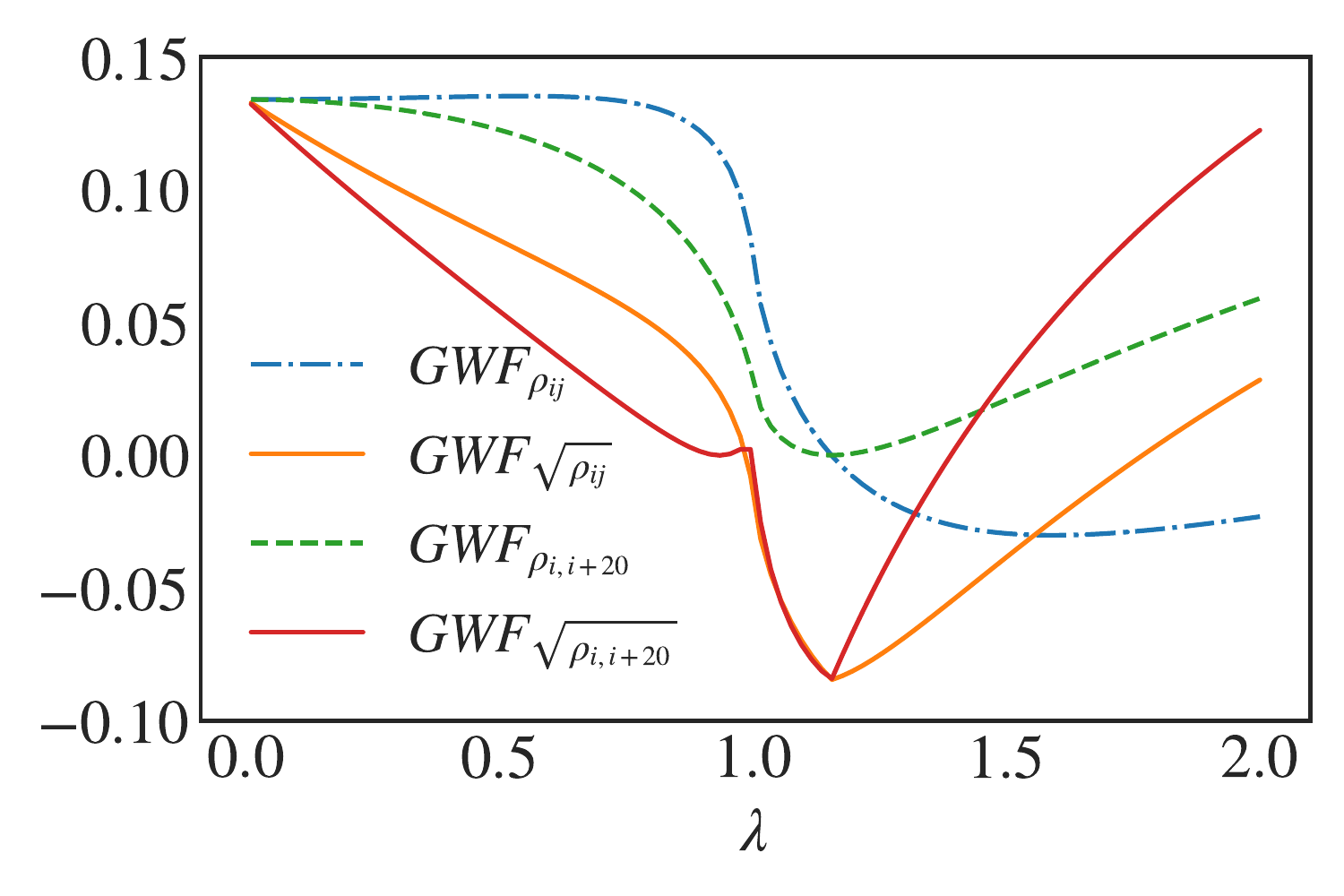}
       }\\
     \subfloat[\label{der_tsgwf}]{%
       \includegraphics[width=0.49\textwidth]{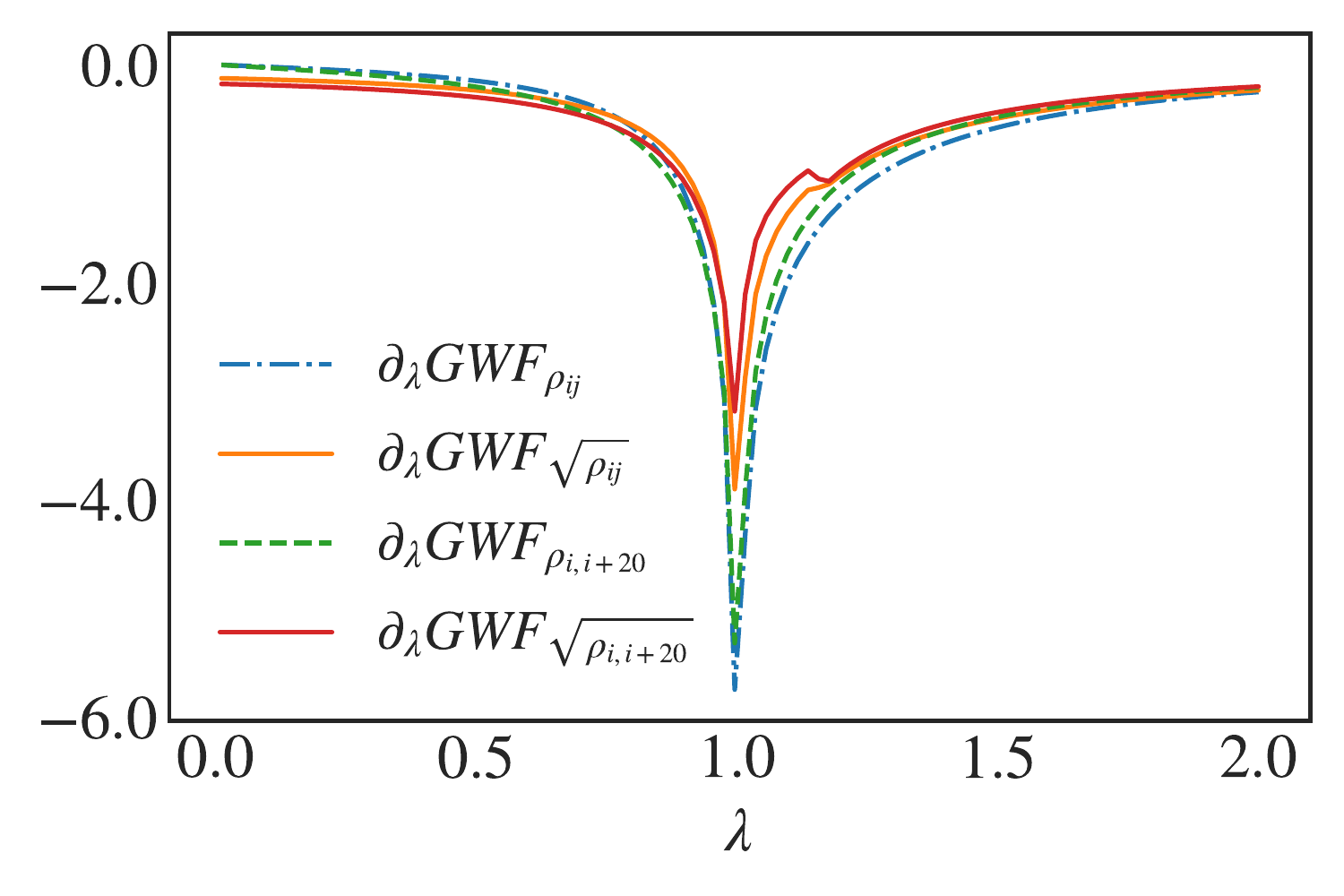}
       }%
     \subfloat[\label{der_sqssgwf}]{%
       \includegraphics[width=0.49\textwidth]{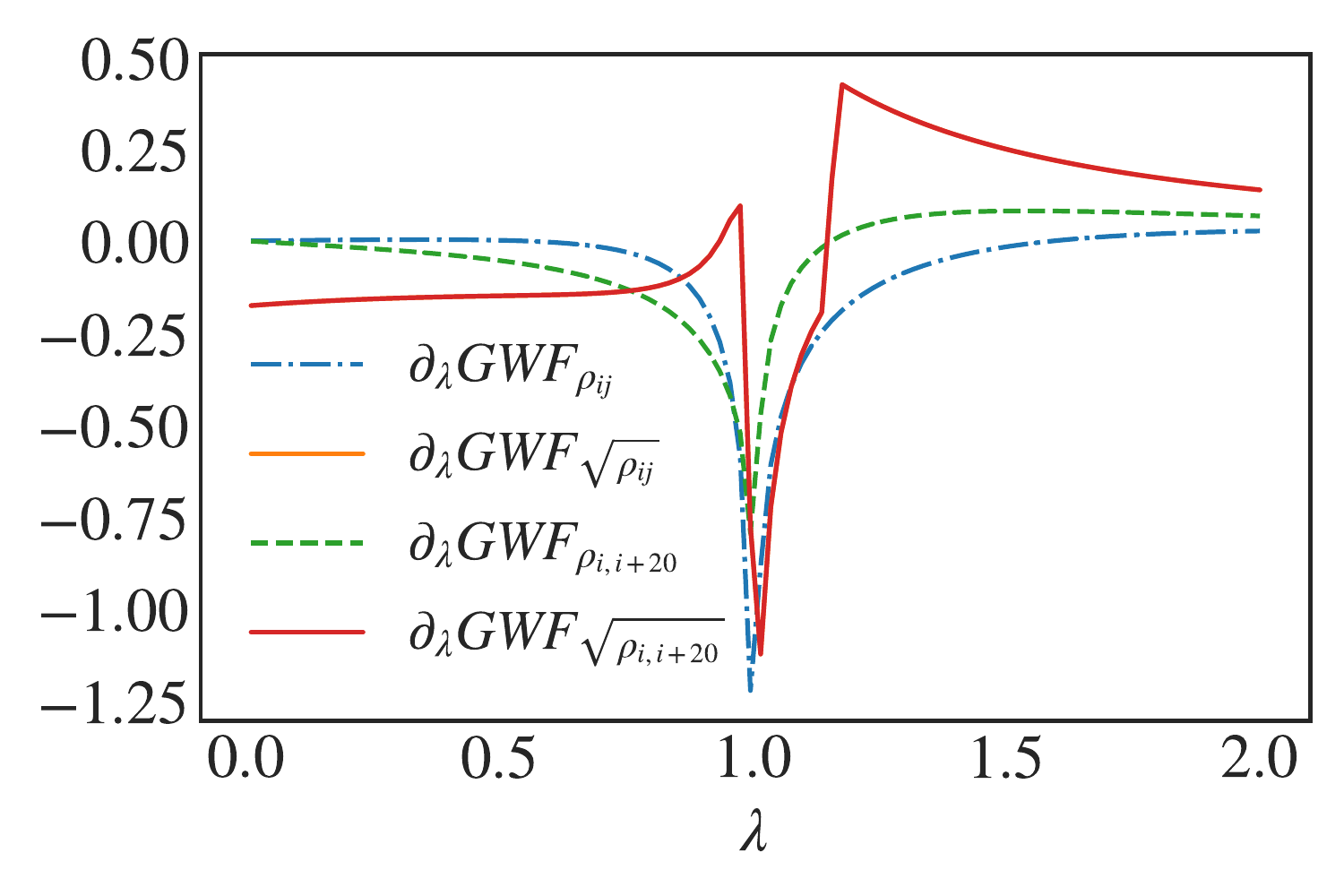}
       }
	\caption{The two sites GWF of the $XY$ model taking $\gamma=0.5$ in~\eqref{gwf_ij} [upper panels] and its first derivative with respect to $\lambda$ [lower panels] for (a) $\theta_i\!=\!\theta_j\!=\!\pi/2$; $\varphi_i\!=\!\varphi_j\!=\!2\pi$ and (b) $ \theta_i\!=\!\theta_j\!=\!\varphi_i\!=\!\varphi_j\!=\!0  $ in the case of nearest and the $20^{\text{th}}$ neighbor using $\rho_{ij}$ and $\sqrt{\rho_{ij}}$.}
	\label{fig_gwf_quantum phase transition_fp}
\end{figure}

\noindent Figure~\eqref{fig_gwf_quantum phase transition_fp} shows the behavior of the two site GWF for the $XY$ model with $\gamma=0.5$ for two angle configurations: $(\theta_i\!=\!\theta_j\!=\!\varphi_i\!=\!\varphi_j\!=\!0)$ and $(\theta_i\!=\!\theta_j\!=\!\pi/2 ; \varphi_i\!=\!\varphi_j\!=\!2\pi)$. Focusing on the upper panels of (a) and (b), we see an inflection point for both configurations for nearest neighbors (blue line) and $20^{\text{th}}$ neighbors (green line) at the critical point $\lambda_c\!=\!1$. As with the DWF, the factorization point can only be directly detected by examining the GWF for ${\sqrt{\rho_{ij}}}$, where a discontinuity appears in the derivative. However, the factorization phenomenon comes with an additional property: the value of the GWF {\it at the factorization point} is constant for any lattice distance $m$ which can be seen clearly for the $\text{GWF}_{\rho_{ij}}$ and $\text{GWF}_{\sqrt{\rho_{ij}}}$ in Figure~\eqref{fig_gwf_quantum phase transition_fp}. Such a behavior was first noted for the quantum discord in the same model~\cite{EPL2011, CampbellPRA2013}. This property can be understood when looking at the energy levels of finite-sizes of the $XY$ chain, where an energy-level crossing between the ground state and the first excited state take place exactly at the factorization point~\cite{CampbellPRA2013, AmicoPRB, GiorgiPRB} which forces the spin-spin correlation functions to have a constant value at any distance $m$. 

\subsection{Three sites}
\begin{figure}[t!]
    \subfloat[\label{3sgwf0}]{%
       \includegraphics[width=0.49\textwidth]{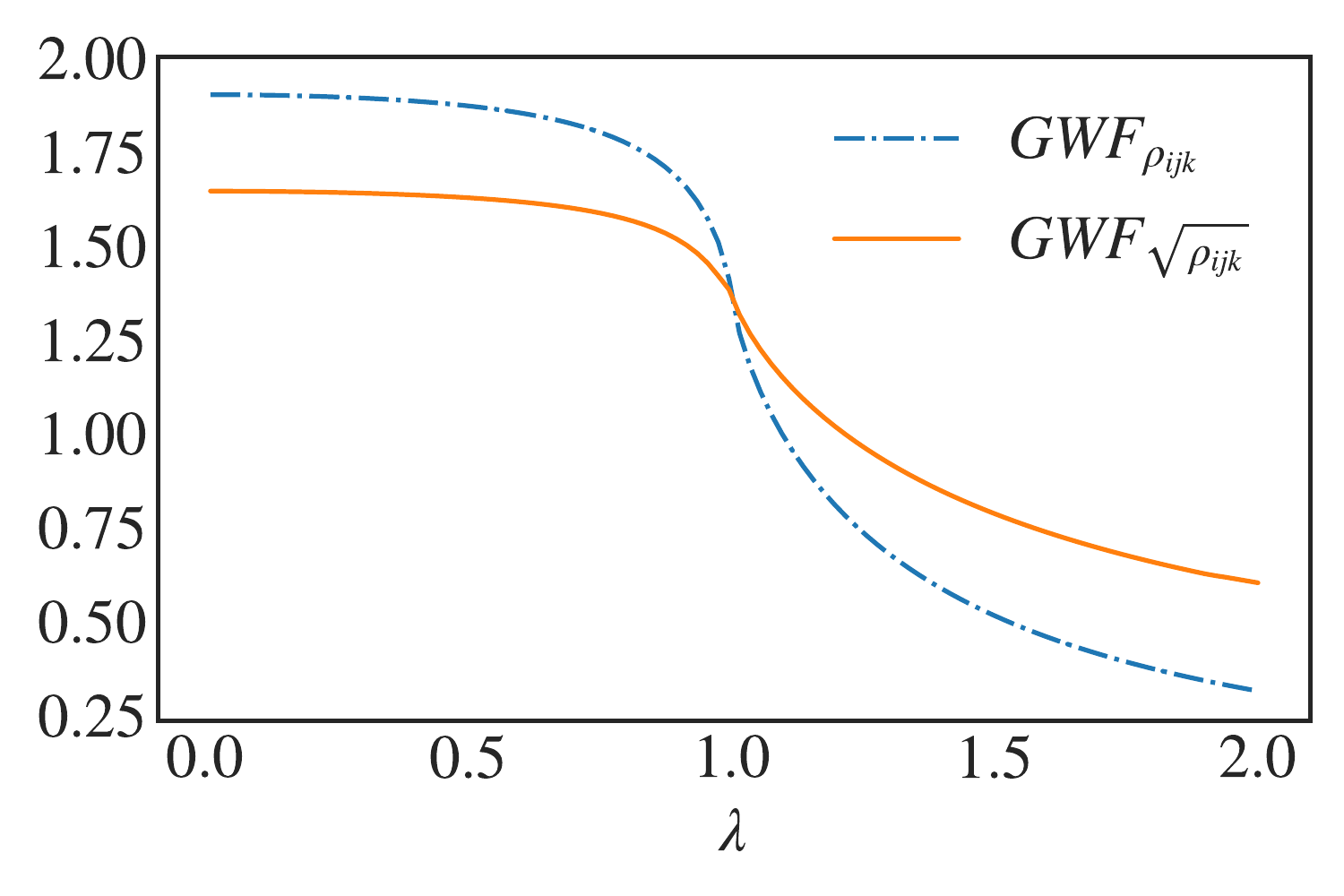}
       }%
     \subfloat[\label{3sgwfpi}]{%
       \includegraphics[width=0.49\textwidth]{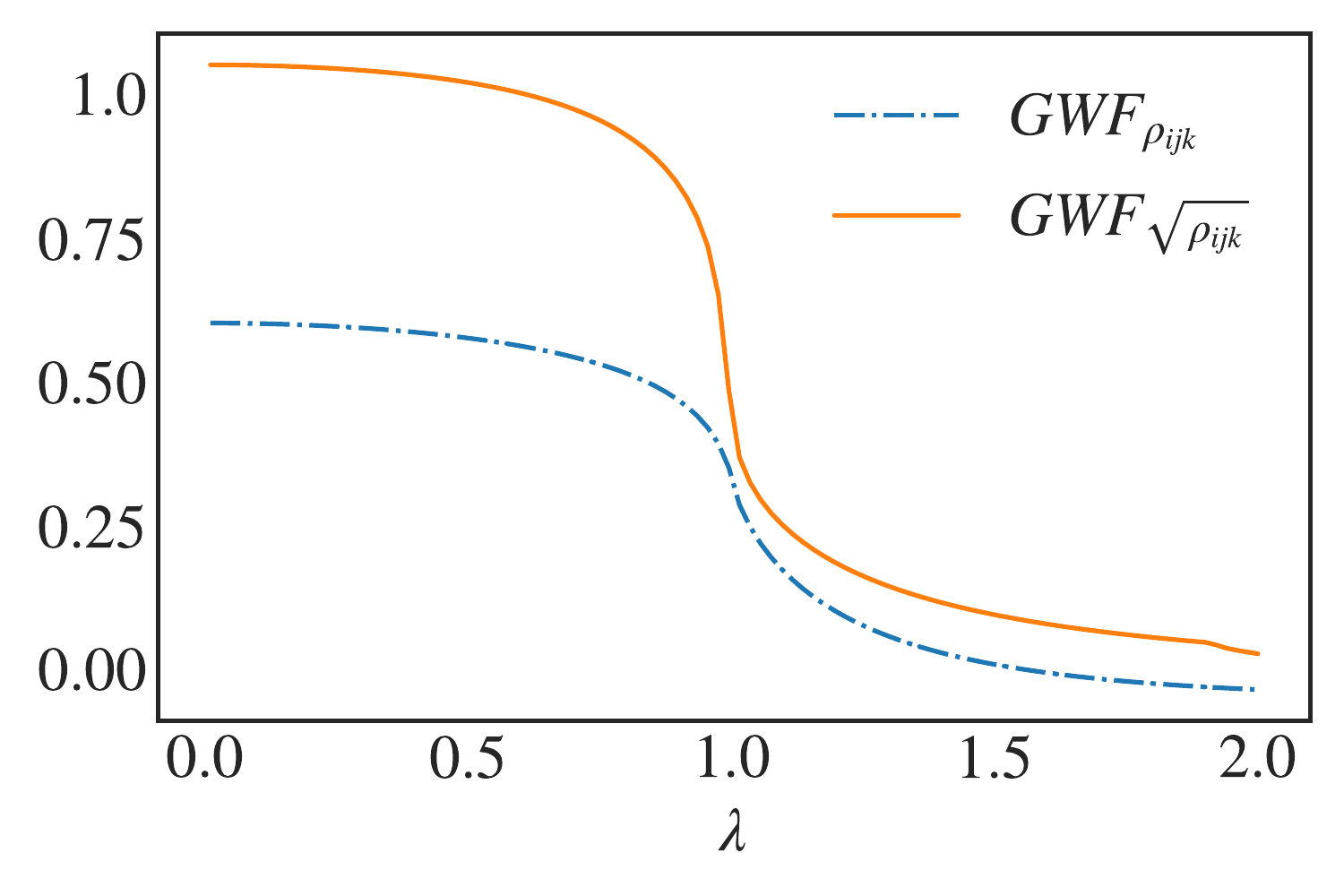}
       }\\
     \subfloat[\label{der_3sgwf0}]{%
       \includegraphics[width=0.49\textwidth]{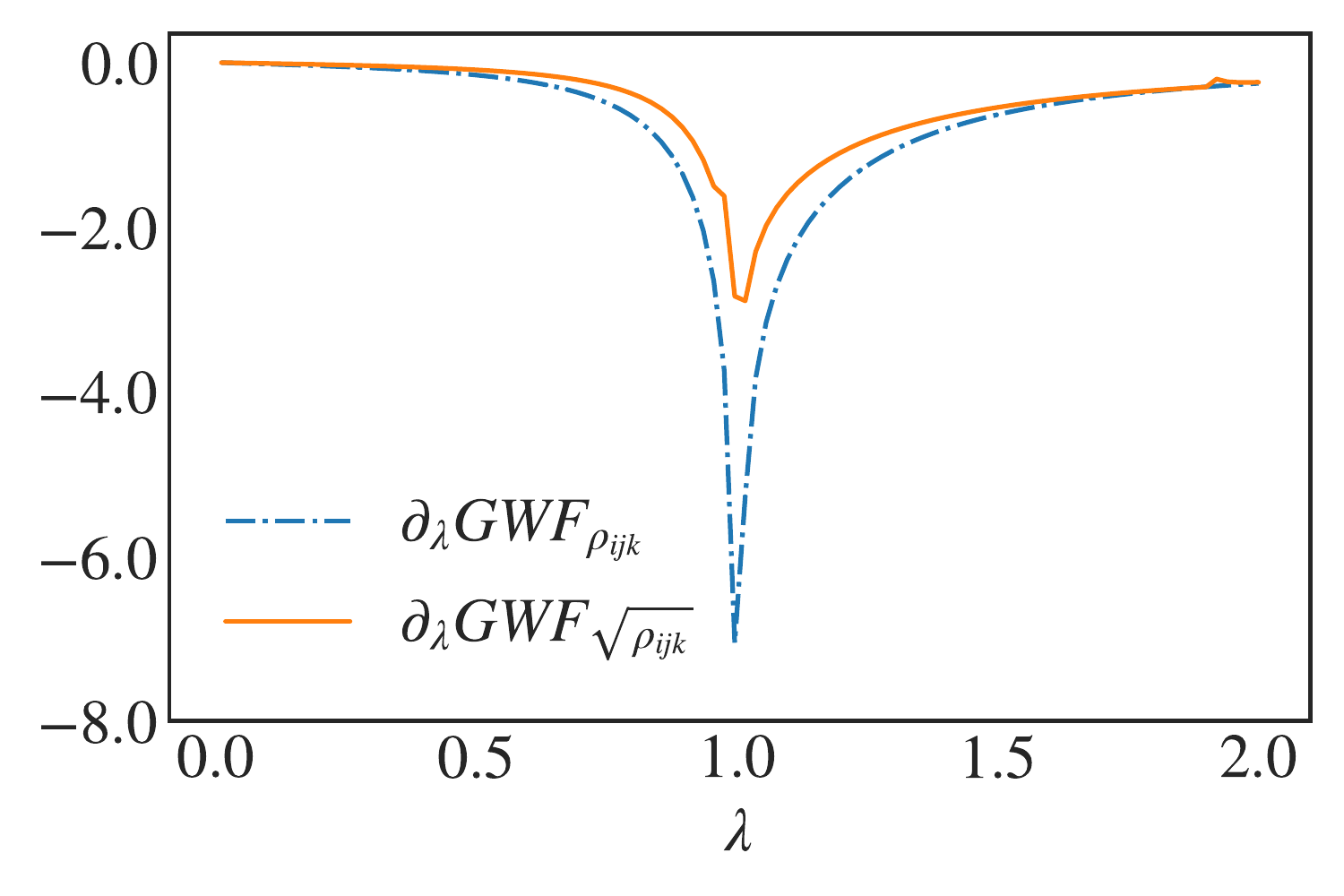}
       }%
     \subfloat[\label{der_3sgwfpi}]{%
       \includegraphics[width=0.49\textwidth]{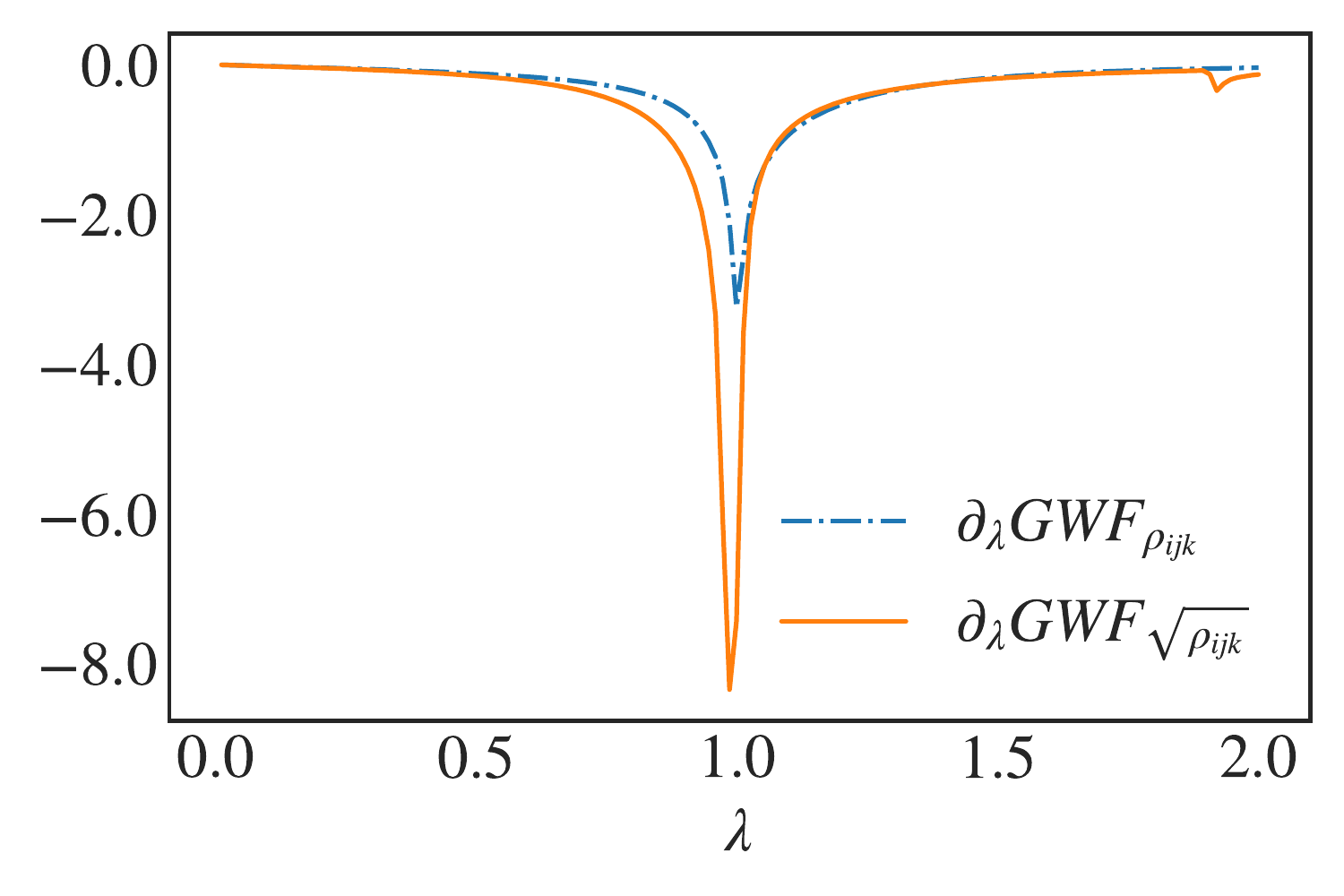}
       }
	\caption{The three sites GWF of the $XY$ model taking $\gamma=0.5$ in~\eqref{gwf_ijk} and its first derivative with respect to $\lambda$ for (a) $\theta_i\!=\!\theta_j\!=\!\pi/2$; $\varphi_i\!=\!\varphi_j\!=\!2\pi$ and (b) $ \theta_i\!=\!\theta_j\!=\!\varphi_i\!=\!\varphi_j\!=\!0  $ in the case of nearest neighbor spins.}
	\label{fig_gwf_ijk}
\end{figure}
A significant advantage of the GWF approach is that it can readily be extended to multipartite systems. Here we examine a three site system of the $XY$ model. The reduced density matrix $\rho_{ijk}$, taken by performing the partial trace over the infinite chain except the sites $(i,j,k)$, expressible as 
\begin{equation}
    \rho_{ijk}=\frac{1}{2^3} \sum_{\alpha,\beta,\gamma=0}^3 \langle \sigma_{i}^{\alpha} \sigma_{j}^{\beta} \sigma_{k}^{\gamma} \rangle \sigma_{i}^{\alpha} \otimes \sigma_{j}^{\beta} \otimes \sigma_{k}^{\gamma}
    \label{rho_ijk}
\end{equation}
and the full expression for the GWF is given below, c.f. formula~\eqref{gwf_ijk}. While in principle one could also consider the DWF for this case, in general the discrete phase space will consist of 64 behaviors, making it difficult to visualize. Furthermore, as we have established from the single and two-site analyses, the DWF corresponds to particular choices for the angles entering into the GWF. In analogy with the analysis of the two sites GWF, we consider two angle configurations: $\theta_i\!=\!\theta_j\!=\!\theta_k\!=\!\pi/2;\varphi_i\!=\!\varphi_j\!=\!\varphi_k\!=\!2\pi$ and $\theta_i\!=\!\theta_j\!=\!\theta_k\!=\!\varphi_i\!=\!\varphi_j\!=\!\varphi_k\!=\!0$ of the three sites GWF. In Figure~\eqref{fig_gwf_ijk} we see that in the multipartite case continues to spotlight the second order quantum phase transition for both sets of angles, however no sign of the factorization point can be seen. The failure of a multipartite non-classicality indicator to witness the factorization point is remarkable and is at variance with the behavior of certain indicators of multipartite entanglement which vanish in the thermodynamic limit~\cite{GiampaoloPRA}.
\begin{small}
	\begin{align}
	&\text{GWF}_{\rho_{ijk}}(\theta_i,\varphi_i,\theta_j,\varphi_j,\theta_k,\varphi_k) = \frac{1}{8} \Big[ 1-\sqrt{3}\left(\cos{2\theta_i} + \cos{2\theta_j} + \cos{2\theta_k} \right) \langle \sigma^z \rangle +
	3 \cos{2\varphi_i}  \nonumber \\& \sin{2\theta_i}\cos{2\varphi_k}\sin{2\theta_k} \langle \sigma^x_i\sigma^x_k \rangle  + 3\left( \cos{2\varphi_i}\sin{2\theta_i}\cos{2\varphi_j}\sin{2\theta_j} + \cos{2\varphi_j}\sin{2\theta_j}\cos{2\varphi_k}\sin{2\theta_k}  \right)  \nonumber \\&  \langle \sigma^x_i\sigma^x_j \rangle  +
	3 \sin{2\theta_i}\sin{2\varphi_i} \sin{2\theta_k}\sin{2\varphi_k} \langle \sigma^y_i\sigma^y_{k} \rangle + 3 \cos{2\theta_i}\cos{2\theta_k}  \langle \sigma^z_i\sigma^z_{k} \rangle + \nonumber \\& 3 \left(\sin{2\theta_i}\sin{2\theta_j}\sin{2\varphi_i}\sin{2\varphi_j} + \sin{2\theta_i}\sin{2\theta_k}\sin{2\varphi_j}\sin{2\varphi_k} \right) \langle \sigma^y_i\sigma^y_{k} \rangle +  \nonumber \\& 
	 +3 \left(\cos{2\theta_i}\cos{2\theta_j} + \cos{2\theta_j}\cos{2\theta_k} \right) \langle \sigma^z_i\sigma^z_{j} \rangle -
	3\sqrt{3} \cos{2\varphi_i}\sin{2\theta_i}\cos{2\varphi_j}\sin{2\theta_j}\cos{2\theta_k}   \langle \sigma^x_i\sigma^x_{j} \rangle  \langle \sigma^z \rangle + \nonumber \\ &
	3\sqrt{3} \cos{2\varphi_i}\sin{2\theta_i}\cos{2\varphi_k}\sin{2\theta_k}\cos{2\theta_j}   \langle \sigma^x_i\sigma^x_{k} \rangle  \langle \sigma^z \rangle - 3\sqrt{3} \sin{2\theta_i}\sin{2\varphi_i} \sin{2\theta_j}\sin{2\varphi_j}\cos{2\theta_k}   \langle \sigma^y_i\sigma^y_{j} \rangle  \langle \sigma^z \rangle + \nonumber \\&
	3\sqrt{3} \sin{2\theta_i}\sin{2\varphi_i} \sin{2\theta_k}\sin{2\varphi_k}\cos{2\theta_j}    \langle \sigma^y_i\sigma^y_{k} \rangle  \langle \sigma^z \rangle -
	3\sqrt{3} \cos{2\theta_i}\cos{2\theta_j}\cos{2\theta_k}  \left( \langle \sigma^z_i\sigma^z_{j} \rangle - \langle \sigma^z_i\sigma^z_{k} \rangle \right)  \langle \sigma^z \rangle
	\Big],
	\label{gwf_ijk}
	\end{align}
\end{small}
\section{The $XXZ$ model}
As a second interesting candidate system, we consider the $XXZ$ model with periodic boundary conditions
\begin{equation}
\mathcal{H}_{XXZ}=\frac{1}{4}\sum_{i=1}^N \sigma_i^x\sigma_{i+1}^x+\sigma_i^y\sigma_{i+1}^y+\Delta \sigma_i^z\sigma_{i+1}^z,
\label{eq6}
\end{equation}
where $\Delta$ is the anisotropy parameter. The phase diagram is split into three regions, separated by two different quantum phase transitions. For $\Delta\!\leq\!-1$, the system is in a ferromagnetic (gapped) phase and at $\Delta\!=\!-1$ a first-order quantum phase transition occurs. For $-1\!<\!\Delta\!<\!1$, the system is in a gapless (Luttinger liquid) phase and at $\Delta\!=\!1$ an infinite-order continuous quantum phase transition occurs, known as the Kosterlitz-Thouless quantum phase transition~\cite{Kosterlitz1973}. Finally, for $\Delta\!>\!1$, the system enters the anti-ferromagnetic (gapped) phase. The equilibrium properties of this model have been well studied, and in particular various measures of bipartite quantum correlations and their behavior across the different quantum phase transitions have been explored~\cite{qptdiscord,JafariPRA2008, SarandyPRA2009, RulliPRA2010, JafariPRA2017}. While entanglement and quantum discord were shown to reveal the critical points, their qualitative behaviors were shown to be strikingly different~\cite{SarandyPRA2009}. Here, by examining the DWF and the GWF we can shed greater light on these behaviors and show that when extremization procedures are employed, features spotlighting criticality become more pronounced.
\begin{table}[t]
    \begin{subtable}{\linewidth}
            \centering
            \tikzmark{t}\\
            \tikzmark{l}
        \begin{tabular}{c   c     c     c c c}
        & & $00$ & $01$&10&11\\[-8pt]
        & \multicolumn{1}{@{}l}{\tikzmark{x}}\\
        00 & & \begingroup \color{magenta}$ \bigstar$ \endgroup&\begingroup \color{red} \textbf{ - - -  } \endgroup&\begingroup \color{red} \textbf{ - - -  } \endgroup&\begingroup \color{magenta} $ \bigstar$ \endgroup\\
        01 & & \begingroup \color{green} $\times$ \endgroup&\begingroup \color{green} $\times$ \endgroup&\begingroup \color{green} $\times$ \endgroup&\begingroup \color{green} $\times$ \endgroup \\
        10 & & \begingroup \color{green} $\times$ \endgroup&\begingroup \color{green} $\times$ \endgroup&\begingroup \color{green} $\times$ \endgroup&\begingroup \color{green} $\times$ \endgroup\\
        11 & & \begingroup \color{magenta} $ \bigstar$ \endgroup&\begingroup \color{red} \textbf{ - - -  } \endgroup&\begingroup \color{red} \textbf{ - - -  } \endgroup&\begingroup \color{magenta}$ \bigstar$ \endgroup \\
        \end{tabular}\tikzmark{r}\\
        \tikzmark{b}
        \tikz[overlay,remember picture] \draw[-triangle 45] (x-|l) -- (x-|r) node[right] {$(p_1,p_2)$};
        \tikz[overlay,remember picture] \draw[-triangle 45] (t-|x) -- (b-|x) node[below] {$(x_1,x_2)$};
        \end{subtable}
        \vskip0.5cm
 \caption{Discrete phase space of the $XXZ$ model~\eqref{eq6}. Each symbol corresponds to one of the three characteristic behaviors shown in Figure~\eqref{fig_XXZ}.}
        \label{tabxxz}
\end{table}

\noindent Due to the form of Hamiltonian~\eqref{eq6} we find that no relevant information about the critical properties of the system can be revealed by studying only the single site density matrix. This is simply due to the fact that the single site density matrix depends only on $\langle \sigma^z \rangle$, which is constant for the $XXZ$ model. Therefore, for the remainder we will focus on the two site setting.
\begin{figure}[b!]
	 \subfloat[\label{xxz1}]{%
       \includegraphics[width=0.33\textwidth]{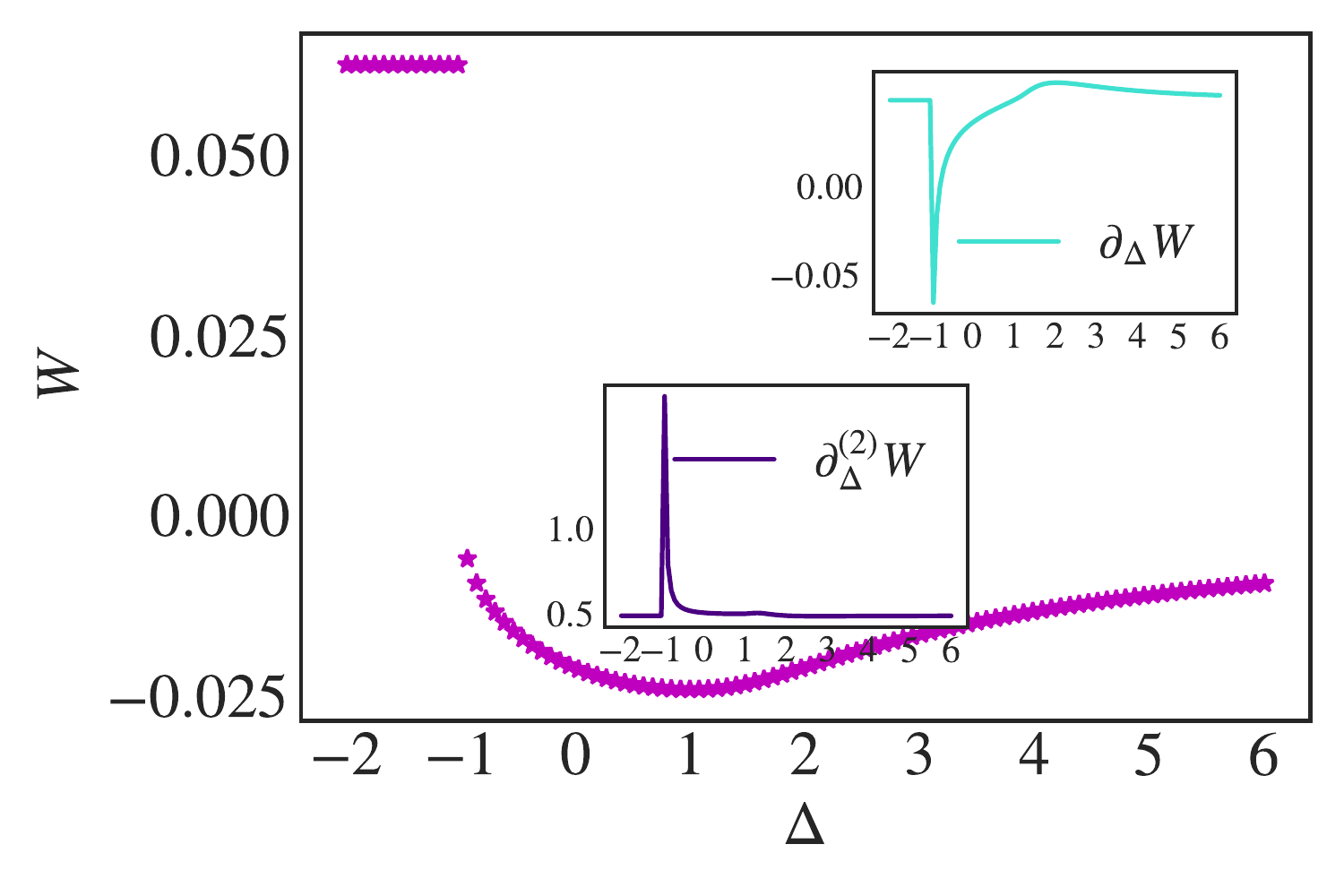}
       }%
     \subfloat[\label{xxz2}]{%
       \includegraphics[width=0.33\textwidth]{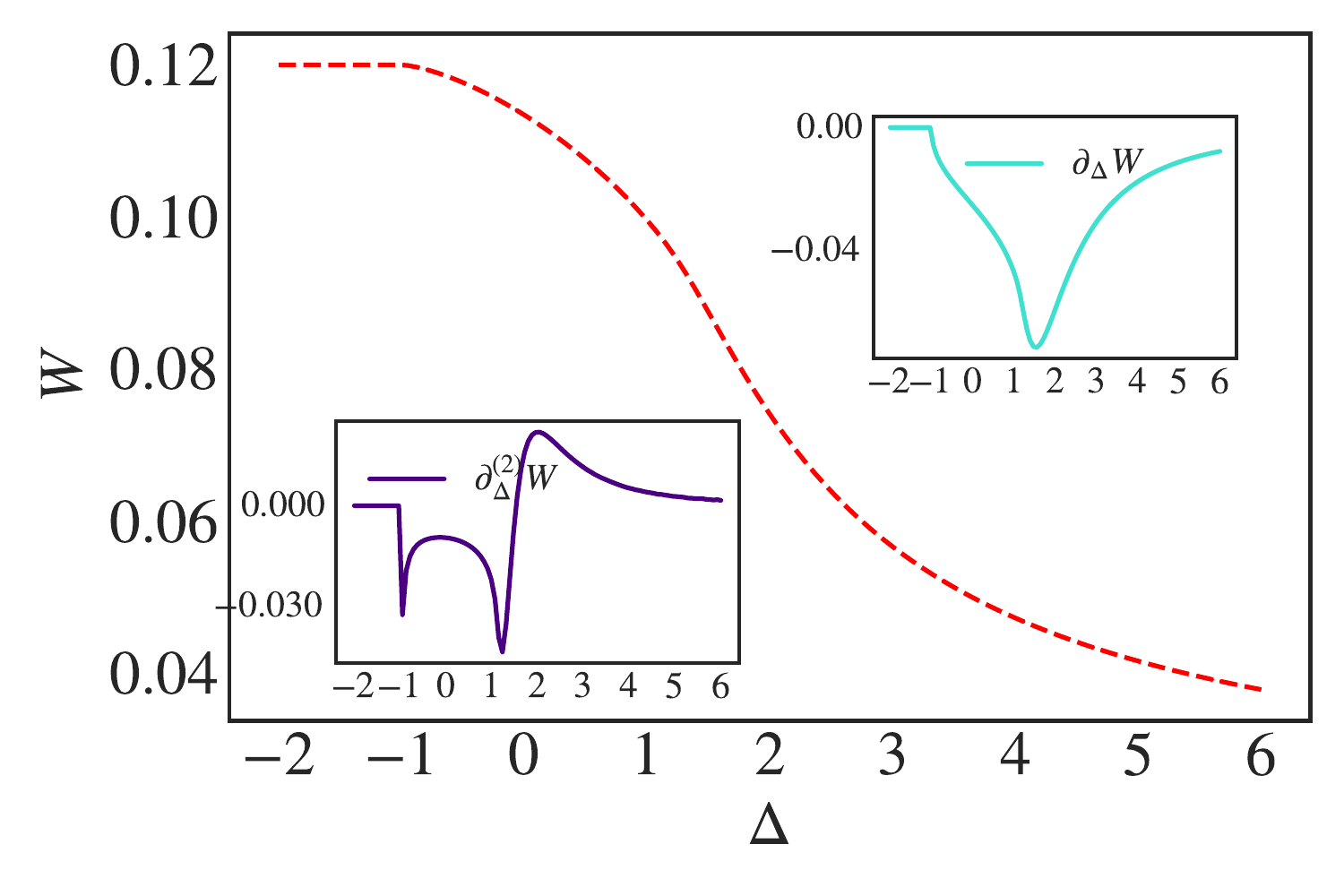}
       }
     \subfloat[\label{xxz3}]{%
       \includegraphics[width=0.33\textwidth]{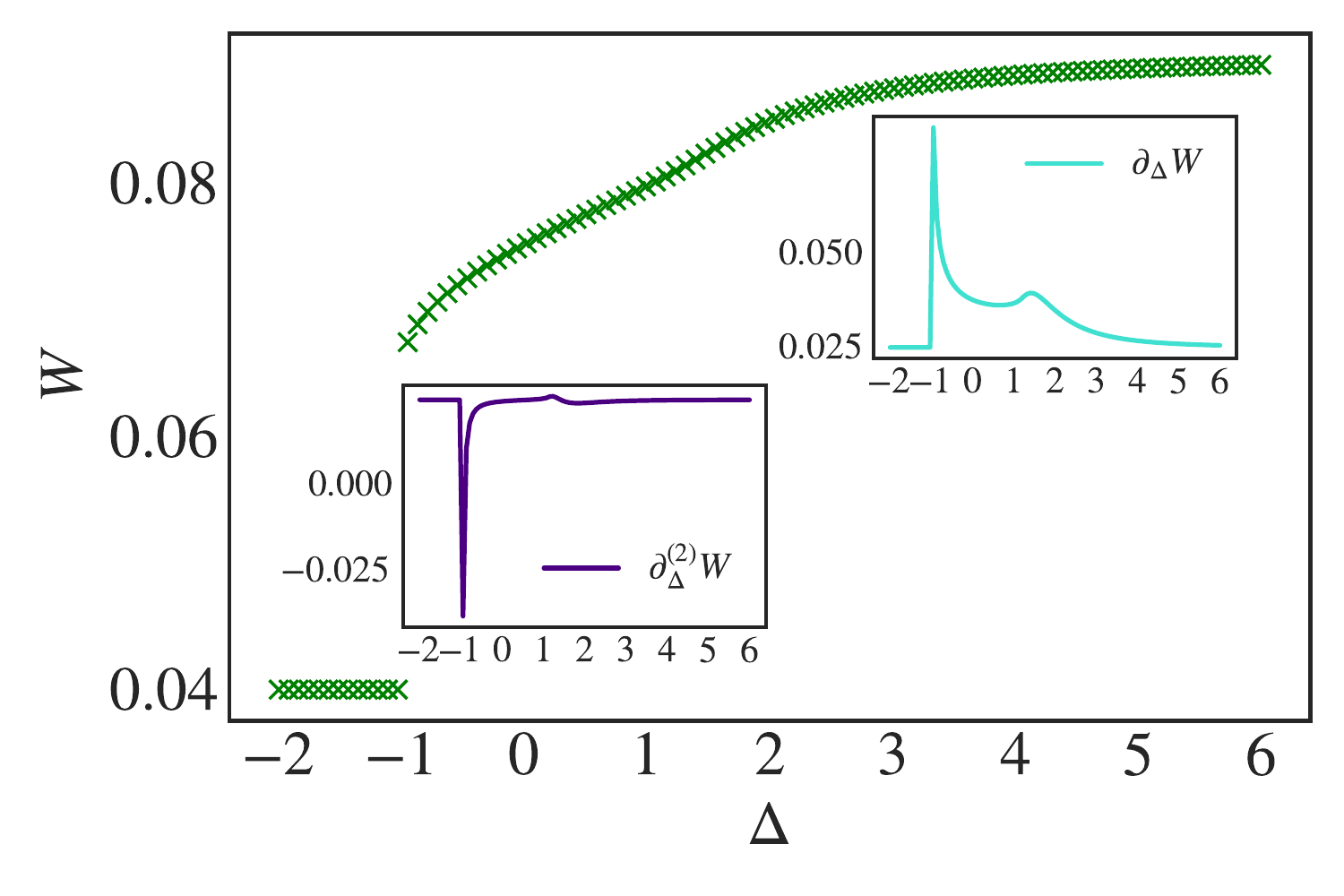}
       }
	\caption{Discrete Wigner function for a pair of nearest neighbors spins for the $XXZ$ model~\eqref{eq6}. The three distinct behaviors correspond to the appropriate phase space points indicated in Table~\eqref{tabxxz}.}
	\label{fig_XXZ}
\end{figure}
\noindent For the $XXZ$ model we find that the two-site DWF calculated following~\eqref{eq4} exhibits three distinct behaviors shown in Figure~\eqref{fig_XXZ} and Table~\eqref{tabxxz} as a function of $\Delta$, and their corresponding first and second derivatives for nearest-neighboring sites. Let us first consider the behavior of the DWF at the corners of the discrete phase space i.e. (00,00), (00,11), (11,00) and (11,11) [cf. Figure~\eqref{xxz1}]. We see that the DWF is discontinuous at the first-order quantum phase transition $\Delta\!\!=\!\!-1$ while it reaches a minimum at the infinite-order quantum phase transition at $\Delta\!\!=\!\!1$, after which the DWF approaches zero with increasing anisotropy. This behavior is qualitatively identical to that of  the entanglement measured via concurrence which in this case is simply $2 |\langle \sigma_i^x \sigma_j^x \rangle|$. The relationship is evident due to the fact that the DWF at these points depends on both $\langle \sigma_i^x \sigma_j^x \rangle$ and $\langle \sigma_i^z \sigma_j^z \rangle$. It is interesting that by direct calculation we confirm that the negativity of the DWF coincides with the presence of entanglement in the state, inline with a negative behavior of the continuous Wigner function implying genuine non-classicality of the state\cite{wfrabat, wfiran}.

\noindent The second significant behavior is located at phase space points (00,01), (00,10), (11,01), and (11,10) shown in Figure~\eqref{xxz2} where, in contrast with the previous cases, signatures of the critical points are less evident immediately in the behavior of the DWF. For $\Delta\!\!<\!\!-1$ these functions are constant and exhibit a sudden change at the first-order quantum phase transition. On inspection we can see a point of inflection around $\Delta\!=\!1.5$. Looking at the first derivative of the DWF we see that it presents an amplitude bump at $\Delta=1.5$, and the second derivative is divergent at $\Delta\!=\!-1$ and around $\Delta\!=\!1$.  The more peculiar behavior seen in this DWF is due to the destructive interference at these phase space points between the two terms that control the DWF which are $1+\langle \sigma_i^z \sigma_j^z \rangle$ and $-2\langle \sigma_i^x \sigma_j^x \rangle$, and the inflection point seen arises from a sudden change in the concavity of $-2\langle \sigma_i^x \sigma_j^x \rangle$. Thus, unlike in the $XY$ model where all DWFs readily witness the quantum phase transition, the DWF in these four points can only easily signal the first-order quantum phase transition exactly, while for the infinite-order quantum phase transition it shows only some anomalies around $\Delta\!\!=\!\!1$. However, we will revisit this behavior in the context of extremization procedures shortly.

\noindent Finally we consider the remaining eight phase space points, Figure~\eqref{xxz3}. Here the DWF depends solely on a single term, $1-\langle \sigma_i^z \sigma_j^z \rangle$, and owing to the fact that spin-spin correlation functions are discontinuous at $\Delta\!\!=\!\!-1$ and that on their own they fail at revealing the infinite-order quantum phase transition at $\Delta\!=\!1$ the DWF at these points inherits these properties from the $\langle \sigma_i^z \sigma_j^z \rangle$ contribution which explains why the DWF is discontinuous and its derivatives are divergent at $\Delta\!=\!-1$, while it does not show any special behavior at the infinite-order quantum phase transition $\Delta\!=\!1$.
\begin{figure}[t!]
	 \subfloat[\label{extremization}]{%
       \includegraphics[width=0.49\textwidth]{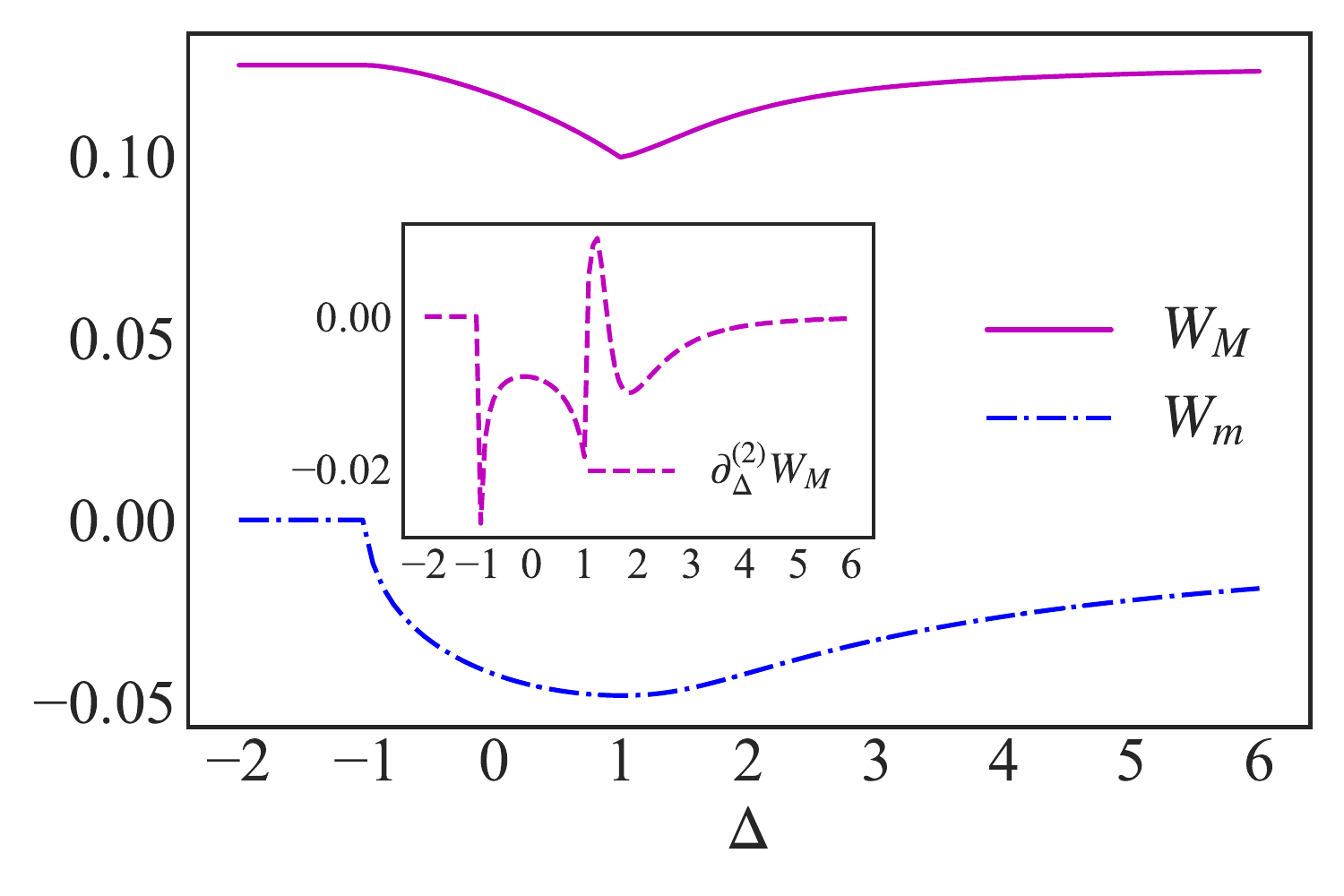}
       }%
     \subfloat[\label{maxmin}]{%
       \includegraphics[width=0.49\textwidth]{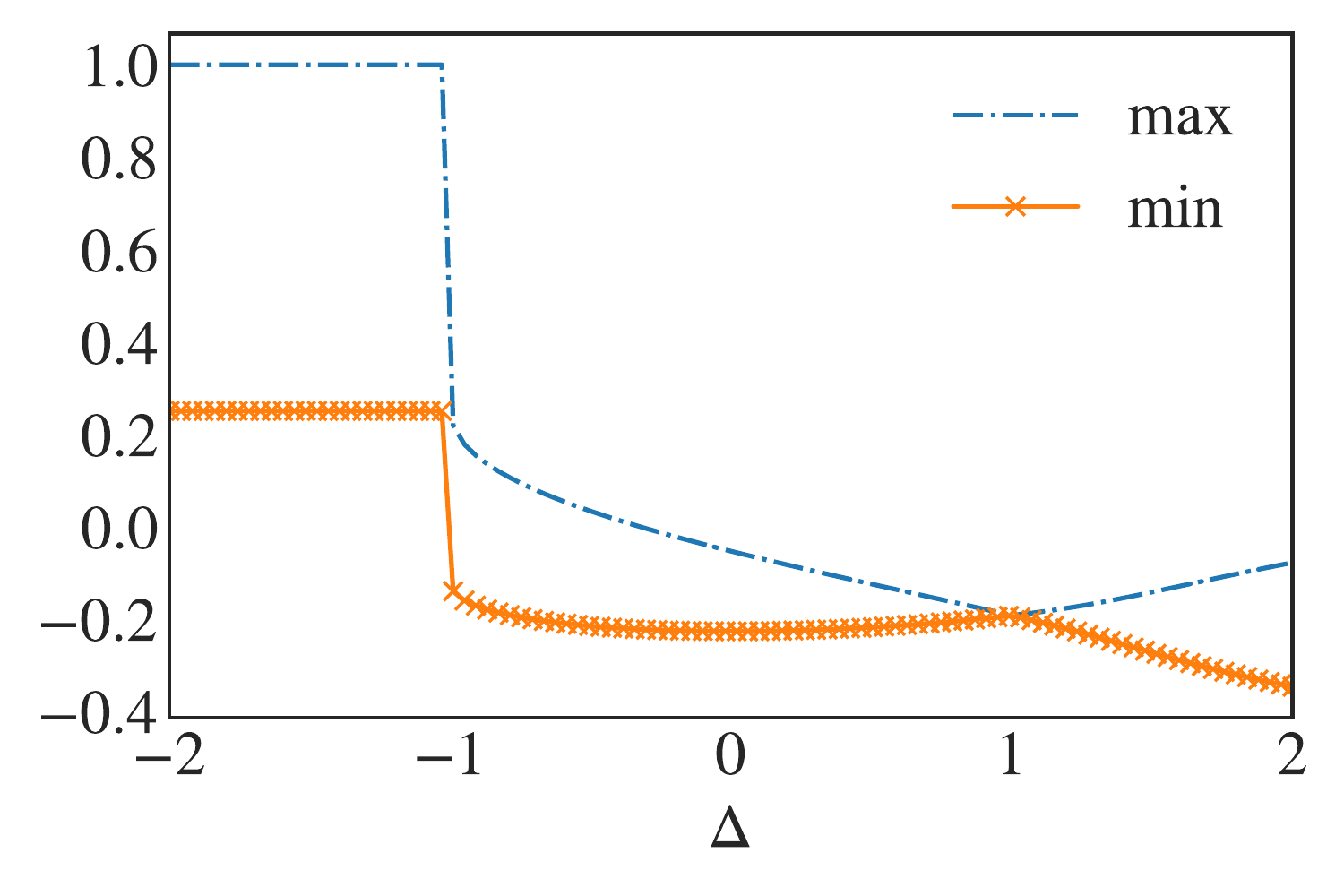}
       }
\caption{(a) Extremization of the DWF following~\eqref{eqExtreme}. The topmost curve shows the maximization $W_M$ and the lower curve shows the minimization $W_m$. The inset shows the second derivative with respect to $\Delta$ of $W_M$. (b) The maximum and the minimum behavior of the GWF for the $XXZ$ model~\eqref{gwf_xxz}.}
\label{fig3}
\end{figure}
\noindent Several correlation measures involve a minimization or maximization to be performed and often such correlation measures stand out as the preferred figures of merit for studying criticality~\cite{EPL2011, SarandyPRA2009, CampbellPRA2013, Werlang2010, CakmakPRB2014}. In this regard it is interesting to consider a similar extremization procedure for the DWF. Let $W_M$ and $W_m$ be the maximized and minimized DWF over the discrete phase space, respectively, given by
\begin{equation}
\begin{aligned}
W_M&=\max(W_{00,00},W_{00,01},W_{01,00}),\\
W_m&=\min(W_{00,00},W_{00,01},W_{01,00}),
\end{aligned}
\label{eqExtreme}
\end{equation}
where we have chosen $W_{00,00}$, $W_{00,01}$, and $W_{01,00}$ to capture the three distinct behaviors exhibited in the discrete phase space. In Figure~\eqref{extremization} we see $W_M$ reveals a cusp exactly at the infinite-order quantum phase transition and thus its first (second) derivative is discontinuous (divergent) at the critical point, $\Delta\!=\!1$, as shown in the inset. This indicates that the DWF could be a good alternative to correlation measures that involve extremization procedures due to the comparative simplicity in its calculation and its easy physical interpretation following~\eqref{eq4}. Information about the entanglement in the two-qubit $XXZ$ model, can be extracted from the minimization of the DWF, $W_m$. On close inspection, we notice from Figure~\eqref{concurrence_xxz} and~\eqref{fig3}, that the concurrence is proportional to $|W_m|$. Looking at $\partial_\Delta^2 W_M$ of the DWF in the inset of Figure~\eqref{extremization} and the corresponding second derivatives of the distinct behaviors in discrete phase space shown in Figure~\eqref{xxz2}, where we have destructive interference between the terms that control the DWF (for example the point (00,01)), we find that both behave quite similarly. Therefore, it appears that to be able to detect \textit{reliably} the infinite-order quantum phase transition, one requires a figure of merit that includes all the spin-spin correlation functions of the quantum system. This is further evidenced by the fact that the other parts of the phase space, where only a single spin-spin correlation term is dominant, are less sensitive to this quantum phase transition. Figure~\eqref{concurrence_xxz} depicts the concurrence~.
\begin{figure}[t!]
    \centering
    \includegraphics[scale=0.7]{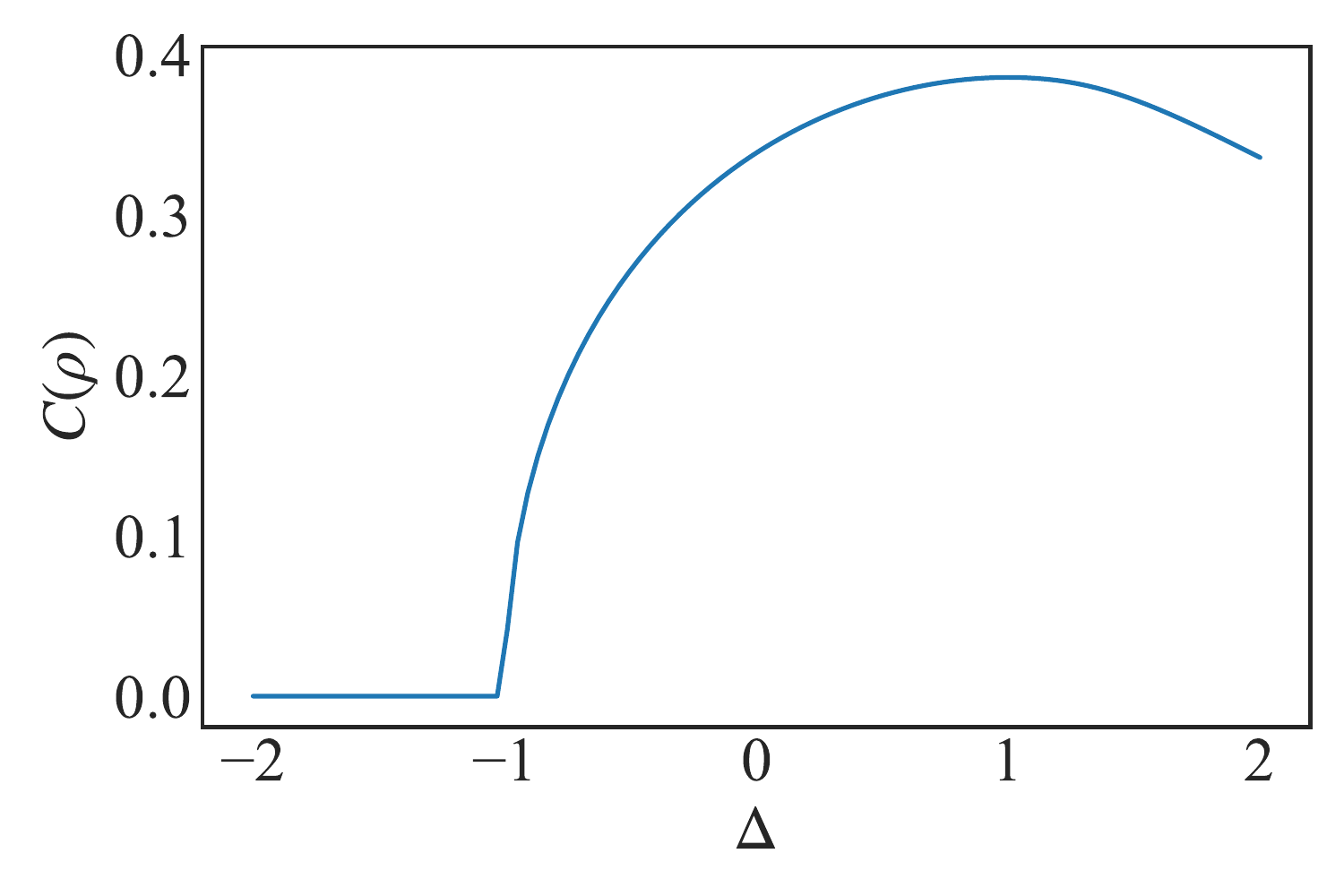}
    \caption{The concurrence in the two-qubit $XXZ$ model with respect to $\Delta$.}
    \label{concurrence_xxz}
\end{figure}

\noindent We finally consider the two site GWF
\begin{align}
\text{GWF}_{\rho_{ij}}&(\theta_i,\varphi_i,\theta_j,\varphi_j)\!=\!\frac{1}{4} \Big( 1+ 3 \cos{2\theta_i}\cos{2\theta_j}  \langle \sigma^z_i\sigma^z_{j} \rangle + \nonumber\\  & 3\sin{2\theta_i}\sin{2\theta_j}\cos{2(\varphi_i-\varphi_j)} \langle \sigma^x_i\sigma^x_{j} \rangle
\Big).
\label{gwf_xxz}
\end{align}
Similarly with the $XY$ model, we are interested in a set of angles $(\theta_i,\varphi_i,\theta_j,\varphi_j)$ that yield the maximum ($\theta_i\!=\!\theta_j\!=\!\pi/2$ for $\Delta\!\leq\!1$ and $\theta_i\!=\!\theta_j\!=\!\pi/4$ for $\Delta\!>\!1$) and the minimum ($\theta_i\!=\!\theta_j\!=\!\pi/4$ for $\Delta\!\leq\!1$ and $\theta_i\!=\!\theta_j\!=\!\pi/2$ for $\Delta\!>\!1$) behavior of the GWF. One may notice that this procedure does not depend on the angles $(\varphi_i,\varphi_j)$, this is because they cancel out in~\eqref{gwf_xxz} when they are bounded by the same interval. Looking at Figure~\eqref{maxmin} we see that the maximum (minimum) behavior of the GWF of the $XXZ$ model is constant when $\Delta \!\!<\!\!-1$ and shows an inherited discontinuity from the spin-spin correlation functions, at the first-order quantum phase transition point $\Delta\!\!=\!\!-1$. Moreover, reaching the point of the infinite-order quantum phase transition $\Delta\!=\!+1$, the angles describing the maximum (minimum) behaviors of~\eqref{gwf_xxz} switch from $\pi/4$ to $\pi/2$ ($\pi/2$ to $\pi/4$) which manifests as a cusp, revealing the infinite-order quantum phase transition at $\Delta\!\!=\!\!1$ and thus by exploiting an extremization procedure we are able to faithfully spotlight the infinite-order quantum phase transition.
\section{Summary}
\noindent Throughout this chapter, we presented an alternative method to study quantum phase transitions from a phase space perspective using two approaches: the discrete Wigner function (DWF) and the generalized Wigner function (GWF). By establishing a connection between the phase space techniques and the thermodynamical quantities of a quantum spin-$\frac{1}{2}$ chain, we have shown the DWF and the GWF to be versatile tools in studying first, second, and infinite-order quantum phase transitions. Furthermore, we have shown that signatures of ground state factorization are only present in bipartite quantities. In addition, our approach may provide a promising tool for the experimental investigation of quantum phase transitions following the procedures proposed in References~\cite{expwf, expdwf2, GWF_entanglement_PRA}. Furthermore, through equations~\eqref{DWF_single_site},~\eqref{gwf_ss},~\eqref{eq4},~\eqref{gwf_ij},~\eqref{gwf_xxz}, and~\eqref{gwf_ijk} a given DWF/GWF is easily physically interpreted and can be generalized to higher dimensional systems which is a task proven to be difficult and complex for quantum correlations measures. Beyond characterizing phase transitions, our approach also provides insight into the behavior of various correlation measures and quantum coherence in such systems. While we have focused on equilibrium systems, we expect our approach to be useful in examining the dynamical properties of such critical systems~\cite{CampbellPRB2016, HeylReview, NJPSchachenmayer, PRASchachenmayer, PRBGasenzer, QST2019, PRXSchachenmayer}. 

\part{Quantum Thermodynamics meets Quantum Phase Transitions}
\chapter{Thermodynamics of Information \label{chap6} }
\section{Foundations of thermodynamics}
Thermodynamics is a theory that aims to describe the behavior of heat and work in physical systems. Historically, understanding the forms of energy was limited to practical needs. In the late $18^{\text{th}}$ century, heat was recognized as a form of energy by Rutherford (also known as Thompson), when he analyzed the amount of heat generated via cannon barrels~\cite{rutherford}. The industrial revolution in the $19^{\text{th}}$ century ignited the development of the theory of thermodynamics, in order to design efficient steam engines. Sadi Carnot is considered the founding father of thermodynamics, as he noticed that the efficiency of steam engines can be optimized by understanding their energy balance between work and heat~\cite{carnot1978reflexions}. Carnot's results were generalized by E. Clapeyron~\cite{clapeyron1834memoire}, and after three decades R. Clausius constructed the mathematical framework of thermodynamics~\cite{clausius}. In principle, the theory of thermodynamics focuses on the transformation of systems from one state of \textit{equilibrium} to another, while satisfying five axioms called the laws of thermodynamics.
\begin{enumerate}
    \item \textbf{Zeroth law;}\\
    when two systems are each in thermal equilibrium with a third system, the first two systems are in thermal equilibrium with each other. This law allows us to use thermometers as the “third system” and to define a temperature scale. The Zeroth Law of thermodynamics defines a state of equilibrium of a system with respect to its environment, which can be described by an equation of state relating the experimentally accessible parameters of the system. For steam engines, the parameters are pressure $P$, temperature $T$, and volume $V$ which are related by
    \begin{equation}
        f(P,V,T)=0,
        \label{eq_state}
    \end{equation}
    where $f$ is a characteristic function. For an ideal gas, the formula~\eqref{eq_state} translates into the well known $PV=Nk_bT$, where $N$ is the number of particles, and $k_b$ is the Boltzmann constant.
    \item \textbf{First law;}\\
    the change in the system's internal energy, $dE$, is equal to the difference between the heat, $\delta Q$, absorbed by the system from its environment, and the work, $\delta W$, done by the system to its environment. Mathematically, this statement can be written as
    \begin{equation}
        dE=\delta Q- \delta W.
    \end{equation}
    Here, $dE$ is an exact differential which means that it does not depend on the path taken in the thermodynamical manifold. In contrast, work $\delta W$ and heat $\delta Q$ are path dependent processes.  
    \item \textbf{Second law;}\\
    heat does not flow spontaneously from a colder bath to a hotter bath, or, equivalently, heat at a given temperature cannot be converted entirely into work. Through this law we define a new quantity called the entropy. For reversible processes it is given by
    \begin{equation}
        \Delta S:=\frac{\delta Q}{T}.
        \label{thermo_entropy}
    \end{equation}
     In general, the second law of thermodynamics expresses the fact that the entropy of the universe increases with time,
     \begin{equation}
         \Delta S_{\text{universe}} \geq 0.
     \end{equation}
     Equality holds for perfect (reversible) processes. 
     \item \textbf{Third law;}\\
     for classical systems, the entropy vanishes in the limit $T\to0$. That is
     \begin{equation}
         \lim_{T \to 0} \Delta S =0.
     \end{equation}
     The third law of thermodynamics is also called Nernst theorem, and its validity for quantum systems is still debated~\cite{oppenheim1, oppenheim2}.
\end{enumerate}
In addition to the traditional and well known laws of thermodynamics given above, there exists a ``Fourth law'' that describes situations away from equilibrium. Through this law, thermodynamics takes a step in complexity, in order to study realistic systems where their physical properties vary in space $\vec{x}$, and time $t$. In this case, the dynamics can be approximated by assuming that the system is described by a state of microscopic local equilibrium in any point in space and time, while the thermodynamic variables, such as the temperature $T(\vec{x},t)$, the local density $n(\vec{x},t)$, and the local energy density $e(\vec{x},t)$, vary weakly on a macroscopic scale. In this case, the rate of change of the dynamical local entropy $s(\vec{x},t)$ is given by
\begin{equation}
    \frac{ds}{dt}=\sum_k \frac{\partial s}{\partial X_k} \frac{dX_k}{dt}.
\end{equation}
where $\{X_k\}_k$ is a set of extensive parameters that vary with time. Hence, we can define the thermodynamic fluxes as the time-derivative of these $\{X_k\}_k$,
\begin{equation}
    J_k :=\frac{dX_k}{dt}.
\end{equation}
Therefore, we have
\begin{equation}
    \frac{ds}{dt}=\sum_k F_k J_k,
    \label{dsdt}
\end{equation}
where $F_k=\frac{\partial s}{\partial X_k}$ represent thermodynamic forces, or affinities. When the system is in a local equilibrium, $F_k$ is small. Therefore, we can expand $J_k$ to leading order in $F_k$, and we can write
\begin{equation}
    J_k=\sum_j L_{j,k} F_j,
    \label{Jk}
\end{equation}
where the kinetic coefficients $L_{j,k}$ are given by
\begin{equation}
    L_{j,k}:=\frac{\partial J_k}{\partial F_j}\bigg|_{F_j =0},
\end{equation}
with $F_j=0$ at equilibrium. The fourth law of thermodynamics states that the matrix of kinetic coefficient $L$ is symmetric, i.e.
\begin{equation}
    L_{j,k}=L_{k,j}.
    \label{Ljk}
\end{equation}
Equation~\eqref{Ljk} is also known as the Onsager theorem~\cite{onsager_thermo}, and is equivalent to Newton's third law. This seems natural when identifying equation~\eqref{Jk} with Newton's second law.

\noindent Thermodynamics has been very successful in describing the average macroscopic behavior of classical systems, which impacted several branches of science from physics (heat engines and black holes), chemistry (chemical compounds and chemical reactions) to biological systems. However, despite this success and universality the theory of thermodynamics is plagued with three major challenges: (i) thermodynamics does not provide a microscopic analysis of physical systems, (ii) the theory is limited to equilibrium states, and only infinitely-slow, quasi-static processes are fully describable; and (iii) the original formulation of thermodynamics is not suitable for quantum systems.
\section{Stochastic thermodynamics}
The above shortcomings of thermodynamics can be resolved by making use of stochastic thermodynamics, where work and heat are treated as stochastic random variables following probability densities~\cite{strasberg_2021}. The fruit of this approach is the invention of various fluctuation theorems, which connect the non-equilibrium properties of a system to its equilibrium counterpart. The most general form of a fluctuation theorem is given by
\begin{equation}
    \frac{\mathcal{P}(\Sigma=-A)}{\mathcal{P}(\Sigma=A)} = \exp{(A)},
    \label{general_fluctuation}
\end{equation}
which relates the probability to find a negative entropy production $\Sigma$, with the probability to find a positive value. Applying Jensen's inequality, $\exp{(-\langle x \rangle)}\geq \langle \exp{(-x)}\rangle $, to the fluctuation theorem~\eqref{general_fluctuation} gives
\begin{equation}
    \langle \Sigma \rangle \geq 0,
\end{equation}
which can be interpreted as an extension of the second law of thermodynamics to the case of out-of-equilibrium systems. In the following we discuss how probability distributions are defined for work, and we derive two important fluctuation relations, called the Tasaki-Crooks relation, and the Jarzynski equality.

\noindent Consider a quantum system governed by a time-dependent Hamiltonian $H(\lambda(t))$, where $\lambda(t)$ is an external \textit{work} parameter, used to drive the system out-of-equilibrium. We initialize the system in a thermal state by allowing it to equilibrate with a thermal reservoir at inverse temperature $\beta$, for a fixed value of the work parameter $\lambda(t<t_i)=\lambda_i$. Hence, the initial state is a Gibbs state, given by
\begin{equation}
    \rho_i(\lambda,\beta)=\frac{e^{-\beta H(\lambda_i)}}{\mathcal{Z}(\lambda_i,\beta)},
    \label{gibbs}
\end{equation}
here, $\mathcal{Z}(\lambda_i,\beta)$ is the partition function. At $t=t_i$ the system is decoupled from the thermal reservoir, and a reversible protocol is applied on the system, by changing the work parameter from $\lambda_i$ to $\lambda_f$ at $t=t_f$. The initial and final Hamiltonian are defined by their respective spectral decompositions
\begin{align}
    H(\lambda_i)&=\sum_n E_n(\lambda_i) \ket{\psi_n}\bra{\psi_n},\\
    H(\lambda_f)&=\sum_m E_m(\lambda_f) \ket{\phi_m}\bra{\phi_m},
\end{align}
where $\ket{\psi_n}$ $(\ket{\phi_m})$ denotes the $n$th ($m$th) eigenstate of the initial (final) Hamiltonian, with the corresponding eigenvalue $E_n(\lambda_i)$ $(E_m(\lambda_f))$. The initial and finale Hamiltonian are connected by the unitary time evolution operator $U(t_f,t_i)$, which takes the following form
\begin{equation}
    U(t_f,t_i)=\boldsymbol{T}_{\rightarrow}\exp{\left[ -i \int_{t_i}^{t_f} H(\lambda(t^{\prime}))dt^{\prime} \right]},
    \label{evolve}
\end{equation}
where $\boldsymbol{T}_{\rightarrow}$ denotes the time ordering operator.

\noindent The work performed on the system is determined by two projective energy measurements. The outcome of the first measurement, where the system is initialized in the Gibbs state~\eqref{gibbs}, is the energy $E_n(\lambda_i)$ with the probability
\begin{equation}
    p(n)=\frac{e^{-\beta E_n(\lambda_i)}}{\mathcal{Z}(\lambda_i)}.
\end{equation}
Then, the system evolves under the evolution operator~\eqref{evolve}, until a second measurement is performed at time $t_f$. It produces the energy outcome $E_m(\lambda_f)$, with the probability
\begin{equation}
    p(m|n)= | \bra{\phi_m} U(t_f,t_i) \ket{\psi_n} |^2.
\end{equation}
Hence, the joint probability of obtaining the outcome $E_n(\lambda_i)$ followed by $E_m(\lambda_f)$ in the final measurement is
\begin{equation}
    p(n,m)=p(n)p(n|m)=\frac{e^{-\beta E_n(\lambda_i)}}{\mathcal{Z}(\lambda_i)} | \bra{\phi_m} U(t_f,t_i) \ket{\psi_n} |^2.
\end{equation}
Therefore, the quantum work probability distribution is defined as
\begin{equation}
    P(W)=\sum_{n,m} p(n,m)\delta \left( W-[E_m(\lambda_f)-E_n(\lambda_i)] \right),
    \label{wpd}
\end{equation}
where $\delta(.)$ is the Dirac delta function. The distribution~\eqref{wpd} is constructed via discrete values of the work $(E_m(\lambda_f)-E_n(\lambda_i))$, weighted by the probability $p(n,m)$. As a consequence, the quantum work distribution represents fluctuations due to thermal statistics in the first measurement, and fluctuations from quantum measurement statistics in the second measurement. 

\noindent The Tasaki-Crooks fluctuation theorem~\cite{tasaki} follows from the work probability distribution~\eqref{wpd}. It is written as
\begin{equation}
    \frac{P(+W)}{P(-W)}=e^{\beta(W-\Delta F)}.
    \label{tasaki}
\end{equation}
Here, $P(+W)$ denotes the work probability distribution of the forward process~\eqref{wpd}, whereas $P(-W)$ represents the distribution of the backward process, in which the system is initialized in the Gibbs state of the final Hamiltonian $H(\lambda_f)$ at $t=0$. The time evolution in this case is dictated by the time reverse protocol $\Theta U(t_f,t_i) \Theta^{\dagger}$, where $\Theta$ is the anti-unitary time reversal operator. The Tasaki-Crooks fluctuation theorem~\eqref{tasaki} shows that for a closed system under non-equilibrium driving, the fluctuation in work are related to the equilibrium free energy difference between the equilibrium states $\rho_i(\lambda_i)$ and $\rho_f(\lambda_f)$,
\begin{equation}
    \Delta F=\frac{1}{\beta} \log\left(\frac{\mathcal{Z}_{\beta}(\lambda_i)}{\mathcal{Z}_{\beta}(\lambda_f)}\right).
\end{equation}
The second prominent fluctuation theorem is Jarzynski equality~\cite{Jarzynski1, Jarzynski2}, which can be derived by integrating the Tasaki-Crooks fluctuation theorem,
\begin{equation}
    \int P(+W)e^{-\beta W} dW=\langle e^{-\beta W} \rangle= e^{-\beta \Delta F}.
    \label{Jarzynski}
\end{equation}
The Jarzynski equality~\eqref{Jarzynski} implies that information about $\Delta F$ can be extracted by measuring the exponential of the work $W$. We note here that the Tasaki-Crooks fluctuation theorem~\eqref{tasaki} is only valid to Markovian process, while the Jarzynski equality~\eqref{Jarzynski} holds also for non-Markovian settings.

\noindent A straightforward application of Jensen's inequality on the Jarzynski equality gives $\langle W \rangle =\Delta F$, from which we can define the dissipated work
\begin{equation}
     \langle W \rangle_{\text{irr}}=\langle W \rangle -\Delta F,
\end{equation}
which quantifies the irreversible nature of non-equilibrium processes, and is always positive $\langle W \rangle_{\text{irr}} \geq 0$ due to the Jarzynski equality~\eqref{Jarzynski}.

\noindent The irreversible entropy change $\langle \Sigma \rangle$ corresponding to the irreversible work $\langle W \rangle_{\text{irr}}$ follows from the Tasaki-Crooks fluctuation theorem~\eqref{tasaki}, by taking the logarithm of both sides and integrating over the forward distribution $P(+W)$,
\begin{equation}
    \langle \Sigma \rangle=\beta(W-\Delta F)=K\left( P(+W) || P(-W) \right),
\end{equation}
where $K(.||.)$ is the classical Kullback-Leibler divergence, which is a statistical distance measuring how the distribution $P(+W)$ is different from $P(-W)$. For a quantum system, the irreversible entropy change is determined by the quantum relative entropy,
\begin{equation}
     \langle \Sigma \rangle=D(\sigma||\rho(\lambda_f,\beta)).
\end{equation}
Here, $\sigma=U(t_f,t_i) \rho(\lambda_i,\beta) U^{\dagger}(t_f,t_i)$ is the out-of-equilibrium state at the end of the protocol.
\section{Quantum work, heat and entropy production}
The original formulation of thermodynamics and the fluctuation theorems can be used alongside quantum theory in order to extend the notion of work, heat and entropy production to quantum systems~\cite{booksteve, thermo_quantum_regime}. In particular, we discuss the case when the quantum system follows a quasi-static process, and the situation of weak (strong) coupling to the environment, where the quantum system is described by Gibbs (non-Gibbsian) equilibrium states.
\begin{enumerate}
    \item \textbf{Quasi-static process.}\\ Consider a quantum system $\Omega$ described by the Hamiltonian $H(\lambda)$, where $\lambda$ is the control parameter e.g. the magnetic field or the volume of a piston. For an infinitely slow quasi-static process, the quantum system $\Omega$ is always in thermal equilibrium with the environment.  The dynamics of the system $\Omega$ is dictated by the Liouville equation $\dot{\rho}=L_{\lambda}(\rho)$, where $L_{\lambda}$ is a superoperator reflecting two contributions to the dynamics. The first is the unitary dynamics due to $H$, and the second is the non-unitary part generated by the interaction with the environment.
    \item \textbf{Gibbs equilibrium states.}\\
    The Gibbs state describes systems interacting very weakly with their environment, it is written as
    \begin{align}
        \rho^{\text{eq}}=\frac{e^{-\beta H}}{\mathcal{Z}}, &&& \text{where} && \mathcal{Z}=\Tr(e^{-\beta H}),
    \end{align}
    and $\beta$ is the inverse temperature of the environment. Therefore, the internal energy is determined via the expectation value of $H$ in the Gibbs state $\rho^{\text{eq}}$, that is $E=\Tr(\rho^{\text{eq}} H)$, whereas the free energy is $F=\frac{1}{\beta}\log(\mathcal{Z})$, so that the thermodynamic entropy is given by
    \begin{equation}
        S=\beta(E-F)=-\Tr\left(\rho^{\text{eq}} \log_2\rho^{\text{eq}}\right).
        \label{thermo_entropy}
    \end{equation}
    The rate of change of the entropy~\eqref{thermo_entropy} in the case of isothermal quasi-static processes is
    \begin{equation}
        dS=\beta \left[ \Tr\left(d\rho^{\text{eq}} H\right)+ (\Tr\left( \rho^{\text{eq}} dH \right) -\Delta F) \right]=\beta \Tr\left(d\rho^{\text{eq}} H\right).
    \end{equation}
    Similarly, the rate of change of the internal energy is given by
    \begin{equation}
        dE= \underbrace{\Tr\left(d\rho^{\text{eq}} H\right)}_{\delta Q=\frac{dS}{\beta}} + \underbrace{\Tr\left(\rho^{\text{eq}} dH \right)}_{\delta W=dF}.
    \end{equation}
    The identification is consistent with the first and second law of thermodynamics. It reflects the fact that heat, $Q$, is the change of internal energy due to the variation of the entropy, while the work, $W$, is the change of the internal energy resulting from the change of the Hamiltonian of the system via external parameters.
    \item \textbf{Non-Gibbsian equilibrium states.}\\
    Realistically, the interaction energy between a quantum system and its environment is not negligible, as a consequence of the strong correlations between them. Therefore, thermodynamics needs to be reformulated in order to take into account the energy cost of the system-environment interaction. In this case the thermodynamic entropy can be written as
    \begin{align}
        \mathcal{S}&=-\Tr\left( \rho^{\text{ss}}\log_2 \rho^{\text{ss}} \right) + \left[ \Tr \left(\rho^{\text{ss}}\log_2 \rho^{\text{ss}}\right)  - \Tr \left(\rho^{\text{ss}}\log_2 \rho^{\text{eq}}\right) \right],\nonumber \\
        &=\beta \left[ E-\left( F+TS( \rho^{\text{ss}} || \rho^{\text{eq}} ) \right)\right] = \beta \left[ E-\mathcal{F} \right].
    \end{align}
    Here, $\rho^{\text{ss}}$ is the non-Gibbsian equilibrium state, and $E=\Tr(\rho^{\text{ss}}H)$ is the internal energy of the system. $\mathcal{F}=F+TS( \rho^{\text{ss}} || \rho^{\text{eq}} )$ is the information free energy, while $S(.||.)$ denotes the quantum relative entropy.
    
    In the same fashion with Gibbs equilibrium states, we write the change of the entropy for quasi-static processes as
    \begin{align}
        d\mathcal{S}&=\beta \left[ \Tr \left( d\rho^{\text{ss}} H \right) + \Tr \left( \rho^{\text{ss}} dH \right) -d\mathcal{F}\right], \nonumber \\
        &=\beta \left( \delta Q_{\text{tot}}-\delta Q_c \right),
    \end{align}
    where $\delta Q_{\text{tot}}=\Tr \left( d\rho^{\text{ss}} H \right)$ is the total heat, and $\delta Q_c=d\mathcal{F}-\Tr \left( \rho^{\text{ss}} dH \right)$ represents the energy cost to maintain coherence and quantum correlations.
    
    We can also associate an entropic cost through the excess heat $\delta Q_{\text{ex}}=d\mathcal{S}/\beta$. Hence the first law of thermodynamics is written as
    \begin{equation}
        dE=\delta W_{\text{ex}}+\delta Q_{\text{ex}},
    \end{equation}
    where $\delta W_{\text{ex}}=\delta W +\delta Q_c$ is the excess work.
\end{enumerate}
Next, we move to the concept of quantum entropy production. We consider a system $\Omega$ initialized in a thermal equilibrium state (which is not necessarily a Gibbs state) with the temperature of the environment, and can be driven by an external parameter $\lambda$. The change of the internal energy $\Delta E$ and thermodynamic entropy $\Delta \mathcal{S}$ are given, respectively, by
\begin{align}
    \Delta E=W+Q, && \Delta \mathcal{S}=\beta Q + \Sigma,
\end{align}
where $\Sigma$ is the entropy production and $Q$ is the amount of heat exchanged with the environment at inverse temperature $\beta$. Therefore, the entropy production is written as,
\begin{equation}
    \Sigma=\Delta \mathcal{S}-\beta( \Delta E - W).
    \label{entropy_prod}
\end{equation}
The internal energy can be expressed in terms of the equilibrium state $\rho^{\text{eq}}$,
\begin{equation}
    \beta E=\beta \Tr \left( \rho^{\text{ss}} H \right) =-\Tr \left( \rho^{\text{ss}} \log_2 \rho^{\text{eq}} \right) + \log \mathcal{Z}.
    \label{bE}
\end{equation}
Therefore, for a protocol driving the system through $\lambda$ from $\lambda_i$ to $\lambda_f$, we have
\begin{align}
    \beta W &=\beta \int_{\lambda_i}^{\lambda_f} \Tr \left( \rho^{\text{ss}}(\lambda) \partial_{\lambda} H(\lambda)\right)  d\lambda,  \nonumber \\
    &=-\int_{\lambda_i}^{\lambda_f} \left[ \Tr \left( \rho^{\text{ss}}(\lambda) \partial_{\lambda} \log_2 \rho^{\text{eq}}(\lambda)\right) -\log_2 Z_f +\log_2 Z_i \right]d\lambda.
    \label{bW}
\end{align}
Injecting equation~\eqref{bE} and~\eqref{bW} in~\eqref{entropy_prod} yields
\begin{equation}
    \Sigma=S(\rho^{\text{ss}}(\lambda_i)||\rho^{\text{eq}}(\lambda_i))-S(\rho^{\text{ss}}(\lambda_f)||\rho^{\text{eq}}(\lambda_f))-\int_{\lambda_i}^{\lambda_f} \Tr \left( \rho^{\text{ss}}(\lambda) \partial_{\lambda} \log_2 \rho^{\text{eq}}(\lambda)\right) d\lambda,
    \label{entropy_prod_non_equ}
\end{equation}
which represents the entropy production along a non-equilibrium path in the thermodynamic manifold.
\section{Resource theory for thermodynamics}
Resource theories allow us to quantify and manage resources using an information-theoretic framework. The basic procedure to construct a resource theory is by defining (i) free operations, (ii) free states, and (iii) the set of allowed state transformations~\cite{gilad2019}. The most prominent example of a resource theory comes from the field of quantum information, where a resource is identified with the entanglement between two quantum states~\cite{rev_entanglement}. In this case, the free operations are the well known ``Local Operations and Classical Communication (LOCC)'', and the free states are the separable states. Whereas the set of allowed state transformations are defined under the entanglement monotones. Another example is the resource theory of coherence, where incoherent operations constitute the free operations, and the free states are any diagonal density matrix $\rho$ in the coherence basis $\ket{i}$. Whereas the coherence monotones, e.g. relative entropy of coherence, sets the state conversion conditions~\cite{coherence_review}. The resource theory for thermodynamics is developed in order to study thermal interactions. The free operations are energy-preserving unitary operators, while the free states are equilibrium states, i.e. Gibbs states or catalysts. In the same fashion with entanglement and coherence, the state conversion conditions are set by a monotone function, which in the case of thermodynamics can be the free energy~\cite{gooldreview}. The advent of the resource theory for thermodynamics inspired the interpretation of information-theoretic results in the context of the thermodynamics of quantum systems.

\noindent The thermodynamic resource theory can be built through different models, which differ in the set of allowed operations. These can only be (i) contact with a thermal bath, and (ii) reversible operations that preserve a thermodynamic quantity. In the following we list some incarnations of these two kinds of operations.
\begin{enumerate}
    \item \textbf{Noisy operations.}\\
    Noisy operations are the simplest resource theory, in which the Hamiltonians are considered fully degenerate. Therefore, the free states are described by maximally mixed states of arbitrary dimensions. Whereas the free operations are the partial trace and all the unitary transformations.
    Noisy operations model the transition $\rho_S \rightarrow \rho^{\prime}_S$, if and only if there exist a $d_R -$dimensional ancillary system $R$ satisfying
    \begin{equation}
        \rho^{\prime}_{S}:=\mathcal{E}_{\text{Noisy}}(\rho_S):=\Tr_R \left( U_{SR} \left[ \rho_S \otimes \frac{\mathbb{I_R}}{d_R} \right] U^{\dagger}_{SR} \right),
        \label{NO}
    \end{equation}
    where $U_{SR}$ is a unitary matrix and $\mathcal{E}_{\text{Noisy}}$ is chosen such that it preserves the maximally mixed state $\rho_S =\frac{\mathbb{I}_S}{d_S}$, i.e. unital map.  Noisy operations aim at describing the information content carried in systems, instead of energy. This information aspect paved the way to use noisy operations in exorcizing the Maxwell demon problem. State transformation follows the Shur monotone functions, which make use of majorization in order to determine the degree of mixture of the state~\cite{gour2015}. Information-theoretical entropy measures, i.e. von Neumann entropy, are examples of the Shur monotone functions.
    \item \textbf{Thermal operations.}\\
    We take a step further in complexity by allowing systems with non-degenerate Hamiltonian $H$, which implies that the distribution of energy levels is non-trivial. Hence, at fixed average energy, the state with maximum entropy (maximally mixed state) corresponds to the thermal Gibbs state given by
    \begin{equation}
        \rho^{\beta}=\frac{e^{-\beta H}}{\mathcal{Z}}.
        \label{gibbs_state}
    \end{equation}
    Gibbs states~\eqref{gibbs_state} have unique features that justify their use as free states. In particular, they are completely passive which means that one cannot increase the mean energy contained in them by unitary operations. Additionally, Gibbs states do not allow arbitrary energy transitions during the transformation $\rho \rightarrow \sigma$, which implies that Gibbs states cannot be used to extract work. Thermal operations (TO) are defined as
    \begin{equation}
        \mathcal{E}_{TO}(\rho_S)=\Tr_R \left[ U_{SR} \left( \rho_S \otimes \rho^{\beta}_R \right) U^{\dagger}_{SR} \right].
        \label{TO}
    \end{equation}
    Here, $U_{SR}$ is a unitary operator reflecting the energy-preserving unitary dynamics across the system $S$ and bath $R$. The energy conservation implies that $U_{SR}$ commutes with the total Hamiltonian $H_{SR}=H_S\otimes \mathbb{I}_R + \mathbb{I}_S \otimes H_R$. In the limit $H_S=\mathbb{I}_S$ and $H_R=\mathbb{I}_R$, the thermal operations~\eqref{TO} reduce to the noisy operations~\eqref{NO}. The monotones for state transformations are different forms of the free energy.
    \item \textbf{Gibbs-preserving maps.}\\
    Gibbs-preserving maps (GPs) constitute a general model for thermal interactions, in which the set of free operations is the set of all quantum maps that preserve the Gibbs state at inverse temperature $\beta$:
    \begin{equation}
        \mathcal{E}_{GP}\left( \rho_S^{\beta} \right)=\rho_S^{\beta}.
    \end{equation}
    For general quantum states, the  Gibbs-preserving maps are less restrictive than thermal operations, and
    may allow for transformations that are not possible using thermal operations, such as creating coherence in the final state of an initial energy-incoherent state. This can be seen by considering a qubit system described by the Hamiltonian $H=e\ket{1}\bra{1}$. The transformation $\ket{1}\rightarrow \ket{+}=\frac{1}{\sqrt{2}}\left( \ket{0}+\ket{1} \right)$ is not possible via thermal operations, because they cannot create coherence; while there exist Gibbs-preserving maps that can achieve this task.
\end{enumerate}
The difference between thermal operations and Gibbs-preserving maps motivated the study of the role of coherence in state transformation. In order to perform the transformation $\ket{1}\rightarrow \ket{+}=\frac{1}{\sqrt{2}}\left( \ket{0}+\ket{1} \right)$, we need to extract coherence of a reservoir, e.g. doubly infinite harmonic oscillator $H=\sum_{n=-\infty}^{\infty} n\Delta \ket{n}\bra{n}$ in a coherent state $\ket{\Psi}=\frac{1}{N} \sum_{n=a}^{a+N} \ket{n}$. In practice, an example of such a reservoir is the laser which can be used in order to apply transformations on quantum systems, e.g. ion traps. Coherence reservoirs can be implemented multiple times in order to perform state transformations, which is due to the catalyst nature of coherence. In fact, after each use of the coherence reservoir its state spreads out, but its ability to implement operations stays invariant. In general, catalysts in thermodynamics can be used to overcome the limits imposed to transform states via thermal operations~\cite{catalytic_coherence}.

\noindent The resource theory of thermodynamics sets energetic limitations on the manipulation of quantum systems, while the resource theories in quantum information introduce bounds for the creation and control of quantum features, i.e. entanglement and quantum coherence. Since thermodynamics and information are intimately connected, we ask whether a thermodynamics price exists for establishing quantum correlations? In the following, we briefly report some prominent results.

\noindent In the case of a two-qubit system, the minimal cost of creating entanglement was found to be
\begin{equation}
    C(E)=\sqrt{\frac{\Delta E}{E} \left( 2-\frac{\Delta E}{E} \right)}, 
    \label{c(e)}
\end{equation}
where $C(.)$ is the concurrence, and $E$ is the energy gap. Equation~\eqref{c(e)} sets the cost for creating entanglement between two qubits, from thermal states~\cite{Huber_2015}. Moreover, for an arbitrary composite system $AB$ it has been shown that the work cost $W_{\text{cost}}$ for creating correlations between $A$ and $B$ is proportional to the mutual information $\mathcal{I}(A:B)$, and the temperature of the system,
\begin{equation}
    W_{\text{cost}}\geq k_bT\mathcal{I}(A:B),
\end{equation}
which defines the work cost for creating correlations~\cite{huber_2015_2}. The quantum advantage in terms of work extraction has been studied. It has been shown that two entangled qubits can store twice the amount of information that can be stored via separable (classical) states. However, for very large system sizes, $N$, the difference between the separably encoded work from correlations $W_{\text{sep}}$, and the
maximally possible work in correlations $W_{\text{max}}$ scales as,
\begin{equation}
\frac{W_{\text{sep}}}{W_{\text{max}}}=1-\mathcal{O}(N^{-1}).
\end{equation}
Hence, the quantum advantage does not hold in the thermodynamic limit $N\to \infty$~\cite{Huber_2015_3}.

\noindent Thermodynamics can be used to reveal the quantumness of a system. In fact it is argued that the energy of the system can be used as an entanglement witness for zero, and non-zero temperature settings~\cite{Vedral_2004}. This is motivated by the observation that a low average energy implies that the density matrix is close to the entangled ground state. Therefore, as this distance becomes small, the witness becomes more efficient in validating the entanglement of the state. On the other hand, it has been established that macroscopic thermodynamic functions, i.e. entropy~\cite{entropy_2015} and magnetic susceptibility~\cite{Wie_niak_2005}, can be used to reveal the entanglement of the system.
\section{Thermodynamics of computation}
Shannon's definition of information, as being a measure of our freedom of choice, remains abstract and does not reflect the physical essence of information. Since information can be encoded in physical systems, it is natural to think of it as a physical quantity. This is confirmed through the Landauer principle, which sets an energetic cost for information processing~\cite{landauer1961}. In particular, Landauer's principle sets a constraint-free minimum cost to erase information. The most famous consequences of Landauer's limit are the exorcism of Maxwell’s Demon~\cite{bennett2003notes}, and the development of fully reversible models of computation~\cite{Bennett1982TheTO}.
\subsection{Landauer's principle}
Rolf Landauer proposed the principle in 1961, while working on the problem of thermodynamical limitations on information processing. The derivation of the limit follows from evaluating the change of entropy after erasure of information. Consider a two-state system $\mathfrak{S}$ containing 1 bit of information, the initial Shannon entropy $H(p)=-\sum_i p_i \log_2 p_i=\log_2 2=1$. Whereas the final entropy after erasure is zero. Suppose the system $\mathfrak{S}$ is in equilibrium, at temperature $T$, with an environment $\mathcal{E}$ in the form of a thermal bath. The second law of thermodynamics states that the total change of entropy of $\mathfrak{S}+\mathcal{E}$ is positive, i.e.
\begin{equation}
    \Delta S_{\text{tot}}=\Delta S_{\mathfrak{S}}+\Delta S_{\mathcal{E}} \geq 0,
\end{equation}
where $\Delta S= S_f - S_i$ denotes the changes of the entropy. Therefore, using Clausius inequality we can write the amount of heat dissipated to the environment (we assume a large thermal bath, so that thermal equilibrium always holds) as,
\begin{equation}
    Q_{\mathcal{E}} \geq -T\Delta S_{\mathfrak{S}}.
\end{equation}
Since the thermodynamic entropy $S$ is linked to Shannon's entropy $H(p)$ by $S=k_bH(p)$, whereas the change of entropy between the pre-erasure and after-erasure states is $\Delta H=-\log_2 2$. Hence, the heat dissipated in the environment is bounded by
\begin{equation}
     Q_{\mathcal{E}} \geq k_bT\Delta H = k_b T \log_2 2.
     \label{landauer}
\end{equation}
Landauer's principle~\eqref{landauer} implies that erasing one bit of information increases the entropy of the environment by \textit{at least} $k_b\log_2 2$. Through this principle, modern computers can be seen as engines consuming free energy to produce heat and mathematical work. Modern computers still operate above the Landauer limit by many degrees of magnitude, however it is predicted by Koomey's law that the limit will be reached around 2050.

\noindent The paradox of Maxwell's demon was solved by Charles Bennett using insights from Landauer's principle~\cite{bennett2003notes}. Maxwell's demon is an intelligent agent in a thought experiment conceptualized by Maxwell, where the demon monitors an ideal gas in a container. The demon inserts a partition, in the middle of the container, with a controllable door which he uses in order to sort the fast particles on one side and the slow particles on the other side of the container. As a consequence, the entropy of the gas decreases, which violates the second law of thermodynamics. In contrast, Bennett showed that the demon will run out of information storage space and must begin to erase the information it has previously gathered, which is a process that increases the entropy of the system and restores the second law of thermodynamics.
\subsection{Quantum Landauer's principle}
Landauer's limit~\eqref{landauer} can be extended to quantum systems, via Reeb and Wolf formulation~\cite{Reeb_2014}. Consider a system $\mathfrak{S}$ in contact with an environment $\mathcal{E}$, the set of minimal assumptions that validate the Landauer's principle are
\begin{enumerate}
    \item $\mathfrak{S}$ and $\mathcal{E}$ are quantum systems living, respectively, in Hilbert space $\mathcal{H}_{\mathfrak{S}}$ and $\mathcal{H}_{\mathcal{E}}$.
    \item The initial state is uncorrelated, i.e. $\rho_{\mathfrak{S}\mathcal{E}}=\rho_{\mathfrak{S}} \otimes \rho_{\mathcal{E}}$.
    \item The environment is modeled by a thermal Gibbs state, i.e. $\rho_{\mathcal{E}}=\exp{\left(-\beta H_{\mathcal{E}}\right)}/\mathcal{Z_E}$, with $\mathcal{Z_E}$ being the partition function.
    \item The system-environment interaction is described by a unitary transformation $U(t)=\exp{\left( -iH_{\text{tot}}t/\hbar\right)}$, where $H_{\text{tot}}$ is the total Hamiltonian, $H_{\text{tot}}=H_{\mathfrak{S}} + H_{\mathcal{E}} + H_{\mathfrak{S}\mathcal{E}}$.
\end{enumerate}
Similarly with the original formulation of the Landauer principle~\eqref{landauer}, we write the change of the entropy for the system (environment) as the difference between the initial and final entropies:
\begin{equation}
    \Delta S_{\mathfrak{S}(\mathcal{E})}=S\left( \rho^0_{\mathfrak{S}(\mathcal{E})} \right)-S\left( \rho^t_{\mathfrak{S}(\mathcal{E})} \right).
\end{equation}
According to the second law of thermodynamics, the entropy production of the process must be positive. Therefore, we have
\begin{align}
    -\Delta S_{\mathfrak{S}}-\Delta S_{\mathcal{E}}&=S(\rho^t_{\mathfrak{S}})-S(\rho^0_{\mathfrak{S}})+S(\rho^t_{\mathcal{E}})-S(\rho^0_{\mathcal{E}}),\nonumber \\
    &=S(\rho^t_{\mathfrak{S}})+S(\rho^t_{\mathcal{E}})-S(\rho^0_{\mathfrak{S}} \otimes \rho^0_{\mathcal{E}}),\nonumber \\
    &=S(\rho^t_{\mathfrak{S}})+S(\rho^t_{\mathcal{E}})-S(\rho^t_{\mathfrak{S}\mathcal{E}}),\nonumber \\
    &=I(\rho^t_{\mathfrak{S}\mathcal{E}})\geq 0.
    \label{eq6.45}
\end{align}
Here, $I(\rho^t_{\mathfrak{S}\mathcal{E}})$ is the mutual information between the system and environment. Equation~\eqref{eq6.45} can be seen as a version of the second law of thermodynamics, due to the non-negativity of the mutual information. The Landauer limit follows by developing~\eqref{eq6.45} as
\begin{align}
    I(\rho^t_{\mathfrak{S}\mathcal{E}})&+\Delta S_{\mathfrak{S}}=-\Delta S_{\mathcal{E}}, \nonumber\\
    &=-\Tr \left(\rho^t_{\mathcal{E}} \log_2 \rho^t_{\mathcal{E}}  \right) + \Tr \left( \rho^0_{\mathcal{E}} \log_2 \left[ \frac{e^{-\beta H_{\mathcal{E}}}}{\Tr e^{-\beta H_{\mathcal{E}}} } \right] \right), \nonumber \\
    &=-\Tr \left(\rho^t_{\mathcal{E}} \log_2 \rho^t_{\mathcal{E}}  \right)-\beta \Tr H_{\mathcal{E}}\rho^0_{\mathcal{E}} - \log_2 \left( \Tr  e^{-\beta H_{\mathcal{E}}} \right) +\beta \Tr \left( H_{\mathcal{E}}\rho^t_{\mathcal{E}} \right) -\beta \Tr \left( H_{\mathcal{E}}\rho^t_{\mathcal{E}} \right), \nonumber \\
    &=\beta \Tr \left( H_{\mathcal{E}} (\rho^t_{\mathcal{E}}-\rho^0_{\mathcal{E}})\right)-\Tr \left( \rho^t_{\mathcal{E}} \log_2 \rho^t_{\mathcal{E}} \right) + \Tr \left( \rho^t_{\mathcal{E}} \log_2 \rho^0_{\mathcal{E}} \right), \nonumber \\
    &=\beta \langle Q_{\mathcal{E}} \rangle -S(\rho^t_{\mathcal{E}}||\rho^0_{\mathcal{E}}).
\end{align}
In the end we have
\begin{equation}
   \beta \langle Q_{\mathcal{E}} \rangle= \Delta S_{\mathfrak{S}} + I(\rho^t_{\mathfrak{S}\mathcal{E}})+S(\rho^t_{\mathcal{E}}||\rho^0_{\mathcal{E}}),
\end{equation}
which represent the equality version of the non-equilibrium Landauer limit. Since the relative entropy $S(.||.)$ and the mutual information $I(.)$ are non-negative quantities, we have
\begin{equation}
    \beta \langle Q_{\mathcal{E}} \rangle \geq \Delta S_{\mathfrak{S}},
\end{equation}
which represents the quantum Landauer limit for non-equilibrium settings~\cite{Reeb_2014}.

\chapter{Work Statistics \& Symmetry Breaking in an Excited State Quantum Phase Transition \label{chap7} }
We examine how the presence of an excited state quantum phase transition (ESQPT) manifests in the dynamics of a many-body system subject to a sudden quench. Focusing on the Lipkin-Meshkov-Glick model initialized in the ground state of the ferromagnetic phase, we demonstrate that the work probability distribution displays non-Gaussian behavior for quenches in the vicinity of the excited state critical point. Furthermore, we show that the entropy of the diagonal ensemble is highly susceptible to critical regions, making it a robust and practical indicator of the associated spectral characteristics. We assess the role that symmetry breaking has on the ensuing dynamics, highlighting that its effect is only present for quenches beyond the critical point. Finally, we show that similar features persist when the system is initialized in an excited state and briefly explore the behavior for initial states in the paramagnetic phase~\cite{mzaouali2021}. 
\section{The LMG model}
The Lipkin-Meshkov-Glick (LMG) model~\cite{lmg1965} describes a set of $N$ spin-$\tfrac{1}{2}$'s with infinite range interaction subject to a transverse field~\cite{prl_vidal, pre_vidal, prb_vidal, prb_castanos, quan_pra, CampbellPRB}. The fully anisotropic Hamiltonian can be written in terms of the Pauli matrices $\sigma^i_{x,z}$ acting on site $i$ as
    \begin{equation}
    \mathcal{H}=-\frac{1}{N} \sum_{i<j} \sigma_x^i \otimes \sigma_x^j  + h\sum_i \sigma_z^i, 
     \label{lmg}
    \end{equation}
where $h\geq 0$ is the strength of the magnetic field in the $z-$direction. It is convenient to recast the LMG model in terms of the total spin operators $S_\alpha\!=\!\sum_{i} \sigma_{\alpha}^i/2$, with $\alpha\!=\!\{ x,y,z\}$, 
\begin{equation}
    \mathcal{H}=-\frac{1}{N}  S_x^2  + h\: \Bigg(S_z +\frac{N}{2} \Bigg),
     \label{lmg_spin}
    \end{equation}
which can be written in a bosonic form by applying the Schwinger representation of spin operators 
\begin{align}
    S^z=t^{\dagger}t-\frac{N}{2}=\hat{n}_t-\frac{N}{2}, \: S^+=t^{\dagger}s=\big(S^-\big)^{\dagger},
\end{align}
where $S^{\pm}\!=\!S_x \pm i\:S_y$ are spin ladder operators. This results in a Hamiltonian describing a system of two species of scalar bosons $s$ and $t$, given by
\begin{equation}
    \mathcal{H}=h\:t^{\dagger}t-\frac{1}{4N}\big( t^{\dagger}s+s^{\dagger}t\big)^2.
    \label{lmg_boson}
\end{equation}
The non-zero elements of the Hamiltonian~\eqref{lmg_boson} in the basis 
\begin{equation}
    \ket{N,n_t} = \frac{\big(t^{\dagger}\big)^{n_t} \big(s^{\dagger}\big)^{N-n_t} }{\sqrt{n_t ! \big(N-n_t\big)!}} \ket{0},
    \label{base}
\end{equation}
are given by
\begin{gather}
    \bra{N,n_t}\mathcal{H} \ket{N,n_t}\!=\! h \: n_t\!-\!\frac{(n_t+1)(N-n_t)+n_t(N-n_t+1)}{4N} ,\nonumber \\
    \bra{N,n_t}\mathcal{H} \ket{N,n_t+2}\!=\!-\frac{\sqrt{(n_t+1)(N-n_t)(n_t+2)(N-n_t-1)}}{4N}, 
    \label{element_matrix}
\end{gather}
where $\ket{0}$ is the vacuum state and $0 \! \leq \! n_t \! \leq \! N $ and therefore the dimension of the Hamiltonian~\eqref{lmg_boson}, is $N\!+\!1$. The LMG model exhibits a $\mathbb{Z}_2$ parity symmetry, given by the operator $\Pi=e^{i\pi(S_z+N/2)}=e^{i\pi t^\dagger t}$ so that $[\mathcal{H},\Pi]=0$. As a consequence, the Hamiltonian can be split into odd and even parity blocks of dimension $D_{\text{odd}}\!=\!N/2\:+1$ and $D_{\text{even}}\!=\!N/2$, respectively.

\begin{figure}[t]
    \includegraphics[width=\columnwidth]{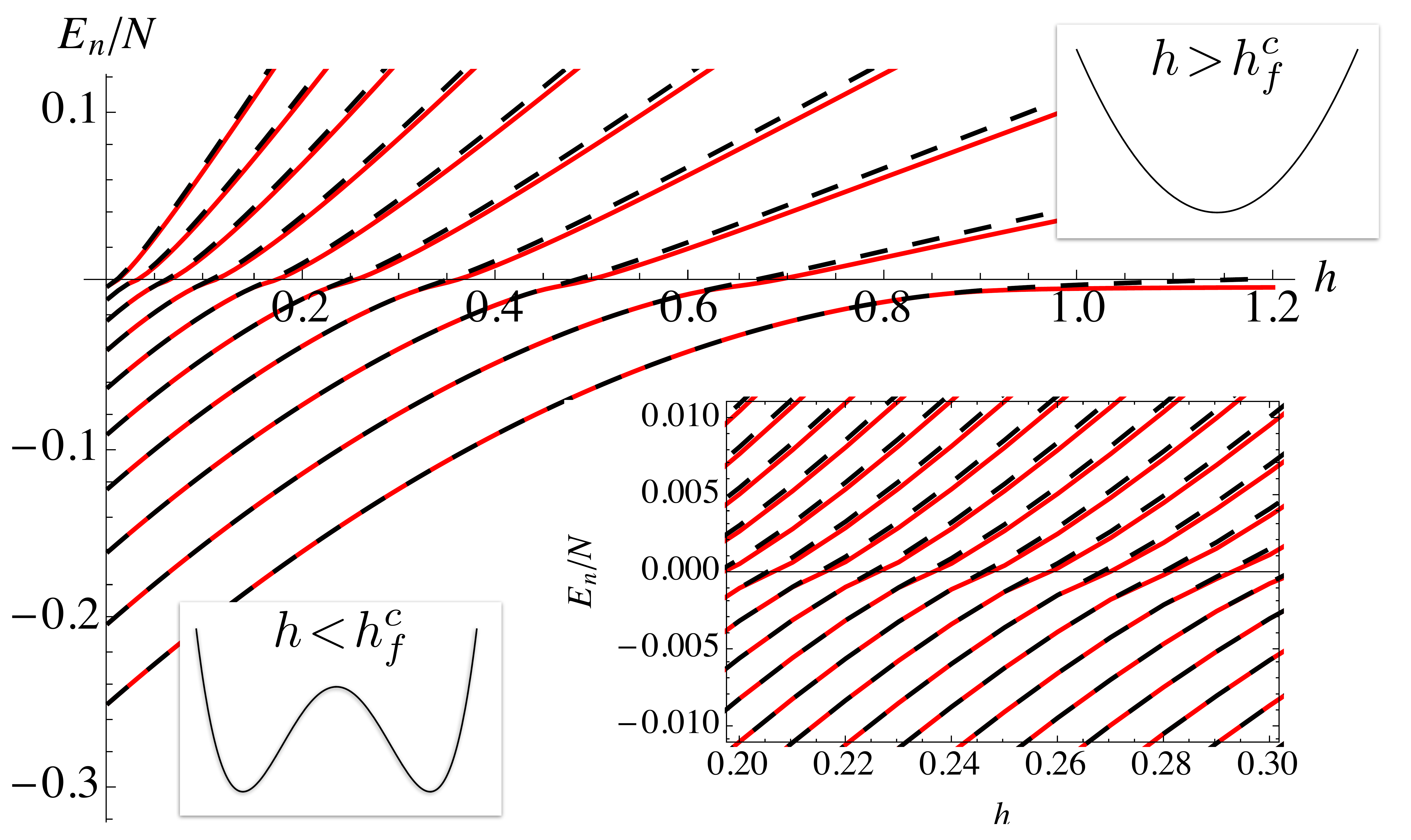}
    \caption{Spectrum of the LMG model~\eqref{lmg_boson}, with respect to the magnetic field $h$. {\it Main:} Energy spectrum for $N\!=\!100$, showing the crossing between the critical line of the ESQPT $E_c\!=\!0$ and the lifting of the degeneracy. Only 1/5 of the total eigenstates are shown for clarity. Solid and dashed lines refer to eigenstates with opposite parity. {\it Lower right inset:} Zoom around the critical energy for $N\!=\!500$ spins. {\it Upper right+lower left insets:} Sketch of the effective potential above and below the ESQPT, respectively.}
    \label{lmg_energies}
    \end{figure}

\noindent In the thermodynamic limit, the model exhibits a spontaneous symmetry-breaking second-order QPT in the ground state at $h_c\!=\!1$~\cite{prl_vidal, pre_vidal, prb_vidal, prb_castanos} between a ferromagnetic phase ($h\!<\!1$) where the spectrum becomes doubly degenerate, and therefore the system is effectively a double well, and a paramagnetic phase ($h\!>\!1$) where all energy levels are distinct and equi-spaced. We qualitatively see the difference between these phases in Figure~\ref{lmg_energies} where we show the eigenenergies of the LMG model~\eqref{lmg_boson} as a function of $h$, for $N\!=\!100$. The double degeneracy in the spectrum leads to another critical feature in the excited-states. For a finite value of $N$ and fixed value of $h\!<\!1$ we see that the spectrum is only doubly degenerate up to a particular energy level which characterizes ESQPT~\cite{Caprio:08}. We can identify ESQPTs either by fixing the energy while varying the control parameter of the model, or equivalently, by increasing the energy at a fixed value of the control parameter. The ESQPT refers to a non-analytical behavior of the density of states, $\nu(E)=\sum_k \delta(E-E_k)$ with $E_k$ the eigenenergies of the Hamiltonian, i.e. $\mathcal{H}=\sum_k E_k\ket{k}\bra{k}$~\cite{Caprio:08,esqpt_2020review}. As the system approaches an ESQPT in the LMG, the density of states develops a logarithmic divergence $\nu(E)\propto -\log|E-E_c|$ due to a concentration of the energy levels at $E_c\!=\!0$~\cite{prl_vidal,pre_vidal,2016_pra_lea} and for $h\!<\!1$ which is the critical region for the ESQPT in the Hamiltonian~\eqref{lmg_boson}.  In what follows we explore how dynamical signatures of the ESQPT are present in the work statistics after a sudden quench and examine the effect of breaking the $\mathbb{Z}_2$ parity symmetry.

\section{Figures of merit}
\noindent We consider a protocol where the Hamiltonian~\eqref{lmg_boson} is initialized in a particular state, $\ket{\psi_i}\!=\!\ket{\psi(0)}$, for a given value of the magnetic field, $h_i$. At $t\!\!=\!\!0$ we abruptly change the magnetic field $h_i\!\to\!h_f$ and study the time evolution of the system under the final Hamiltonian, $\mathcal{H}_f$, according to $\ket{\psi(t)}\!=\!e^{-i\mathcal{H}_f t} \ket{\psi(0)}$. In what follows, we consider $\mathbb{Z}_2$ symmetric and $\mathbb{Z}_2$  symmetry-broken ground states and excited states. A key figure of merit for studying the dynamical response of a system to such a sudden perturbation is captured by the time-dependent fidelity or survival probability which has been extensively used in studying the critical features of spin models~\cite{RMP_Silva, JoPA_Jafari, prb_najafi, prb_bose, scirep_campo, pra_fazio, prl_zanardi, prl_jafari, PR_Gorin, qian_pra, quan_pra, CampbellPRB}. Assuming the system begins in an eigenstate, it is defined as
\begin{equation}
\mathcal{L}(t)=\vert \chi(t) \vert^2
\label{echo}
\end{equation}
where
\begin{equation}
\chi(t)=\bra{\psi(0)} \psi(t) \rangle=\bra{\psi(0)} e^{-i\mathcal{H}_f t} \ket{\psi(0)}
\label{echo2}
\end{equation}
is the characteristic function of the work distribution in the case of a sudden quench and given by~\cite{prl_silva}
\begin{equation}
P_W=\sum_{m} p_{m\vert n}^\tau \delta \left( W - (E_m - E_n) \right),
\label{pw}
\end{equation}
with $E_m (E_n)$ the energy of the corresponding eigenstate of the final (initial) Hamiltonian. Unless otherwise stated, we will assume the system begins in the ground state, $n\!=\!0$, and therefore $p_{m\vert 0}$ is the conditional probability of measuring $E_m$ after the quench. The moments of the work distribution due to the sudden quench can be readily determined~\cite{prx_fusco}:
\begin{equation}
    \langle W^l \rangle = \sum_{m} \Big( E_m^f - E_0^i \Big)^l |\langle\psi^i_0 \vert \psi^f_m \rangle |^2\equiv (-i)^l \partial_{t}^l \chi(t) \vert_{t\to0},
    \label{work}
\end{equation}
with the first and second moments corresponding to the average work and variance, respectively. Recent proposals have demonstrated that the distribution~\eqref{pw} is experimentally accessible~\cite{WorkDistPRL}. Under these conditions, namely initial ground state and sudden quench, the work distribution is mathematically equivalent to the infinite time average of the quantum state, i.e. the diagonal ensemble. Therefore, we have all the information necessary to determine the entropy of the diagonal ensemble~\cite{prb_goold_1, prb_goold_2, cakan2020_DE}, which is simply given by the Shannon entropy of $P_W$:
\begin{equation}
S_W=-\sum_{W}\ P_W \log_2 P_W.
\label{entropy}
\end{equation}
\section{Quench from the ferromagnetic phase}
Before analyzing the impact of symmetry breaking and ESQPT in the work statistics, it is convenient to find the critical value of the magnetic field $h_f^c$ for which the initially prepared ground state at $h_i$ is brought to the critical energy $E_c$ at which the ESQPT takes place. For that we rely on a semiclassical approximation. We first compute the semiclassical energy using a spin coherent representation, $\ket{\alpha}=(1+\alpha^2)^{-J}e^{\alpha S^+}\ket{J,-J}$ with $\alpha\in \mathbb{R}$ and where $\ket{J,m_J}$ denotes the standard basis of $\{S^2,S_z\}$ for the Dicke states, such that $S_z\ket{J,m_J}=m_J\ket{J,m_J}$. The energy  is
\begin{align}\label{eq:Ealpha}
E(\alpha,h)=\lim_{N\rightarrow \infty} \bra{\alpha}\mathcal{H}\ket{\alpha}=\frac{(\alpha^4-1)h-2\alpha^2}{(1+\alpha^2)^2},
\end{align}
where we have neglected the irrelevant constant energy contribution $hN/2$, which does not modify the double-well structure of $E(\alpha,h)$. 
The ground state parameter under this spin-coherent representation is achieved by minimization of $E(\alpha,h)$, which yields
\begin{align}
\alpha_{\rm gs}(h)= \begin{cases}0 \qquad \qquad h>1 \\ \pm \sqrt{\frac{1-h}{1+h}} \quad 0\leq h\leq 1 \end{cases}.
\end{align}
The solution $\alpha=0$ that ensures $dE(\alpha,h)/d\alpha=0$ becomes a local maximum for $h>1$,  which is precisely the ESQPT. For the energy functional as given in~\eqref{eq:Ealpha}, the ESQPT takes place at the critical energy $E_c=-h$ for $0\leq h\leq 1$. The two equivalent solutions for $0\leq h\leq 1$ signal the spontaneous symmetry breaking. 

\noindent In this manner, we can compute the energy of the quenched initial state as $E_q(h_i,h_f)\equiv E(\alpha_{\rm gs}(h_i),h_f)=\lim_{N\rightarrow \infty} \langle \alpha_{\rm gs}(h_i)|\mathcal{H}(h_f) |\alpha_{\rm gs}(h_i)\rangle $ where we have explicitly written the dependence of the Hamiltonian on the final parameter $h_f$. The critical quench strength follows from $E_q(h_i,h_f^c)=E_c$ which for $0\leq h_i\leq 1$ results in the simple expression
\begin{align}
h_f^c=\frac{1+h_i}{2}.
\label{eq:hfc}
  \end{align}
That is, for $h_f<h_f^c$ the ground state of $\mathcal{H}$ at $h_i$ is confined within the symmetry-broken phase, while a quench $h_f>h_f^c$ provides sufficient energy so that the quenched state is brought above the ESQPT where the degeneracy is lifted (cf. Figure~\ref{lmg_energies}).

\noindent Note however that if $h_i>1$, then $E(h_i,h_f)=-h_f$ which corresponds to the ground state energy for $h_f\geq 1$ and the critical energy of the ESQPT for $h_f<1$. Hence, the impact of the QPT can be captured by quenching from the paramagnetic phase ($h_i>1$) to the critical point $h_f=h_c=1$. 

\begin{figure*}[t]
    \subfloat[\label{echo_symm}]{%
       \includegraphics[width=0.33\textwidth]{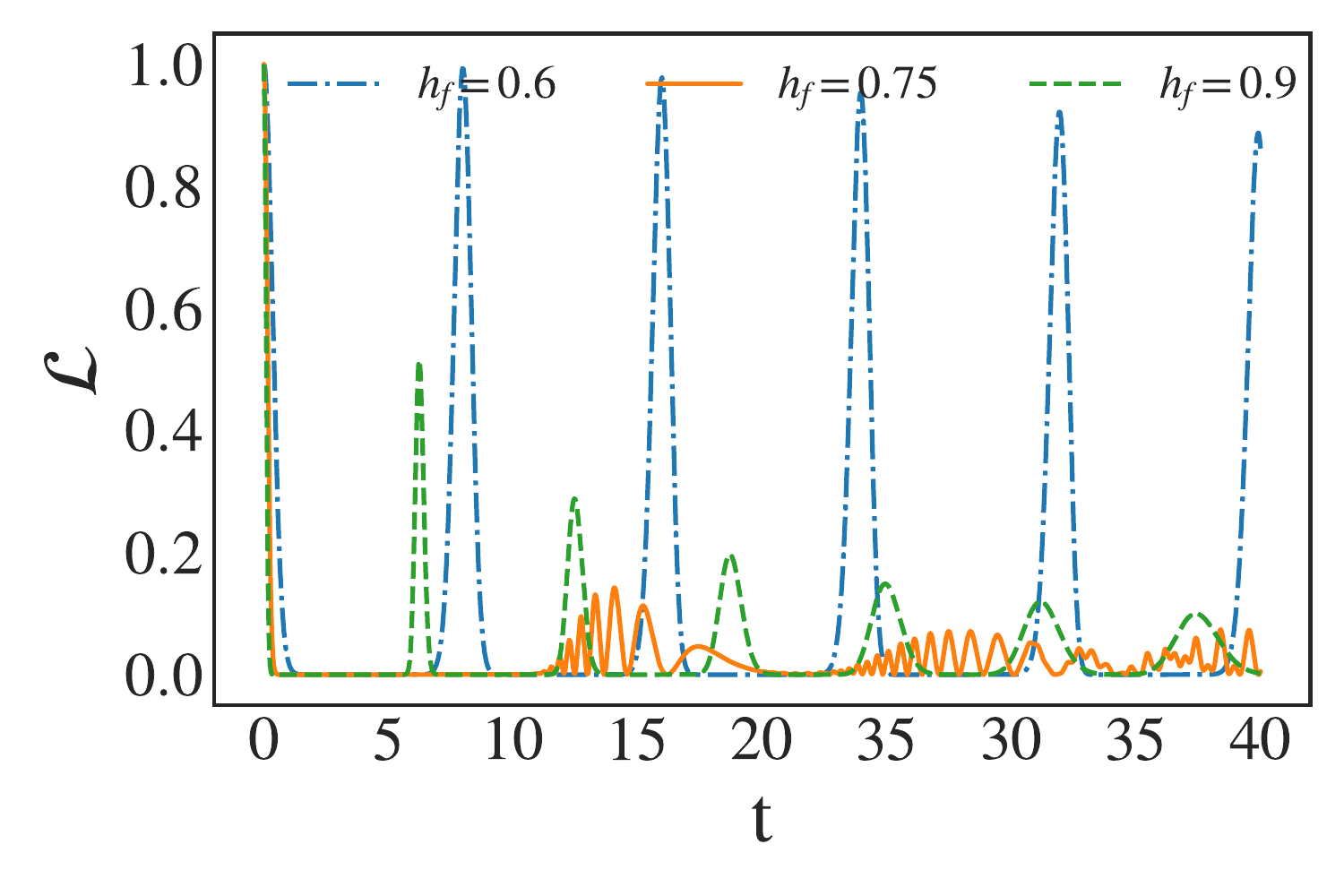}
       }%
     \subfloat[\label{pw_symm}]{%
       \includegraphics[width=0.33\textwidth]{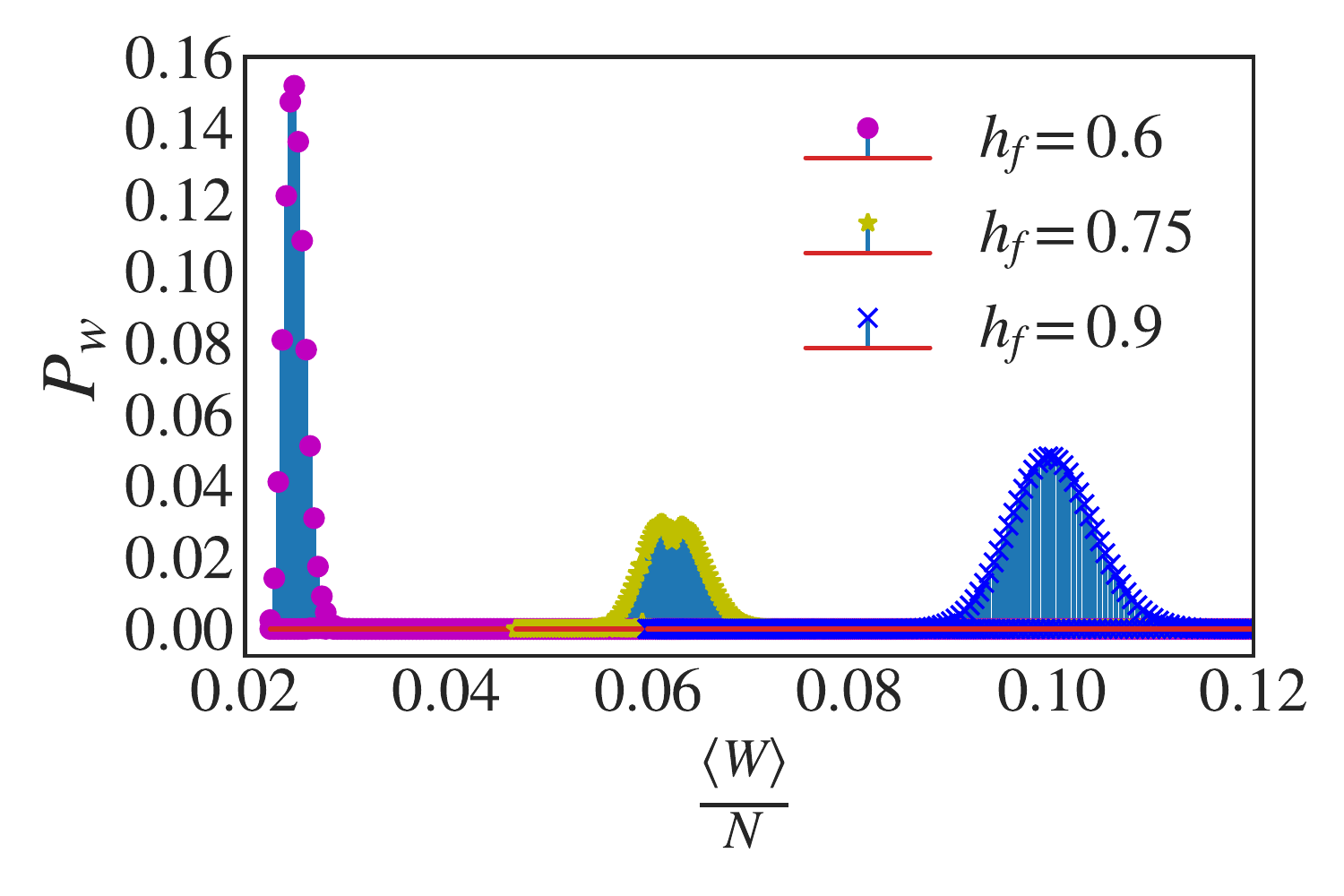}
       }
     \subfloat[\label{sw_symm}]{%
       \includegraphics[width=0.33\textwidth]{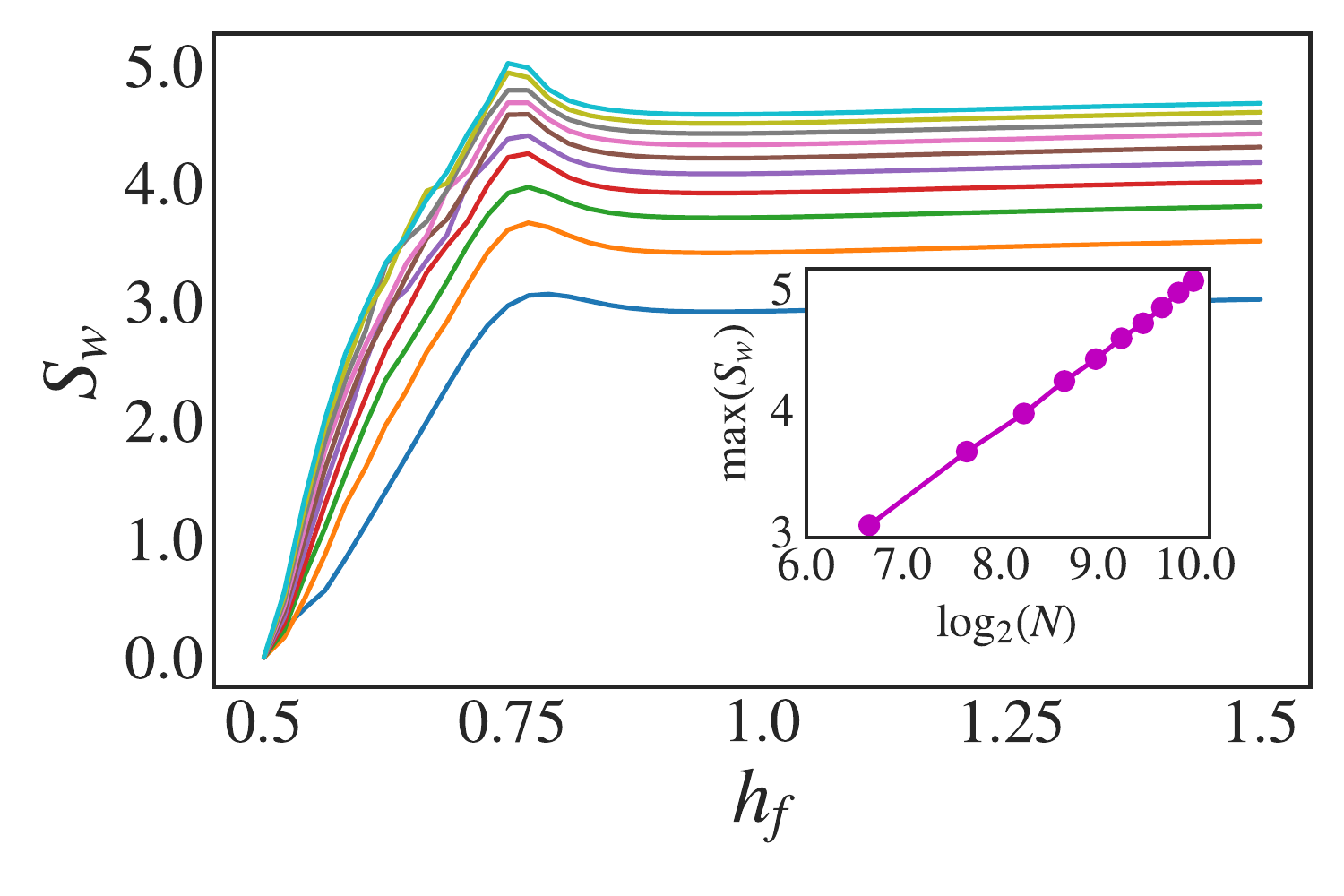}
       }
\caption{{\bf Symmetric ground state.} (a)+(b) The survival probability~\eqref{echo}, and the work probability distribution~\eqref{pw}, respectively for a system size $N\!=\!2000$ and initial magnetic field $h_i=0.5$, when the quench is performed to below ($h_f\!=\!0.6$), above ($h_f\!=\!0.9$) and at the ESQPT critical point ($h_f\!=\!0.75$). (c) The Shannon entropy~\eqref{entropy} with respect to the magnetic field $h_f$ for various system sizes $N\!=\!100 [\text{bottom,~blue}]\!\to\!1000[\text{top,~cyan}]$, the inset show the scaling of the maximum of $S_W$ with respect to $\log_2(N)$.}
\label{symb_gs}
\end{figure*}

\subsection{Symmetric ground state}
We begin our analysis by initializing our system in the ground state of the ferromagnetic phase ($h\!<\!1$) of the LMG model~\eqref{lmg_boson} and perform a sudden quench from $h_i\!=\!0.5$. The ESQPT corresponds to the point at which the energy of the final Hamiltonian, measured in the initial energy basis, crosses the critical line of the ESQPT $E_c\!=\!0$, cf. Figure~\ref{lmg_energies}. As explained above, the energy of the post-quenched Hamiltonian crosses $E_c\!=\!0$  at $h_f^c\!=(1+h_i)/2=0.75$ where the ESQPT occurs.  We consider quenches to three different values of $h_f$: (i) below the ESQPT, $h_f\!<\!0.75$, (ii) to the ESQPT, $h_f\!=\! 0.75$, and (iii) beyond the ESQPT, $h_f\!>\!0.75$. We remark that qualitatively similar results hold for other choices of $h_i$ with the caveat that the location of the ESQPT is shifted accordingly as dictated by equation~\eqref{eq:hfc}.

\noindent Figure~\eqref{echo_symm} depicts the survival probability~\eqref{echo} for a system size $N\!=\!2000$ and a quench starting from $h_i\!=\!0.5$. We find that quenches either sufficiently below or above the ESQPT point show qualitatively similar behaviors. In particular, for $h_f\!=\!0.6$ we find strong periodic revivals with the system almost perfectly returning to the initial state, while for $h_f\!=\!0.9$ the system still exhibits sharp revivals between periods of dynamical orthogonality, albeit with the revivals decaying in amplitude. This qualitative behavior persists for other values of $h_f$, including when quenching to and beyond the second order ground state QPT $h_c\!=\!1$~\cite{CampbellPRB} with the notable exception in the vicinity of the ESQPT. For quenches to the ESQPT point we see the survival probability no longer exhibits such a clear periodic behavior, but instead remains dynamically close to a fully orthogonal state. The sensitivity to the presence of the ESQPT is further reflected in the work probability distribution shown in Figure~\eqref{pw_symm}, where we find $P_W$ is generally Gaussian for quenches to arbitrary values of $h_f$, except in the vicinity of the ESQPT, $h_f\!=\! h_f^c\!=\!0.75$, where the shape of $P_W$ changes to a double peak with the emergence of a dip, reflecting the effect of the presence of the ESQPT, as previously discussed~\cite{2011_relano}. 

\begin{figure}[t!]
    \subfloat[First moment $\langle W \rangle$.\label{work1_symm}]{%
       \includegraphics[width=0.49\textwidth]{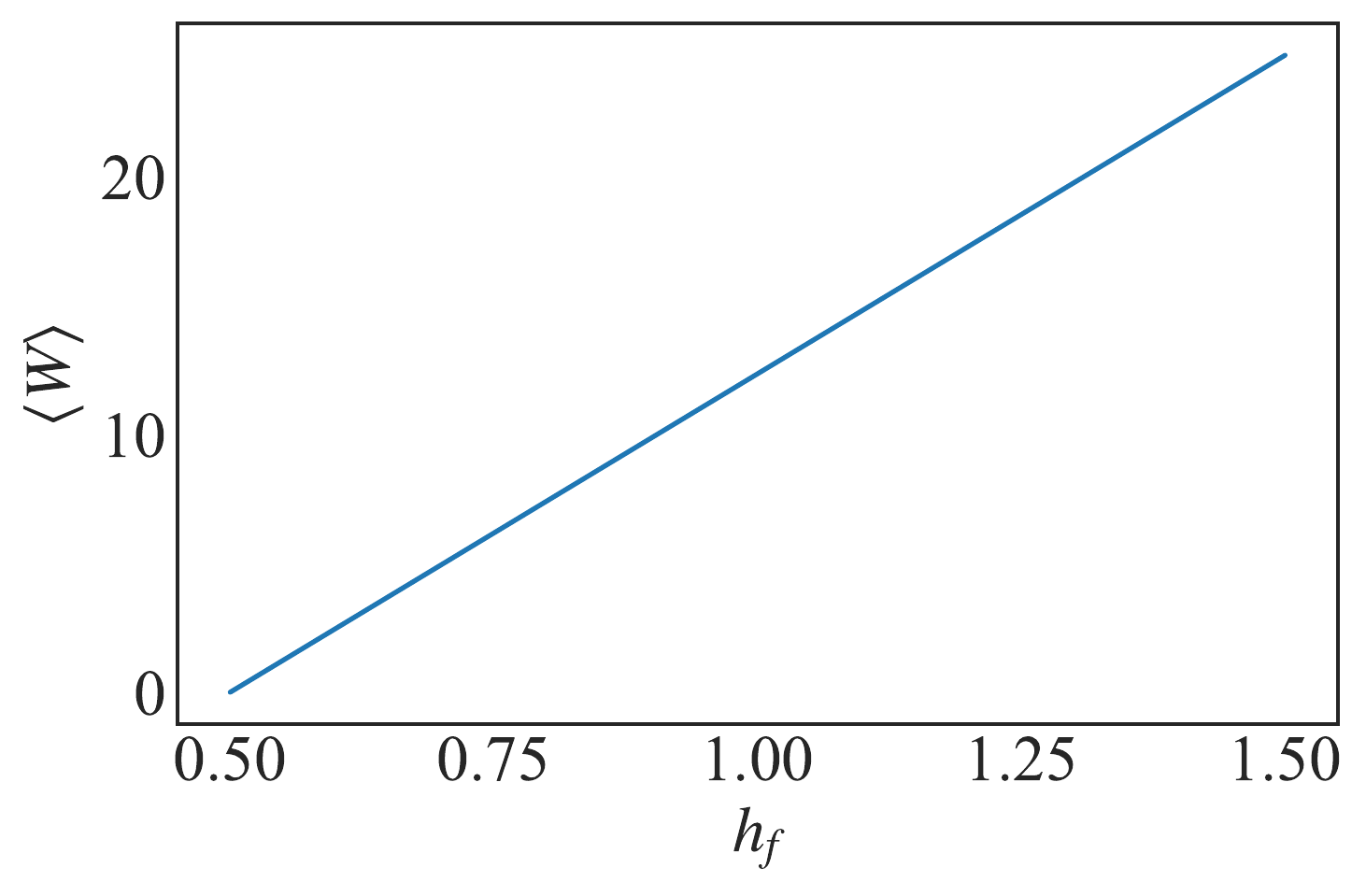}
       }%
     \subfloat[Second moment $\langle W^2 \rangle$.\label{work2_symm}]{%
       \includegraphics[width=0.49\textwidth]{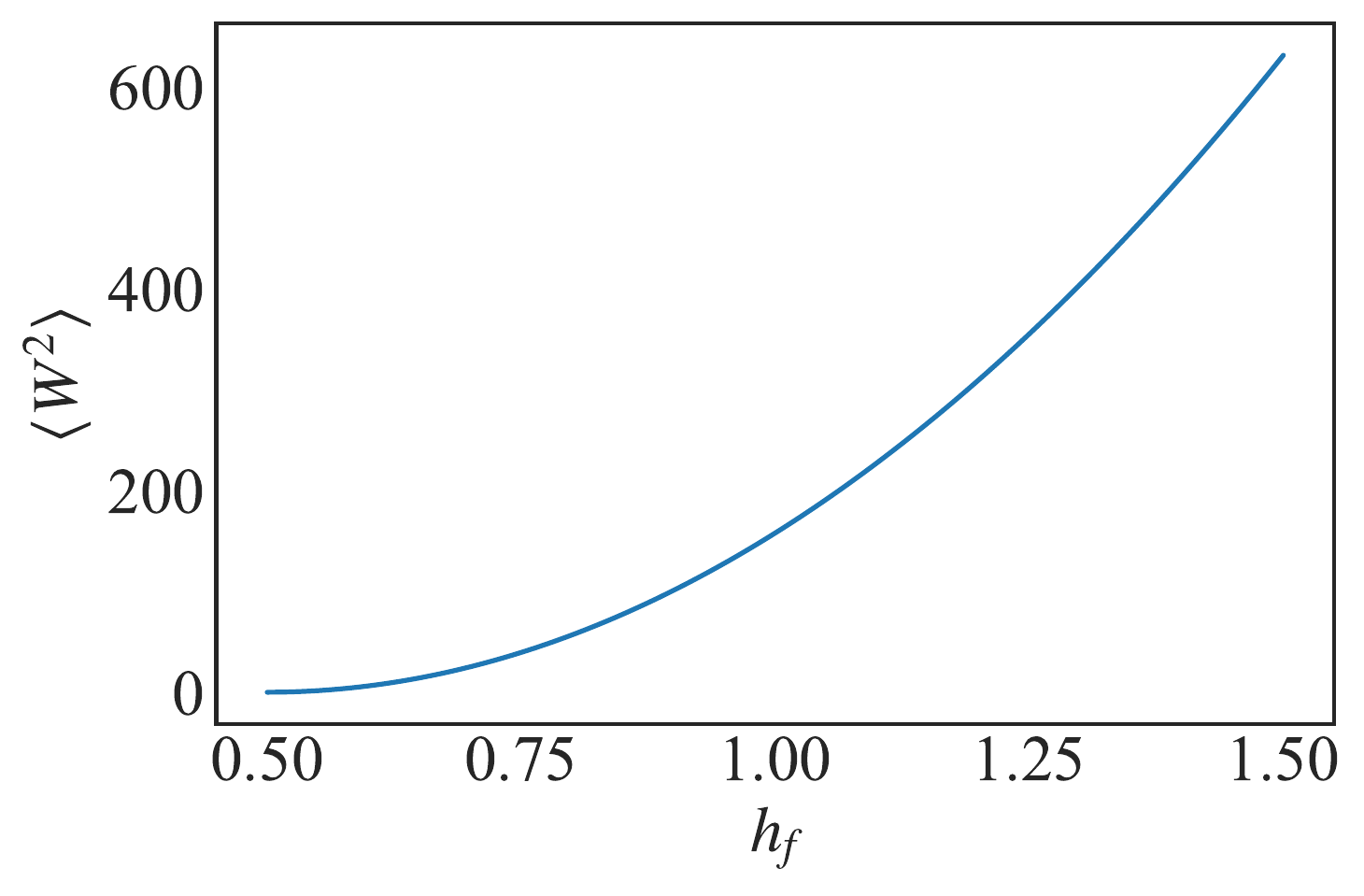}
       }
    \caption{Moments of the average work~\eqref{work} with respect to the final magnetic field $h_f$, for a system size $N=100$ and initial ground state at $h_i=0.5$. In (a) we represent the first moment, and in (b) the second moment.}
    \label{fig_mom_work}
\end{figure}

\noindent We next examine the entropy of the diagonal ensemble~\eqref{entropy}, in Figure~\eqref{sw_symm} for $h_i\!=\!0.5$ as a function of the quench amplitude, $h_f$, for various system sizes. We immediately see the emergence of a peak in the entropy at the ESQPT point. We observe a logarithmic scaling of $S_W\propto \log_2(N)$ as the system size is increased, as demonstrated in the inset where we show this explicitly for the peak, however we remark that this scaling holds for any value of $h_f$. Finally, we note that the moments of the work distribution are also readily accessible. However they exhibit no sensitivity to the presence of the ESQPT, with the first (second) moments scaling linearly (quadratically) with the quench amplitude, as depicted in Figure~\eqref{fig_mom_work}.

\subsection{Symmetry-broken ground state}
As previously mentioned, in the thermodynamic limit and for $h<1$ the LMG undergoes a spontaneous $\mathbb{Z}_2$ symmetry breaking. For finite systems, any small perturbation in $S_x$ leads to a symmetry breaking in the ferromagnetic phase ($h<1$), while it does not alter the paramagnetic phase ($h>1$). For that reason, we introduce a small perturbation $|\epsilon|\!<\!<\!1$ in $S_x$, such that it does not affect the critical features of the model, i.e.
\begin{equation}
    \mathcal{H}=-\frac{1}{N}  S_x^2  + h\: \Bigg(S_z +\frac{N}{2} \Bigg) + \epsilon\: S_x.
     \label{lmg_epsilon}
\end{equation}
In this case, the non-zero elements of the Hamiltonian in the basis~\eqref{base}, are given by~\eqref{element_matrix} and
\begin{equation}
\bra{N,n_t}\mathcal{H} \ket{N,n_t+1}\!=\!\frac{\epsilon}{2} \sqrt{(N-n_t)(n_t+1)}.
\end{equation}
When $h\!<\!1$ the ground state of the Hamiltonian~\eqref{lmg_epsilon}, is a fully symmetry broken (FSB) ground state, i.e. a superposition of the two degenerate fully symmetric ground states with opposite parity, $\ket{\varphi_{\pm}}$ such that $\Pi\ket{\varphi_{\pm}}=\pm \ket{\varphi_{\pm}}$.  In particular, the FSB states can be written as $\ket{\varphi_{\rm FSB,\pm}}=(\ket{\varphi_{+}}\pm\ket{\varphi_-})/\sqrt{2}$ which yield a maximum value of the symmetry-breaking order parameter, $|\langle S_x \rangle|$, and are only degenerated up to an energy factor $|\epsilon|\ll 1$. 
\begin{figure}[t!]
    \subfloat[\label{echo_symm_br}]{%
       \includegraphics[width=0.33\textwidth]{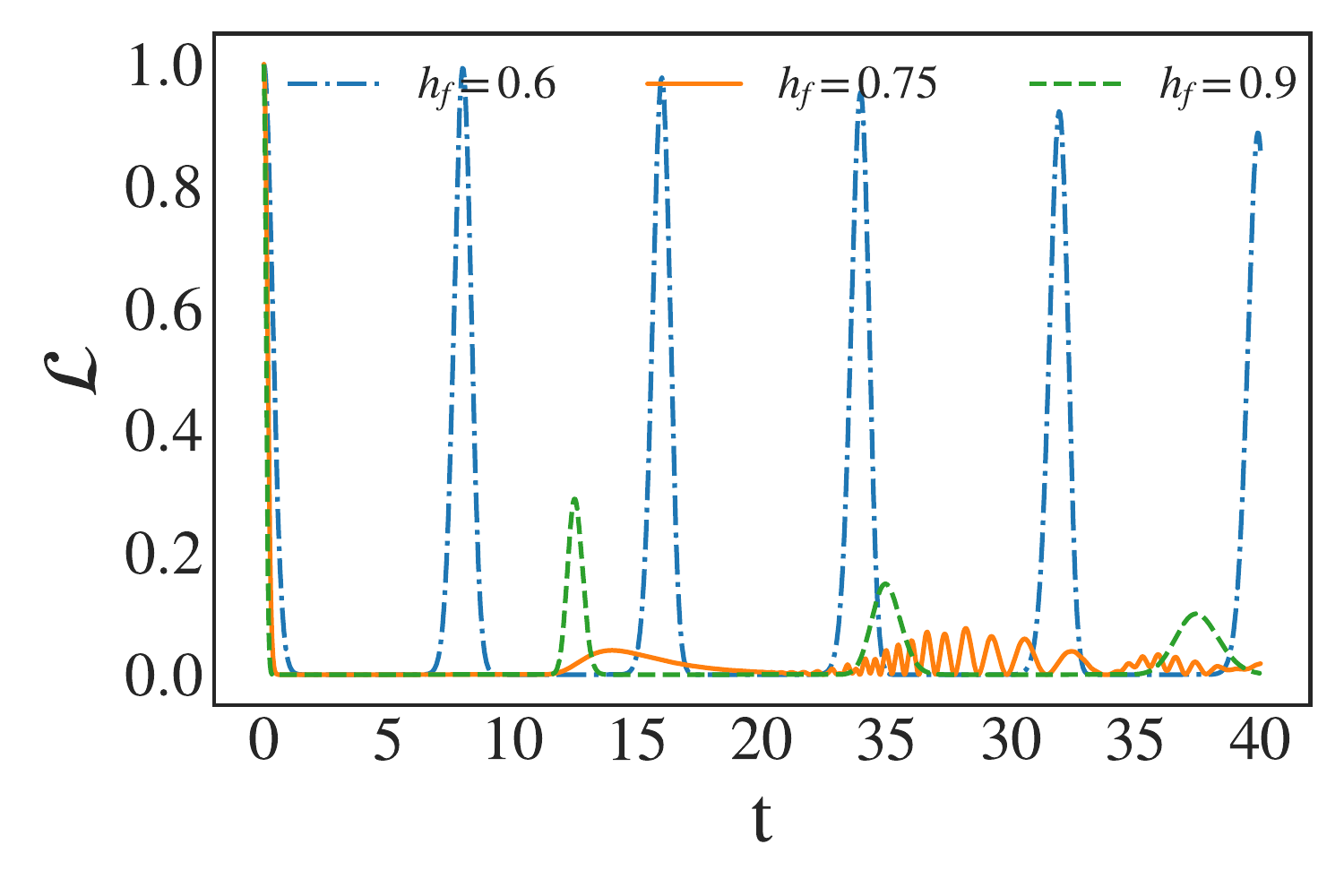}
       }%
     \subfloat[\label{pw_symm_br}]{%
       \includegraphics[width=0.33\textwidth]{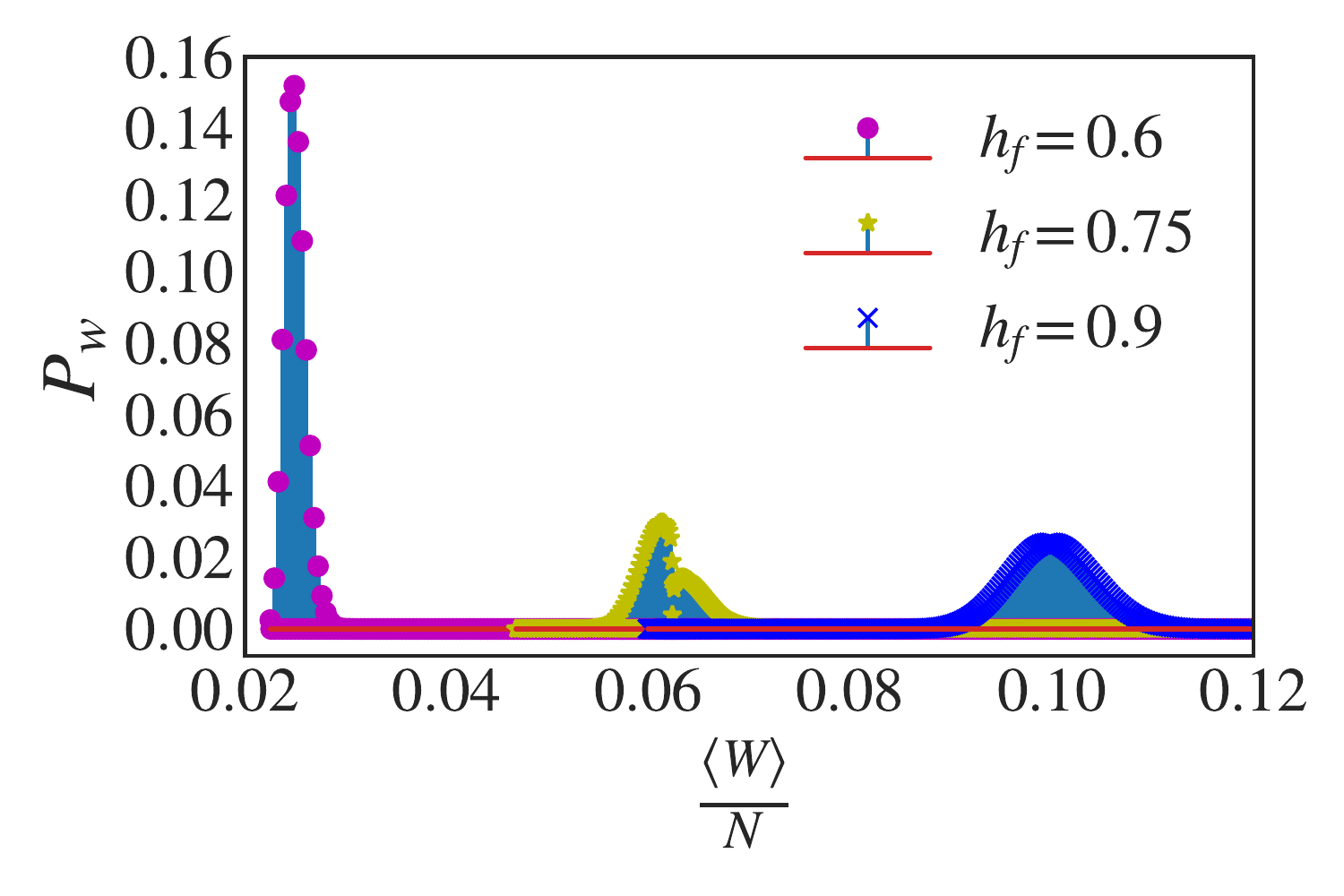}
       }
     \subfloat[\label{sw_symm_br}]{%
       \includegraphics[width=0.33\textwidth]{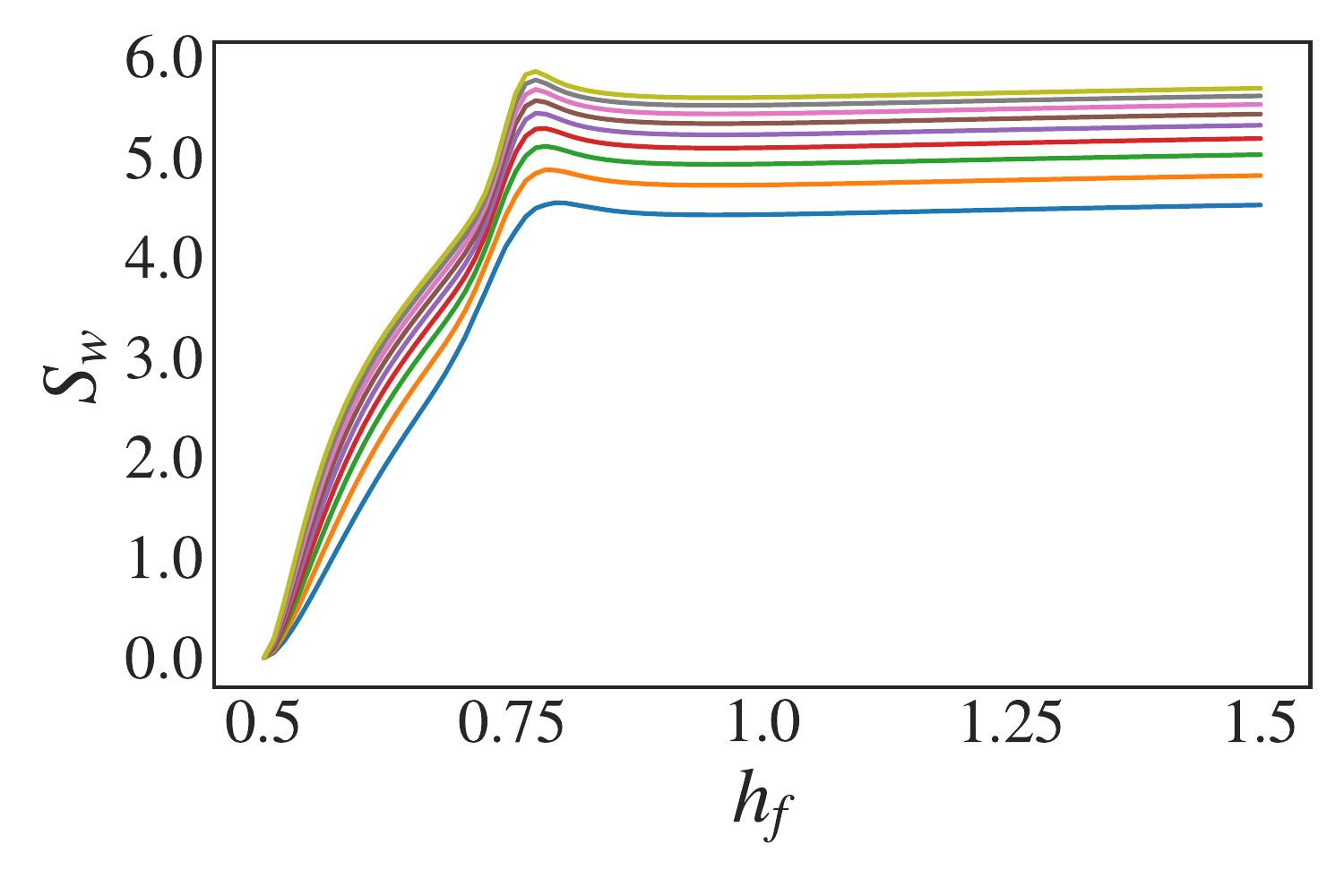}
       }
\caption{{\bf Fully symmetry broken ground state.} (a)+(b) The survival probability~\eqref{echo}, and the work probability distribution\eqref{pw}, respectively for a system size $N\!=\!2000$ and initial ground state at $h_i=0.5$, when the quench is performed to below ($h_f\!=\!0.6$), above ($h_f\!=\!0.9$) and at the ESQPT critical point ($h_f\!=\!0.75$). (c) The Shannon entropy~\eqref{entropy} with respect to the magnetic field $h_f$ for various system sizes $N\!=\!100 [\text{bottom,~blue}]\!\to\!1000[\text{top,~yellow}]$. 
}
\label{fsb}
\end{figure}
We can now examine the effect that breaking the $\mathbb{Z}_2$ symmetry has on the figures of merit. Considering the same quench parameters as before, in Figure~\eqref{echo_symm_br} we show the behavior of the survival probability. While largely consistent with the previous case, we nevertheless see some qualitative differences appearing. For quenches below the ESQPT, we see that symmetry breaking has no effect and the survival probability is identical in both instances, as can be seen by comparing the blue dot-dashed curves in Figure~\eqref{echo_symm} and~\eqref{echo_symm_br}.  For quenches beyond the ESQPT, breaking the symmetry leads to a change in the period of the revivals in the survival probability. In fact, the period of these revivals doubles in the symmetry broken case with respect to the symmetric ground state due to the fact that the $\mathbb{Z}_2$ symmetry is no longer conserved. In this case, the absent peaks in the survival probability with respect to the symmetric ground state (cf. Figure~\eqref{echo_symm}) correspond to the overlap $|\langle \varphi_{\rm FSB,-}|e^{-i\mathcal{H}_f t}|\varphi_{\rm FSB,+}\rangle|^2$, which is intimately related to the emergence of a dynamical quantum phase transition. Quenches to the ESQPT point are notably affected by breaking the symmetry, with the system remaining closer to orthogonality throughout. 

\noindent The work distribution is similarly affected, showing evidence of the symmetry breaking only when the quench is sufficiently strong. As shown in Figure~\eqref{pw_symm_br}, the distribution is the same for both the symmetric and symmetry broken initial states when the quench is below the ESQPT as the energies in this region are not symmetry dependent (cf. Figure~\eqref{lmg_energies}). However, while $P_W$ retains its Gaussian profile for quenches far above/below the ESQPT point, when quenching beyond the ESQPT the distribution peak is halved, which is due to the spreading of $P_W$ over both parity subspaces of the model. The effect of symmetry breaking is most notable in the work distribution when the system is quenched to the ESQPT. Once again the distribution loses the Gaussian profile and exhibits a dip similar to the symmetric case. We now find that to the left of the dip, corresponding to states below the critical energy, both the symmetric and symmetry broken initial states show the same distribution, however to the right of the dip the amplitude of the probabilities is halved again due to involvement of both parity subspaces.

\begin{figure}[t!]
    \subfloat[First moment $\langle W \rangle$.\label{work1_symm_br}]{%
       \includegraphics[width=0.49\textwidth]{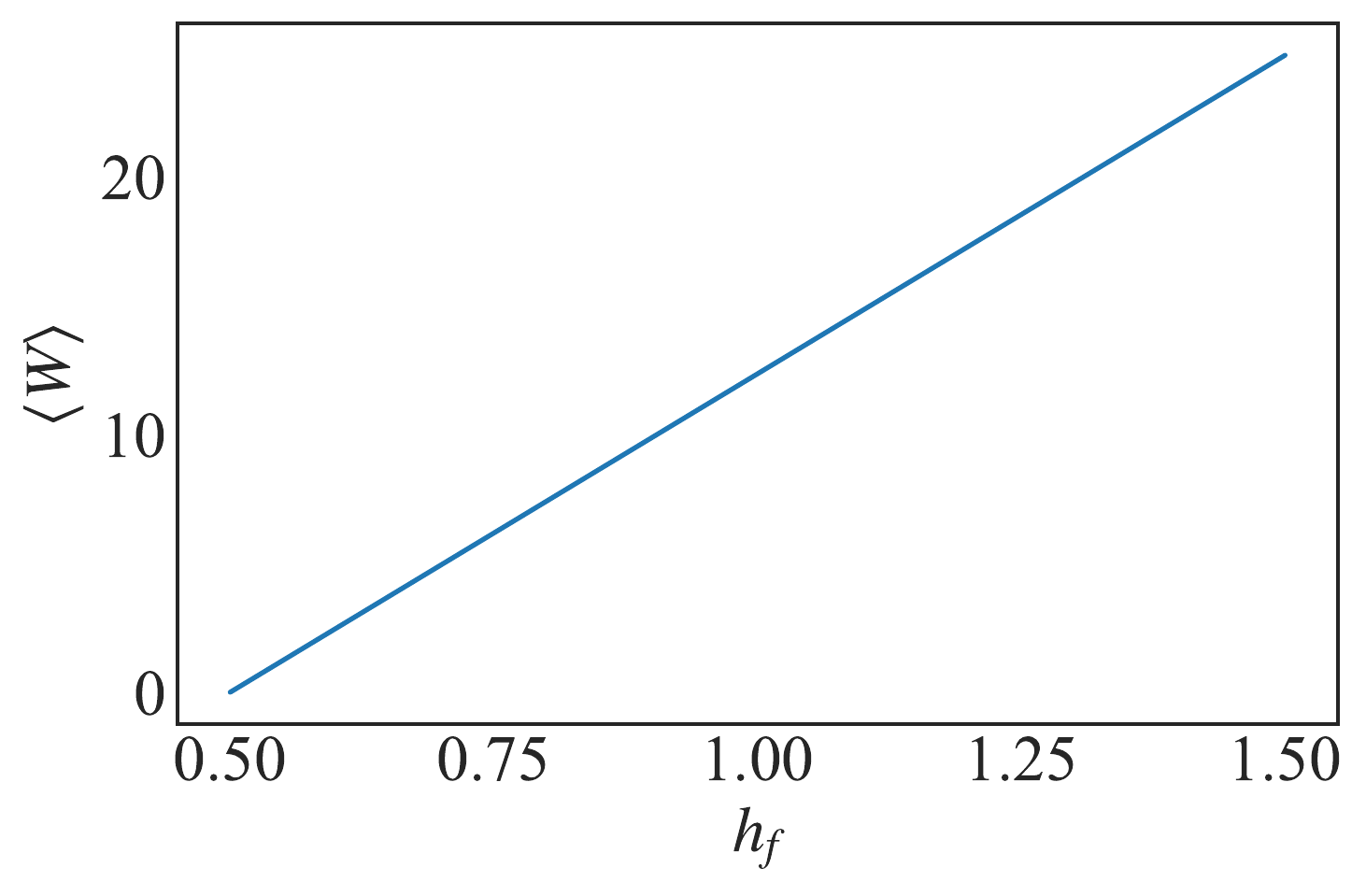}
       }%
     \subfloat[Second moment $\langle W^2 \rangle$.\label{work2_symm_br}]{%
       \includegraphics[width=0.49\textwidth]{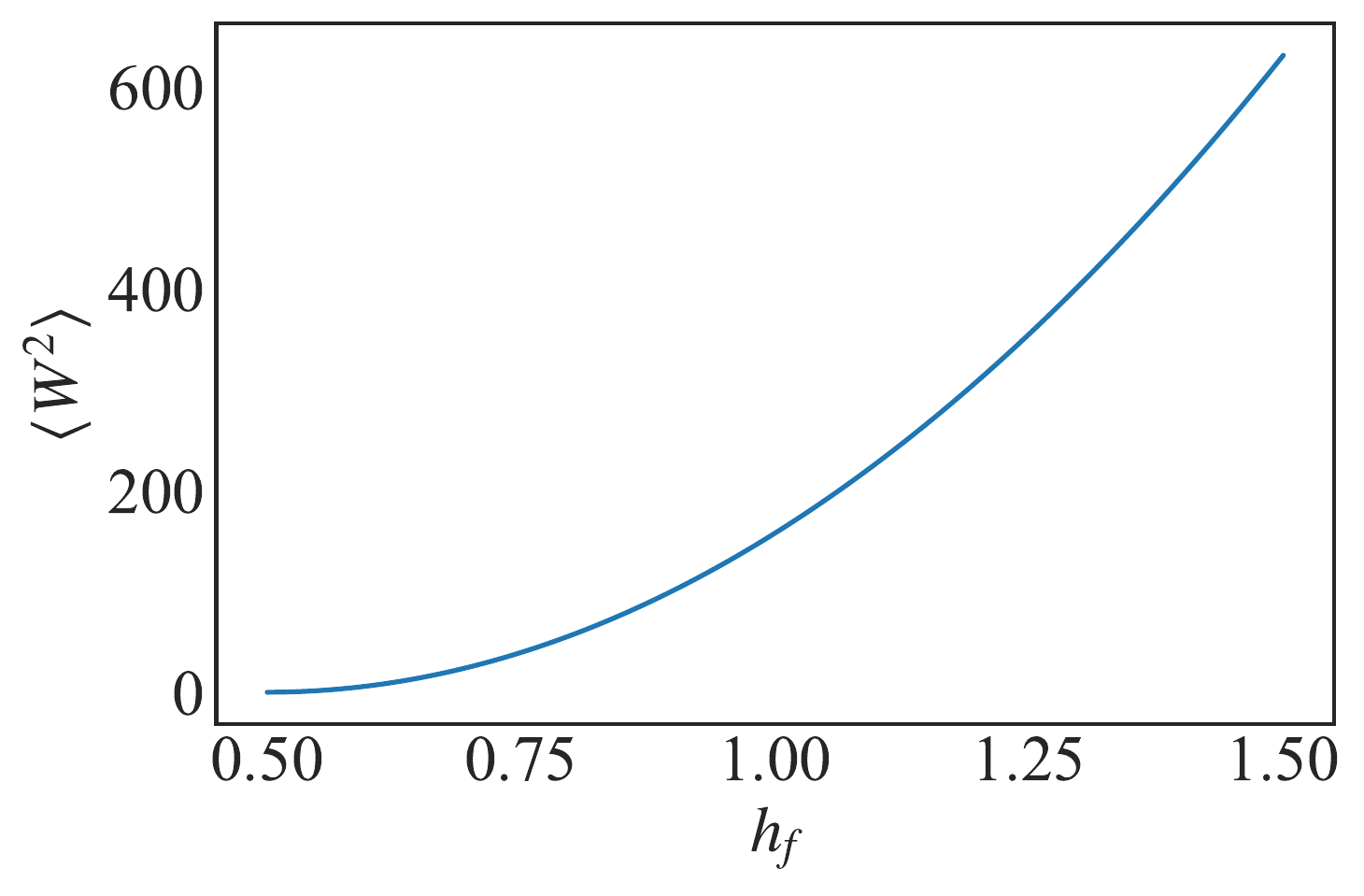}
       }
    \caption{Moments of the average work~\eqref{work} with respect to the final magnetic field $h_f$, for a system size $N=100$ and initial symmetry-broken ground state at $h_i=0.5$. In (a) we represent the first moment, and in (b) the second moment.}
    \label{fig_mom_work_br}
\end{figure}

\noindent The behavior of the entropy of the diagonal ensemble~\eqref{entropy}, when we break the $\mathbb{Z}_2$ symmetry is shown in Figure~\eqref{sw_symm_br} for various system sizes $N\!=\!100 [\text{blue}]\!\to\!1000[\text{yellow}] $ and is consistent with the behavior of the fully symmetric case shown in Figure~\eqref{sw_symm}. As in the symmetric case, $S_W$ increases quickly, peaking at the ESQPT point, $h_f^c\!=\!0.75$. We remark that, while a symmetric ground state can only populate a single parity subspace, an initial symmetry-broken ground state populates both subspaces. As a consequence, the entropy $S_W$ is larger in the symmetry-broken case by a factor $\log_2(2)=1$ when $h_f>h_f^c$, reflecting the spreading of $P_W$ over the two parity subspaces.  Finally, in contrast to these figures of merit, the first and second moments of the work distribution~\eqref{work} are unaffected by symmetry breaking, as shown in Figure~\eqref{fig_mom_work_br}.

\subsection{Weighted superposition}
\begin{figure}[t]
    \subfloat[\label{echo_super}]{%
       \includegraphics[width=0.49\textwidth]{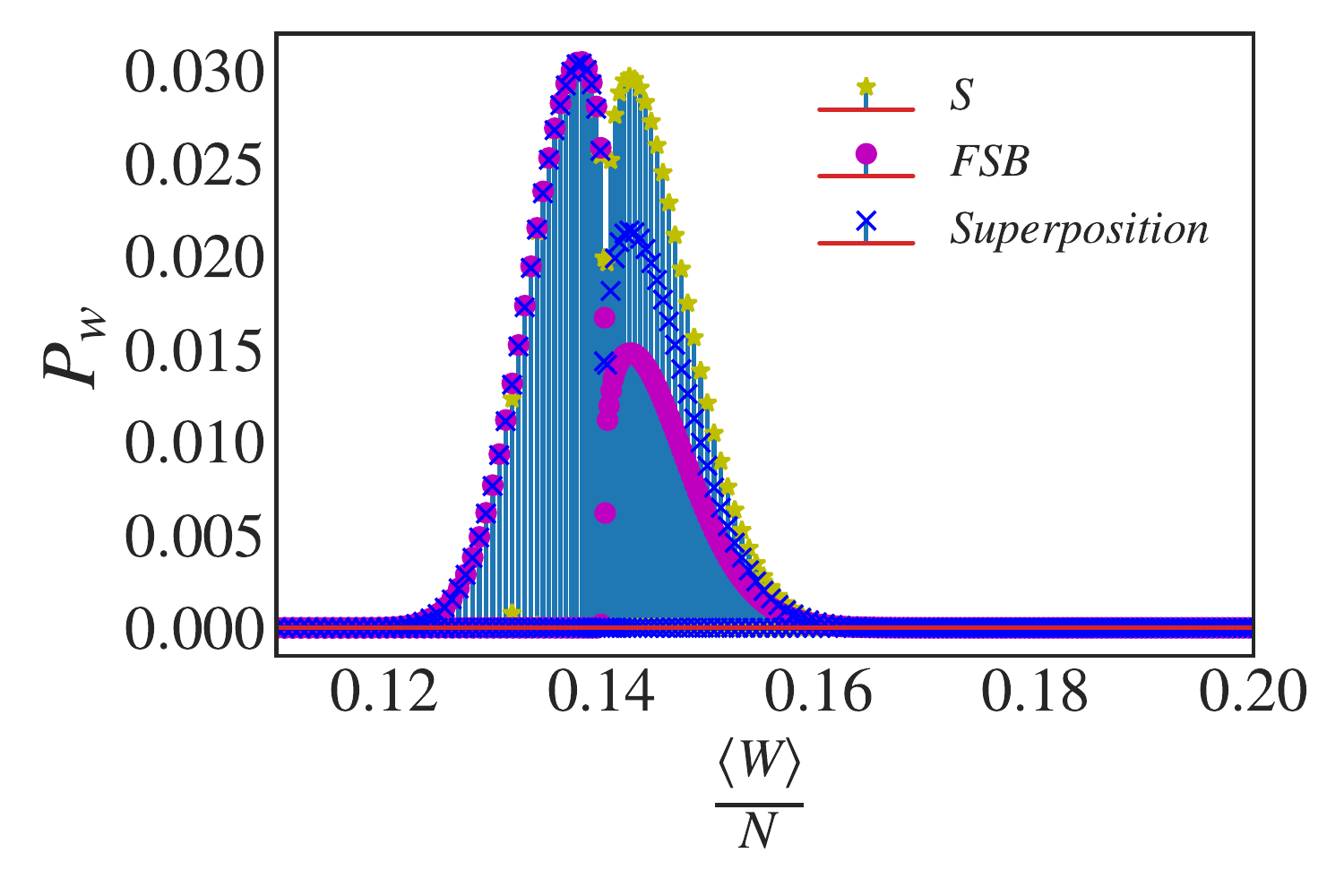}
       }%
     \subfloat[\label{pw_super}]{%
       \includegraphics[width=0.49\textwidth]{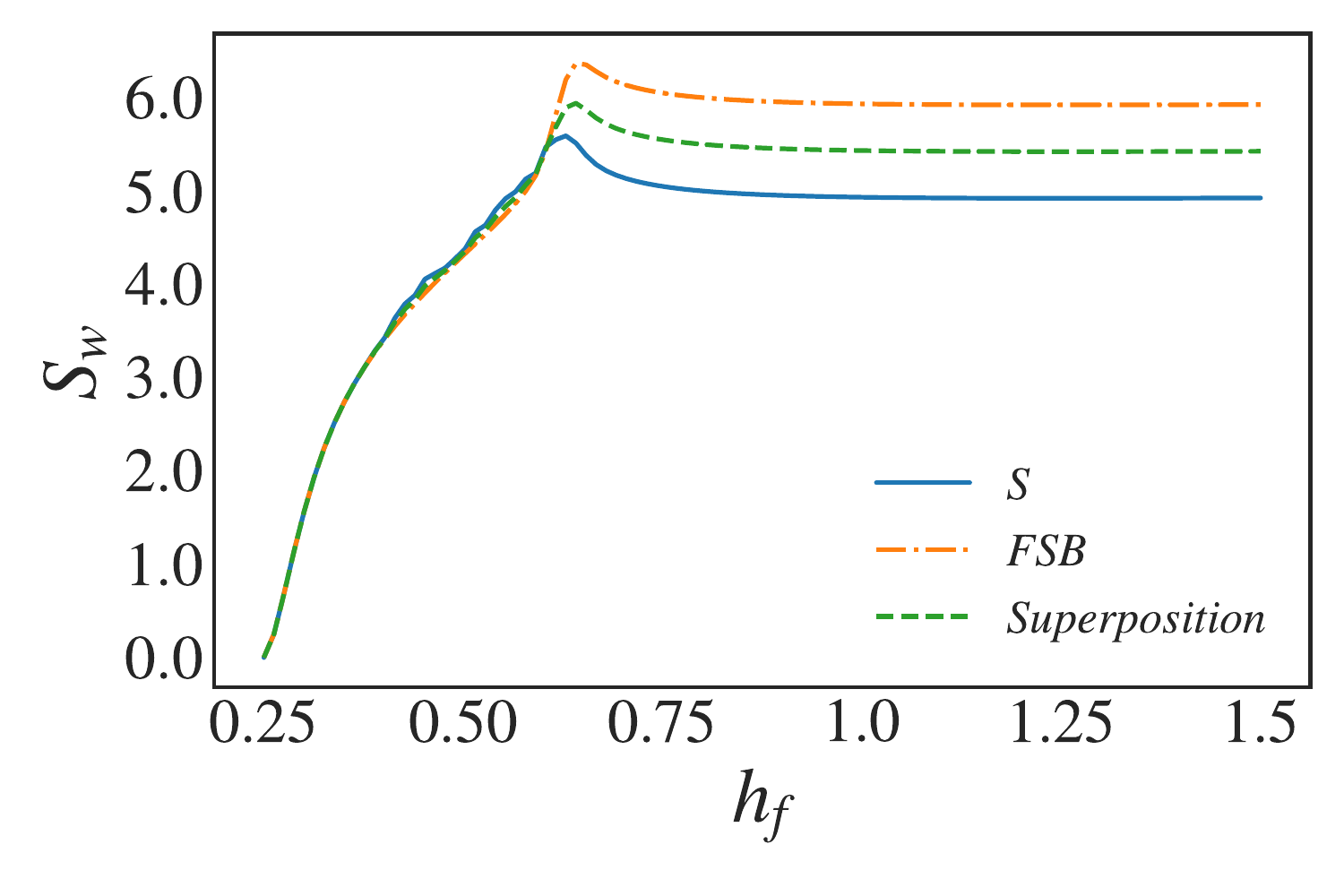}
       }
\caption{(a) The work probability distribution $P_W$~\eqref{pw}, and (b) The Shannon entropy $S_W$~\eqref{entropy}, with respect to $h_f$ in the LMG model~\eqref{lmg_epsilon}. In both panels we quench from $h_i\!=\!0.25$ in a system of size $N\!=\!1000$, initialized in the symmetric ground state (S), fully symmetry broken (FSB) ground state and in superposition between the two ground states. Note the peak at $h_f\approx 0.625$ which corresponds the critical value $h_f^c=(1+h_i)/2$ in this case (cf. Equation~\eqref{eq:hfc}).}
\label{superposition}
\end{figure}

Having discussed the features of symmetric and fully symmetry broken ground states, we complete the picture by considering the case of an initial state which does not maximise the value of the symmetry-breaking order parameter $|\langle S_x\rangle|$ but still breaks the $\mathbb{Z}_2$ parity symmetry. As an example, we choose $|\varphi_{\rm sup}\rangle\propto 2\ket{\varphi_{+}}+\ket{\varphi_{-}}$ and we (arbitrarily) fix $h_i\!=\!0.25$ with $N\!=\!1000$. In Figure~\eqref{echo_super} we show the work probability distribution for all three initial states when the quench is exactly to the ESQPT critical point, $h_f^c\!=\!0.625$. All distributions exhibit the same double-peaked behavior and, furthermore, the distributions are identical to the left of the cusp. It is only for values of the work above the cusp, which corresponds to those states of the final eigenspectrum that are above the critical energy, that show the effects of symmetry breaking. Indeed, by taking a suitable superposition we can smoothly transition between the two extreme cases shown above. The entropy of the diagonal ensemble similarly reflects the effect of taking such a superposition, as shown in Figure~\eqref{pw_super}, where $S_W$ also interpolates between the two extreme behaviors, and nevertheless clearly spotlights the presence of the ESQPT. 

\subsection{Quenching from excited states}

\begin{figure}[t]
    \subfloat[\label{echo_exc}]{%
       \includegraphics[width=0.49\textwidth]{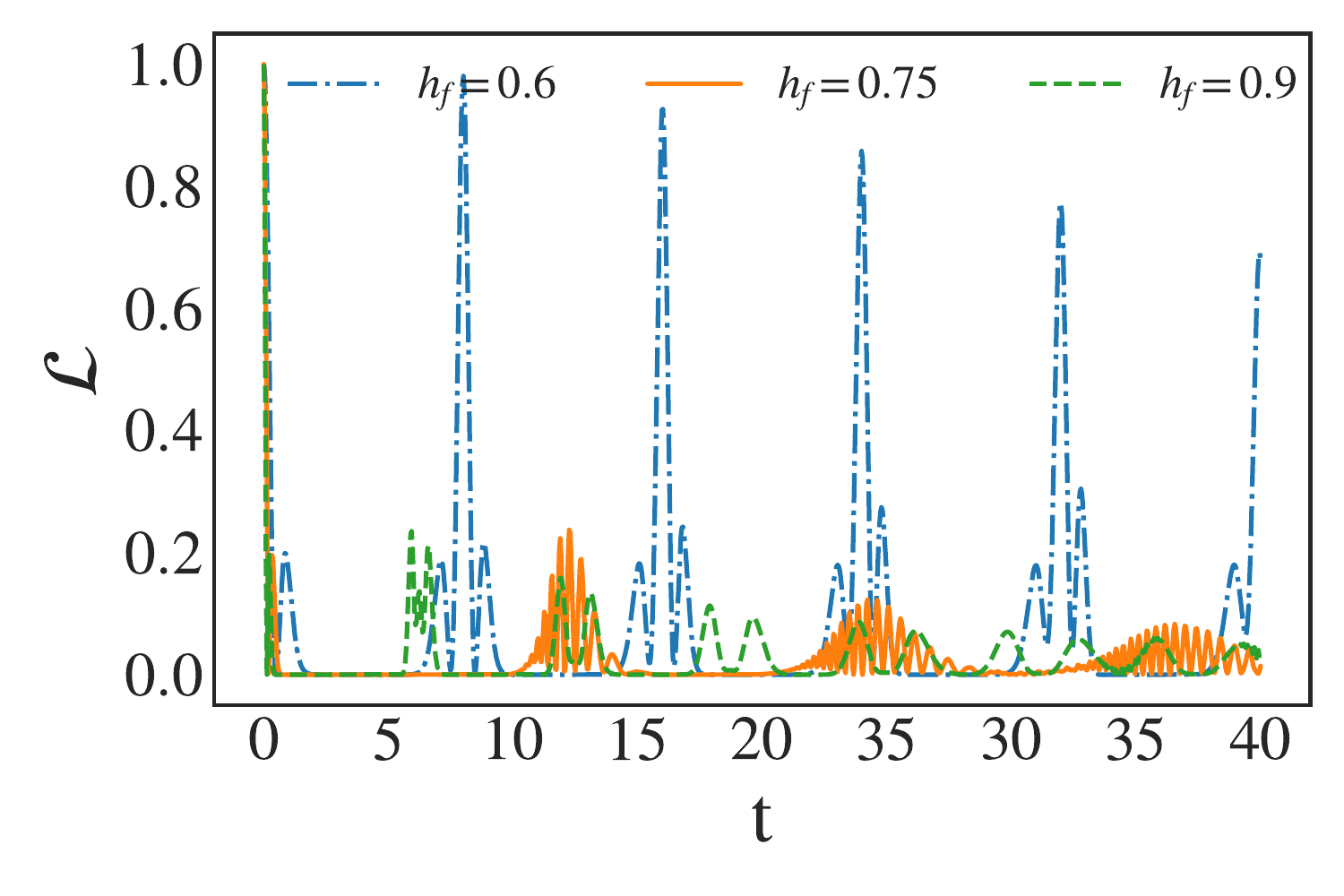}
       }%
     \subfloat[\label{pw_exc}]{%
       \includegraphics[width=0.49\textwidth]{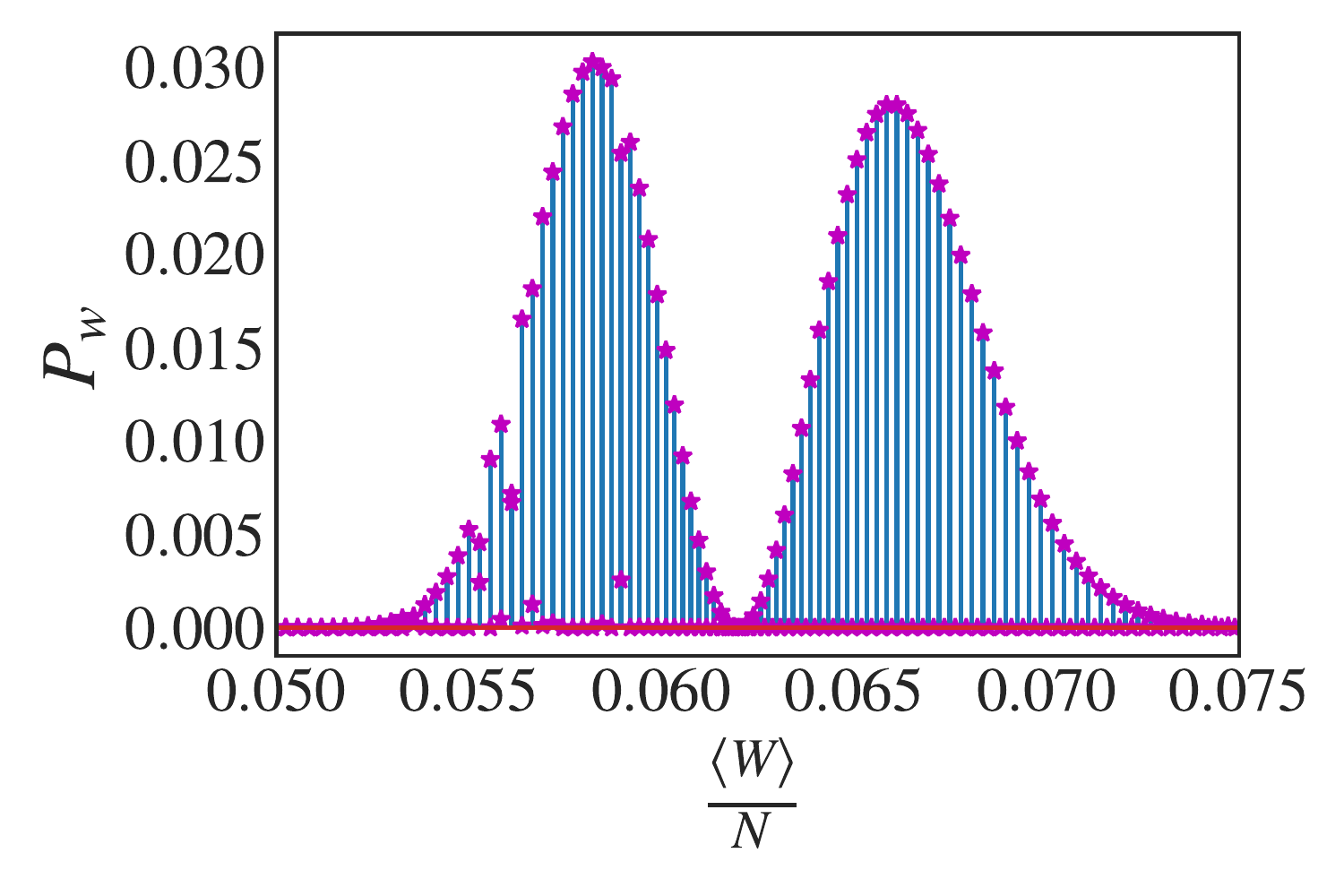}
       }
\caption{{\bf Symmetric excited state.} (a) The survival probability~\eqref{echo}, for a system size $N\!=\!2000$, when the quench is performed from $h_i\!=\! 0.5$ to below ($h_f\!=\!0.6$), above ($h_f\!=\!0.9$) and to the ESQPT critical point ($h_f\!=\!0.75$). (b) The work probability distribution~\eqref{pw}, for a system of size $N\!=\!2000$ quenched from $h_i\!=\! 0.5$ to the ESQPT critical point ($h_f\!=\!0.75$).}
\label{lmg_es}
\end{figure}

An interesting feature of the model is that the double degeneracy occurring in the ferromagnetic phase is not restricted to the ground and first excited states and, in fact, extends to higher excited states up to the critical energy for $0\leq h\leq 1$. These higher excited states exhibit the same critical features and therefore here we examine whether signatures of the ESQPT are also present in the dynamics and work statistics for systems initialized in their excited state conserving the $\mathbb{Z}_2$ parity symmetry. To this end, we fix $h_i\!=\!0.5$, $N\!=\!2000$ and initialize the system in the second excited state. In Figure~\eqref{echo_exc} we show the survival probability when the quench is performed to below ($h_f\!=\!0.6$), above ($h_f\!=\!0.9$) and to the ESQPT critical point $h_f^c\!=\!0.75$. Note that although $h_f^c$ corresponds to the critical value when quenching from the ground state, for low lying excited states it ensures that the quenched state has an energy in the vicinity  of the ESQPT. While the behavior is in keeping with the ground state cases, a remarkable feature emerging is the presence of higher frequencies in the revivals of the survival probability for quenches both below and above the ESQPT point and quenching to the critical energy again ensures the system remains close to orthogonality throughout the dynamics. A consequence of these higher frequencies is directly exhibited in the work probability distribution cf. Figure~\eqref{pw_exc}, where a bi-modal shape emerges. For quenches to the ESQPT point we find that there is still a cusp appearing in the distribution, similarly as for the ground state case. In addition, it is worth commenting that the same phenomenology of symmetry breaking applies to this scenario too.

\section{Quench from the paramagnetic phase}
\begin{figure*}[t]
    \subfloat[\label{echo_para}]{%
       \includegraphics[width=0.33\textwidth]{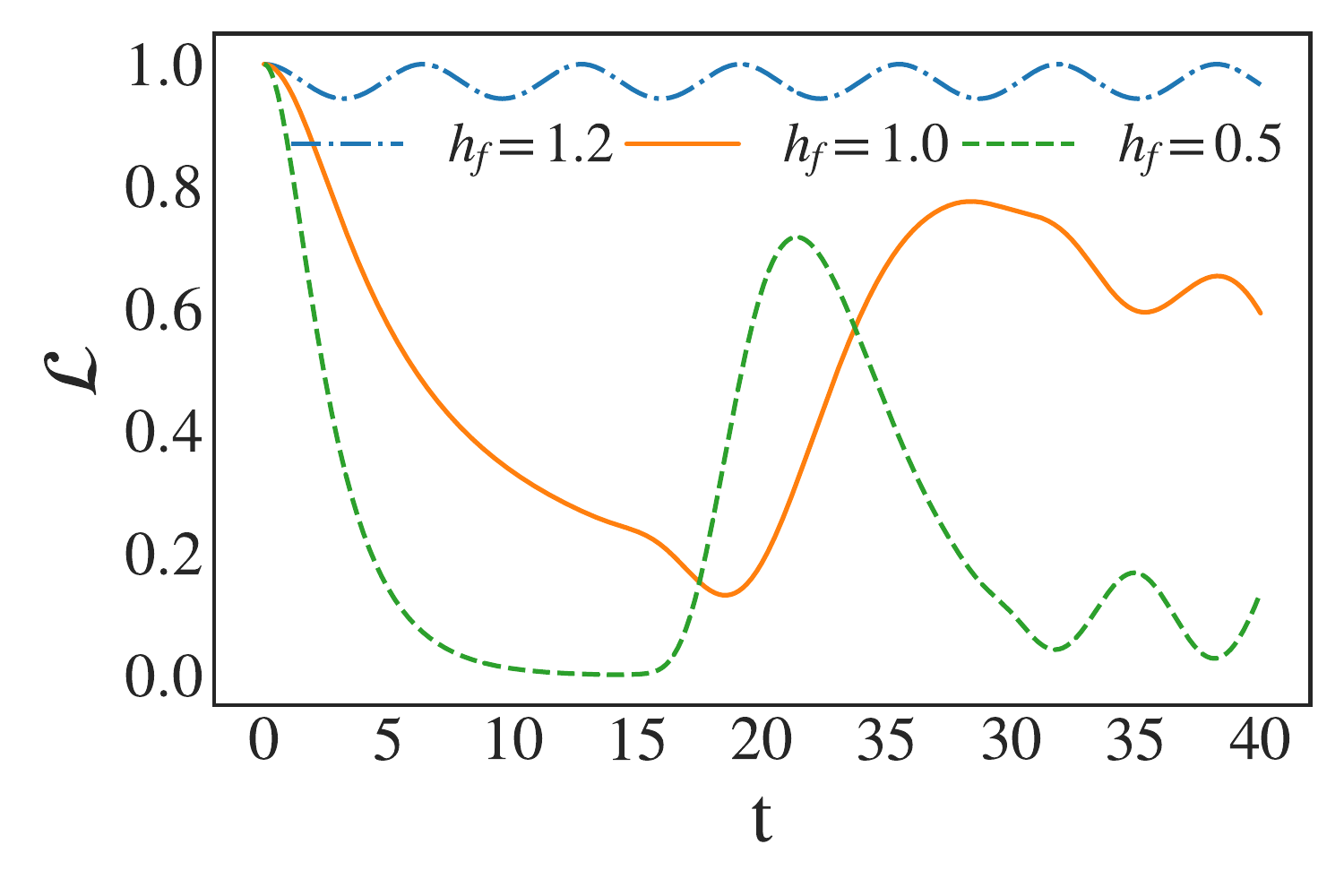}
       }%
     \subfloat[\label{pw_para}]{%
       \includegraphics[width=0.33\textwidth]{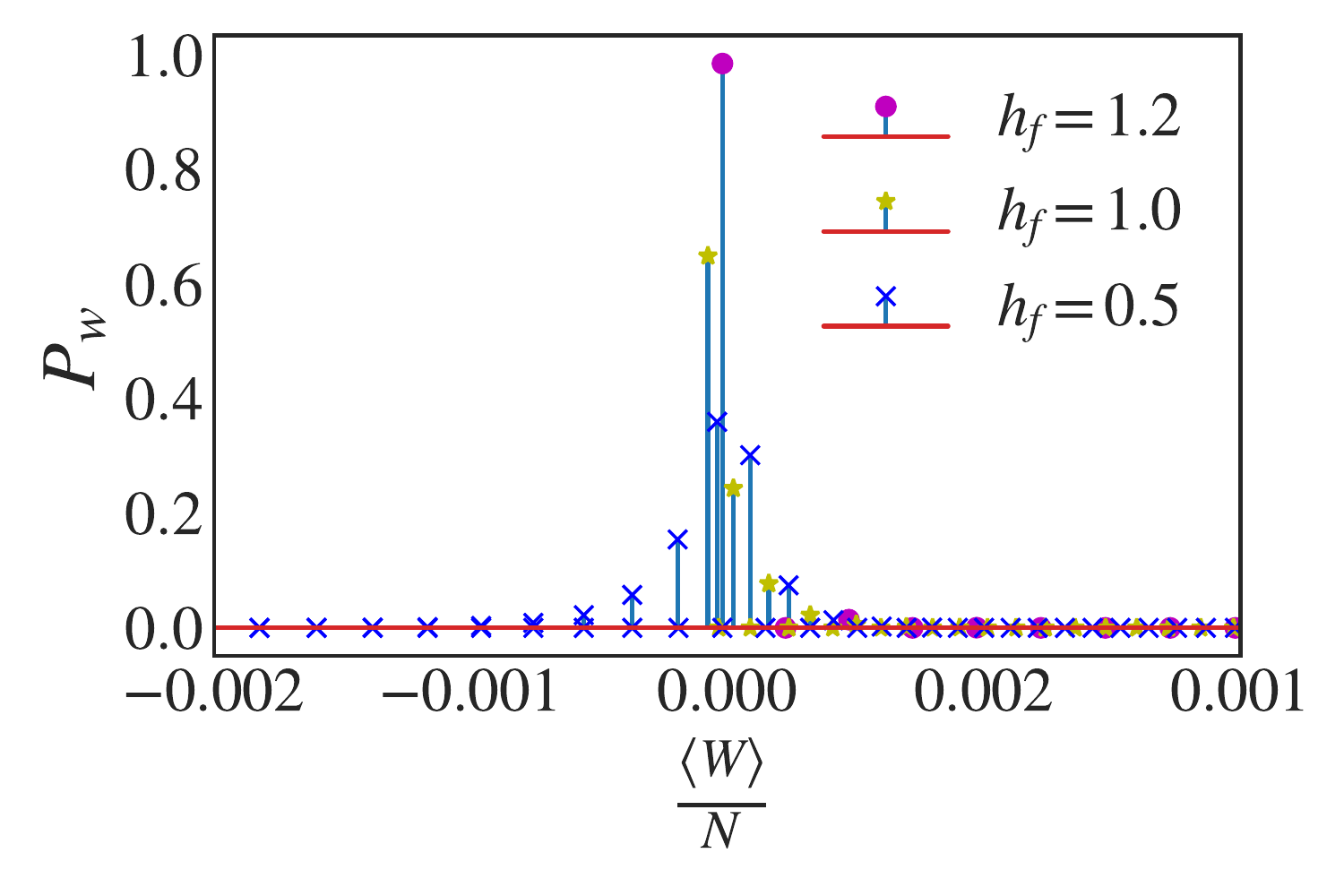}
       }
       \subfloat[\label{sw_para}]{%
       \includegraphics[width=0.33\textwidth]{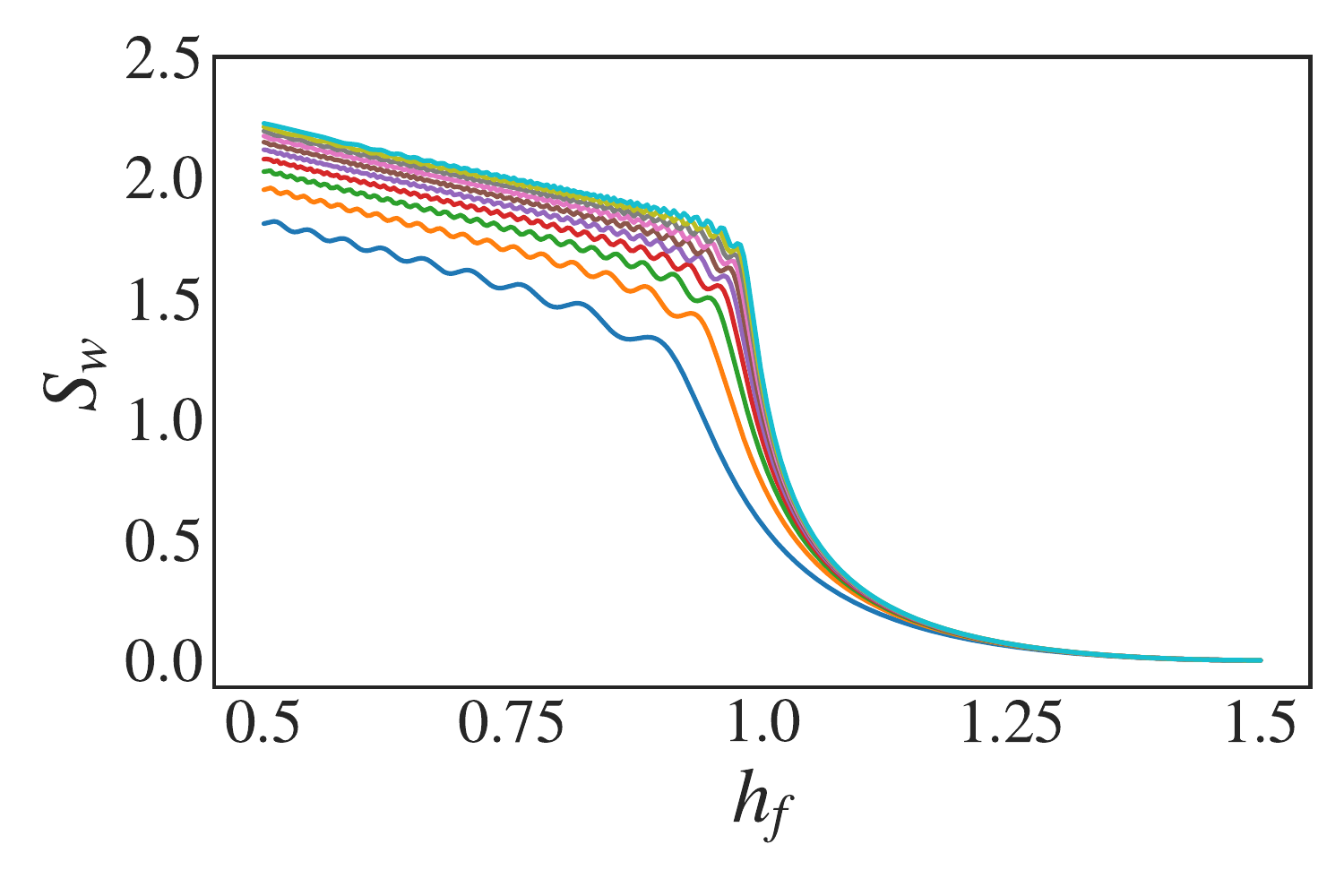}
       }
\caption{{\bf Paramagnetic ground state.} (a)+(b) The survival probability~\eqref{echo}, and the work probability distribution~\eqref{pw}, respectively for a system size $N\!=\!2000$, when the quench is performed to below ($h_f\!=\!1.2$), above ($h_f\!=\!0.5$) and the second order QPT critical point ($h_f\!=\!1.0$). (c) The Shannon entropy~\eqref{entropy} with respect to the magnetic field $h_f$ for various system sizes $N\!=\!100 [\text{bottom,~blue}]\!\to\!1000[\text{top,~cyan}]$.
}
\label{para}
\end{figure*}

For completeness we also consider the case of a system initialized in the paramagnetic phase ($h\!>\!1$). Unlike from the ferromagnetic phase, it is not possible to cross the ESQPT by quenching the ground state with $h_i>1$. To the contrary, the quench will be able to signal the QPT when $h_f=h_c=1$, while for $h_f<1$ the state is brought to the critical energy of the ESQPT. In Figure~\eqref{echo_para} the survival probability is shown for system size $N\!=\!2000$ and a quench starting from $h_i\!=\!1.5$. Constraining the quench to within the same phase, that is $h_f\!=\!1.2$ (blue dot-dashed curve), shows small oscillations with perfect revivals. Conversely, quenching either to the second order QPT $h_f\!=\! h_c\!=\!1$ (orange solid line) or beyond $h_f=0.5$ (dashed green line) initially drives the state far from equilibrium and the dynamics is no longer oscillatory. Furthermore, large quenches beyond the QPT drive the system to orthogonal states~\cite{CampbellPRB, CampbellPRL2020}. The work probability distribution reflects the results found for the survival probability as shown in Figure~\eqref{pw_para}. $P_W$ is dominated by a single value of the work when the quench is confined within the same phase, this is a consequence of the fact that the energy levels in the paramagnetic phase are equidistant. Quenching to the QPT, we see that the distribution is still ruled by one value of the work, however other contributions are starting to emerge. A large quench crossing the QPT results in a broader probability distribution reflecting the irreversible nature of the dynamics when crossing a critical point.

\noindent Turning our attention to the entropy of the diagonal ensemble, Figure~\eqref{sw_para} shows that the entropy is small in the paramagnetic phase $h\!>\!1$, reflecting the fact that the dynamics is reversible and dominated by a single eigenstate. As we approach the QPT, the entropy sharply increases and a cusp appears tending to $h_f\!\to\!h_c=1$ as $N\!\to\!\infty$ thus indicating that the entropy of the diagonal ensemble is a faithful indicator of the ground state QPT in this case. We remark this is in contrast to the case of initial states in the ferromagnetic phase discussed previously where the presence of the ESQPT and its crossing was succinctly captured by the diagonal entropy, regardless of the presence of absence of symmetry breaking, but it was not sensitive to the ground state QPT at $h_f\!=\!1$.
\section{Summary}
\noindent In this chapter we have examined the dual effect of symmetry breaking and excited state quantum phase transitions have on the dynamics of a many-body system. Focusing on the LMG model we have demonstrated that while the average work, and higher moments of the distributions, are indifferent to either the presence of an ESQPT or the effect of symmetry breaking, the distribution itself is acutely sensitive to both. Furthermore, we have established that the entropy of the diagonal ensemble is a favorable figure of merit for pinpointing and studying ESQPTs and the symmetry breaking effects~\cite{2020_wang_arxiv}. The qualitative features exhibited when the system is initialized in the ground state were shown to largely extend to initially excited states, with some notable changes, in particular, the emergence of a bimodal distribution for the work that is nevertheless sensitive to quenches to the ESQPT. Finally, we examined the behavior for quenches that start in the paramagnetic phase, where the only critical features exist in the ground state and demonstrated that the entropy of the diagonal ensemble continues to be a useful tool for spotlighting the underlying critical features of the spectrum.

\intotoc*{\cleardoublepage}{Epilogue}
\chapter*{Epilogue}
\noindent We focused throughout this thesis on studying the equilibrium and non-equilibrium properties of critical quantum systems, using the formalism of quantum information theory and quantum thermodynamics. After an overview on the physics of classical and quantum phase transitions, we shed light on the theory of classical and quantum information. Then, equipped with this background we focused on analyzing the equilibrium properties of critical quantum systems, where we evaluated analytically and numerically the long range entanglement, quantum discord and quantum coherence in the critical Heisenberg $XX$ model, in order to demonstrate the versatility of the different forms of quantum information-theoretic measures in revealing the salient critical properties of quantum systems. Furthermore, we have analyzed the resilience of these measures with respect to thermal fluctuations, where we reported the robustness of quantum coherence and quantum discord, and the weakness of quantum entanglement to temperature.

\noindent To fully explore the possible representations of physical systems in quantum mechanics, we have discussed the phase space formalism for infinite-, and finite-dimensional systems, with special focus on the later where we introduced the Wootters and Stratonovich Wigner functions. Through these measures we have introduced an analytic novel way to represent, detect, and distinguish first-, second- and infinite-order equilibrium quantum phase transitions in the critical quantum spin chains. We have developed a general analytical formula relating the Wootters and Stratonovich Wigner function and the thermodynamic quantities of spin models, which apply to single, bipartite and multipartite systems governed by the $XY$ and the $XXZ$ models. Additionally, we have shown, numerically, that the factorization of the ground state of the $XY$ model is only detectable using quantities based on the square root of the bipartite reduced density matrix. Hence, we established that phase space techniques provide a simple, experimentally promising tool in the study of many-body systems and we showed their relation with measures of entanglement and quantum coherence. 

\noindent Christopher Jarzynski once said: ``If we shift our focus away from equilibrium states, we find a rich universe of non-equilibrium behavior''~\cite{JarNat2015}. Motivated by this statement and equipped with the previous results, we shifted our attention toward addressing the non-equilibrium features of critical quantum systems, using the quantum thermodynamics approach. We provided a birds eye view on the phenomenological theory of classical thermodynamics and its extension to non-equilibrium settings via the stochastic theory of thermodynamics. Furthermore, we discussed how thermodynamics can be used as a resource for state manipulation, how it can be extended to quantum settings, and its interplay with classical and quantum computing. In the end, we provided a study on the work statistics in non-equilibrium critical quantum systems. In particular, we examined how the presence of an excited state quantum phase transition manifests in the dynamics of the Lipkin-Meshkov-Glick spin model subject to a sudden quench. By initializing the system in the ground state of the ferromagnetic phase we showed, numerically, that the work probability distribution displays non-Gaussian behavior for quenches in the vicinity of the excited state critical point. Furthermore, we reported that the entropy of the diagonal ensemble is highly susceptible to critical regions, making it a robust and practical indicator of the associated spectral characteristics. In the end, we evaluated the role that symmetry breaking has on the ensuing dynamics, highlighting that its effect is only present for quenches beyond the critical point; and we argued that similar features persist when the system is initialized in an excited state, while briefly exploring the behavior for initial states in the paramagnetic phase.

\noindent An interesting line of research is the exploitation of the features of quantum systems in order to design efficient quantum technologies. Indeed, entanglement and quantum coherence have been used in order to process information efficiently and communicate safely. Recently, there has been an upsurge in extending thermodynamics to the quantum regime in order to design
quantum-based thermal machines to overcome some limitations of their classical counterparts. One way to achieve this is by modeling the working fluid by a critical quantum system, and driving it across the critical point. At the level of equilibrium quantum phase transitions, it has been shown that crossing a second-order quantum phase transition can enhance the efficiency of quantum heat engines. Here we are interested in showing the impact of the excited state quantum phase transition on the efficiency of quantum heat engines. To achieve the goal, we will model the working medium by an LMG model, driven across the excited state quantum phase transition through different protocols, e.g. adiabatic and non-adiabatic driving.

\noindent Another exciting and promising research field is quantum computing, which is the most prominent application of quantum technologies, and is believed to offer a computational advantage by being able to solve certain computational problems exponentially faster than classical computers. An important aspect in building new technology is the resource consumption, i.e its energy efficiency, which in the case of quantum computing is lacking consensus. The theoretical framework to address this issue is through quantum thermodynamics which provides the necessary tools to quantify and characterize the efficiency of emerging quantum technologies, and therefore is crucial in laying a road map to scalable devices. Following the recent push toward a quantum energy initiative~\cite{Auffeves:2021paz}, we are interested in studying the energetic cost of quantum computing by investigating the thermodynamics of quantum computing paradigms, such as: quantum annealers. These architectures are a promising quantum algorithm for solving optimization tasks, it consist of mapping the optimization problem onto finding the ground state of the Ising spin glass. Therefore, we are interested in investigating the dynamics and thermodynamics of the Ising spin glass, in order to evaluate how work, entropy and information behave upon external perturbation, to locate regions of minimal and high efficiency.


\cleardoublepage

\backmatter
\intotoc*{\cleardoublepage}{\bibname}
\bibliography{Biblio/ref}

\begin{thebibliography}{250}
\expandafter\ifx\csname natexlab\endcsname\relax\def\natexlab#1{#1}\fi
\providecommand{\url}[1]{\texttt{#1}}
\providecommand{\href}[2]{#2}
\providecommand{\path}[1]{#1}
\providecommand{\DOIprefix}{doi:}
\providecommand{\ArXivprefix}{arXiv:}
\providecommand{\URLprefix}{URL: }
\providecommand{\Pubmedprefix}{pmid:}
\providecommand{\doi}[1]{\href{http://dx.doi.org/#1}{\path{#1}}}
\providecommand{\Pubmed}[1]{\href{pmid:#1}{\path{#1}}}
\providecommand{\bibinfo}[2]{#2}
\ifx\xfnm\relax \def\xfnm[#1]{\unskip,\space#1}\fi
\bibitem[{Harari(2014)}]{harari2014sapiens}
\bibinfo{author}{Y.~N. Harari}, \bibinfo{title}{Sapiens: A brief history of
  humankind}, \bibinfo{publisher}{Random House}, \bibinfo{year}{2014}.
\bibitem[{Chappin(2012)}]{chappin2012review}
\bibinfo{author}{E.~Chappin},
\newblock \bibinfo{title}{Review of phase transitions (primers in complex
  systems)}  (\bibinfo{year}{2012}).
\bibitem[{Mathis et~al.(2017)Mathis, Bhattacharya, and
  Walker}]{mathis2017emergence}
\bibinfo{author}{C.~Mathis}, \bibinfo{author}{T.~Bhattacharya},
  \bibinfo{author}{S.~I. Walker},
\newblock \bibinfo{title}{The emergence of life as a first-order phase
  transition},
\newblock \bibinfo{journal}{Astrobiology} \bibinfo{volume}{17}
  (\bibinfo{year}{2017}) \bibinfo{pages}{266--276}.
\bibitem[{Goldenfeld(2018)}]{goldenfeld2018lectures}
\bibinfo{author}{N.~Goldenfeld}, \bibinfo{title}{Lectures on phase transitions
  and the renormalization group}, \bibinfo{publisher}{CRC Press},
  \bibinfo{year}{2018}.
\bibitem[{Sachdev(2011)}]{sachdev2011quantum}
\bibinfo{author}{S.~Sachdev}, \bibinfo{title}{Quantum phase transitions},
  \bibinfo{publisher}{Cambridge university press}, \bibinfo{year}{2011}.
\bibitem[{Vojta(2003)}]{vojta2003quantum}
\bibinfo{author}{M.~Vojta},
\newblock \bibinfo{title}{Quantum phase transitions},
\newblock \bibinfo{journal}{Reports on Progress in Physics}
  \bibinfo{volume}{66} (\bibinfo{year}{2003}) \bibinfo{pages}{2069}.
\bibitem[{Curie(1895)}]{curie1895proprietes}
\bibinfo{author}{P.~Curie}, \bibinfo{title}{Propri{\'e}t{\'e}s magn{\'e}tiques
  des corps a diverses temp{\'e}ratures}, \bibinfo{number}{4},
  \bibinfo{publisher}{Gauthier-Villars et fils}, \bibinfo{year}{1895}.
\bibitem[{Kittel(1949)}]{kittel1949physical}
\bibinfo{author}{C.~Kittel},
\newblock \bibinfo{title}{Physical theory of ferromagnetic domains},
\newblock \bibinfo{journal}{Reviews of modern Physics} \bibinfo{volume}{21}
  (\bibinfo{year}{1949}) \bibinfo{pages}{541}.
\bibitem[{Onnes(1913)}]{onnes1913investigations}
\bibinfo{author}{H.~K. Onnes},
\newblock \bibinfo{title}{Investigations into the properties of substances at
  low temperatures, which have led, amongst other things, to the preparation of
  liquid helium},
\newblock \bibinfo{journal}{Nobel lecture} \bibinfo{volume}{4}
  (\bibinfo{year}{1913}) \bibinfo{pages}{306--336}.
\bibitem[{Ising(1925)}]{Ising:1925em}
\bibinfo{author}{E.~Ising},
\newblock \bibinfo{title}{{Contribution to the Theory of Ferromagnetism}},
\newblock \bibinfo{journal}{Z. Phys.} \bibinfo{volume}{31}
  (\bibinfo{year}{1925}) \bibinfo{pages}{253--258}.
\bibitem[{Helmut~Kronmüller(2007)}]{magnetismhandbook}
\bibinfo{author}{S.~P. Helmut~Kronmüller}, \bibinfo{title}{Handbook of
  magnetism and advanced magnetic materials}, \bibinfo{edition}{1} ed.,
  \bibinfo{publisher}{Wiley-Interscience}, \bibinfo{year}{2007}.
\bibitem[{K.H.J~Buschow(2003)}]{physicsmagnetism}
\bibinfo{author}{F.~d.~B. K.H.J~Buschow}, \bibinfo{title}{Physics of Magnetism
  and Magnetic Materials}, \bibinfo{edition}{1} ed.,
  \bibinfo{publisher}{Springer}, \bibinfo{year}{2003}.
\bibitem[{Cohen-Tannoudji et~al.(1986{\natexlab{a}})Cohen-Tannoudji, Diu, and
  Laloe}]{cohen1986quantum}
\bibinfo{author}{C.~Cohen-Tannoudji}, \bibinfo{author}{B.~Diu},
  \bibinfo{author}{F.~Laloe},
\newblock \bibinfo{title}{Quantum mechanics, volume 1},
\newblock \bibinfo{journal}{Quantum Mechanics} \bibinfo{volume}{1}
  (\bibinfo{year}{1986}{\natexlab{a}}) \bibinfo{pages}{898}.
\bibitem[{Cohen-Tannoudji et~al.(1986{\natexlab{b}})Cohen-Tannoudji, Diu, and
  Laloe}]{cohen1986quantum2}
\bibinfo{author}{C.~Cohen-Tannoudji}, \bibinfo{author}{B.~Diu},
  \bibinfo{author}{F.~Laloe},
\newblock \bibinfo{title}{Quantum mechanics, volume 2},
\newblock \bibinfo{journal}{Quantum Mechanics} \bibinfo{volume}{2}
  (\bibinfo{year}{1986}{\natexlab{b}}) \bibinfo{pages}{626}.
\bibitem[{Cohen-Tannoudji et~al.(2019)Cohen-Tannoudji, Diu, and
  Lalo{\"e}}]{cohen2019quantum3}
\bibinfo{author}{C.~Cohen-Tannoudji}, \bibinfo{author}{B.~Diu},
  \bibinfo{author}{F.~Lalo{\"e}}, \bibinfo{title}{Quantum Mechanics, Volume 3:
  Fermions, Bosons, Photons, Correlations, and Entanglement},
  \bibinfo{publisher}{John Wiley \& Sons}, \bibinfo{year}{2019}.
\bibitem[{Bloch(1928)}]{bloch1928quantum}
\bibinfo{author}{F.~Bloch},
\newblock \bibinfo{title}{Quantum mechanics of electrons in crystal lattices},
\newblock \bibinfo{journal}{Z. Phys} \bibinfo{volume}{52}
  (\bibinfo{year}{1928}) \bibinfo{pages}{555--600}.
\bibitem[{Dirac(1928)}]{dirac1928quantum}
\bibinfo{author}{P.~A.~M. Dirac},
\newblock \bibinfo{title}{The quantum theory of the electron},
\newblock \bibinfo{journal}{Proceedings of the Royal Society of London. Series
  A, Containing Papers of a Mathematical and Physical Character}
  \bibinfo{volume}{117} (\bibinfo{year}{1928}) \bibinfo{pages}{610--624}.
\bibitem[{{N\'eel, M. Louis}(1948)}]{neel}
\bibinfo{author}{{N\'eel, M. Louis}},
\newblock \bibinfo{title}{Propri\'et\'es magn\'etiques des ferrites ;
  ferrimagn\'etisme et antiferromagn\'etisme},
\newblock \bibinfo{journal}{Ann. Phys.} \bibinfo{volume}{12}
  (\bibinfo{year}{1948}) \bibinfo{pages}{137--198}.
\bibitem[{Poole(2007)}]{diamagnetism}
\bibinfo{author}{C.~Poole}, \bibinfo{title}{Superconductivity},
  \bibinfo{edition}{2} ed., \bibinfo{publisher}{Academic Press},
  \bibinfo{year}{2007}.
\bibitem[{Landau(1936)}]{landau1936theory}
\bibinfo{author}{L.~Landau},
\newblock \bibinfo{title}{The theory of phase transitions},
\newblock \bibinfo{journal}{Nature} \bibinfo{volume}{138}
  (\bibinfo{year}{1936}) \bibinfo{pages}{840--841}.
\bibitem[{Ginzburg and Landau(2009)}]{landausuperc}
\bibinfo{author}{V.~L. Ginzburg}, \bibinfo{author}{L.~D. Landau},
\newblock \bibinfo{title}{On the theory of superconductivity},
\newblock in: \bibinfo{booktitle}{On superconductivity and superfluidity},
  \bibinfo{publisher}{Springer}, \bibinfo{year}{2009}, pp.
  \bibinfo{pages}{113--137}.
\bibitem[{Landau(2018)}]{landausuperf}
\bibinfo{author}{L.~Landau},
\newblock \bibinfo{title}{The theory of superfluidity of helium ii},
\newblock in: \bibinfo{booktitle}{An Introduction to the Theory of
  Superfluidity}, \bibinfo{publisher}{CRC Press}, \bibinfo{year}{2018}, pp.
  \bibinfo{pages}{185--204}.
\bibitem[{Cooper(1956)}]{bcs1}
\bibinfo{author}{L.~N. Cooper},
\newblock \bibinfo{title}{Bound electron pairs in a degenerate fermi gas},
\newblock \bibinfo{journal}{Phys. Rev.} \bibinfo{volume}{104}
  (\bibinfo{year}{1956}) \bibinfo{pages}{1189--1190}.
\bibitem[{Bardeen et~al.(1957{\natexlab{a}})Bardeen, Cooper, and
  Schrieffer}]{bcs2}
\bibinfo{author}{J.~Bardeen}, \bibinfo{author}{L.~N. Cooper},
  \bibinfo{author}{J.~R. Schrieffer},
\newblock \bibinfo{title}{Microscopic theory of superconductivity},
\newblock \bibinfo{journal}{Phys. Rev.} \bibinfo{volume}{106}
  (\bibinfo{year}{1957}{\natexlab{a}}) \bibinfo{pages}{162--164}.
\bibitem[{Bardeen et~al.(1957{\natexlab{b}})Bardeen, Cooper, and
  Schrieffer}]{bcs3}
\bibinfo{author}{J.~Bardeen}, \bibinfo{author}{L.~N. Cooper},
  \bibinfo{author}{J.~R. Schrieffer},
\newblock \bibinfo{title}{Theory of superconductivity},
\newblock \bibinfo{journal}{Phys. Rev.} \bibinfo{volume}{108}
  (\bibinfo{year}{1957}{\natexlab{b}}) \bibinfo{pages}{1175--1204}.
\bibitem[{Anderson(1958)}]{andresonlocal}
\bibinfo{author}{P.~W. Anderson},
\newblock \bibinfo{title}{Absence of diffusion in certain random lattices},
\newblock \bibinfo{journal}{Phys. Rev.} \bibinfo{volume}{109}
  (\bibinfo{year}{1958}) \bibinfo{pages}{1492--1505}.
\bibitem[{Wilson and Kogut(1974)}]{wilson1974renormalization}
\bibinfo{author}{K.~G. Wilson}, \bibinfo{author}{J.~Kogut},
\newblock \bibinfo{title}{The renormalization group and the $\epsilon$
  expansion},
\newblock \bibinfo{journal}{Physics reports} \bibinfo{volume}{12}
  (\bibinfo{year}{1974}) \bibinfo{pages}{75--199}.
\bibitem[{Wilson(1983)}]{wilson1983}
\bibinfo{author}{K.~G. Wilson},
\newblock \bibinfo{title}{The renormalization group and critical phenomena},
\newblock \bibinfo{journal}{Rev. Mod. Phys.} \bibinfo{volume}{55}
  (\bibinfo{year}{1983}) \bibinfo{pages}{583--600}.
\bibitem[{Bennett and Shor(1998)}]{bennett1998quantum}
\bibinfo{author}{C.~H. Bennett}, \bibinfo{author}{P.~W. Shor},
\newblock \bibinfo{title}{Quantum information theory},
\newblock \bibinfo{journal}{IEEE transactions on information theory}
  \bibinfo{volume}{44} (\bibinfo{year}{1998}) \bibinfo{pages}{2724--2742}.
\bibitem[{Wilde(2013)}]{wilde2013quantum}
\bibinfo{author}{M.~M. Wilde}, \bibinfo{title}{Quantum information theory},
  \bibinfo{publisher}{Cambridge University Press}, \bibinfo{year}{2013}.
\bibitem[{Preskill(2000)}]{preskill2000quantum}
\bibinfo{author}{J.~Preskill},
\newblock \bibinfo{title}{Quantum information and physics: some future
  directions},
\newblock \bibinfo{journal}{Journal of Modern Optics} \bibinfo{volume}{47}
  (\bibinfo{year}{2000}) \bibinfo{pages}{127--137}.
\bibitem[{Nielsen and Chuang(2002)}]{nielsen2002quantum}
\bibinfo{author}{M.~A. Nielsen}, \bibinfo{author}{I.~Chuang},
  \bibinfo{title}{Quantum computation and quantum information},
  \bibinfo{year}{2002}.
\bibitem[{Petz(2007)}]{petz2007quantum}
\bibinfo{author}{D.~Petz}, \bibinfo{title}{Quantum information theory and
  quantum statistics}, \bibinfo{publisher}{Springer Science \& Business Media},
  \bibinfo{year}{2007}.
\bibitem[{Amico et~al.(2008)Amico, Fazio, Osterloh, and Vedral}]{fazioreview}
\bibinfo{author}{L.~Amico}, \bibinfo{author}{R.~Fazio},
  \bibinfo{author}{A.~Osterloh}, \bibinfo{author}{V.~Vedral},
\newblock \bibinfo{title}{Entanglement in many-body systems},
\newblock \bibinfo{journal}{Rev. Mod. Phys.} \bibinfo{volume}{80}
  (\bibinfo{year}{2008}) \bibinfo{pages}{517--576}.
\bibitem[{Dutta et~al.(2015)Dutta, Aeppli, Chakrabarti, Divakaran, Rosenbaum,
  and Sen}]{dutta2015quantum}
\bibinfo{author}{A.~Dutta}, \bibinfo{author}{G.~Aeppli}, \bibinfo{author}{B.~K.
  Chakrabarti}, \bibinfo{author}{U.~Divakaran}, \bibinfo{author}{T.~F.
  Rosenbaum}, \bibinfo{author}{D.~Sen}, \bibinfo{title}{Quantum phase
  transitions in transverse field spin models: from statistical physics to
  quantum information}, \bibinfo{publisher}{Cambridge University Press},
  \bibinfo{year}{2015}.
\bibitem[{Osborne and Nielsen(2002)}]{niel_osb}
\bibinfo{author}{T.~J. Osborne}, \bibinfo{author}{M.~A. Nielsen},
\newblock \bibinfo{title}{Entanglement in a simple quantum phase transition},
\newblock \bibinfo{journal}{Phys. Rev. A} \bibinfo{volume}{66}
  (\bibinfo{year}{2002}) \bibinfo{pages}{032110}.
\bibitem[{Osterloh et~al.(2002)Osterloh, Amico, Falci, and Fazio}]{rozario}
\bibinfo{author}{A.~Osterloh}, \bibinfo{author}{L.~Amico},
  \bibinfo{author}{G.~Falci}, \bibinfo{author}{R.~Fazio},
\newblock \bibinfo{title}{Scaling of entanglement close to a quantum phase
  transition},
\newblock \bibinfo{journal}{Nature} \bibinfo{volume}{416}
  (\bibinfo{year}{2002}) \bibinfo{pages}{608 EP}.
\bibitem[{Wu et~al.(2004)Wu, Sarandy, and Lidar}]{QPT2004}
\bibinfo{author}{L.-A. Wu}, \bibinfo{author}{M.~S. Sarandy},
  \bibinfo{author}{D.~A. Lidar},
\newblock \bibinfo{title}{Quantum phase transitions and bipartite
  entanglement},
\newblock \bibinfo{journal}{Phys. Rev. Lett.} \bibinfo{volume}{93}
  (\bibinfo{year}{2004}) \bibinfo{pages}{250404}.
\bibitem[{Dillenschneider(2008)}]{qptdiscord}
\bibinfo{author}{R.~Dillenschneider},
\newblock \bibinfo{title}{Quantum discord and quantum phase transition in spin
  chains},
\newblock \bibinfo{journal}{Phys. Rev. B} \bibinfo{volume}{78}
  (\bibinfo{year}{2008}) \bibinfo{pages}{224413}.
\bibitem[{Campbell et~al.(2013)Campbell, Richens, Lo~Gullo, and
  Busch}]{CampbellPRA2013}
\bibinfo{author}{S.~Campbell}, \bibinfo{author}{J.~Richens},
  \bibinfo{author}{N.~Lo~Gullo}, \bibinfo{author}{T.~Busch},
\newblock \bibinfo{title}{Criticality, factorization, and long-range
  correlations in the anisotropic $xy$ model},
\newblock \bibinfo{journal}{Phys. Rev. A} \bibinfo{volume}{88}
  (\bibinfo{year}{2013}) \bibinfo{pages}{062305}.
\bibitem[{Irons et~al.(2017)Irons, Quintanilla, Perring, Amico, and
  Aeppli}]{AmicoPRB}
\bibinfo{author}{H.~R. Irons}, \bibinfo{author}{J.~Quintanilla},
  \bibinfo{author}{T.~G. Perring}, \bibinfo{author}{L.~Amico},
  \bibinfo{author}{G.~Aeppli},
\newblock \bibinfo{title}{Control of entanglement transitions in quantum spin
  clusters},
\newblock \bibinfo{journal}{Phys. Rev. B} \bibinfo{volume}{96}
  (\bibinfo{year}{2017}) \bibinfo{pages}{224408}.
\bibitem[{Sarandy(2009)}]{SarandyPRA2009}
\bibinfo{author}{M.~S. Sarandy},
\newblock \bibinfo{title}{Classical correlation and quantum discord in critical
  systems},
\newblock \bibinfo{journal}{Phys. Rev. A} \bibinfo{volume}{80}
  (\bibinfo{year}{2009}) \bibinfo{pages}{022108}.
\bibitem[{Werlang et~al.(2010)Werlang, Trippe, Ribeiro, and
  Rigolin}]{Werlang2010}
\bibinfo{author}{T.~Werlang}, \bibinfo{author}{C.~Trippe},
  \bibinfo{author}{G.~A.~P. Ribeiro}, \bibinfo{author}{G.~Rigolin},
\newblock \bibinfo{title}{Quantum correlations in spin chains at finite
  temperatures and quantum phase transitions},
\newblock \bibinfo{journal}{Phys. Rev. Lett.} \bibinfo{volume}{105}
  (\bibinfo{year}{2010}) \bibinfo{pages}{095702}.
\bibitem[{Karpat et~al.(2014)Karpat, \ifmmode~\mbox{\c{C}}\else
  \c{C}\fi{}akmak, and Fanchini}]{CakmakPRB2014}
\bibinfo{author}{G.~Karpat}, \bibinfo{author}{B.~\ifmmode~\mbox{\c{C}}\else
  \c{C}\fi{}akmak}, \bibinfo{author}{F.~F. Fanchini},
\newblock \bibinfo{title}{Quantum coherence and uncertainty in the anisotropic
  xy chain},
\newblock \bibinfo{journal}{Phys. Rev. B} \bibinfo{volume}{90}
  (\bibinfo{year}{2014}) \bibinfo{pages}{104431}.
\bibitem[{Li and Lin(2016)}]{qptcoherence}
\bibinfo{author}{Y.-C. Li}, \bibinfo{author}{H.-Q. Lin},
\newblock \bibinfo{title}{Quantum coherence and quantum phase transitions},
\newblock \bibinfo{journal}{Sci. Rep.} \bibinfo{volume}{6}
  (\bibinfo{year}{2016}) \bibinfo{pages}{26365}.
\bibitem[{Lorenzo et~al.(2017)Lorenzo, Marino, Plastina, Palma, and
  Apollaro}]{TonySciRep}
\bibinfo{author}{S.~Lorenzo}, \bibinfo{author}{J.~Marino},
  \bibinfo{author}{F.~Plastina}, \bibinfo{author}{G.~M. Palma},
  \bibinfo{author}{T.~J.~G. Apollaro},
\newblock \bibinfo{title}{Quantum critical scaling under periodic driving},
\newblock \bibinfo{journal}{Sci. Rep.} \bibinfo{volume}{7}
  (\bibinfo{year}{2017}) \bibinfo{pages}{5672}.
\bibitem[{Power et~al.(2015)Power, Campbell, Moreno-Cardoner, and
  De~Chiara}]{CampbellPRB2015}
\bibinfo{author}{M.~J.~M. Power}, \bibinfo{author}{S.~Campbell},
  \bibinfo{author}{M.~Moreno-Cardoner}, \bibinfo{author}{G.~De~Chiara},
\newblock \bibinfo{title}{Nonclassicality and criticality in symmetry-protected
  magnetic phases},
\newblock \bibinfo{journal}{Phys. Rev. B} \bibinfo{volume}{91}
  (\bibinfo{year}{2015}) \bibinfo{pages}{214411}.
\bibitem[{Malvezzi et~al.(2016)Malvezzi, Karpat, \ifmmode~\mbox{\c{C}}\else
  \c{C}\fi{}akmak, Fanchini, Debarba, and Vianna}]{CakmakPRB2016}
\bibinfo{author}{A.~L. Malvezzi}, \bibinfo{author}{G.~Karpat},
  \bibinfo{author}{B.~\ifmmode~\mbox{\c{C}}\else \c{C}\fi{}akmak},
  \bibinfo{author}{F.~F. Fanchini}, \bibinfo{author}{T.~Debarba},
  \bibinfo{author}{R.~O. Vianna},
\newblock \bibinfo{title}{Quantum correlations and coherence in spin-1
  heisenberg chains},
\newblock \bibinfo{journal}{Phys. Rev. B} \bibinfo{volume}{93}
  (\bibinfo{year}{2016}) \bibinfo{pages}{184428}.
\bibitem[{Stasi\ifmmode~\acute{n}\else \'{n}\fi{}ska
  et~al.(2014)Stasi\ifmmode~\acute{n}\else \'{n}\fi{}ska, Rogers, Paternostro,
  De~Chiara, and Sanpera}]{RogersPRA2014}
\bibinfo{author}{J.~Stasi\ifmmode~\acute{n}\else \'{n}\fi{}ska},
  \bibinfo{author}{B.~Rogers}, \bibinfo{author}{M.~Paternostro},
  \bibinfo{author}{G.~De~Chiara}, \bibinfo{author}{A.~Sanpera},
\newblock \bibinfo{title}{Long-range multipartite entanglement close to a
  first-order quantum phase transition},
\newblock \bibinfo{journal}{Phys. Rev. A} \bibinfo{volume}{89}
  (\bibinfo{year}{2014}) \bibinfo{pages}{032330}.
\bibitem[{Hofmann et~al.(2014)Hofmann, Osterloh, and G\"uhne}]{HofmannPRB}
\bibinfo{author}{M.~Hofmann}, \bibinfo{author}{A.~Osterloh},
  \bibinfo{author}{O.~G\"uhne},
\newblock \bibinfo{title}{Scaling of genuine multiparticle entanglement close
  to a quantum phase transition},
\newblock \bibinfo{journal}{Phys. Rev. B} \bibinfo{volume}{89}
  (\bibinfo{year}{2014}) \bibinfo{pages}{134101}.
\bibitem[{Giampaolo and Hiesmayr(2013)}]{GiampaoloPRA}
\bibinfo{author}{S.~M. Giampaolo}, \bibinfo{author}{B.~C. Hiesmayr},
\newblock \bibinfo{title}{Genuine multipartite entanglement in the $xy$ model},
\newblock \bibinfo{journal}{Phys. Rev. A} \bibinfo{volume}{88}
  (\bibinfo{year}{2013}) \bibinfo{pages}{052305}.
\bibitem[{Campbell et~al.(2013)Campbell, Mazzola, Chiara, Apollaro, Plastina,
  Busch, and Paternostro}]{NJPCampbell}
\bibinfo{author}{S.~Campbell}, \bibinfo{author}{L.~Mazzola},
  \bibinfo{author}{G.~D. Chiara}, \bibinfo{author}{T.~J.~G. Apollaro},
  \bibinfo{author}{F.~Plastina}, \bibinfo{author}{T.~Busch},
  \bibinfo{author}{M.~Paternostro},
\newblock \bibinfo{title}{Global quantum correlations in finite-size spin
  chains},
\newblock \bibinfo{journal}{New J. Phys.} \bibinfo{volume}{15}
  (\bibinfo{year}{2013}) \bibinfo{pages}{043033}.
\bibitem[{Bayat(2017)}]{BayatPRL2017}
\bibinfo{author}{A.~Bayat},
\newblock \bibinfo{title}{Scaling of tripartite entanglement at impurity
  quantum phase transitions},
\newblock \bibinfo{journal}{Phys. Rev. Lett.} \bibinfo{volume}{118}
  (\bibinfo{year}{2017}) \bibinfo{pages}{036102}.
\bibitem[{Justino and de~Oliveira(2012)}]{BellIneqPRA2012}
\bibinfo{author}{L.~Justino}, \bibinfo{author}{T.~R. de~Oliveira},
\newblock \bibinfo{title}{Bell inequalities and entanglement at quantum phase
  transitions in the $xxz$ model},
\newblock \bibinfo{journal}{Phys. Rev. A} \bibinfo{volume}{85}
  (\bibinfo{year}{2012}) \bibinfo{pages}{052128}.
\bibitem[{Mahdavifar et~al.(2017)Mahdavifar, Mahdavifar, and
  Jafari}]{JafariPRA2017}
\bibinfo{author}{S.~Mahdavifar}, \bibinfo{author}{S.~Mahdavifar},
  \bibinfo{author}{R.~Jafari},
\newblock \bibinfo{title}{Magnetic quantum correlations in the one-dimensional
  transverse-field $xxz$ model},
\newblock \bibinfo{journal}{Phys. Rev. A} \bibinfo{volume}{96}
  (\bibinfo{year}{2017}) \bibinfo{pages}{052303}.
\bibitem[{Kargarian et~al.(2008)Kargarian, Jafari, and Langari}]{JafariPRA2008}
\bibinfo{author}{M.~Kargarian}, \bibinfo{author}{R.~Jafari},
  \bibinfo{author}{A.~Langari},
\newblock \bibinfo{title}{Renormalization of entanglement in the anisotropic
  heisenberg $(xxz)$ model},
\newblock \bibinfo{journal}{Phys. Rev. A} \bibinfo{volume}{77}
  (\bibinfo{year}{2008}) \bibinfo{pages}{032346}.
\bibitem[{Rulli and Sarandy(2010)}]{RulliPRA2010}
\bibinfo{author}{C.~C. Rulli}, \bibinfo{author}{M.~S. Sarandy},
\newblock \bibinfo{title}{Entanglement and local extremes at an infinite-order
  quantum phase transition},
\newblock \bibinfo{journal}{Phys. Rev. A} \bibinfo{volume}{81}
  (\bibinfo{year}{2010}) \bibinfo{pages}{032334}.
\bibitem[{Mzaouali and El~Baz(2019)}]{Zakaria2019}
\bibinfo{author}{Z.~Mzaouali}, \bibinfo{author}{M.~El~Baz},
\newblock \bibinfo{title}{Long range quantum coherence, quantum \& classical
  correlations in heisenberg xx chain},
\newblock \bibinfo{journal}{Physica A} \bibinfo{volume}{518}
  (\bibinfo{year}{2019}) \bibinfo{pages}{119}.
\bibitem[{Mzaouali et~al.(2019)Mzaouali, Campbell, and El~Baz}]{mzaouali2019}
\bibinfo{author}{Z.~Mzaouali}, \bibinfo{author}{S.~Campbell},
  \bibinfo{author}{M.~El~Baz},
\newblock \bibinfo{title}{Discrete and generalized phase space techniques in
  critical quantum spin chains},
\newblock \bibinfo{journal}{Physics Letters A} \bibinfo{volume}{383}
  (\bibinfo{year}{2019}) \bibinfo{pages}{125932}.
\bibitem[{Ahami and El~Baz(2021)}]{ahami2021thermal}
\bibinfo{author}{N.~Ahami}, \bibinfo{author}{M.~El~Baz},
\newblock \bibinfo{title}{Thermal entanglement in a mixed spin heisenberg xxx
  chain with dm interaction},
\newblock \bibinfo{journal}{International Journal of Quantum Information}
  \bibinfo{volume}{19} (\bibinfo{year}{2021}) \bibinfo{pages}{2150021}.
\bibitem[{Abaach et~al.(2021)Abaach, El~Baz, and Faqir}]{abaach2021pairwise}
\bibinfo{author}{S.~Abaach}, \bibinfo{author}{M.~El~Baz},
  \bibinfo{author}{M.~Faqir},
\newblock \bibinfo{title}{Pairwise quantum correlations in four-level quantum
  dot systems},
\newblock \bibinfo{journal}{Physics Letters A} \bibinfo{volume}{391}
  (\bibinfo{year}{2021}) \bibinfo{pages}{127140}.
\bibitem[{Mansour et~al.(2020)Mansour, Siyouri, Faqir, and
  El~Baz}]{mansour2020quantum}
\bibinfo{author}{H.~A. Mansour}, \bibinfo{author}{F.~Siyouri},
  \bibinfo{author}{M.~Faqir}, \bibinfo{author}{M.~El~Baz},
\newblock \bibinfo{title}{Quantum correlations dynamics in two coupled
  semiconductor inas quantum dots},
\newblock \bibinfo{journal}{Physica Scripta} \bibinfo{volume}{95}
  (\bibinfo{year}{2020}) \bibinfo{pages}{095101}.
\bibitem[{El~Bir and El~Baz(2020)}]{el2020quantum}
\bibinfo{author}{O.~El~Bir}, \bibinfo{author}{M.~El~Baz},
\newblock \bibinfo{title}{Quantum correlations under the effect of a thermal
  environment in a triangular optomechanical cavity},
\newblock \bibinfo{journal}{JOSA B} \bibinfo{volume}{37} (\bibinfo{year}{2020})
  \bibinfo{pages}{A237--A244}.
\bibitem[{DeChiara and Sanpera(2018)}]{DeChiaraReview}
\bibinfo{author}{G.~DeChiara}, \bibinfo{author}{A.~Sanpera},
\newblock \bibinfo{title}{Genuine quantum correlations in quantum many-body
  systems: a review of recent progress},
\newblock \bibinfo{journal}{Rep. Prog. Phys.} \bibinfo{volume}{81}
  (\bibinfo{year}{2018}) \bibinfo{pages}{074002}.
\bibitem[{Kadowaki and Nishimori(1998)}]{nishimori}
\bibinfo{author}{T.~Kadowaki}, \bibinfo{author}{H.~Nishimori},
\newblock \bibinfo{title}{Quantum annealing in the transverse ising model},
\newblock \bibinfo{journal}{Phys. Rev. E} \bibinfo{volume}{58}
  (\bibinfo{year}{1998}) \bibinfo{pages}{5355--5363}.
\bibitem[{Heim et~al.(2015)Heim, R{\o}nnow, Isakov, and
  Troyer}]{heim2015annealer}
\bibinfo{author}{B.~Heim}, \bibinfo{author}{T.~F. R{\o}nnow},
  \bibinfo{author}{S.~V. Isakov}, \bibinfo{author}{M.~Troyer},
\newblock \bibinfo{title}{Quantum versus classical annealing of ising spin
  glasses},
\newblock \bibinfo{journal}{Science} \bibinfo{volume}{348}
  (\bibinfo{year}{2015}) \bibinfo{pages}{215--217}.
\bibitem[{Yan and Sinitsyn(2021)}]{yan2021annealer}
\bibinfo{author}{B.~Yan}, \bibinfo{author}{N.~A. Sinitsyn},
\newblock \bibinfo{title}{Analytical solution for ground state estimation by a
  nonadiabatic quantum annealing to arbitrary ising spin hamiltonian},
\newblock \bibinfo{journal}{arXiv preprint arXiv:2110.12354}
  (\bibinfo{year}{2021}).
\bibitem[{Bose(2007)}]{bose2007quantumcommu}
\bibinfo{author}{S.~Bose},
\newblock \bibinfo{title}{Quantum communication through spin chain dynamics: an
  introductory overview},
\newblock \bibinfo{journal}{Contemporary Physics} \bibinfo{volume}{48}
  (\bibinfo{year}{2007}) \bibinfo{pages}{13--30}.
\bibitem[{Nakatani(2018)}]{dmrgtensor}
\bibinfo{author}{N.~Nakatani},
\newblock \bibinfo{title}{Matrix product states and density matrix
  renormalization group algorithm},
\newblock in: \bibinfo{booktitle}{Reference Module in Chemistry, Molecular
  Sciences and Chemical Engineering}, \bibinfo{publisher}{Elsevier},
  \bibinfo{year}{2018}.
\bibitem[{Perrot(1998)}]{perrot1998z}
\bibinfo{author}{P.~Perrot}, \bibinfo{title}{A to Z of Thermodynamics},
  \bibinfo{publisher}{Oxford University Press on Demand}, \bibinfo{year}{1998}.
\bibitem[{Carnot(1978)}]{carnot1978reflexions}
\bibinfo{author}{S.~Carnot}, \bibinfo{title}{R{\'e}flexions sur la puissance
  motrice du feu}, \bibinfo{number}{26}, \bibinfo{publisher}{Vrin},
  \bibinfo{year}{1978}.
\bibitem[{Reif(1998)}]{reif}
\bibinfo{author}{F.~Reif},
\newblock \bibinfo{title}{Fundamentals of statistical and thermal physics},
\newblock \bibinfo{journal}{American Journal of Physics} \bibinfo{volume}{66}
  (\bibinfo{year}{1998}) \bibinfo{pages}{164--167}.
\bibitem[{Binder et~al.(2018)Binder, Correa, Gogolin, Anders, and
  Adesso}]{thermo_quantum_regime}
\bibinfo{author}{F.~Binder}, \bibinfo{author}{L.~Correa},
  \bibinfo{author}{C.~Gogolin}, \bibinfo{author}{J.~Anders},
  \bibinfo{author}{G.~Adesso}, \bibinfo{title}{Thermodynamics in the Quantum
  Regime Fundamental Aspects and New Directions: Fundamental Aspects and New
  Directions}, \bibinfo{year}{2018}.
\bibitem[{Deffner and Campbell(2019)}]{booksteve}
\bibinfo{author}{S.~Deffner}, \bibinfo{author}{S.~Campbell},
  \bibinfo{title}{Quantum Thermodynamics}, 2053-2571,
  \bibinfo{publisher}{Morgan \& Claypool Publishers}, \bibinfo{year}{2019}.
\bibitem[{Bennett(2003)}]{bennett2003notes}
\bibinfo{author}{C.~H. Bennett}, \bibinfo{title}{Notes on landauer's principle,
  reversible computation and maxwell's demon}, \bibinfo{year}{2003}.
  \href{http://arxiv.org/abs/physics/0210005}{{\tt arXiv:physics/0210005}}.
\bibitem[{{Szilard}(1929)}]{szilard}
\bibinfo{author}{L.~{Szilard}},
\newblock \bibinfo{title}{{{\"u}ber die Entropieverminderung in einem
  thermodynamischen System bei Eingriffen intelligenter Wesen}},
\newblock \bibinfo{journal}{Zeitschrift fur Physik} \bibinfo{volume}{53}
  (\bibinfo{year}{1929}) \bibinfo{pages}{840--856}.
\bibitem[{Toyabe et~al.(2010)Toyabe, Sagawa, Ueda, Muneyuki, and
  Sano}]{szilardexp}
\bibinfo{author}{S.~Toyabe}, \bibinfo{author}{T.~Sagawa},
  \bibinfo{author}{M.~Ueda}, \bibinfo{author}{E.~Muneyuki},
  \bibinfo{author}{M.~Sano},
\newblock \bibinfo{title}{Experimental demonstration of information-to-energy
  conversion and validation of the generalized jarzynski equality},
\newblock \bibinfo{journal}{Nature Physics} \bibinfo{volume}{6}
  (\bibinfo{year}{2010}) \bibinfo{pages}{988–992}.
\bibitem[{Bennett(1982)}]{Bennett1982TheTO}
\bibinfo{author}{C.~H. Bennett},
\newblock \bibinfo{title}{The thermodynamics of computation—a review},
\newblock \bibinfo{journal}{International Journal of Theoretical Physics}
  \bibinfo{volume}{21} (\bibinfo{year}{1982}) \bibinfo{pages}{905--940}.
\bibitem[{Goold et~al.(2016)Goold, Huber, Riera, Rio, and
  Skrzypczyk}]{gooldreview}
\bibinfo{author}{J.~Goold}, \bibinfo{author}{M.~Huber},
  \bibinfo{author}{A.~Riera}, \bibinfo{author}{L.~d. Rio},
  \bibinfo{author}{P.~Skrzypczyk},
\newblock \bibinfo{title}{The role of quantum information in thermodynamics—a
  topical review},
\newblock \bibinfo{journal}{Journal of Physics A: Mathematical and Theoretical}
  \bibinfo{volume}{49} (\bibinfo{year}{2016}) \bibinfo{pages}{143001}.
\bibitem[{Campbell(2016)}]{CampbellPRB2016}
\bibinfo{author}{S.~Campbell},
\newblock \bibinfo{title}{Criticality revealed through quench dynamics in the
  lipkin-meshkov-glick model},
\newblock \bibinfo{journal}{Phys. Rev. B} \bibinfo{volume}{94}
  (\bibinfo{year}{2016}) \bibinfo{pages}{184403}.
\bibitem[{Mzaouali et~al.(2021)Mzaouali, Puebla, Goold, El~Baz, and
  Campbell}]{mzaouali2021}
\bibinfo{author}{Z.~Mzaouali}, \bibinfo{author}{R.~Puebla},
  \bibinfo{author}{J.~Goold}, \bibinfo{author}{M.~El~Baz},
  \bibinfo{author}{S.~Campbell},
\newblock \bibinfo{title}{Work statistics and symmetry breaking in an
  excited-state quantum phase transition},
\newblock \bibinfo{journal}{Physical Review E} \bibinfo{volume}{103}
  (\bibinfo{year}{2021}).
\bibitem[{Hammam et~al.(2021)Hammam, Hassouni, Fazio, and
  Manzano}]{Hammam_2021}
\bibinfo{author}{K.~Hammam}, \bibinfo{author}{Y.~Hassouni},
  \bibinfo{author}{R.~Fazio}, \bibinfo{author}{G.~Manzano},
\newblock \bibinfo{title}{Optimizing autonomous thermal machines powered by
  energetic coherence},
\newblock \bibinfo{journal}{New Journal of Physics} \bibinfo{volume}{23}
  (\bibinfo{year}{2021}) \bibinfo{pages}{043024}.
\bibitem[{Giorgi and Campbell(2015)}]{Giorgi_2015iop}
\bibinfo{author}{G.~L. Giorgi}, \bibinfo{author}{S.~Campbell},
\newblock \bibinfo{title}{Correlation approach to work extraction from finite
  quantum systems},
\newblock \bibinfo{journal}{Journal of Physics B: Atomic, Molecular and Optical
  Physics} \bibinfo{volume}{48} (\bibinfo{year}{2015}) \bibinfo{pages}{035501}.
\bibitem[{Kamin et~al.(2020)Kamin, Tabesh, Salimi, and Santos}]{batt2020}
\bibinfo{author}{F.~H. Kamin}, \bibinfo{author}{F.~T. Tabesh},
  \bibinfo{author}{S.~Salimi}, \bibinfo{author}{A.~C. Santos},
\newblock \bibinfo{title}{Entanglement, coherence, and charging process of
  quantum batteries},
\newblock \bibinfo{journal}{Phys. Rev. E} \bibinfo{volume}{102}
  (\bibinfo{year}{2020}) \bibinfo{pages}{052109}.
\bibitem[{Liu et~al.(2021)Liu, Shi, Shi, Wang, and Yang}]{batt2021}
\bibinfo{author}{J.-X. Liu}, \bibinfo{author}{H.-L. Shi},
  \bibinfo{author}{Y.-H. Shi}, \bibinfo{author}{X.-H. Wang},
  \bibinfo{author}{W.-L. Yang},
\newblock \bibinfo{title}{Entanglement and work extraction in the central-spin
  quantum battery},
\newblock \bibinfo{journal}{Phys. Rev. B} \bibinfo{volume}{104}
  (\bibinfo{year}{2021}) \bibinfo{pages}{245418}.
\bibitem[{Fadaie et~al.(2018)Fadaie, Yunt, and M\"ustecapl\ifmmode \imath \else
  \i \fi{}o\ifmmode~\breve{g}\else \u{g}\fi{}lu}]{topo_qhe}
\bibinfo{author}{M.~Fadaie}, \bibinfo{author}{E.~Yunt}, \bibinfo{author}{O.~E.
  M\"ustecapl\ifmmode \imath \else \i \fi{}o\ifmmode~\breve{g}\else
  \u{g}\fi{}lu},
\newblock \bibinfo{title}{Topological phase transition in quantum-heat-engine
  cycles},
\newblock \bibinfo{journal}{Phys. Rev. E} \bibinfo{volume}{98}
  (\bibinfo{year}{2018}) \bibinfo{pages}{052124}.
\bibitem[{Fogarty and Busch(2020)}]{Fogarty_2020qhe}
\bibinfo{author}{T.~Fogarty}, \bibinfo{author}{T.~Busch},
\newblock \bibinfo{title}{A many-body heat engine at criticality},
\newblock \bibinfo{journal}{Quantum Science and Technology} \bibinfo{volume}{6}
  (\bibinfo{year}{2020}) \bibinfo{pages}{015003}.
\bibitem[{Solfanelli et~al.(2020)Solfanelli, Falsetti, and
  Campisi}]{qhe_campsi2020}
\bibinfo{author}{A.~Solfanelli}, \bibinfo{author}{M.~Falsetti},
  \bibinfo{author}{M.~Campisi},
\newblock \bibinfo{title}{Nonadiabatic single-qubit quantum otto engine},
\newblock \bibinfo{journal}{Phys. Rev. B} \bibinfo{volume}{101}
  (\bibinfo{year}{2020}) \bibinfo{pages}{054513}.
\bibitem[{Myers and Deffner(2020)}]{myers_qhe}
\bibinfo{author}{N.~M. Myers}, \bibinfo{author}{S.~Deffner},
\newblock \bibinfo{title}{Bosons outperform fermions: The thermodynamic
  advantage of symmetry},
\newblock \bibinfo{journal}{Phys. Rev. E} \bibinfo{volume}{101}
  (\bibinfo{year}{2020}) \bibinfo{pages}{012110}.
\bibitem[{Ma et~al.(2017)Ma, Su, and Sun}]{qpt_qhe2017}
\bibinfo{author}{Y.-H. Ma}, \bibinfo{author}{S.-H. Su}, \bibinfo{author}{C.-P.
  Sun},
\newblock \bibinfo{title}{Quantum thermodynamic cycle with quantum phase
  transition},
\newblock \bibinfo{journal}{Phys. Rev. E} \bibinfo{volume}{96}
  (\bibinfo{year}{2017}) \bibinfo{pages}{022143}.
\bibitem[{Çakmak et~al.(2016)Çakmak, Altintas, and
  E.~M\"{u}stecaplıo\~{g}lu}]{lmg_qhe2016}
\bibinfo{author}{S.~Çakmak}, \bibinfo{author}{F.~Altintas},
  \bibinfo{author}{O.~E.~M\"{u}stecaplıo\~{g}lu},
\newblock \bibinfo{title}{Lipkin-meshkov-glick model in a quantum otto cycle},
\newblock \bibinfo{journal}{The European Physical Journal Plus}
  \bibinfo{volume}{131} (\bibinfo{year}{2016}).
\bibitem[{Lieb(1976)}]{stab_matter}
\bibinfo{author}{E.~H. Lieb},
\newblock \bibinfo{title}{The stability of matter},
\newblock \bibinfo{journal}{Rev. Mod. Phys.} \bibinfo{volume}{48}
  (\bibinfo{year}{1976}) \bibinfo{pages}{553--569}.
\bibitem[{Baxter(1982)}]{baxterbook}
\bibinfo{author}{R.~Baxter}, \bibinfo{title}{Exactly solved models in
  statistical mechanics}, \bibinfo{edition}{illustrated edition} ed.,
  \bibinfo{publisher}{Academic Press}, \bibinfo{year}{1982}.
\bibitem[{Franchini(2017)}]{franchini2017}
\bibinfo{author}{F.~Franchini},
\newblock \bibinfo{title}{An introduction to integrable techniques for
  one-dimensional quantum systems},
\newblock \bibinfo{journal}{Lecture Notes in Physics}  (\bibinfo{year}{2017}).
\bibitem[{Onsager(1944)}]{onsager1}
\bibinfo{author}{L.~Onsager},
\newblock \bibinfo{title}{Crystal statistics. i. a two-dimensional model with
  an order-disorder transition},
\newblock \bibinfo{journal}{Phys. Rev.} \bibinfo{volume}{65}
  (\bibinfo{year}{1944}) \bibinfo{pages}{117--149}.
\bibitem[{Kaufman(1949)}]{onsager2}
\bibinfo{author}{B.~Kaufman},
\newblock \bibinfo{title}{Crystal statistics. ii. partition function evaluated
  by spinor analysis},
\newblock \bibinfo{journal}{Phys. Rev.} \bibinfo{volume}{76}
  (\bibinfo{year}{1949}) \bibinfo{pages}{1232--1243}.
\bibitem[{Kaufman and Onsager(1949)}]{onsager3}
\bibinfo{author}{B.~Kaufman}, \bibinfo{author}{L.~Onsager},
\newblock \bibinfo{title}{Crystal statistics. iii. short-range order in a
  binary ising lattice},
\newblock \bibinfo{journal}{Phys. Rev.} \bibinfo{volume}{76}
  (\bibinfo{year}{1949}) \bibinfo{pages}{1244--1252}.
\bibitem[{C.~/~Green(1972)}]{griffith}
\bibinfo{author}{M.~S. E.~D. C.~/~Green}, \bibinfo{title}{Exact results}, Phase
  transitions and critical phenomena 1, \bibinfo{publisher}{Academic Press
  London.}, \bibinfo{year}{1972}.
\bibitem[{{Weiss, Pierre}(1907)}]{weiss-curie}
\bibinfo{author}{{Weiss, Pierre}},
\newblock \bibinfo{title}{L'hypoth\`ese du champ mol\'eculaire et la
  propri\'et\'e ferromagn\'etique},
\newblock \bibinfo{journal}{J. Phys. Theor. Appl.} \bibinfo{volume}{6}
  (\bibinfo{year}{1907}) \bibinfo{pages}{661--690}.
\bibitem[{Bitko et~al.(1996)Bitko, Rosenbaum, and Aeppli}]{bitko1996}
\bibinfo{author}{D.~Bitko}, \bibinfo{author}{T.~F. Rosenbaum},
  \bibinfo{author}{G.~Aeppli},
\newblock \bibinfo{title}{Quantum critical behavior for a model magnet},
\newblock \bibinfo{journal}{Phys. Rev. Lett.} \bibinfo{volume}{77}
  (\bibinfo{year}{1996}) \bibinfo{pages}{940--943}.
\bibitem[{Coleman(1999)}]{coleman1999}
\bibinfo{author}{P.~Coleman},
\newblock \bibinfo{title}{Theories of non-fermi liquid behavior in heavy
  fermions},
\newblock \bibinfo{journal}{Physica B: Condensed Matter}
  \bibinfo{volume}{259–261} (\bibinfo{year}{1999})
  \bibinfo{pages}{353–358}.
\bibitem[{von Löhneysen(1996)}]{L_hneysen_1996}
\bibinfo{author}{H.~von Löhneysen},
\newblock \bibinfo{title}{Non-fermi-liquid behaviour in the heavy-fermion
  system},
\newblock \bibinfo{journal}{Journal of Physics: Condensed Matter}
  \bibinfo{volume}{8} (\bibinfo{year}{1996}) \bibinfo{pages}{9689--9706}.
\bibitem[{Dagotto(1994)}]{dagotto1994}
\bibinfo{author}{E.~Dagotto},
\newblock \bibinfo{title}{Correlated electrons in high-temperature
  superconductors},
\newblock \bibinfo{journal}{Rev. Mod. Phys.} \bibinfo{volume}{66}
  (\bibinfo{year}{1994}) \bibinfo{pages}{763--840}.
\bibitem[{Orenstein and Millis(2000)}]{orenstein2000}
\bibinfo{author}{J.~Orenstein}, \bibinfo{author}{A.~J. Millis},
\newblock \bibinfo{title}{Advances in the physics of high-temperature
  superconductivity},
\newblock \bibinfo{journal}{Science} \bibinfo{volume}{288}
  (\bibinfo{year}{2000}) \bibinfo{pages}{468--474}.
\bibitem[{Sachdev(2000)}]{sachdev2000}
\bibinfo{author}{S.~Sachdev},
\newblock \bibinfo{title}{Quantum criticality: Competing ground states in low
  dimensions},
\newblock \bibinfo{journal}{Science} \bibinfo{volume}{288}
  (\bibinfo{year}{2000}) \bibinfo{pages}{475--480}.
\bibitem[{Kravchenko et~al.(1995)Kravchenko, Mason, Bowker, Furneaux, Pudalov,
  and D'Iorio}]{krav1995}
\bibinfo{author}{S.~V. Kravchenko}, \bibinfo{author}{W.~E. Mason},
  \bibinfo{author}{G.~E. Bowker}, \bibinfo{author}{J.~E. Furneaux},
  \bibinfo{author}{V.~M. Pudalov}, \bibinfo{author}{M.~D'Iorio},
\newblock \bibinfo{title}{Scaling of an anomalous metal-insulator transition in
  a two-dimensional system in silicon at b=0},
\newblock \bibinfo{journal}{Phys. Rev. B} \bibinfo{volume}{51}
  (\bibinfo{year}{1995}) \bibinfo{pages}{7038--7045}.
\bibitem[{Carr(2010)}]{under_qpt}
\bibinfo{author}{L.~Carr}, \bibinfo{title}{Understanding Quantum Phase
  Transitions}, Condensed Matter Physics, \bibinfo{edition}{1} ed.,
  \bibinfo{publisher}{CRC}, \bibinfo{year}{2010}.
\bibitem[{Chakraborty(1999)}]{quantumdot}
\bibinfo{author}{T.~Chakraborty}, \bibinfo{title}{Quantum dots: a survey of the
  properties of artificial atoms}, \bibinfo{edition}{1st ed} ed.,
  \bibinfo{publisher}{Elsevier}, \bibinfo{year}{1999}.
\bibitem[{Hiura et~al.(2019)Hiura, Itabashi, Takeishi, Takayama, Kiba, and
  Murayama}]{qdots_spin_trans}
\bibinfo{author}{S.~Hiura}, \bibinfo{author}{K.~Itabashi},
  \bibinfo{author}{K.~Takeishi}, \bibinfo{author}{J.~Takayama},
  \bibinfo{author}{T.~Kiba}, \bibinfo{author}{A.~Murayama},
\newblock \bibinfo{title}{Quantum spin transport to semiconductor quantum dots
  through superlattice},
\newblock \bibinfo{journal}{Applied Physics Letters} \bibinfo{volume}{114}
  (\bibinfo{year}{2019}) \bibinfo{pages}{072406}.
\bibitem[{Eckern and Wysokiński(2021)}]{qdot_heat_transp}
\bibinfo{author}{U.~Eckern}, \bibinfo{author}{K.~I. Wysokiński},
\newblock \bibinfo{title}{Charge and heat transport through quantum dots with
  local and correlated-hopping interactions},
\newblock \bibinfo{journal}{Physical Review Research} \bibinfo{volume}{3}
  (\bibinfo{year}{2021}).
\bibitem[{Phirouznia et~al.(2012)Phirouznia, Bourkheili, and
  Hesari}]{spin-cond_qdot}
\bibinfo{author}{A.~Phirouznia}, \bibinfo{author}{S.~H. Bourkheili},
  \bibinfo{author}{V.~Z. Hesari},
\newblock \bibinfo{title}{Controllable non-equilibrium spin injection of
  conduction electrons in magnetic quantum dots},
\newblock \bibinfo{journal}{Physica E: Low-dimensional Systems and
  Nanostructures} \bibinfo{volume}{44} (\bibinfo{year}{2012})
  \bibinfo{pages}{1304--1308}.
\bibitem[{Oreg and Goldhaber-Gordon(2003)}]{oreg2003}
\bibinfo{author}{Y.~Oreg}, \bibinfo{author}{D.~Goldhaber-Gordon},
\newblock \bibinfo{title}{Two-channel kondo effect in a modified single
  electron transistor},
\newblock \bibinfo{journal}{Phys. Rev. Lett.} \bibinfo{volume}{90}
  (\bibinfo{year}{2003}) \bibinfo{pages}{136602}.
\bibitem[{Andrei et~al.(1988)Andrei, Deville, Glattli, Williams, Paris, and
  Etienne}]{wigcrys1988}
\bibinfo{author}{E.~Y. Andrei}, \bibinfo{author}{G.~Deville},
  \bibinfo{author}{D.~C. Glattli}, \bibinfo{author}{F.~I.~B. Williams},
  \bibinfo{author}{E.~Paris}, \bibinfo{author}{B.~Etienne},
\newblock \bibinfo{title}{Observation of a magnetically induced wigner solid},
\newblock \bibinfo{journal}{Phys. Rev. Lett.} \bibinfo{volume}{60}
  (\bibinfo{year}{1988}) \bibinfo{pages}{2765--2768}.
\bibitem[{Santos et~al.(1992)Santos, Suen, Shayegan, Li, Engel, and
  Tsui}]{2d1992}
\bibinfo{author}{M.~B. Santos}, \bibinfo{author}{Y.~W. Suen},
  \bibinfo{author}{M.~Shayegan}, \bibinfo{author}{Y.~P. Li},
  \bibinfo{author}{L.~W. Engel}, \bibinfo{author}{D.~C. Tsui},
\newblock \bibinfo{title}{Observation of a reentrant insulating phase near the
  1/3 fractional quantum hall liquid in a two-dimensional hole system},
\newblock \bibinfo{journal}{Phys. Rev. Lett.} \bibinfo{volume}{68}
  (\bibinfo{year}{1992}) \bibinfo{pages}{1188--1191}.
\bibitem[{Maciej~Lewenstein(2012)}]{ultracoldbook}
\bibinfo{author}{V.~A. Maciej~Lewenstein, Anna~Sanpera},
  \bibinfo{title}{Ultracold Atoms in Optical Lattices: Simulating quantum
  many-body systems}, \bibinfo{publisher}{Oxford University Press},
  \bibinfo{year}{2012}.
\bibitem[{Greiner et~al.(2002)Greiner, Mandel, Esslinger, Haensch, and
  Bloch}]{bloch2002}
\bibinfo{author}{M.~Greiner}, \bibinfo{author}{O.~Mandel},
  \bibinfo{author}{T.~Esslinger}, \bibinfo{author}{T.~Haensch},
  \bibinfo{author}{I.~Bloch},
\newblock \bibinfo{title}{Quantum phase transition from a superfluid to a mott
  insulator in a gas of ultracold atoms},
\newblock \bibinfo{journal}{Nature} \bibinfo{volume}{415}
  (\bibinfo{year}{2002}) \bibinfo{pages}{39--44}.
\bibitem[{Claude E~Shannon(1963)}]{shannon}
\bibinfo{author}{W.~W. Claude E~Shannon}, \bibinfo{title}{The Mathematical
  Theory of Communication}, \bibinfo{publisher}{U. Illinois},
  \bibinfo{year}{1963}.
\bibitem[{Bennett et~al.(1993)Bennett, Brassard, Cr\'epeau, Jozsa, Peres, and
  Wootters}]{telepo}
\bibinfo{author}{C.~H. Bennett}, \bibinfo{author}{G.~Brassard},
  \bibinfo{author}{C.~Cr\'epeau}, \bibinfo{author}{R.~Jozsa},
  \bibinfo{author}{A.~Peres}, \bibinfo{author}{W.~K. Wootters},
\newblock \bibinfo{title}{Teleporting an unknown quantum state via dual
  classical and einstein-podolsky-rosen channels},
\newblock \bibinfo{journal}{Phys. Rev. Lett.} \bibinfo{volume}{70}
  (\bibinfo{year}{1993}) \bibinfo{pages}{1895--1899}.
\bibitem[{Shor(1997)}]{shor}
\bibinfo{author}{P.~W. Shor},
\newblock \bibinfo{title}{Polynomial-time algorithms for prime factorization
  and discrete logarithms on a quantum computer},
\newblock \bibinfo{journal}{SIAM Journal on Computing} \bibinfo{volume}{26}
  (\bibinfo{year}{1997}) \bibinfo{pages}{1484–1509}.
\bibitem[{Bennett and Brassard(2014)}]{crypto}
\bibinfo{author}{C.~H. Bennett}, \bibinfo{author}{G.~Brassard},
\newblock \bibinfo{title}{Quantum cryptography: Public key distribution and
  coin tossing},
\newblock \bibinfo{journal}{Theoretical Computer Science} \bibinfo{volume}{560}
  (\bibinfo{year}{2014}) \bibinfo{pages}{7–11}.
\bibitem[{M.(2006)}]{quantum_comm_book}
\bibinfo{author}{P.~M.}, \bibinfo{title}{Quantum computation and quantum
  communication: theory and experiments}, \bibinfo{publisher}{Springer},
  \bibinfo{year}{2006}.
\bibitem[{Einstein et~al.(1935)Einstein, Podolsky, and Rosen}]{EPR}
\bibinfo{author}{A.~Einstein}, \bibinfo{author}{B.~Podolsky},
  \bibinfo{author}{N.~Rosen},
\newblock \bibinfo{title}{Can quantum-mechanical description of physical
  reality be considered complete?},
\newblock \bibinfo{journal}{Phys. Rev.} \bibinfo{volume}{47}
  (\bibinfo{year}{1935}) \bibinfo{pages}{777--780}.
\bibitem[{Ekert(1991)}]{ekert91}
\bibinfo{author}{A.~K. Ekert},
\newblock \bibinfo{title}{Quantum cryptography based on bell's theorem},
\newblock \bibinfo{journal}{Phys. Rev. Lett.} \bibinfo{volume}{67}
  (\bibinfo{year}{1991}) \bibinfo{pages}{661--663}.
\bibitem[{Bennett and Wiesner(1992)}]{weisner92}
\bibinfo{author}{C.~H. Bennett}, \bibinfo{author}{S.~J. Wiesner},
\newblock \bibinfo{title}{Communication via one- and two-particle operators on
  einstein-podolsky-rosen states},
\newblock \bibinfo{journal}{Phys. Rev. Lett.} \bibinfo{volume}{69}
  (\bibinfo{year}{1992}) \bibinfo{pages}{2881--2884}.
\bibitem[{Bouwmeester et~al.(1997)Bouwmeester, Pan, Mattle, Eibl, Weinfurter,
  and Zeilinger}]{exp_telep}
\bibinfo{author}{D.~Bouwmeester}, \bibinfo{author}{J.-W. Pan},
  \bibinfo{author}{K.~Mattle}, \bibinfo{author}{M.~Eibl},
  \bibinfo{author}{H.~Weinfurter}, \bibinfo{author}{A.~Zeilinger},
\newblock \bibinfo{title}{Experimental quantum teleportation},
\newblock \bibinfo{journal}{Nature} \bibinfo{volume}{390}
  (\bibinfo{year}{1997}) \bibinfo{pages}{575–579}.
\bibitem[{Boschi et~al.(1998)Boschi, Branca, De~Martini, Hardy, and
  Popescu}]{exp_telep_popescu}
\bibinfo{author}{D.~Boschi}, \bibinfo{author}{S.~Branca},
  \bibinfo{author}{F.~De~Martini}, \bibinfo{author}{L.~Hardy},
  \bibinfo{author}{S.~Popescu},
\newblock \bibinfo{title}{Experimental realization of teleporting an unknown
  pure quantum state via dual classical and einstein-podolsky-rosen channels},
\newblock \bibinfo{journal}{Phys. Rev. Lett.} \bibinfo{volume}{80}
  (\bibinfo{year}{1998}) \bibinfo{pages}{1121--1125}.
\bibitem[{Horodecki et~al.(2009)Horodecki, Horodecki, Horodecki, and
  Horodecki}]{rev_entanglement}
\bibinfo{author}{R.~Horodecki}, \bibinfo{author}{P.~Horodecki},
  \bibinfo{author}{M.~Horodecki}, \bibinfo{author}{K.~Horodecki},
\newblock \bibinfo{title}{Quantum entanglement},
\newblock \bibinfo{journal}{Rev. Mod. Phys.} \bibinfo{volume}{81}
  (\bibinfo{year}{2009}) \bibinfo{pages}{865--942}.
\bibitem[{Wootters(1998)}]{wootters_concurrence}
\bibinfo{author}{W.~K. Wootters},
\newblock \bibinfo{title}{Entanglement of formation of an arbitrary state of
  two qubits},
\newblock \bibinfo{journal}{Phys. Rev. Lett.} \bibinfo{volume}{80}
  (\bibinfo{year}{1998}) \bibinfo{pages}{2245--2248}.
\bibitem[{Ollivier and Zurek(2001)}]{discord1}
\bibinfo{author}{H.~Ollivier}, \bibinfo{author}{W.~H. Zurek},
\newblock \bibinfo{title}{Quantum discord: A measure of the quantumness of
  correlations},
\newblock \bibinfo{journal}{Phys. Rev. Lett.} \bibinfo{volume}{88}
  (\bibinfo{year}{2001}) \bibinfo{pages}{017901}.
\bibitem[{Henderson and Vedral(2001)}]{discord2}
\bibinfo{author}{L.~Henderson}, \bibinfo{author}{V.~Vedral},
\newblock \bibinfo{title}{Classical, quantum and total correlations},
\newblock \bibinfo{journal}{Journal of Physics A: Mathematical and General}
  \bibinfo{volume}{34} (\bibinfo{year}{2001}) \bibinfo{pages}{6899–6905}.
\bibitem[{Bera et~al.(2017)Bera, Das, Sadhukhan, Singha~Roy, Sen(De), and
  Sen}]{discord3}
\bibinfo{author}{A.~Bera}, \bibinfo{author}{T.~Das},
  \bibinfo{author}{D.~Sadhukhan}, \bibinfo{author}{S.~Singha~Roy},
  \bibinfo{author}{A.~Sen(De)}, \bibinfo{author}{U.~Sen},
\newblock \bibinfo{title}{Quantum discord and its allies: a review of recent
  progress},
\newblock \bibinfo{journal}{Reports on Progress in Physics}
  \bibinfo{volume}{81} (\bibinfo{year}{2017}) \bibinfo{pages}{024001}.
\bibitem[{Gheorghiu et~al.(2015)Gheorghiu, de~Oliveira, and
  Sanders}]{barry_discord}
\bibinfo{author}{V.~Gheorghiu}, \bibinfo{author}{M.~C. de~Oliveira},
  \bibinfo{author}{B.~C. Sanders},
\newblock \bibinfo{title}{Nonzero classical discord},
\newblock \bibinfo{journal}{Physical Review Letters} \bibinfo{volume}{115}
  (\bibinfo{year}{2015}).
\bibitem[{Horodecki et~al.(2005)Horodecki, Oppenheim, and
  Winter}]{quantum_state_merg}
\bibinfo{author}{M.~Horodecki}, \bibinfo{author}{J.~Oppenheim},
  \bibinfo{author}{A.~Winter},
\newblock \bibinfo{title}{Partial quantum information},
\newblock \bibinfo{journal}{Nature} \bibinfo{volume}{436}
  (\bibinfo{year}{2005}) \bibinfo{pages}{673–676}.
\bibitem[{Gu et~al.(2012)Gu, Chrzanowski, Assad, Symul, Modi, Ralph, Vedral,
  and Lam}]{quantum_channel}
\bibinfo{author}{M.~Gu}, \bibinfo{author}{H.~M. Chrzanowski},
  \bibinfo{author}{S.~M. Assad}, \bibinfo{author}{T.~Symul},
  \bibinfo{author}{K.~Modi}, \bibinfo{author}{T.~C. Ralph},
  \bibinfo{author}{V.~Vedral}, \bibinfo{author}{P.~K. Lam},
\newblock \bibinfo{title}{Observing the operational significance of discord
  consumption},
\newblock \bibinfo{journal}{Nature Physics} \bibinfo{volume}{8}
  (\bibinfo{year}{2012}) \bibinfo{pages}{671–675}.
\bibitem[{Girolami et~al.(2013)Girolami, Tufarelli, and Adesso}]{metrology1}
\bibinfo{author}{D.~Girolami}, \bibinfo{author}{T.~Tufarelli},
  \bibinfo{author}{G.~Adesso},
\newblock \bibinfo{title}{Characterizing nonclassical correlations via local
  quantum uncertainty},
\newblock \bibinfo{journal}{Phys. Rev. Lett.} \bibinfo{volume}{110}
  (\bibinfo{year}{2013}) \bibinfo{pages}{240402}.
\bibitem[{Girolami et~al.(2014)Girolami, Souza, Giovannetti, Tufarelli,
  Filgueiras, Sarthour, Soares-Pinto, Oliveira, and Adesso}]{metrology2}
\bibinfo{author}{D.~Girolami}, \bibinfo{author}{A.~M. Souza},
  \bibinfo{author}{V.~Giovannetti}, \bibinfo{author}{T.~Tufarelli},
  \bibinfo{author}{J.~G. Filgueiras}, \bibinfo{author}{R.~S. Sarthour},
  \bibinfo{author}{D.~O. Soares-Pinto}, \bibinfo{author}{I.~S. Oliveira},
  \bibinfo{author}{G.~Adesso},
\newblock \bibinfo{title}{Quantum discord determines the interferometric power
  of quantum states},
\newblock \bibinfo{journal}{Phys. Rev. Lett.} \bibinfo{volume}{112}
  (\bibinfo{year}{2014}) \bibinfo{pages}{210401}.
\bibitem[{Correa et~al.(2013)Correa, Palao, Adesso, and Alonso}]{discord_refr}
\bibinfo{author}{L.~A. Correa}, \bibinfo{author}{J.~P. Palao},
  \bibinfo{author}{G.~Adesso}, \bibinfo{author}{D.~Alonso},
\newblock \bibinfo{title}{Performance bound for quantum absorption
  refrigerators},
\newblock \bibinfo{journal}{Physical Review E} \bibinfo{volume}{87}
  (\bibinfo{year}{2013}).
\bibitem[{Br\'adler et~al.(2010)Br\'adler, Wilde, Vinjanampathy, and
  Uskov}]{bio_discord}
\bibinfo{author}{K.~Br\'adler}, \bibinfo{author}{M.~M. Wilde},
  \bibinfo{author}{S.~Vinjanampathy}, \bibinfo{author}{D.~B. Uskov},
\newblock \bibinfo{title}{Identifying the quantum correlations in
  light-harvesting complexes},
\newblock \bibinfo{journal}{Phys. Rev. A} \bibinfo{volume}{82}
  (\bibinfo{year}{2010}) \bibinfo{pages}{062310}.
\bibitem[{Baumgratz et~al.(2014)Baumgratz, Cramer, and
  Plenio}]{quant_coherence}
\bibinfo{author}{T.~Baumgratz}, \bibinfo{author}{M.~Cramer},
  \bibinfo{author}{M.~B. Plenio},
\newblock \bibinfo{title}{Quantifying coherence},
\newblock \bibinfo{journal}{Phys. Rev. Lett.} \bibinfo{volume}{113}
  (\bibinfo{year}{2014}) \bibinfo{pages}{140401}.
\bibitem[{Streltsov et~al.(2017)Streltsov, Adesso, and
  Plenio}]{coherence_review}
\bibinfo{author}{A.~Streltsov}, \bibinfo{author}{G.~Adesso},
  \bibinfo{author}{M.~B. Plenio},
\newblock \bibinfo{title}{Colloquium : Quantum coherence as a resource},
\newblock \bibinfo{journal}{Reviews of Modern Physics} \bibinfo{volume}{89}
  (\bibinfo{year}{2017}).
\bibitem[{Lin(1991)}]{QJSD}
\bibinfo{author}{J.~Lin},
\newblock \bibinfo{title}{Divergence measures based on the shannon entropy},
\newblock \bibinfo{journal}{IEEE Trans. Inf. Theory} \bibinfo{volume}{37}
  (\bibinfo{year}{1991}) \bibinfo{pages}{145--151}.
\bibitem[{Radhakrishnan et~al.(2016)Radhakrishnan, Parthasarathy, Jambulingam,
  and Byrnes}]{coherencePRL}
\bibinfo{author}{C.~Radhakrishnan}, \bibinfo{author}{M.~Parthasarathy},
  \bibinfo{author}{S.~Jambulingam}, \bibinfo{author}{T.~Byrnes},
\newblock \bibinfo{title}{Distribution of quantum coherence in multipartite
  systems},
\newblock \bibinfo{journal}{Phys. Rev. Lett.} \bibinfo{volume}{116}
  (\bibinfo{year}{2016}) \bibinfo{pages}{150504}.
\bibitem[{P. and E.(1928)}]{jordan-wigner}
\bibinfo{author}{J.~P.}, \bibinfo{author}{W.~E.},
\newblock \bibinfo{title}{Über das paulische Äquivalenzverbot},
\newblock \bibinfo{journal}{Zeitschrift für Physik} \bibinfo{volume}{47}
  (\bibinfo{year}{1928}) \bibinfo{pages}{631 – 651}. \bibinfo{note}{Cited by:
  1376}.
\bibitem[{Weinberg and Bukov(2017)}]{quspin1}
\bibinfo{author}{P.~Weinberg}, \bibinfo{author}{M.~Bukov},
\newblock \bibinfo{title}{{QuSpin: a Python Package for Dynamics and Exact
  Diagonalisation of Quantum Many Body Systems part I: spin chains}},
\newblock \bibinfo{journal}{SciPost Phys.} \bibinfo{volume}{2}
  (\bibinfo{year}{2017}) \bibinfo{pages}{003}.
\bibitem[{Weinberg and Bukov(2019)}]{quspin2}
\bibinfo{author}{P.~Weinberg}, \bibinfo{author}{M.~Bukov},
\newblock \bibinfo{title}{{QuSpin: a Python Package for Dynamics and Exact
  Diagonalisation of Quantum Many Body Systems. Part II: bosons, fermions and
  higher spins}},
\newblock \bibinfo{journal}{SciPost Phys.} \bibinfo{volume}{7}
  (\bibinfo{year}{2019}) \bibinfo{pages}{20}.
\bibitem[{Wells(2014)}]{pauli_basis_exp}
\bibinfo{author}{H.~J. Wells}, \bibinfo{title}{Quantum spin chains and random
  matrix theory}, \bibinfo{year}{2014}.
  \href{http://arxiv.org/abs/1410.1666}{{\tt arXiv:1410.1666}}.
\bibitem[{Wick(1950)}]{wick1950}
\bibinfo{author}{G.~C. Wick},
\newblock \bibinfo{title}{The evaluation of the collision matrix},
\newblock \bibinfo{journal}{Phys. Rev.} \bibinfo{volume}{80}
  (\bibinfo{year}{1950}) \bibinfo{pages}{268--272}.
\bibitem[{Fumani et~al.(2016)Fumani, Nemati, Mahdavifar, and
  Darooneh}]{fumani2018}
\bibinfo{author}{F.~K. Fumani}, \bibinfo{author}{S.~Nemati},
  \bibinfo{author}{S.~Mahdavifar}, \bibinfo{author}{A.~H. Darooneh},
\newblock \bibinfo{title}{Magnetic entanglement in spin-1/2 xy chains},
\newblock \bibinfo{journal}{Physica A: Statistical Mechanics and its
  Applications} \bibinfo{volume}{445} (\bibinfo{year}{2016})
  \bibinfo{pages}{256--263}.
\bibitem[{Yano and Nishimori(2005)}]{nishimori2005_ent}
\bibinfo{author}{H.~Yano}, \bibinfo{author}{H.~Nishimori},
\newblock \bibinfo{title}{{Ground State Entanglement in Spin Systems}},
\newblock \bibinfo{journal}{Progress of Theoretical Physics Supplement}
  \bibinfo{volume}{157} (\bibinfo{year}{2005}) \bibinfo{pages}{164--167}.
\bibitem[{Sachdev(2011)}]{Sachdev}
\bibinfo{author}{S.~Sachdev}, \bibinfo{title}{{Quantum Phase Transitions}},
  \bibinfo{edition}{2nd} ed., \bibinfo{publisher}{Cambridge University Press},
  \bibinfo{address}{Cambridge, UK}, \bibinfo{year}{2011}.
\bibitem[{Wang et~al.(2010)Wang, Li, Nie, and Li}]{discord_X_state}
\bibinfo{author}{C.-Z. Wang}, \bibinfo{author}{C.-X. Li},
  \bibinfo{author}{L.-Y. Nie}, \bibinfo{author}{J.-F. Li},
\newblock \bibinfo{title}{Classical correlation and quantum discord mediated by
  cavity in two coupled qubits},
\newblock \bibinfo{journal}{Journal of Physics B: Atomic, Molecular and Optical
  Physics} \bibinfo{volume}{44} (\bibinfo{year}{2010}) \bibinfo{pages}{015503}.
\bibitem[{Maziero et~al.(2010)Maziero, Guzman, C\'eleri, Sarandy, and
  Serra}]{Maziero2010}
\bibinfo{author}{J.~Maziero}, \bibinfo{author}{H.~C. Guzman},
  \bibinfo{author}{L.~C. C\'eleri}, \bibinfo{author}{M.~S. Sarandy},
  \bibinfo{author}{R.~M. Serra},
\newblock \bibinfo{title}{Quantum and classical thermal correlations in the
  $xy$ spin-$\frac{1}{2}$ chain},
\newblock \bibinfo{journal}{Phys. Rev. A} \bibinfo{volume}{82}
  (\bibinfo{year}{2010}) \bibinfo{pages}{012106}.
\bibitem[{Radhakrishnan et~al.(2017{\natexlab{a}})Radhakrishnan, Parthasarathy,
  Segar, and Byrnes}]{coherence_india1}
\bibinfo{author}{C.~Radhakrishnan}, \bibinfo{author}{M.~Parthasarathy},
  \bibinfo{author}{J.~Segar}, \bibinfo{author}{T.~Byrnes},
\newblock \bibinfo{title}{Quantum coherence of the heisenberg spin models with
  dzyaloshinsky-moriya interactions},
\newblock \bibinfo{journal}{Scientific Reports} \bibinfo{volume}{7}
  (\bibinfo{year}{2017}{\natexlab{a}}).
\bibitem[{Radhakrishnan et~al.(2017{\natexlab{b}})Radhakrishnan, Ermakov, and
  Byrnes}]{coherence_india2}
\bibinfo{author}{C.~Radhakrishnan}, \bibinfo{author}{I.~Ermakov},
  \bibinfo{author}{T.~Byrnes},
\newblock \bibinfo{title}{Quantum coherence of planar spin models with
  dzyaloshinsky-moriya interaction},
\newblock \bibinfo{journal}{Phys. Rev. A} \bibinfo{volume}{96}
  (\bibinfo{year}{2017}{\natexlab{b}}) \bibinfo{pages}{012341}.
\bibitem[{Zhang et~al.(2012)Zhang, Chen, Liu, and Li}]{zhang2012classification}
\bibinfo{author}{T.~Zhang}, \bibinfo{author}{P.-X. Chen},
  \bibinfo{author}{W.-T. Liu}, \bibinfo{author}{C.-Z. Li},
  \bibinfo{title}{Classification of entanglement and quantum phase transition
  in xx model}, \bibinfo{year}{2012}.
  \href{http://arxiv.org/abs/1206.4246}{{\tt arXiv:1206.4246}}.
\bibitem[{Styer et~al.(2002{\natexlab{a}})Styer, Balkin, Becker, Burns, Dudley,
  Forth, Gaumer, Kramer, Oertel, Park, Rinkoski, Smith, and
  Wotherspoon}]{quantum_formulation}
\bibinfo{author}{D.~F. Styer}, \bibinfo{author}{M.~S. Balkin},
  \bibinfo{author}{K.~M. Becker}, \bibinfo{author}{M.~R. Burns},
  \bibinfo{author}{C.~E. Dudley}, \bibinfo{author}{S.~T. Forth},
  \bibinfo{author}{J.~S. Gaumer}, \bibinfo{author}{M.~A. Kramer},
  \bibinfo{author}{D.~C. Oertel}, \bibinfo{author}{L.~H. Park},
  \bibinfo{author}{M.~T. Rinkoski}, \bibinfo{author}{C.~T. Smith},
  \bibinfo{author}{T.~D. Wotherspoon},
\newblock \bibinfo{title}{Nine formulations of quantum mechanics},
\newblock \bibinfo{journal}{American Journal of Physics} \bibinfo{volume}{70}
  (\bibinfo{year}{2002}{\natexlab{a}}) \bibinfo{pages}{288--297}.
\bibitem[{Styer et~al.(2002{\natexlab{b}})Styer, Balkin, Becker, Burns, Dudley,
  Forth, Gaumer, Kramer, Oertel, Park, Rinkoski, Smith, and
  Wotherspoon}]{wigner_review_1}
\bibinfo{author}{D.~F. Styer}, \bibinfo{author}{M.~S. Balkin},
  \bibinfo{author}{K.~M. Becker}, \bibinfo{author}{M.~R. Burns},
  \bibinfo{author}{C.~E. Dudley}, \bibinfo{author}{S.~T. Forth},
  \bibinfo{author}{J.~S. Gaumer}, \bibinfo{author}{M.~A. Kramer},
  \bibinfo{author}{D.~C. Oertel}, \bibinfo{author}{L.~H. Park},
  \bibinfo{author}{M.~T. Rinkoski}, \bibinfo{author}{C.~T. Smith},
  \bibinfo{author}{T.~D. Wotherspoon},
\newblock \bibinfo{title}{Nine formulations of quantum mechanics},
\newblock \bibinfo{journal}{American Journal of Physics} \bibinfo{volume}{70}
  (\bibinfo{year}{2002}{\natexlab{b}}) \bibinfo{pages}{288--297}.
\bibitem[{Rundle and Everitt(2021)}]{wigner_review_2}
\bibinfo{author}{R.~P. Rundle}, \bibinfo{author}{M.~J. Everitt},
\newblock \bibinfo{title}{Overview of the phase space formulation of quantum
  mechanics with application to quantum technologies},
\newblock \bibinfo{journal}{Advanced Quantum Technologies} \bibinfo{volume}{4}
  (\bibinfo{year}{2021}) \bibinfo{pages}{2100016}.
\bibitem[{Wigner(1932)}]{wigner1932}
\bibinfo{author}{E.~Wigner},
\newblock \bibinfo{title}{On the quantum correction for thermodynamic
  equilibrium},
\newblock \bibinfo{journal}{Phys. Rev.} \bibinfo{volume}{40}
  (\bibinfo{year}{1932}) \bibinfo{pages}{749--759}.
\bibitem[{Moyal(1949)}]{moyal_1949}
\bibinfo{author}{J.~E. Moyal},
\newblock \bibinfo{title}{Quantum mechanics as a statistical theory},
\newblock \bibinfo{journal}{Mathematical Proceedings of the Cambridge
  Philosophical Society} \bibinfo{volume}{45} (\bibinfo{year}{1949})
  \bibinfo{pages}{99–124}.
\bibitem[{Groenewold(1946)}]{GROENEWOLD}
\bibinfo{author}{H.~Groenewold},
\newblock \bibinfo{title}{On the principles of elementary quantum mechanics},
\newblock \bibinfo{journal}{Physica} \bibinfo{volume}{12}
  (\bibinfo{year}{1946}) \bibinfo{pages}{405--460}.
\bibitem[{Glauber(1963)}]{glauber1963}
\bibinfo{author}{R.~J. Glauber},
\newblock \bibinfo{title}{Coherent and incoherent states of the radiation
  field},
\newblock \bibinfo{journal}{Phys. Rev.} \bibinfo{volume}{131}
  (\bibinfo{year}{1963}) \bibinfo{pages}{2766--2788}.
\bibitem[{Zhang et~al.(1990)Zhang, Feng, and Gilmore}]{review_coh_state1990}
\bibinfo{author}{W.-M. Zhang}, \bibinfo{author}{D.~H. Feng},
  \bibinfo{author}{R.~Gilmore},
\newblock \bibinfo{title}{Coherent states: Theory and some applications},
\newblock \bibinfo{journal}{Rev. Mod. Phys.} \bibinfo{volume}{62}
  (\bibinfo{year}{1990}) \bibinfo{pages}{867--927}.
\bibitem[{Tilma et~al.(2016)Tilma, Everitt, Samson, Munro, and
  Nemoto}]{GWF2016}
\bibinfo{author}{T.~Tilma}, \bibinfo{author}{M.~J. Everitt},
  \bibinfo{author}{J.~H. Samson}, \bibinfo{author}{W.~J. Munro},
  \bibinfo{author}{K.~Nemoto},
\newblock \bibinfo{title}{Wigner functions for arbitrary quantum systems},
\newblock \bibinfo{journal}{Phys. Rev. Lett.} \bibinfo{volume}{117}
  (\bibinfo{year}{2016}) \bibinfo{pages}{180401}.
\bibitem[{Dirac(1942)}]{dirac1942}
\bibinfo{author}{P.~A.~M. Dirac},
\newblock \bibinfo{title}{Bakerian lecture - the physical interpretation of
  quantum mechanics},
\newblock \bibinfo{journal}{Proceedings of the Royal Society of London. Series
  A. Mathematical and Physical Sciences} \bibinfo{volume}{180}
  (\bibinfo{year}{1942}) \bibinfo{pages}{1--40}.
\bibitem[{Barrow(1988)}]{feynamn_neg_prob}
\bibinfo{author}{J.~D. Barrow},
\newblock \bibinfo{title}{Quantum implications: Essays in honour of david
  bohm},
\newblock \bibinfo{journal}{Physics Bulletin} \bibinfo{volume}{39}
  (\bibinfo{year}{1988}) \bibinfo{pages}{120--120}.
\bibitem[{Arkhipov et~al.(2018)Arkhipov, Barasiński, and
  Svozilík}]{negativity2018}
\bibinfo{author}{I.~I. Arkhipov}, \bibinfo{author}{A.~Barasiński},
  \bibinfo{author}{J.~Svozilík},
\newblock \bibinfo{title}{Negativity volume of the generalized wigner function
  as an entanglement witness for hybrid bipartite states},
\newblock \bibinfo{journal}{Scientific Reports} \bibinfo{volume}{8}
  (\bibinfo{year}{2018}).
\bibitem[{Hudson(1974)}]{hudson}
\bibinfo{author}{R.~Hudson},
\newblock \bibinfo{title}{When is the wigner quasi-probability density
  non-negative?},
\newblock \bibinfo{journal}{Reports on Mathematical Physics}
  \bibinfo{volume}{6} (\bibinfo{year}{1974}) \bibinfo{pages}{249--252}.
\bibitem[{Ferry and Nedjalkov(2018)}]{ferry_wigner_book}
\bibinfo{author}{D.~K. Ferry}, \bibinfo{author}{M.~Nedjalkov},
  \bibinfo{title}{The Wigner Function in Science and Technology}, 2053-2563,
  \bibinfo{publisher}{IOP Publishing}, \bibinfo{year}{2018}.
\bibitem[{Praxmeyer et~al.(2016)Praxmeyer, Chen, Yang, Yang, and Lee}]{frog}
\bibinfo{author}{L.~Praxmeyer}, \bibinfo{author}{C.-C. Chen},
  \bibinfo{author}{P.~Yang}, \bibinfo{author}{S.-D. Yang},
  \bibinfo{author}{R.-K. Lee},
\newblock \bibinfo{title}{Direct measurement of time-frequency analogs of
  sub-planck structures},
\newblock \bibinfo{journal}{Phys. Rev. A} \bibinfo{volume}{93}
  (\bibinfo{year}{2016}) \bibinfo{pages}{053835}.
\bibitem[{Lutterbach and Davidovich(1997)}]{direct_mesurement_wigner}
\bibinfo{author}{L.~G. Lutterbach}, \bibinfo{author}{L.~Davidovich},
\newblock \bibinfo{title}{Method for direct measurement of the wigner function
  in cavity qed and ion traps},
\newblock \bibinfo{journal}{Phys. Rev. Lett.} \bibinfo{volume}{78}
  (\bibinfo{year}{1997}) \bibinfo{pages}{2547--2550}.
\bibitem[{Tufarelli et~al.(2011)Tufarelli, Kim, and
  Bose}]{direct_mesurement_weyl1}
\bibinfo{author}{T.~Tufarelli}, \bibinfo{author}{M.~S. Kim},
  \bibinfo{author}{S.~Bose},
\newblock \bibinfo{title}{Oscillator state reconstruction via tunable qubit
  coupling in markovian environments},
\newblock \bibinfo{journal}{Phys. Rev. A} \bibinfo{volume}{83}
  (\bibinfo{year}{2011}) \bibinfo{pages}{062120}.
\bibitem[{Tufarelli et~al.(2012)Tufarelli, Ferraro, Kim, and
  Bose}]{direct_mesurement_weyl2}
\bibinfo{author}{T.~Tufarelli}, \bibinfo{author}{A.~Ferraro},
  \bibinfo{author}{M.~S. Kim}, \bibinfo{author}{S.~Bose},
\newblock \bibinfo{title}{Reconstructing the quantum state of oscillator
  networks with a single qubit},
\newblock \bibinfo{journal}{Phys. Rev. A} \bibinfo{volume}{85}
  (\bibinfo{year}{2012}) \bibinfo{pages}{032334}.
\bibitem[{Sanchez-Soto(2008)}]{DWFbook}
\bibinfo{author}{L.~Sanchez-Soto},
\newblock \bibinfo{title}{Chapter 7 the discrete wigner function},
\newblock \bibinfo{journal}{Progress in Optics - PROG OPTICS}
  \bibinfo{volume}{51} (\bibinfo{year}{2008}) \bibinfo{pages}{469--516}.
\bibitem[{Rundle et~al.(2017)Rundle, Mills, Tilma, Samson, and
  Everitt}]{GWF_entanglement_PRA}
\bibinfo{author}{R.~P. Rundle}, \bibinfo{author}{P.~W. Mills},
  \bibinfo{author}{T.~Tilma}, \bibinfo{author}{J.~H. Samson},
  \bibinfo{author}{M.~J. Everitt},
\newblock \bibinfo{title}{Simple procedure for phase-space measurement and
  entanglement validation},
\newblock \bibinfo{journal}{Phys. Rev. A} \bibinfo{volume}{96}
  (\bibinfo{year}{2017}) \bibinfo{pages}{022117}.
\bibitem[{Walter~Greiner(1994)}]{gell-mann}
\bibinfo{author}{B.~M. Walter~Greiner}, \bibinfo{title}{Quantum mechanics:
  Symmetries}, Greiner, Walter//Theoretical Physics, \bibinfo{edition}{2nd/rev}
  ed., \bibinfo{publisher}{Springer}, \bibinfo{year}{1994}.
\bibitem[{Wootters(1987)}]{wootters}
\bibinfo{author}{W.~K. Wootters},
\newblock \bibinfo{title}{A wigner-function formulation of finite-state quantum
  mechanics},
\newblock \bibinfo{journal}{Ann. Phys.} \bibinfo{volume}{176}
  (\bibinfo{year}{1987}) \bibinfo{pages}{1 -- 21}.
\bibitem[{Signoles et~al.(2014)Signoles, Facon, Grosso, Dotsenko, Haroche,
  Raimond, Brune, and Gleyzes}]{rydb_exp_wigner}
\bibinfo{author}{A.~Signoles}, \bibinfo{author}{A.~Facon},
  \bibinfo{author}{D.~Grosso}, \bibinfo{author}{I.~Dotsenko},
  \bibinfo{author}{S.~Haroche}, \bibinfo{author}{J.-M. Raimond},
  \bibinfo{author}{M.~Brune}, \bibinfo{author}{S.~Gleyzes},
\newblock \bibinfo{title}{Confined quantum zeno dynamics of a watched atomic
  arrow},
\newblock \bibinfo{journal}{Nature Physics} \bibinfo{volume}{10}
  (\bibinfo{year}{2014}) \bibinfo{pages}{715–719}.
\bibitem[{Leiner et~al.(2017)Leiner, Zeier, and Glaser}]{drops1}
\bibinfo{author}{D.~Leiner}, \bibinfo{author}{R.~Zeier}, \bibinfo{author}{S.~J.
  Glaser},
\newblock \bibinfo{title}{Wigner tomography of multispin quantum states},
\newblock \bibinfo{journal}{Phys. Rev. A} \bibinfo{volume}{96}
  (\bibinfo{year}{2017}) \bibinfo{pages}{063413}.
\bibitem[{Leiner and Glaser(2018)}]{drops2}
\bibinfo{author}{D.~Leiner}, \bibinfo{author}{S.~J. Glaser},
\newblock \bibinfo{title}{Wigner process tomography: Visualization of spin
  propagators and their spinor properties},
\newblock \bibinfo{journal}{Phys. Rev. A} \bibinfo{volume}{98}
  (\bibinfo{year}{2018}) \bibinfo{pages}{012112}.
\bibitem[{Tian et~al.(2018)Tian, Wang, Zhang, Li, Li, and Zhang}]{caesium}
\bibinfo{author}{Y.~Tian}, \bibinfo{author}{Z.~Wang},
  \bibinfo{author}{P.~Zhang}, \bibinfo{author}{G.~Li}, \bibinfo{author}{J.~Li},
  \bibinfo{author}{T.~Zhang},
\newblock \bibinfo{title}{Measurement of complete and continuous wigner
  functions for discrete atomic systems},
\newblock \bibinfo{journal}{Phys. Rev. A} \bibinfo{volume}{97}
  (\bibinfo{year}{2018}) \bibinfo{pages}{013840}.
\bibitem[{Chen et~al.(2019)Chen, Geng, Zhou, Song, Shen, and Xu}]{nitrogen}
\bibinfo{author}{B.~Chen}, \bibinfo{author}{J.~Geng},
  \bibinfo{author}{F.~Zhou}, \bibinfo{author}{L.~Song},
  \bibinfo{author}{H.~Shen}, \bibinfo{author}{N.~Xu},
\newblock \bibinfo{title}{Quantum state tomography of a single electron spin in
  diamond with wigner function reconstruction},
\newblock \bibinfo{journal}{Applied Physics Letters} \bibinfo{volume}{114}
  (\bibinfo{year}{2019}) \bibinfo{pages}{041102}.
\bibitem[{Kafatos(1989)}]{ghz}
\bibinfo{author}{M.~C. Kafatos},
\newblock \bibinfo{title}{Bell's theorem, quantum theory and conceptions of the
  universe},
\newblock \bibinfo{year}{1989}.
\bibitem[{Agarwal et~al.(1997)Agarwal, Puri, and Singh}]{cat}
\bibinfo{author}{G.~S. Agarwal}, \bibinfo{author}{R.~R. Puri},
  \bibinfo{author}{R.~P. Singh},
\newblock \bibinfo{title}{Atomic schr\"odinger cat states},
\newblock \bibinfo{journal}{Phys. Rev. A} \bibinfo{volume}{56}
  (\bibinfo{year}{1997}) \bibinfo{pages}{2249--2254}.
\bibitem[{Dicke(1954)}]{dicke}
\bibinfo{author}{R.~H. Dicke},
\newblock \bibinfo{title}{Coherence in spontaneous radiation processes},
\newblock \bibinfo{journal}{Phys. Rev.} \bibinfo{volume}{93}
  (\bibinfo{year}{1954}) \bibinfo{pages}{99--110}.
\bibitem[{Giorgi(2009)}]{GiorgiPRB}
\bibinfo{author}{G.~L. Giorgi},
\newblock \bibinfo{title}{Ground-state factorization and quantum phase
  transition in dimerized spin chains},
\newblock \bibinfo{journal}{Phys. Rev. B} \bibinfo{volume}{79}
  (\bibinfo{year}{2009}) \bibinfo{pages}{060405}.
\bibitem[{Paz(2002)}]{expdwf2}
\bibinfo{author}{J.~P. Paz},
\newblock \bibinfo{title}{Discrete wigner functions and the phase-space
  representation of quantum teleportation},
\newblock \bibinfo{journal}{Phys. Rev. A} \bibinfo{volume}{65}
  (\bibinfo{year}{2002}) \bibinfo{pages}{062311}.
\bibitem[{Tomasello et~al.(2011)Tomasello, Rossini, Hamma, and Amico}]{EPL2011}
\bibinfo{author}{B.~Tomasello}, \bibinfo{author}{D.~Rossini},
  \bibinfo{author}{A.~Hamma}, \bibinfo{author}{L.~Amico},
\newblock \bibinfo{title}{Ground-state factorization and correlations with
  broken symmetry},
\newblock \bibinfo{journal}{EPL} \bibinfo{volume}{96} (\bibinfo{year}{2011})
  \bibinfo{pages}{27002}.
\bibitem[{Kosterlitz and Thouless(1973)}]{Kosterlitz1973}
\bibinfo{author}{J.~M. Kosterlitz}, \bibinfo{author}{D.~J. Thouless},
\newblock \bibinfo{title}{{Ordering, metastability and phase transitions in
  two-dimensional systems}},
\newblock \bibinfo{journal}{J. Phys.} \bibinfo{volume}{C6}
  (\bibinfo{year}{1973}) \bibinfo{pages}{1181--1203}.
  \bibinfo{note}{[,349(1973)]}.
\bibitem[{Siyouri et~al.(2016)Siyouri, El~Baz, and Hassouni}]{wfrabat}
\bibinfo{author}{F.~Siyouri}, \bibinfo{author}{M.~El~Baz},
  \bibinfo{author}{Y.~Hassouni},
\newblock \bibinfo{title}{The negativity of wigner function as a measure of
  quantum correlations},
\newblock \bibinfo{journal}{Quantum Inf. Process.} \bibinfo{volume}{15}
  (\bibinfo{year}{2016}) \bibinfo{pages}{4237--4252}.
\bibitem[{Taghiabadi et~al.(2016)Taghiabadi, Akhtarshenas, and
  Sarbishaei}]{wfiran}
\bibinfo{author}{R.~Taghiabadi}, \bibinfo{author}{S.~J. Akhtarshenas},
  \bibinfo{author}{M.~Sarbishaei},
\newblock \bibinfo{title}{Revealing quantum correlation by negativity of the
  wigner function},
\newblock \bibinfo{journal}{Quantum Inf. Process.} \bibinfo{volume}{15}
  (\bibinfo{year}{2016}) \bibinfo{pages}{1999--2020}.
\bibitem[{Dunn et~al.(1995)Dunn, Walmsley, and Mukamel}]{expwf}
\bibinfo{author}{T.~J. Dunn}, \bibinfo{author}{I.~A. Walmsley},
  \bibinfo{author}{S.~Mukamel},
\newblock \bibinfo{title}{Experimental determination of the quantum-mechanical
  state of a molecular vibrational mode using fluorescence tomography},
\newblock \bibinfo{journal}{Phys. Rev. Lett.} \bibinfo{volume}{74}
  (\bibinfo{year}{1995}) \bibinfo{pages}{884--887}.
\bibitem[{Heyl(2018)}]{HeylReview}
\bibinfo{author}{M.~Heyl},
\newblock \bibinfo{title}{Dynamical quantum phase transitions: a review},
\newblock \bibinfo{journal}{Rep. Prog. Phys.} \bibinfo{volume}{81}
  (\bibinfo{year}{2018}) \bibinfo{pages}{054001}.
\bibitem[{Schachenmayer et~al.(2015)Schachenmayer, Pikovski, and
  Rey}]{NJPSchachenmayer}
\bibinfo{author}{J.~Schachenmayer}, \bibinfo{author}{A.~Pikovski},
  \bibinfo{author}{A.~M. Rey},
\newblock \bibinfo{title}{Dynamics of correlations in two-dimensional quantum
  spin models with long-range interactions: a phase-space monte-carlo study},
\newblock \bibinfo{journal}{New J. Phys.} \bibinfo{volume}{17}
  (\bibinfo{year}{2015}) \bibinfo{pages}{065009}.
\bibitem[{Acevedo et~al.(2017)Acevedo, Safavi-Naini, Schachenmayer, Wall,
  Nandkishore, and Rey}]{PRASchachenmayer}
\bibinfo{author}{O.~L. Acevedo}, \bibinfo{author}{A.~Safavi-Naini},
  \bibinfo{author}{J.~Schachenmayer}, \bibinfo{author}{M.~L. Wall},
  \bibinfo{author}{R.~Nandkishore}, \bibinfo{author}{A.~M. Rey},
\newblock \bibinfo{title}{Exploring many-body localization and thermalization
  using semiclassical methods},
\newblock \bibinfo{journal}{Phys. Rev. A} \bibinfo{volume}{96}
  (\bibinfo{year}{2017}) \bibinfo{pages}{033604}.
\bibitem[{Czischek et~al.(2018)Czischek, G\"arttner, and
  Gasenzer}]{PRBGasenzer}
\bibinfo{author}{S.~Czischek}, \bibinfo{author}{M.~G\"arttner},
  \bibinfo{author}{T.~Gasenzer},
\newblock \bibinfo{title}{Quenches near ising quantum criticality as a
  challenge for artificial neural networks},
\newblock \bibinfo{journal}{Phys. Rev. B} \bibinfo{volume}{98}
  (\bibinfo{year}{2018}) \bibinfo{pages}{024311}.
\bibitem[{Czischek et~al.(2019)Czischek, G\"{a}rttner, Oberthaler, Kastner, and
  Gasenzer}]{QST2019}
\bibinfo{author}{S.~Czischek}, \bibinfo{author}{M.~G\"{a}rttner},
  \bibinfo{author}{M.~Oberthaler}, \bibinfo{author}{M.~Kastner},
  \bibinfo{author}{T.~Gasenzer},
\newblock \bibinfo{title}{Quenches near criticality of the quantum ising chain
  -- power and limitations of the discrete truncated wigner approximation},
\newblock \bibinfo{journal}{Quantum Sci. Technol.} \bibinfo{volume}{4}
  (\bibinfo{year}{2019}) \bibinfo{pages}{014006}.
\bibitem[{Schachenmayer et~al.(2015)Schachenmayer, Pikovski, and
  Rey}]{PRXSchachenmayer}
\bibinfo{author}{J.~Schachenmayer}, \bibinfo{author}{A.~Pikovski},
  \bibinfo{author}{A.~M. Rey},
\newblock \bibinfo{title}{Many-body quantum spin dynamics with monte carlo
  trajectories on a discrete phase space},
\newblock \bibinfo{journal}{Phys. Rev. X} \bibinfo{volume}{5}
  (\bibinfo{year}{2015}) \bibinfo{pages}{011022}.
\bibitem[{Thompson(1798)}]{rutherford}
\bibinfo{author}{B.~Thompson},
\newblock \bibinfo{title}{Iv. an inquiry concerning the source of the heat
  which is excited by friction},
\newblock \bibinfo{journal}{Philosophical Transactions of the Royal Society of
  London} \bibinfo{volume}{88} (\bibinfo{year}{1798}) \bibinfo{pages}{80--102}.
\bibitem[{Clapeyron(1834)}]{clapeyron1834memoire}
\bibinfo{author}{E.~Clapeyron}, \bibinfo{title}{M\'emoire sur la puissance
  motrice de la chaleur}, \bibinfo{publisher}{Bachelier}, \bibinfo{year}{1834}.
\bibitem[{Clausius(????)}]{clausius}
\bibinfo{author}{R.~Clausius}, \bibinfo{title}{The Nature of the Motion which
  we call Heat}, ????, pp. \bibinfo{pages}{111--134}.
\bibitem[{Masanes and Oppenheim(2017)}]{oppenheim1}
\bibinfo{author}{L.~Masanes}, \bibinfo{author}{J.~Oppenheim},
\newblock \bibinfo{title}{A general derivation and quantification of the third
  law of thermodynamics},
\newblock \bibinfo{journal}{Nature Communications} \bibinfo{volume}{8}
  (\bibinfo{year}{2017}).
\bibitem[{Horodecki and Oppenheim(2013)}]{oppenheim2}
\bibinfo{author}{M.~Horodecki}, \bibinfo{author}{J.~Oppenheim},
\newblock \bibinfo{title}{Fundamental limitations for quantum and nanoscale
  thermodynamics},
\newblock \bibinfo{journal}{Nature Communications} \bibinfo{volume}{4}
  (\bibinfo{year}{2013}).
\bibitem[{Onsager(1931)}]{onsager_thermo}
\bibinfo{author}{L.~Onsager},
\newblock \bibinfo{title}{Reciprocal relations in irreversible processes. i.},
\newblock \bibinfo{journal}{Phys. Rev.} \bibinfo{volume}{37}
  (\bibinfo{year}{1931}) \bibinfo{pages}{405--426}.
\bibitem[{Strasberg(2021)}]{strasberg_2021}
\bibinfo{author}{P.~Strasberg}, \bibinfo{title}{Quantum Stochastic
  Thermodynamics: Foundations and Selected Applications (Oxford Graduate
  Texts)}, \bibinfo{publisher}{Oxford University Press}, \bibinfo{year}{2021}.
\bibitem[{Crooks(1999)}]{tasaki}
\bibinfo{author}{G.~E. Crooks},
\newblock \bibinfo{title}{Entropy production fluctuation theorem and the
  nonequilibrium work relation for free energy differences},
\newblock \bibinfo{journal}{Phys. Rev. E} \bibinfo{volume}{60}
  (\bibinfo{year}{1999}) \bibinfo{pages}{2721--2726}.
\bibitem[{Jarzynski(1997{\natexlab{a}})}]{Jarzynski1}
\bibinfo{author}{C.~Jarzynski},
\newblock \bibinfo{title}{Nonequilibrium equality for free energy differences},
\newblock \bibinfo{journal}{Phys. Rev. Lett.} \bibinfo{volume}{78}
  (\bibinfo{year}{1997}{\natexlab{a}}) \bibinfo{pages}{2690--2693}.
\bibitem[{Jarzynski(1997{\natexlab{b}})}]{Jarzynski2}
\bibinfo{author}{C.~Jarzynski},
\newblock \bibinfo{title}{Equilibrium free-energy differences from
  nonequilibrium measurements: A master-equation approach},
\newblock \bibinfo{journal}{Phys. Rev. E} \bibinfo{volume}{56}
  (\bibinfo{year}{1997}{\natexlab{b}}) \bibinfo{pages}{5018--5035}.
\bibitem[{Chitambar and Gour(2019)}]{gilad2019}
\bibinfo{author}{E.~Chitambar}, \bibinfo{author}{G.~Gour},
\newblock \bibinfo{title}{Quantum resource theories},
\newblock \bibinfo{journal}{Reviews of Modern Physics} \bibinfo{volume}{91}
  (\bibinfo{year}{2019}).
\bibitem[{Gour et~al.(2015)Gour, Müller, Narasimhachar, Spekkens, and {Yunger
  Halpern}}]{gour2015}
\bibinfo{author}{G.~Gour}, \bibinfo{author}{M.~P. Müller},
  \bibinfo{author}{V.~Narasimhachar}, \bibinfo{author}{R.~W. Spekkens},
  \bibinfo{author}{N.~{Yunger Halpern}},
\newblock \bibinfo{title}{The resource theory of informational nonequilibrium
  in thermodynamics},
\newblock \bibinfo{journal}{Physics Reports} \bibinfo{volume}{583}
  (\bibinfo{year}{2015}) \bibinfo{pages}{1--58}. \bibinfo{note}{The resource
  theory of informational nonequilibrium in thermodynamics}.
\bibitem[{\AA{}berg(2014)}]{catalytic_coherence}
\bibinfo{author}{J.~\AA{}berg},
\newblock \bibinfo{title}{Catalytic coherence},
\newblock \bibinfo{journal}{Phys. Rev. Lett.} \bibinfo{volume}{113}
  (\bibinfo{year}{2014}) \bibinfo{pages}{150402}.
\bibitem[{Huber et~al.(2015)Huber, Perarnau-Llobet, Hovhannisyan, Skrzypczyk,
  Klöckl, Brunner, and Ac{\'{\i}}n}]{Huber_2015}
\bibinfo{author}{M.~Huber}, \bibinfo{author}{M.~Perarnau-Llobet},
  \bibinfo{author}{K.~V. Hovhannisyan}, \bibinfo{author}{P.~Skrzypczyk},
  \bibinfo{author}{C.~Klöckl}, \bibinfo{author}{N.~Brunner},
  \bibinfo{author}{A.~Ac{\'{\i}}n},
\newblock \bibinfo{title}{Thermodynamic cost of creating correlations},
\newblock \bibinfo{journal}{New Journal of Physics} \bibinfo{volume}{17}
  (\bibinfo{year}{2015}) \bibinfo{pages}{065008}.
\bibitem[{Bruschi et~al.(2015)Bruschi, Perarnau-Llobet, Friis, Hovhannisyan,
  and Huber}]{huber_2015_2}
\bibinfo{author}{D.~E. Bruschi}, \bibinfo{author}{M.~Perarnau-Llobet},
  \bibinfo{author}{N.~Friis}, \bibinfo{author}{K.~V. Hovhannisyan},
  \bibinfo{author}{M.~Huber},
\newblock \bibinfo{title}{Thermodynamics of creating correlations: Limitations
  and optimal protocols},
\newblock \bibinfo{journal}{Phys. Rev. E} \bibinfo{volume}{91}
  (\bibinfo{year}{2015}) \bibinfo{pages}{032118}.
\bibitem[{Perarnau-Llobet et~al.(2015)Perarnau-Llobet, Hovhannisyan, Huber,
  Skrzypczyk, Brunner, and Ac\'{\i}n}]{Huber_2015_3}
\bibinfo{author}{M.~Perarnau-Llobet}, \bibinfo{author}{K.~V. Hovhannisyan},
  \bibinfo{author}{M.~Huber}, \bibinfo{author}{P.~Skrzypczyk},
  \bibinfo{author}{N.~Brunner}, \bibinfo{author}{A.~Ac\'{\i}n},
\newblock \bibinfo{title}{Extractable work from correlations},
\newblock \bibinfo{journal}{Phys. Rev. X} \bibinfo{volume}{5}
  (\bibinfo{year}{2015}) \bibinfo{pages}{041011}.
\bibitem[{Vedral(2004)}]{Vedral_2004}
\bibinfo{author}{V.~Vedral},
\newblock \bibinfo{title}{High-temperature macroscopic entanglement},
\newblock \bibinfo{journal}{New Journal of Physics} \bibinfo{volume}{6}
  (\bibinfo{year}{2004}) \bibinfo{pages}{102--102}.
\bibitem[{Bäuml et~al.(2015)Bäuml, Bru{\ss}, Huber, Kampermann, and
  Winter}]{entropy_2015}
\bibinfo{author}{S.~Bäuml}, \bibinfo{author}{D.~Bru{\ss}},
  \bibinfo{author}{M.~Huber}, \bibinfo{author}{H.~Kampermann},
  \bibinfo{author}{A.~Winter},
\newblock \bibinfo{title}{Witnessing entanglement by proxy},
\newblock \bibinfo{journal}{New Journal of Physics} \bibinfo{volume}{18}
  (\bibinfo{year}{2015}) \bibinfo{pages}{015002}.
\bibitem[{Wie{\'{s}}niak et~al.(2005)Wie{\'{s}}niak, Vedral, and
  Brukner}]{Wie_niak_2005}
\bibinfo{author}{M.~Wie{\'{s}}niak}, \bibinfo{author}{V.~Vedral},
  \bibinfo{author}{{\v{C}}.~Brukner},
\newblock \bibinfo{title}{Magnetic susceptibility as a macroscopic entanglement
  witness},
\newblock \bibinfo{journal}{New Journal of Physics} \bibinfo{volume}{7}
  (\bibinfo{year}{2005}) \bibinfo{pages}{258--258}.
\bibitem[{Landauer(1961)}]{landauer1961}
\bibinfo{author}{R.~Landauer},
\newblock \bibinfo{title}{Irreversibility and heat generation in the computing
  process},
\newblock \bibinfo{journal}{IBM Journal of Research and Development}
  \bibinfo{volume}{5} (\bibinfo{year}{1961}) \bibinfo{pages}{183--191}.
\bibitem[{Reeb and Wolf(2014)}]{Reeb_2014}
\bibinfo{author}{D.~Reeb}, \bibinfo{author}{M.~M. Wolf},
\newblock \bibinfo{title}{An improved landauer principle with finite-size
  corrections},
\newblock \bibinfo{journal}{New Journal of Physics} \bibinfo{volume}{16}
  (\bibinfo{year}{2014}) \bibinfo{pages}{103011}.
\bibitem[{Lipkin et~al.(1965)Lipkin, Meshkov, and Glick}]{lmg1965}
\bibinfo{author}{H.~J. Lipkin}, \bibinfo{author}{N.~Meshkov},
  \bibinfo{author}{A.~J. Glick},
\newblock \bibinfo{title}{Validity of many-body approximation methods for a
  solvable model: (i). exact solutions and perturbation theory},
\newblock \bibinfo{journal}{Nuclear Physics} \bibinfo{volume}{62}
  (\bibinfo{year}{1965}) \bibinfo{pages}{188 -- 198}.
\bibitem[{Ribeiro et~al.(2007)Ribeiro, Vidal, and Mosseri}]{prl_vidal}
\bibinfo{author}{P.~Ribeiro}, \bibinfo{author}{J.~Vidal},
  \bibinfo{author}{R.~Mosseri},
\newblock \bibinfo{title}{Thermodynamical limit of the lipkin-meshkov-glick
  model},
\newblock \bibinfo{journal}{Phys. Rev. Lett.} \bibinfo{volume}{99}
  (\bibinfo{year}{2007}) \bibinfo{pages}{050402}.
\bibitem[{Ribeiro et~al.(2008)Ribeiro, Vidal, and Mosseri}]{pre_vidal}
\bibinfo{author}{P.~Ribeiro}, \bibinfo{author}{J.~Vidal},
  \bibinfo{author}{R.~Mosseri},
\newblock \bibinfo{title}{Exact spectrum of the lipkin-meshkov-glick model in
  the thermodynamic limit and finite-size corrections},
\newblock \bibinfo{journal}{Phys. Rev. E} \bibinfo{volume}{78}
  (\bibinfo{year}{2008}) \bibinfo{pages}{021106}.
\bibitem[{Dusuel and Vidal(2005)}]{prb_vidal}
\bibinfo{author}{S.~Dusuel}, \bibinfo{author}{J.~Vidal},
\newblock \bibinfo{title}{Continuous unitary transformations and finite-size
  scaling exponents in the lipkin-meshkov-glick model},
\newblock \bibinfo{journal}{Phys. Rev. B} \bibinfo{volume}{71}
  (\bibinfo{year}{2005}) \bibinfo{pages}{224420}.
\bibitem[{Casta\~nos et~al.(2006)Casta\~nos, L\'opez-Pe\~na, Hirsch, and
  L\'opez-Moreno}]{prb_castanos}
\bibinfo{author}{O.~Casta\~nos}, \bibinfo{author}{R.~L\'opez-Pe\~na},
  \bibinfo{author}{J.~G. Hirsch}, \bibinfo{author}{E.~L\'opez-Moreno},
\newblock \bibinfo{title}{Classical and quantum phase transitions in the
  lipkin-meshkov-glick model},
\newblock \bibinfo{journal}{Phys. Rev. B} \bibinfo{volume}{74}
  (\bibinfo{year}{2006}) \bibinfo{pages}{104118}.
\bibitem[{Quan et~al.(2007)Quan, Wang, and Sun}]{quan_pra}
\bibinfo{author}{H.~T. Quan}, \bibinfo{author}{Z.~D. Wang},
  \bibinfo{author}{C.~P. Sun},
\newblock \bibinfo{title}{Quantum critical dynamics of a qubit coupled to an
  isotropic lipkin-meshkov-glick bath},
\newblock \bibinfo{journal}{Phys. Rev. A} \bibinfo{volume}{76}
  (\bibinfo{year}{2007}) \bibinfo{pages}{012104}.
\bibitem[{Campbell(2016)}]{CampbellPRB}
\bibinfo{author}{S.~Campbell},
\newblock \bibinfo{title}{{Criticality revealed through quench dynamics in the
  Lipkin-Meshkov-Glick model}},
\newblock \bibinfo{journal}{Phys. Rev. B} \bibinfo{volume}{94}
  (\bibinfo{year}{2016}) \bibinfo{pages}{184403}.
\bibitem[{Caprio et~al.(2008)Caprio, Cejnar, and Iachello}]{Caprio:08}
\bibinfo{author}{M.~Caprio}, \bibinfo{author}{P.~Cejnar},
  \bibinfo{author}{F.~Iachello},
\newblock \bibinfo{title}{Excited state quantum phase transitions in many-body
  systems},
\newblock \bibinfo{journal}{Ann. Phys. (N. Y.)} \bibinfo{volume}{323}
  (\bibinfo{year}{2008}) \bibinfo{pages}{1106 -- 1135}.
\bibitem[{Cejnar et~al.(2021)Cejnar, Str\'ansk\'y, Macek, and
  Kloc}]{esqpt_2020review}
\bibinfo{author}{P.~Cejnar}, \bibinfo{author}{P.~Str\'ansk\'y},
  \bibinfo{author}{M.~Macek}, \bibinfo{author}{M.~Kloc},
\newblock \bibinfo{title}{Excited-state quantum phase transitions},
\newblock \bibinfo{journal}{J. Phys. A: Math. Theor.} \bibinfo{volume}{54}
  (\bibinfo{year}{2021}) \bibinfo{pages}{133001}.
\bibitem[{Santos et~al.(2016)Santos, T\'avora, and
  P\'erez-Bernal}]{2016_pra_lea}
\bibinfo{author}{L.~F. Santos}, \bibinfo{author}{M.~T\'avora},
  \bibinfo{author}{F.~P\'erez-Bernal},
\newblock \bibinfo{title}{Excited-state quantum phase transitions in many-body
  systems with infinite-range interaction: Localization, dynamics, and
  bifurcation},
\newblock \bibinfo{journal}{Phys. Rev. A} \bibinfo{volume}{94}
  (\bibinfo{year}{2016}) \bibinfo{pages}{012113}.
\bibitem[{Polkovnikov et~al.(2011)Polkovnikov, Sengupta, Silva, and
  Vengalattore}]{RMP_Silva}
\bibinfo{author}{A.~Polkovnikov}, \bibinfo{author}{K.~Sengupta},
  \bibinfo{author}{A.~Silva}, \bibinfo{author}{M.~Vengalattore},
\newblock \bibinfo{title}{Colloquium: Nonequilibrium dynamics of closed
  interacting quantum systems},
\newblock \bibinfo{journal}{Rev. Mod. Phys.} \bibinfo{volume}{83}
  (\bibinfo{year}{2011}) \bibinfo{pages}{863--883}.
\bibitem[{Jafari(2016)}]{JoPA_Jafari}
\bibinfo{author}{R.~Jafari},
\newblock \bibinfo{title}{Quench dynamics and ground state fidelity of the
  one-dimensional extended quantum compass model in a transverse field},
\newblock \bibinfo{journal}{J. Phys. A: Math. Theor.} \bibinfo{volume}{49}
  (\bibinfo{year}{2016}) \bibinfo{pages}{185004}.
\bibitem[{Najafi and Rajabpour(2017)}]{prb_najafi}
\bibinfo{author}{K.~Najafi}, \bibinfo{author}{M.~A. Rajabpour},
\newblock \bibinfo{title}{On the possibility of complete revivals after quantum
  quenches to a critical point},
\newblock \bibinfo{journal}{Phys. Rev. B} \bibinfo{volume}{96}
  (\bibinfo{year}{2017}) \bibinfo{pages}{014305}.
\bibitem[{Bayat et~al.(2015)Bayat, Bose, Johannesson, and Sodano}]{prb_bose}
\bibinfo{author}{A.~Bayat}, \bibinfo{author}{S.~Bose},
  \bibinfo{author}{H.~Johannesson}, \bibinfo{author}{P.~Sodano},
\newblock \bibinfo{title}{Universal single-frequency oscillations in a quantum
  impurity system after a local quench},
\newblock \bibinfo{journal}{Phys. Rev. B} \bibinfo{volume}{92}
  (\bibinfo{year}{2015}) \bibinfo{pages}{155141}.
\bibitem[{Chenu et~al.(2018)Chenu, Egusquiza, Molina-Vilaplana, and del
  Campo}]{scirep_campo}
\bibinfo{author}{A.~Chenu}, \bibinfo{author}{I.~L. Egusquiza},
  \bibinfo{author}{J.~Molina-Vilaplana}, \bibinfo{author}{A.~del Campo},
\newblock \bibinfo{title}{Quantum work statistics, loschmidt echo and
  information scrambling},
\newblock \bibinfo{journal}{Sci. Rep.} \bibinfo{volume}{8}
  (\bibinfo{year}{2018}) \bibinfo{pages}{12634}.
\bibitem[{Rossini et~al.(2007)Rossini, Calarco, Giovannetti, Montangero, and
  Fazio}]{pra_fazio}
\bibinfo{author}{D.~Rossini}, \bibinfo{author}{T.~Calarco},
  \bibinfo{author}{V.~Giovannetti}, \bibinfo{author}{S.~Montangero},
  \bibinfo{author}{R.~Fazio},
\newblock \bibinfo{title}{Decoherence induced by interacting quantum spin
  baths},
\newblock \bibinfo{journal}{Phys. Rev. A} \bibinfo{volume}{75}
  (\bibinfo{year}{2007}) \bibinfo{pages}{032333}.
\bibitem[{Quan et~al.(2006)Quan, Song, Liu, Zanardi, and Sun}]{prl_zanardi}
\bibinfo{author}{H.~T. Quan}, \bibinfo{author}{Z.~Song}, \bibinfo{author}{X.~F.
  Liu}, \bibinfo{author}{P.~Zanardi}, \bibinfo{author}{C.~P. Sun},
\newblock \bibinfo{title}{Decay of loschmidt echo enhanced by quantum
  criticality},
\newblock \bibinfo{journal}{Phys. Rev. Lett.} \bibinfo{volume}{96}
  (\bibinfo{year}{2006}) \bibinfo{pages}{140604}.
\bibitem[{Jafari and Johannesson(2017)}]{prl_jafari}
\bibinfo{author}{R.~Jafari}, \bibinfo{author}{H.~Johannesson},
\newblock \bibinfo{title}{Loschmidt echo revivals: Critical and noncritical},
\newblock \bibinfo{journal}{Phys. Rev. Lett.} \bibinfo{volume}{118}
  (\bibinfo{year}{2017}) \bibinfo{pages}{015701}.
\bibitem[{Gorin et~al.(2006)Gorin, Prosen, Seligman, and Žnidarič}]{PR_Gorin}
\bibinfo{author}{T.~Gorin}, \bibinfo{author}{T.~Prosen}, \bibinfo{author}{T.~H.
  Seligman}, \bibinfo{author}{M.~Žnidarič},
\newblock \bibinfo{title}{Dynamics of loschmidt echoes and fidelity decay},
\newblock \bibinfo{journal}{Phys. Rep.} \bibinfo{volume}{435}
  (\bibinfo{year}{2006}) \bibinfo{pages}{33 -- 156}.
\bibitem[{Wang et~al.(2015)Wang, Wang, Yang, and Wang}]{qian_pra}
\bibinfo{author}{Q.~Wang}, \bibinfo{author}{P.~Wang},
  \bibinfo{author}{Y.~Yang}, \bibinfo{author}{W.-g. Wang},
\newblock \bibinfo{title}{Decay of quantum loschmidt echo and fidelity in the
  broken phase of the lipkin-meshkov-glick model},
\newblock \bibinfo{journal}{Phys. Rev. A} \bibinfo{volume}{91}
  (\bibinfo{year}{2015}) \bibinfo{pages}{042102}.
\bibitem[{Silva(2008)}]{prl_silva}
\bibinfo{author}{A.~Silva},
\newblock \bibinfo{title}{Statistics of the work done on a quantum critical
  system by quenching a control parameter},
\newblock \bibinfo{journal}{Phys. Rev. Lett.} \bibinfo{volume}{101}
  (\bibinfo{year}{2008}) \bibinfo{pages}{120603}.
\bibitem[{Fusco et~al.(2014)Fusco, Pigeon, Apollaro, Xuereb, Mazzola, Campisi,
  Ferraro, Paternostro, and De~Chiara}]{prx_fusco}
\bibinfo{author}{L.~Fusco}, \bibinfo{author}{S.~Pigeon},
  \bibinfo{author}{T.~J.~G. Apollaro}, \bibinfo{author}{A.~Xuereb},
  \bibinfo{author}{L.~Mazzola}, \bibinfo{author}{M.~Campisi},
  \bibinfo{author}{A.~Ferraro}, \bibinfo{author}{M.~Paternostro},
  \bibinfo{author}{G.~De~Chiara},
\newblock \bibinfo{title}{Assessing the nonequilibrium thermodynamics in a
  quenched quantum many-body system via single projective measurements},
\newblock \bibinfo{journal}{Phys. Rev. X} \bibinfo{volume}{4}
  (\bibinfo{year}{2014}) \bibinfo{pages}{031029}.
\bibitem[{Batalh\~ao et~al.(2014)Batalh\~ao, Souza, Mazzola, Auccaise,
  Sarthour, Oliveira, Goold, De~Chiara, Paternostro, and Serra}]{WorkDistPRL}
\bibinfo{author}{T.~B. Batalh\~ao}, \bibinfo{author}{A.~M. Souza},
  \bibinfo{author}{L.~Mazzola}, \bibinfo{author}{R.~Auccaise},
  \bibinfo{author}{R.~S. Sarthour}, \bibinfo{author}{I.~S. Oliveira},
  \bibinfo{author}{J.~Goold}, \bibinfo{author}{G.~De~Chiara},
  \bibinfo{author}{M.~Paternostro}, \bibinfo{author}{R.~M. Serra},
\newblock \bibinfo{title}{Experimental reconstruction of work distribution and
  study of fluctuation relations in a closed quantum system},
\newblock \bibinfo{journal}{Phys. Rev. Lett.} \bibinfo{volume}{113}
  (\bibinfo{year}{2014}) \bibinfo{pages}{140601}.
\bibitem[{Goold et~al.(2015)Goold, Gogolin, Clark, Eisert, Scardicchio, and
  Silva}]{prb_goold_1}
\bibinfo{author}{J.~Goold}, \bibinfo{author}{C.~Gogolin},
  \bibinfo{author}{S.~R. Clark}, \bibinfo{author}{J.~Eisert},
  \bibinfo{author}{A.~Scardicchio}, \bibinfo{author}{A.~Silva},
\newblock \bibinfo{title}{Total correlations of the diagonal ensemble herald
  the many-body localization transition},
\newblock \bibinfo{journal}{Phys. Rev. B} \bibinfo{volume}{92}
  (\bibinfo{year}{2015}) \bibinfo{pages}{180202}.
\bibitem[{Pietracaprina et~al.(2017)Pietracaprina, Gogolin, and
  Goold}]{prb_goold_2}
\bibinfo{author}{F.~Pietracaprina}, \bibinfo{author}{C.~Gogolin},
  \bibinfo{author}{J.~Goold},
\newblock \bibinfo{title}{Total correlations of the diagonal ensemble as a
  generic indicator for ergodicity breaking in quantum systems},
\newblock \bibinfo{journal}{Phys. Rev. B} \bibinfo{volume}{95}
  (\bibinfo{year}{2017}) \bibinfo{pages}{125118}.
\bibitem[{\ifmmode~\mbox{\c{C}}\else \c{C}\fi{}akan
  et~al.(2021)\ifmmode~\mbox{\c{C}}\else \c{C}\fi{}akan, Cirac, and
  Ba\~nuls}]{cakan2020_DE}
\bibinfo{author}{A.~\ifmmode~\mbox{\c{C}}\else \c{C}\fi{}akan},
  \bibinfo{author}{J.~I. Cirac}, \bibinfo{author}{M.~C. Ba\~nuls},
\newblock \bibinfo{title}{Approximating the long time average of the density
  operator: Diagonal ensemble},
\newblock \bibinfo{journal}{Phys. Rev. B} \bibinfo{volume}{103}
  (\bibinfo{year}{2021}) \bibinfo{pages}{115113}.
\bibitem[{P\'erez-Fern\'andez et~al.(2011)P\'erez-Fern\'andez, Cejnar, Arias,
  Dukelsky, Garc\'{\i}a-Ramos, and Rela\~no}]{2011_relano}
\bibinfo{author}{P.~P\'erez-Fern\'andez}, \bibinfo{author}{P.~Cejnar},
  \bibinfo{author}{J.~M. Arias}, \bibinfo{author}{J.~Dukelsky},
  \bibinfo{author}{J.~E. Garc\'{\i}a-Ramos}, \bibinfo{author}{A.~Rela\~no},
\newblock \bibinfo{title}{Quantum quench influenced by an excited-state phase
  transition},
\newblock \bibinfo{journal}{Phys. Rev. A} \bibinfo{volume}{83}
  (\bibinfo{year}{2011}) \bibinfo{pages}{033802}.
\bibitem[{Fogarty et~al.(2020)Fogarty, Deffner, Busch, and
  Campbell}]{CampbellPRL2020}
\bibinfo{author}{T.~Fogarty}, \bibinfo{author}{S.~Deffner},
  \bibinfo{author}{T.~Busch}, \bibinfo{author}{S.~Campbell},
\newblock \bibinfo{title}{Orthogonality catastrophe as a consequence of the
  quantum speed limit},
\newblock \bibinfo{journal}{Phys. Rev. Lett.} \bibinfo{volume}{124}
  (\bibinfo{year}{2020}) \bibinfo{pages}{110601}.
\bibitem[{Wang and P\'erez-Bernal(2021)}]{2020_wang_arxiv}
\bibinfo{author}{Q.~Wang}, \bibinfo{author}{F.~P\'erez-Bernal},
\newblock \bibinfo{title}{Characterizing the lipkin-meshkov-glick model
  excited-state quantum phase transition using dynamical and statistical
  properties of the diagonal entropy},
\newblock \bibinfo{journal}{Phys. Rev. E} \bibinfo{volume}{103}
  (\bibinfo{year}{2021}) \bibinfo{pages}{032109}.
\bibitem[{Jarzynski(2015)}]{JarNat2015}
\bibinfo{author}{C.~Jarzynski},
\newblock \bibinfo{title}{Diverse phenomena, common themes},
\newblock \bibinfo{journal}{Nature Physics} \bibinfo{volume}{11}
  (\bibinfo{year}{2015}) \bibinfo{pages}{105--107}.
\bibitem[{Auff\`eves(2021)}]{Auffeves:2021paz}
\bibinfo{author}{A.~Auff\`eves},
\newblock \bibinfo{title}{{Quantum technologies need a quantum energy
  initiative}}  (\bibinfo{year}{2021}).
  \href{http://arxiv.org/abs/2111.09241}{{\tt arXiv:2111.09241}}.

\end{thebibliography}
\end{document}